\documentclass[]{tADP2e-arxiv}
\usepackage{multirow}
\pdfoutput=1

\newcommand{\be}{\begin{equation}}
\newcommand{\ee}{\end{equation}}
\newcommand{\bea}{\begin{eqnarray}}
\newcommand{\eea}{\end{eqnarray}}
\newcommand{\vap}{\varepsilon}

\DeclareMathOperator{\sgn}{sgn}
\DeclareMathOperator{\Cl}{Cl_{2}}

\DeclareMathOperator{\Real}{Re}
\DeclareMathOperator{\Li}{Li_{2}}
\DeclareMathOperator{\atan}{arctan}

\begin{document}


\markboth{M. J. Bowick and L. Giomi}{Two-Dimensional Matter: Order, Curvature and Defects}

\title{Two-Dimensional Matter: Order, Curvature and Defects}

\author{%
Mark J. Bowick and Luca Giomi\\\vspace{6pt}
Department of Physics, Syracuse University, Syracuse (NY), 13244
\\\vspace{6pt}
}

\maketitle

\begin{abstract}

\noindent Many systems in nature and the synthetic world involve ordered arrangements of units on two-dimensional surfaces. We review here the fundamental role payed by both the topology of the underlying surface and its Gaussian curvature. Topology dictates certain broad features of the defect structure of the ground state but curvature-driven energetics controls the detailed structured of ordered phases. Among the surprises are the appearance in the ground state of structures that would normally be thermal excitations and thus prohibited at zero temperature. Examples include excess dislocations in the form of grain boundary scars for spherical crystals above a minimal system size, dislocation unbinding for toroidal hexatics, interstitial fractionalization in spherical crystals and the appearance of well-separated disclinations for toroidal crystals. Much of the analysis leads to universal predictions that do not depend on the details of the microscopic interactions that lead to order in the first place. These predictions are subject to test by the many experimental soft and hard matter systems that lead to curved ordered structures such as colloidal particles self-assembling on droplets of one liquid in a second liquid. The defects themselves may be functionalized to create ligands with directional bonding. Thus nano to meso scale superatoms may be designed with specific valency for use in building supermolecules and novel bulk materials.  Parameters such as particle number, geometrical aspect ratios and anisotropy of elastic moduli permit the tuning of the precise architecture of the  superatoms and associated supermolecules. Thus the field has tremendous potential from both a fundamental and materials science/supramolecular chemistry viewpoint.\bigskip\bigskip

\centerline{\bfseries Contents}\medskip

\noindent{\ref{sec:1}.} Order and geometry in condensed matter\\
\hspace*{7pt} {\ref{sec:1a}.} Introduction\\
\hspace*{7pt} {\ref{sec:1b}.} Ordered structures in two-dimensional matter\\
\hspace*{24pt} {\ref{sec:1c}.} Amphiphilic membranes\\
\hspace*{24pt} {\ref{sec:1d}.} Colloidosomes\\
\hspace*{24pt} {\ref{sec:1e}.} Viral capsids\\
\hspace*{24pt} {\ref{sec:1f}.} Carbon nanotubes and related materials\\
{\ref{sec:2}} Interacting topological defects in curved media\\
\hspace*{7pt} {\ref{sec:2a}.} Geometrical frustration\\
\hspace*{7pt} {\ref{sec:2b}.} Mathematical preliminaries and notation\\
\hspace*{7pt} {\ref{sec:2c}.} Elasticity of defects in the plane\\
\hspace*{7pt} {\ref{sec:2d}.} Coupling mechanisms between curvature and defects\\
{\ref{sec:3}.} Crystalline and nematic order on the sphere\\
\hspace*{7pt} {\ref{sec:3a}.} Introduction\\
\hspace*{7pt} {\ref{sec:3b}.} Crystals of point particles\\
\hspace*{24pt} {\ref{sec:3c}.} Planar crystals\\
\hspace*{24pt} {\ref{sec:3d}.} Continuum free energy\\
\hspace*{24pt} {\ref{sec:3e}.} Spherical crystals\\
\hspace*{7pt} {\ref{sec:3f}.} Geometric formalism on the sphere\\
\hspace*{24pt} {\ref{sec:3g}.} The energy of spherical crystals\\
\hspace*{24pt} {\ref{sec:3h}.} Energies of icosadeltahedral configurations\\
\hspace*{24pt} {\ref{sec:3i}.} The energy difference of the $(n,m)$ lattices\\
\hspace*{7pt} {\ref{sec:3j}.} Thomson problem with a continuous distribution of dislocations\\
\hspace*{24pt} {\ref{sec:3k}.} The intermediate regime\\
\hspace*{7pt} {\ref{sec:3l}.} Interstitial fractionalization of the sphere\\
\hspace*{24pt} {\ref{sec:3m}.} Introduction\\
\hspace*{24pt} {\ref{sec:3n}.} Interstitial fractionalization\\
\hspace*{24pt} {\ref{sec:3o}.} Interstitial defect energies\\
\hspace*{24pt} {\ref{sec:3p}.} Position dependece of interstitial defect energies\\
\hspace*{7pt} {\ref{sec:3q}.} Spherical nematics\\
{\ref{sec:4}.} Crystalline order on surfaces with variable Gaussian curvature and boundary\\
\hspace*{7pt} {\ref{sec:4a}.} Introduction\\
\hspace*{7pt} {\ref{sec:4b}.} Surface of revolution and conformal mapping\\
\hspace*{7pt} {\ref{sec:4c}.} Crystalline order on the Gaussian bump\\
\hspace*{7pt} {\ref{sec:4d}.} Paraboloidal crystals\\
\hspace*{7pt} {\ref{sec:4e}.} Experimental realizations of paraboloidal crystals\\
{\ref{sec:5}.} Crystalline and $p-$atic order in toroidal geometries\\
\hspace*{7pt} {\ref{sec:5a}.} Introduction\\
\hspace*{7pt} {\ref{sec:5b}.} Geometry of the torus\\
\hspace*{7pt} {\ref{sec:5c}.} Defect-free $p-$atic textures and \emph{genera} transitions\\
\hspace*{7pt} {\ref{sec:5d}.} Defective ground states in hexatics\\
\hspace*{7pt} {\ref{sec:5e}.} Toroidal crystals\\
\hspace*{24pt} {\ref{sec:5f}.} Geometry of toroidal polyhedra\\
\hspace*{24pt} {\ref{sec:5g}.} Elasticity of defects on the torus\\
\hspace*{24pt} {\ref{sec:5h}.} The fat torus limit\\
{\ref{sec:6}.} Conclusion and discussion
\end{abstract}

\section{\label{sec:1}Order and geometry in condensed matter}

\subsection{\label{sec:1a}Introduction}
More than 200 years ago, in his treatise on the resistance of fluids, d'Alembert wrote: ``Geometry, which should always follow physics when used to describe nature, sometimes commands it'' \cite{DAlembert}. Since then the physics community has explored the power of geometry not only to describe, but also to explain structures and their properties. In the past 20 years soft condensed matter physics has provided many examples of how the geometry of matter is not a quiescent background for some microscopic degrees of freedom, but instead plays a major role in determining structural and mechanical properties and designing the phase diagram of materials such as colloids, liquid crystals, membranes, glasses and carbon nanostructure. 

Geometric models of condensed matter systems have been developed for a wide class of materials since the pioneering work of Bernal and Finney \cite{Bernal:1960,Bernal:1964,BernalFinney:1967a,BernalFinney:1967b,Finney:1970}. In a series of classic papers they suggested that several properties of liquids have their geometric counterpart in randomly packed arrays of spheres. The difference in density between the solid and the liquid phase of a simple monoatomic substance, for instance, is approximately the same between periodically and randomly packed hard spheres (roughly $15\%-16\%$). Also the radial distribution function of randomly packed spheres corresponds well with that determined by X-ray and neutron diffraction for rare-gas liquids. After Bernal, a significant amount of work has been done on random close packing and, even though the legitimacy of the notion of random close packing itself has been questioned frequently in recent years \cite{TorquatoEtAl:2000}, it is now established that many features of the liquid state have in fact a purely geometrical nature.

After the discovery of icosahedral order in metallic glasses \cite{ShechtmanEtAl:1984,LevineSteinhardt:1984,Dyre:2006} the idea of \emph{geometrical frustration} (the geometric impossibility of establishing a preferred local order everywhere in space, see Sec. \ref{sec:2}), became a fundamental concept for the characterization of amorphous solids. Farges and coworkers \cite{FargesEtAl:1975,FargesEtAl:1977,FargesEtAl:1981} were the first to show, by electron diffraction experiments and computer simulations, that the first atoms of small aggregates of rare gases condensed in ultra-high vacum form regular tetrahedra, which later organize in the form of small icosahedral clusters. Since icosahedra do not fill three-dimensional Euclidean space $\mathbb{R}^{3}$, the structure resulting from the aggregation of these icosahedral building blocks does not exhibit long range translational order. The lack of crystallization in covalent glasses is also rooted in the geometrical frustration associated with the constant coordination number of their constituents. Tetravalent monoatomic materials, for instance, cannot form a constant angle between bonds incident at the same atom and organize in a regular network at the same time. In multiatomic glass-forming materials the situation is more involved and the route to the formation of amorphous structures is related to the fact that crystallization would require complex activated phenomena and too large a decrease in entropy.

A breakthrough in the geometrical description of amorphous solids came in 1979 when Kl\'eman and Sadoc first observed that a number of continuous random lattices can be classified as specific mappings of ordered lattices in spaces of constant curvature onto $\mathbb{R}^{3}$ \cite{KlemanSadoc:1979}. This idea was inspired by the observation that whereas regular tetrahedra do not fill three-dimensional Euclidean space, they do regularly tile the three-dimensional sphere $\mathbb{S}^{3}$ (the manifold described by the equation $\sum_{i=1}^{4}x_{i}^{4}=R^{2}$ in $\mathbb{R}^{4}$) on which they build a regular polytope (Schl\"afli symbol $\{3,3,5\}$) with 120 vertices of coordination number 12. This idea, later developed by several others (see Sadoc and Mosseri \cite{SadocMosseri:1982}, Steinhardt \emph{et al}. \cite{SteinhardtEtAl:1981}, Nelson \cite{Nelson:1983}, Kl\'eman \cite{Kleman:1989}), paved the way for a new approach to spatial disorder based on the interplay between order and geometry on three-dimensional manifolds of constant Gaussian curvature.

In-plane order on two-dimensional manifolds has been the subject of much research since the discovery of the ordered phases $L_{\beta}$ and $P_{\beta}$ of phospholipid membranes and is now a rich and mature chapter of condensed matter physics \cite{MembranesAndSurfaces}. After the seminal work of Nelson and Peliti \cite{NelsonPeliti:1987} on shape fluctuations in membranes with crystalline and hexatic order, much work has been done elucidating the intimate relation between in-plane order and the geometry of the underlying substrate with many striking results and even more open questions. The fundamental role of topological defects in two-dimensional systems, first elucidated in a series of pioneering papers by Berezinskii \cite{Berezinskii:1970}, Kosterlitz and Thouless \cite{KosterlitzThouless:1972,KosterlitzThouless:1973,Kosterlitz:1974}, is enhanced in the presence of Gaussian curvature in the underlying medium, and leads to structures in the ground state that would normally be highly suppressed in flat systems. The goal of this article is to review the most recent developments in the study of the ground state properties of two-dimensional order on curved media; that is the structure and the mechanics of ordered phases on two-dimensional substrates equipped with non-zero Gaussian curvature, in a regime where thermal fluctuations are negligible in comparison with other energy scales in the system. While focusing on ground states we note that finite temperature physics on curved spaces is a rich source of open problems.  

This review is organized as follows. In Sec. \ref{sec:2} we discuss the concept of geometrical frustration and review some fundamentals of the elasticity of topological defects in flat and curved spaces. The study of crystalline and orientational order on the sphere is the basis of most of our current knowledge of order in curved space and will be reviewed in Sec. \ref{sec:3}. Sec. \ref{sec:4} is dedicated to crystalline order on surfaces with variable Gaussian curvature and boundary. The existence of defective ground states in toroidal monolayers, with intrinsic crystalline or hexatic order, is a recent development  in the study of order on curved surfaces and is reviewed in Sec. \ref{sec:5}. Finally, in Sec. \ref{sec:6}, we discuss some current and potential applications of defective structures to materials science and nano-engineering.

\subsection{Ordered structures in two-dimensional matter}\label{sec:1b}

Before discussing more technical aspects of the physics of ordered structures on curved surfaces we want to recall some salient features of physical systems with in-pane order and spatial curvature.

\subsection{\label{sec:1c}Amphiphilic Membranes}

\begin{figure}[h!]
\centering
\includegraphics[scale=0.25]{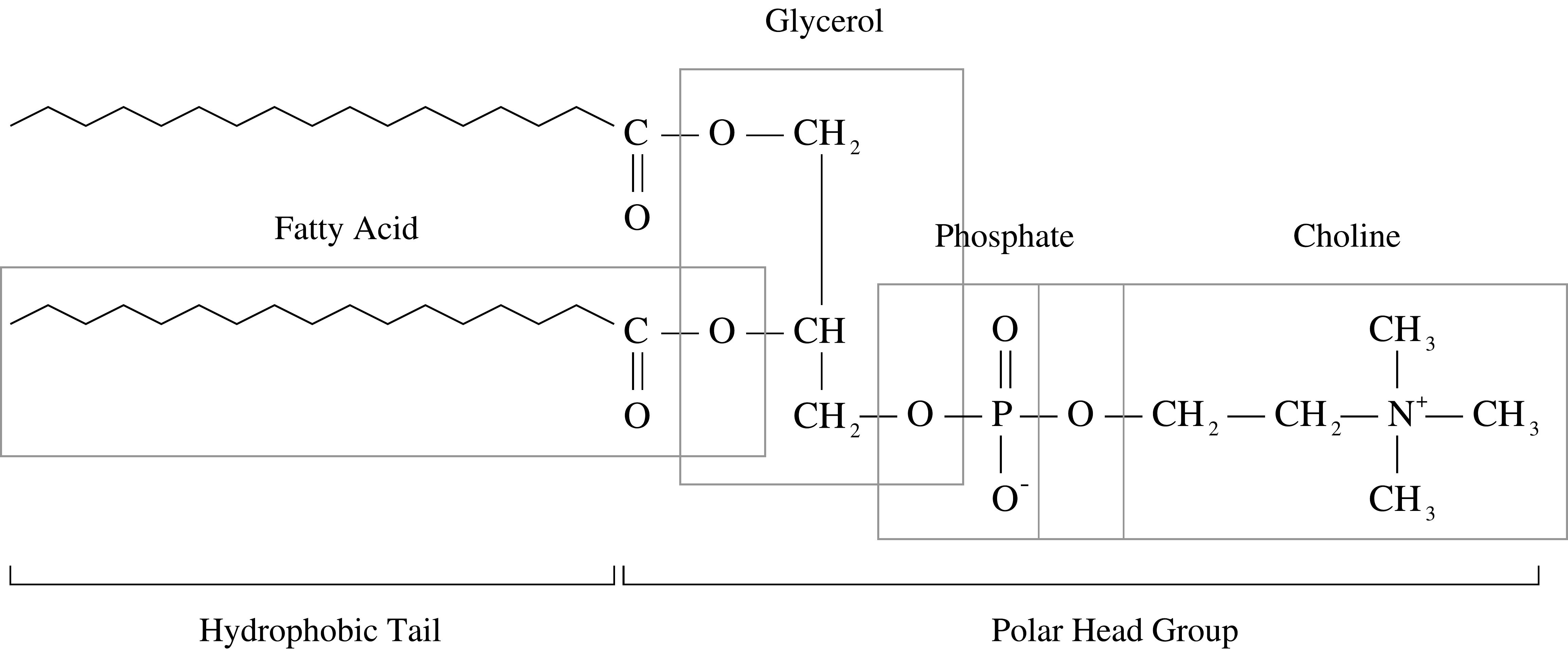}
\caption{\label{fig:sec1-pc}The structure of phosphatidylcholine.}
\end{figure}

\begin{figure}[t]
\centering
\includegraphics[width=0.4\textwidth]{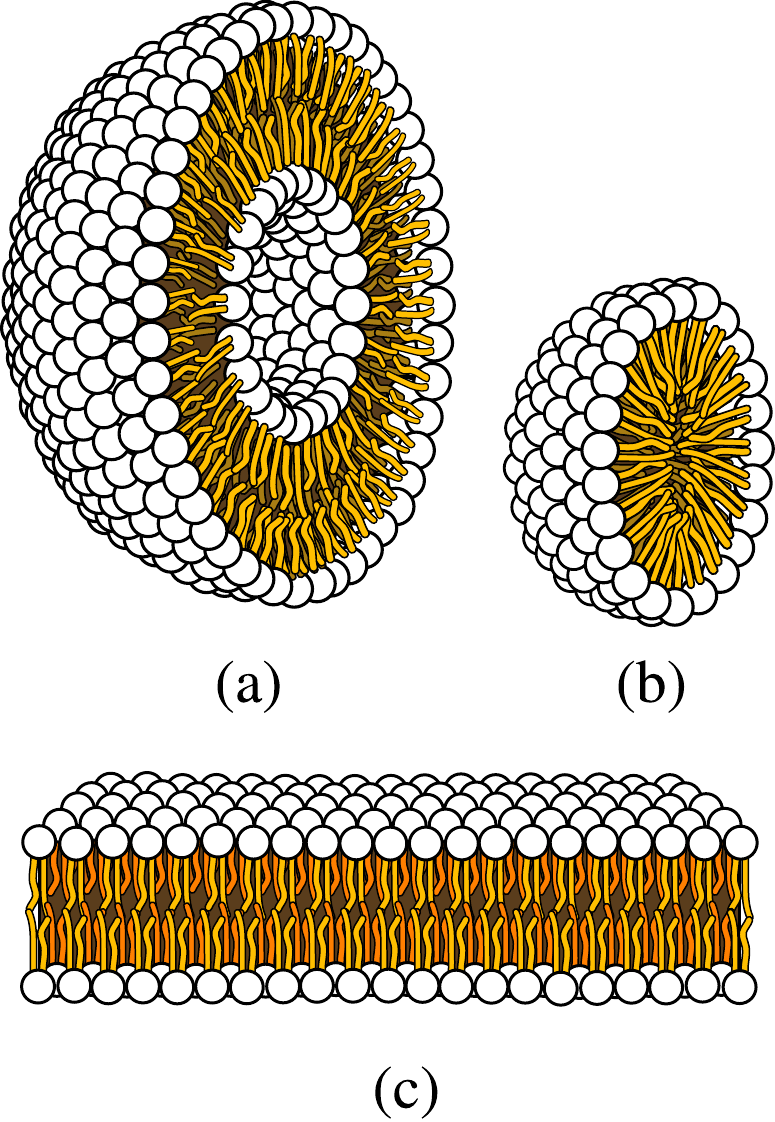}
\caption{\label{fig:sec1-phospholipids}Example of structures formed from self-assembly of amphiphilic molecules. (a) a vesicle, (b) a micelle and (c) a bilayer. [Courtesy of Mariana Ruiz Villarreal].}
\end{figure}

Amphiphilic membranes are thin sheets ($50-100$ \AA) of amphiphilic molecules immersed in a fluid and organized in the form of a \emph{bilayer} (see Fig. \ref{fig:sec1-phospholipids}). The most common constituents of biological membranes are phospholipids consisting of a polar head group and a hydrophobic tail made up of two fatty acyl chains (see Fig. \ref{fig:sec1-pc}). Tails have typical length $14-20$ carbon atoms and regulate the thickness and the stability of the bilayer. The polar head group contains one or more phosphate groups -POOH-O-R. Most phospholipid head groups belong to the phosphoglycerides, which contain glycerol joining the head and the tail.  Examples of phosphoglycerides include phosphatidylcholine (PC), phosphatidylethanolamine (PE) and phosphatidylserine (PS). These are distinguished by the residue R carried by the phosphate group (choline: R=-CH$_{2}$-CH$_{2}$-N$^{+}$-(CH$_{3}$)$_{3}$ and ethanolamine: R=-CH$_{2}$-CH$_{2}$-NH$_{2}$). The fatty acyl chain in biomembranes usually contains an even number of carbon atoms. They may be saturated  (neighboring C atoms are all connected by single bonds) or unsaturated  (some neighboring C atoms are connected by double bonds). 

At low temperature pure phospholipids crystallize and form a bilayer with all the tails in the \emph{trans} configuration and the heads parallel to the bilayer surface and firmly linked into a lattice (mostly by hydrogen bonds). At higher temperature the order is disrupted and the bilayer becomes fluid. 
In solution lipid bilayers can be found in a number of phases with differing degrees of order among the hydrocarbon chains. For phospholipids in the PC family these phases are usually called $L_{\alpha}$, $L_{\beta}$ and $P_{\beta}$. In the $L_{\alpha}$ phase, the tails are liquid and disordered. $L_{\beta}$ is a gel-like phase in which hydrocarbon tails are ordered and diffusion is severely reduced. The $L_{\beta}\rightarrow L_{\alpha}$ transition is usually referred to as \emph{main} transition. The main transition temperature $T_{m}$ increases with the length of the molecules due to a stronger Van der Waals attraction between adjacent molecules. The degree of unsaturation of the hydrocarbon tails also affects $T_{m}$. An unsaturated double bond can produce a kink in the alkane chain, disrupting the regular periodic structure and thus lowering the transition temperature. For bilayers in the PC family, $T_{m}$ ranges form $-60$ to $80$ $\left.^{\circ}\mathrm{C}\right.$, depending on the tail length and the number of double bonds. The order of the hydrocarbon chains also implies a larger thickness of the bilayer. The two lamellar phases can be separated by an intermediate ``rippled'' phase $P_{\beta}$ in which the bilayer exhibits an undulated structure and almost solid-like diffusion properties. Hydrocarbon chains can also appear tilted with respect to the bilayer plane. Tilted phases are generally denoted as $L_{\beta}'$ and $P_{\beta}'$. There are in fact several $L_{\beta}$ phases characterized by differing amounts of tilt and in-plane orientational order \cite{SmithEtAl:1988}. 

Upon changing their concentration, amphiphiles in solution aggregate in a wide variety of structures in addition to bilayers. Above the \emph{critical micelle concentration} (which is of the order of $10^{-3}$ mol/$\ell$) spherical micelles appear (see Fig. \ref{fig:sec1-phospholipids}). Their formation occurs more readily for single-chain amphiphiles (e.g. monoglycerids) and is favored by the presence of large head groups. At higher amphiphile concentrations spherical micelles are replaced by non-spherical ones and eventually by cylindrical rods. Spherical and cylindrical micelles can themselves organize in higher order structures such as cubic lattices or hexagonally packed rod piles. More exotic phases can be obtained by adding oil to the solution. Once the oil is dispersed in water, amphiphiles can form a monolayer across the water-oil interface and self-assemble in complex tubular structures known under the common name of \emph{plumber`s nightmare} \cite{AmphiphilicLayers}. 

\subsubsection{\label{sec:1d}Colloidosomes}

\begin{figure}
\centering
\includegraphics[scale=0.3]{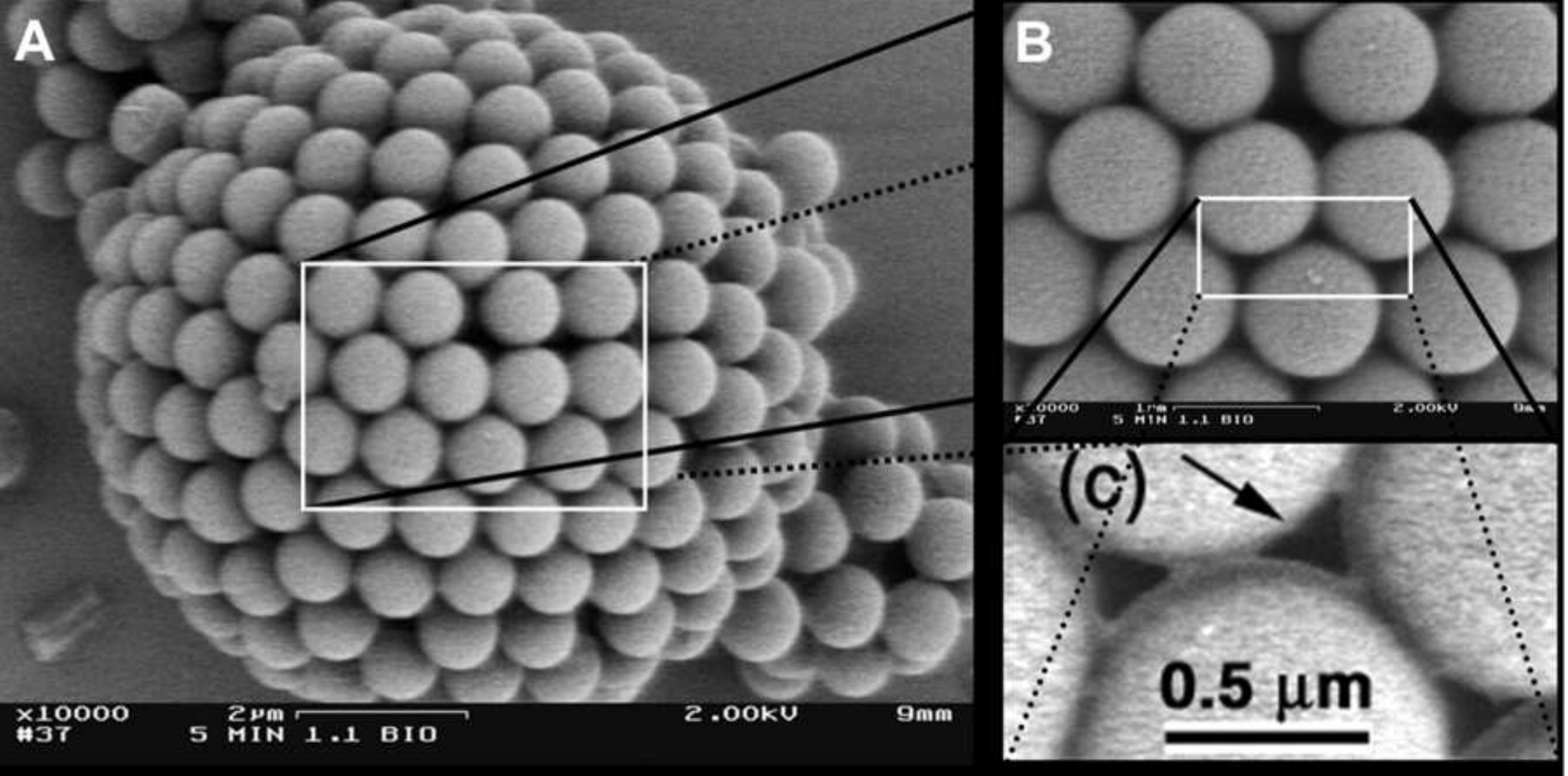}
\caption{\label{fig:sec1-colloidosome}Scanning electron microscope image of a colloidosome from Dinsmore \emph{et al}. \cite{DinsmoreEtAl:2002}. The colloidosome is composed of 0.9 $\mu$m diameter polystyrene spheres sintered at $105^{\circ}\mathrm{C}$. The close-up on the left shows the effect of sintering at the contact points of neighboring spheres.}
\end{figure}

The name \emph{colloidosome} was coined by Dinsmore \emph{et al}. to indicate microcapsules consisting of a shell of coagulated or partially fused colloidal particles surrounding a liquid core \cite{DinsmoreEtAl:2002}. Because of their controllable size, elasticity and permeability, colloidosomes have been recognized to form a promising class of ``soft devices'' for the encapsulation and delivery of active ingredients with a variety of potential applications for the development of novel drug and vaccine delivery vehicles and for the slow release of cosmetic and food supplements. Their major advantage relies on the fact that the permeability of the shell depends mainly on the size of the gaps between neighboring colloidal particles which can be tuned by controlling their size, interactions and degree of fusion.  Velev \emph{et al}. \cite{VelevEtAl:1996a,VelevEtAl:1996b,VelevEtAl:1997} were the first to report a method for the preparation of colloidosomes by templating octanol-in-water emulsions stabilized by latex particles and subsequently removing the octanol core by dissolution in ethanol. Structures of similar architecture have been obtained by templating water-in-oil emulsions \cite{YiEtAl:2002,WangCaruso:2002}. Multilayer shells consisting of alternating positive and negative polyelectrolytes and/or nanoparticles have also been prepared by using layer-by-layer assembly techniques, with the final hollow shells being obtained by removal of the central, sacrificial colloidal particles \cite{KumaraswamyEtAl:2002,FangEtA:2002}. Loxley and Vincent \cite{LoxleyVincent:1998} developed a new way of preparing polymeric capsules with liquid cores based on a phase separation of the polymer within the templated emulsion. The colloidosomes produced by Dinsmore \emph{et al}. \cite{DinsmoreEtAl:2002} were obtained by the assembly of latex particles into shells around water-in-oil emulsion drops, followed by thermal fusion of the particles in the shell and centrifugal transfer into water through a planar oil-water interface (see Fig. \ref{fig:sec1-colloidosome}).

The coverage of emulsified droplets by colloidal particles takes place by self-assembly. The particles dispersed in the fluid spontaneously adsorb on the interface provided the surface energy between the fluid on the inside and the outside of the droplet ($\sigma_{i,o}$) is larger than the difference of those between the particles and the internal fluid $(\sigma_{p,i})$ and the particles and the external fluid $(\sigma_{p,o})$. Thus $\sigma_{i,o}>|\sigma_{p,i}-\sigma_{p,o}|$. A similar mechanism is used in Pickering emulsions, which are stabilized by surface adsorption of colloidal particles. Once adsorbed at the interface, interacting particles distribute evenly, assuring a full and uniform coverage of the droplet. The interactions depend on the type of colloidal particles used as well as the liquids. Coated polymethylmethacrylate (PMMA) or polystyrene spheres, for instance, acquire a permanent electric dipole moment at the interface between the two fluids, probably because of the dissociation of charges on the hydrated surface similar to what happens at a water-air interface \cite{Pieranski:1980}. The resultant dipolar interaction stabilizes the particles and allows full coverage of the droplet. 

The colloidal particles comprising the shell are then locked together to achieve the desired permeability and robustness of the colloidosome. Several techniques are available to achieve this. By sintering polystyrene particles at a temperature slightly above the glass transition ($T_{g}\approx 100$ $\left.^{\circ}\mathrm{C}\right.$), it is possible to achieve a partial fusion of neighboring particles at the contact points. This process also allows one to control precisely the size of the gaps between particles and therefore the permeability of the colloidosome. Another method consists of binding the particles with a polyelectrolyte of opposite charge which can bridge neighboring particles and immobilize them at the interface. Particle locking clearly enhances the toughness of the colloidal shell and increases its rupture stress. For sintered polystyrene particles the latter can be tuned within the range $1-100$ MPa. Colloidosomes locked with polyelectrolytes are even more deformable and can withstand strains of order 50 \% before rupturing. 

The crystalline arrangement of charged colloids on a hemispherical droplet has been recently studied by Irvine and Chaikin who fabricated colloidal suspensions of PMMA spheres at the interface between water and cyclohexyl bromide (CHB) \cite{IrvineChaikin}. Because of the large difference in dielectric constants, ions from the oil strongly partition into the water phase \cite{LeunissenEtAl:2007}. When nearly 100\% PMMA particles are dispersed into the oil phase they form a Wigner crystal and a monolayer near the interface separated by a zone depleted of particles. These last two features are due the ion partitioning that provides the water phase with mobile charges. The net charge of water can be controlled through the pH, shifting the hydrolysis and ion partitioning equilibria. An electrically neutral water droplet acts as a conductor and attracts PMMA particles at the interface through an image charge mechanism. As a result the particles are barely wet by the water phase and organize in a perfect monolayer at (and not across) the interface. 

\begin{figure}
\centering
\includegraphics[width=0.5\textwidth]{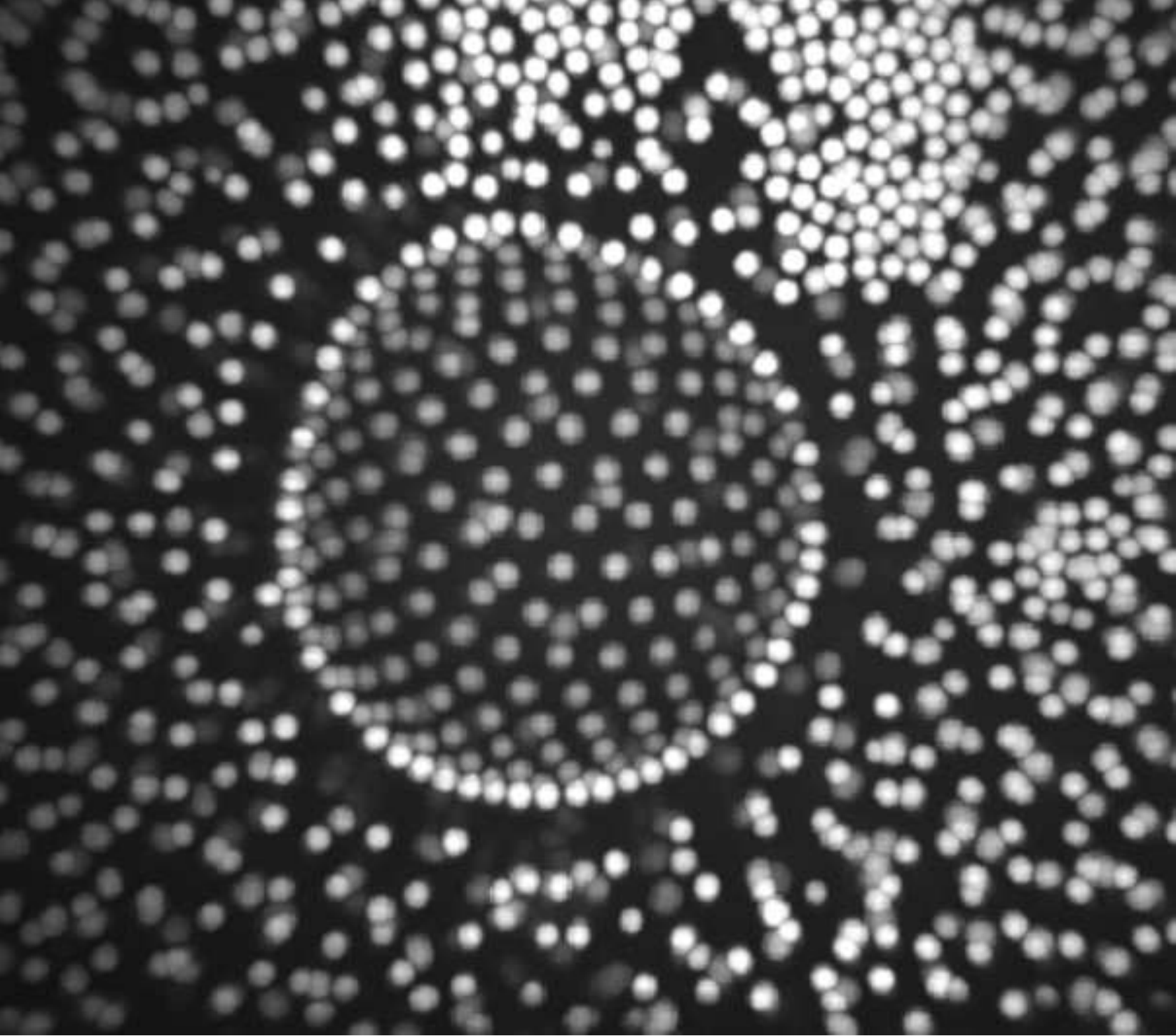}
\caption{\label{fig:sec1-irvine}PMMA colloids sitting at the hemispherical interface between water and cyclohexyl bromide. The particles are positively charged and interact via a screened Coulomb interaction with a Debye screening length proportional to the concentration of ions in the solvent. [Courtesy of W. Irvine and P. M. Chaikin, New York University, New York, NY].}
\end{figure}

\subsubsection{\label{sec:1e}Viral capsids}

Viral capsids are protein shells that enclose the genetic material of a virus and protect it from enzymatic digestion. Capsid proteins are expressed from the DNA or RNA genome of the virus and in physiological conditions self-assemble in very efficient structures which can withstand high forces (their Young modulus is $\sim 2$ MPa) and at the same time effectively disassemble to allow the viral genome to be released in the host cell. Most viral capsids have spherical or rod-like shape, but less standard shapes, such as conical or toroidal, also occur. 

The crystallographic structures of spherical viruses have been extensively investigated and, thanks to the modern techniques of X-ray spectroscopy and cryo-transmission electron microscopy, form part of the core knowledge of modern virology. In most  cases the capsid proteins are grouped in subunits called \emph{capsomers}, oligomers made of either five (pentamer) or six (hexamer) proteins. Spherical viruses typically posses icosahedral symmetry with twelve pentamers located at the vertices of a regular icosahedron. The number of hexamers that complete the capsids is given by $10(T-1)$, where $T$, the triangulation number, takes values from a sequence of ``magic numbers'' (i.e. $T=1,\,3,\,4,\,7\ldots$) associated with the lattice structure of the capsid, as brilliantly explained by Caspar and Klug (CK) in a seminal 1962 \cite{CasparKlug:1962} (the CK construction of icosahedral lattices will be reviewed in Sec. \ref{sec:3}). The diameters of spherical viruses span the range from $10$ to $100$ nm. While small capsids are almost perfectly spherical, large viruses, such as the bacteriophage HK97 or the phycodnavirus, typical exhibit a faceted geometry with nearly flat portions separated by ridges and sharp corners corresponding to the twelve pentamers. This morphological difference was explained by Lidmar \emph{et al} \cite{LindmarEtAl:2003} as a \emph{buckling transition} resulting from a balance between the stretching energy associated with the pentamers in capsomer lattices and the bending elasticity of the viral capsid. 

Non-icosahedral capsids of spherocylindrical shape are common among bacteriophages such as some $T$-even phages as well as the $\phi$CBK and the $\phi$29. In this case the capsid has the form of a cylindrical tube composed of a ring of hexamers closed at the ends by two half-icosahedral caps. This structure is also found in a variant of the $T=7$ papovavirus and can be induced in other icosahedral viruses by point mutation in the capsid proteins \cite{KiselevKlug:1969,DongEtAl:1998}. Of special interest are polymorphic viruses, which can appear in either spherical or spherocylindrical conformations. Polymorphism has been observed in the polyoma/SV40 animal virus \cite{SalunkeEtAl:1989} and the cowpea chlorotic mottle virus (CCMV) \cite{Bancroft:1970} and, for the latter case, appears to be related to the pH and salt concentration of the environment. The human immunodeficiency virus (HIV) also shows broad polymorphism in its capsid shape, including cone-like structures in addition to tubular and spherical ones and has been well studied in recent years \cite{NguyenBruinsmaGelbart:2005}.

\begin{figure}[t]
\centering
\includegraphics[scale=0.5]{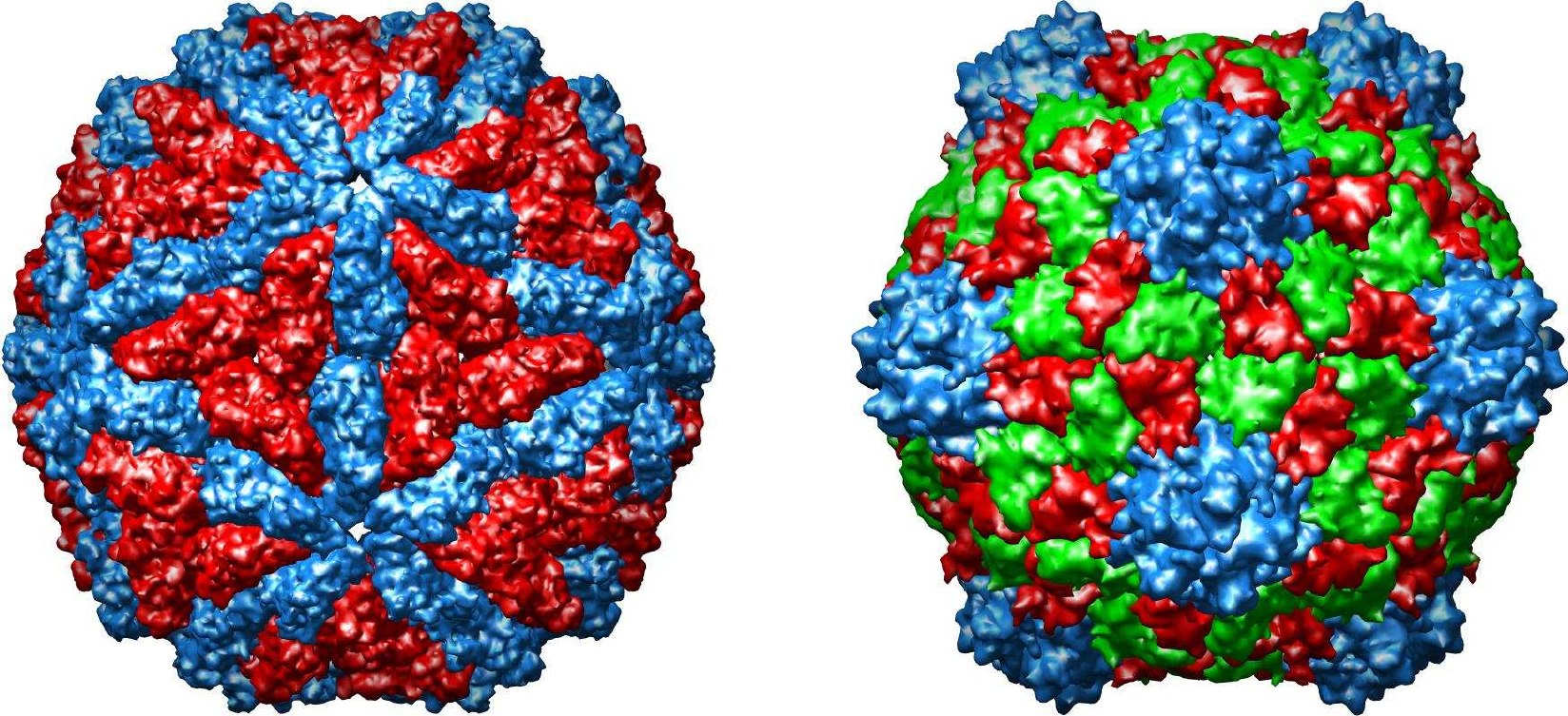}
\caption{\label{fig:sec1-virus}(Color online) Two examples of viral capsids. The L-A virus (on the left) and the cowpea mosaic virus (CPMV) (on the right). Both have 60 capsomers, but the latter has a marked faceted geometry. From \emph{VIPERdb} \cite{VIPERdb}.}
\end{figure}

\subsubsection{Carbon nanotubes and related materials}\label{sec:1f}

The science of carbon nano-materials has experienced a period of phenomenal growth since carbon nanotubes (CNTs) were found by Iijima in 1991 \cite{Iijima:1991} and, since then, a large number of similar structures, including helix-shaped graphitic nanotubes \cite{AmelinckxEtAl:1994}, nanotori \cite{LiuEtAl:1997}, graphitic nanocones \cite{KrishnanEtAl:1997} and nanoflowers \cite{LiuEtAl:2005} have been reported in the literature. The excitement in this field stems from certain exceptional properties that make CNTs potentially useful in many applications in nanotechnology, electronics, optics and other fields of materials science. They exhibit extraordinary strength, unique electrical properties and are efficient conductors of heat. Several methods for the preparation of graphitic nanomaterials have been developed, including laser pyrolysis, arc discharge, and electron irradiation. Recently, metal-catalyzed methods have also been used to synthesize carbon nanomaterials.

Most single wall nanotubes (SWNT) have diameter of about 1 nm, while the length is often of the order of microns. The lattice structure of a SWNT can be obtained from that of a graphene plane by assigning a pair of indices $(n,m)$ which specify how the graphene lattice is rolled up into a seamless cylinder (see Sec. \ref{sec:5}). $n=0$ nanotubes are referred to as ``zigzag'', while $n=m$ tubules are called ``armchair''. The generic name ``chiral'' is used otherwise \cite{Saito}. In terms of tensile deformations, SWNT are the stiffest materials known with a Young modulus in the range 1-5 TPa and a tensile strength of 13-53 GPa. This strength results from the covalent $sp^{2}$ bonds formed between the individual carbon atoms. 

The lattice structure of a nanotube strongly affects its electrical properties because of the interplay between the unique electronic structure of graphene and the tubular geometry.  For a given $(n,m)$ structure, a nanotube can be a conductor (if $n=m$), a small-gap semiconductor (if $n-m$ is a multiple of 3) or a standard semiconductor (otherwise). Thus all armchair ($n=m$) nanotubes are metallic, and nanotubes with $(n,m)$ equal (5,0), (6,4), (9,1), etc. are semiconducting. 

Topological defects may occur on the side-wall of carbon nanotubes and related materials in the form of atomic vacancies, $5-7$ dislocations and Stone-Wales defects \cite{StoneWales:1986} (i.e. quadrupoles consisting of two pairs of $5$-membered and $7$-membered rings) and are believed to significantly change the mechanical and transport properties of carbon nanotubes. The latter, in particular, were suggested to serve as possible nucleation centers for the formation of dislocations in the original ideal graphene network and constitute the onset of possible plastic deformations \cite{NardelliEtAl:1998}. Electronic transport is affected by the presence of defects by lowering the conductivity through the defective region of the tube. Defect formation in armchair-type tubes (which can conduct electricity) can cause the region surrounding that defect to become semiconducting. Furthermore, single monoatomic vacancies induce magnetic properties. Phonon scattering by defective regions heavily affects the thermal properties of carbon nanotubes and leads to an overall reduction of the thermal conductivity. Phonon transport simulations indicate that substitutional defects such as nitrogen or boron will primarily lead to scattering of high frequency optical phonons. Larger scale defects, however, such as Stone Wales defects, cause phonon scattering over a wide range of frequencies, leading to a greater reduction in thermal conductivity \cite{MingoEtAl:2008}. On the other hand, defective regions appear to be natural places for chemical functionalization. Numerical simulations have shown how the presence of Stone-Wales defects considerably enhances the adsorption of carboxyl groups (COOH) which can then bind to molecules with amide and ester bonds \cite{OuYangEtAl:2008}. 

\section{\label{sec:2}Interacting topological defects in curved media}

\subsection{\label{sec:2a}Geometrical frustration}

\begin{figure}[t]
\centering
\includegraphics[width=0.8\textwidth]{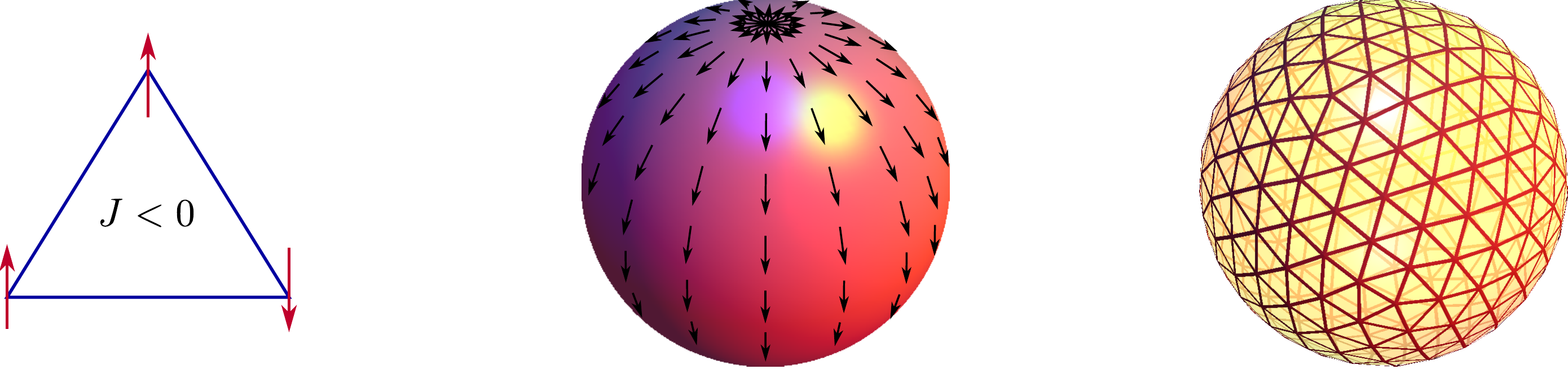}
\caption{\label{fig:sec2-geometrical-frustration}(Color online) Examples of geometrical frustration. (Left) Ising antiferromagnet on a triangular lattice. Because of the topology of the underlying lattice, no arrangement of the three spins can minimize the Hamiltonian $H=-J\sum_{\langle ij \rangle}S_{i}S_{j}$. (Center) Vector field of a sphere. As a consequence of the Poincar\'e-Hopf theorem, a vector field must vanish at least at two points, corresponding to the north and south pole of the sphere in this example. (Right) Triangulation of the sphere. As required by the Euler theorem, any triangulation of the sphere must contain some vertices of coordination number different from six (i.e. twelve $5-$fold vertices in this case).}
\end{figure}

The notion of \emph{geometrical frustration} was introduced to describe situations where certain types of local order, favoured by physical interactions, cannot propagate throughout a system \cite{SadocMosseri}. The expression was used for the first time by Toulouse in 1977 \cite{Toulouse:1977} to describe certain particular magnetic systems with nearest-neighbours interactions which cannot all be satisfied simultaneously. A textbook example of frustration in magnetic models is represented by a system of Ising spins on a triangular lattice with antiferromagnetic bonds: while a perfect antiferromagnetic alignment would minimize all terms in the Ising Hamiltonian, such an alignment is not allowed by the topology of the underlying lattice so that for any triangular plaquette there is always at least one unsatisfied bond (see Fig. \ref{fig:sec2-geometrical-frustration}). This concept can be extended naturally to any system where interactions impose a local order, but the most favoured local configuration is geometrically incompatible with the structure of the embedding space. 

Two-dimensional manifolds equipped with some microscopic field for which a notion of local order can be defined unambiguously provide a paradigm for systems exhibiting geometrical frustration. Consider for instance an assembly of identical particles interacting with a spherically symmetric pair potential $V_{ij}=V(|\bm{x}_{i}-\bm{x}_{j}|)$,  with $\bm{x}_{i}$ the position vector of the $i$th particle in a suitable coordinate system. In flat two-dimensional space, particles almost always pack in triangular lattices, unless the interaction potential is carefully tuned to select some some other lattice topology. Endowing the medium with a non-planar topology introduces frustration in the sense that the energetically favoured $6-$fold orientational order can no longer be established everywhere in the system. Such geometrical frustration arises at the microscopic level from the celebrated Euler theorem of topology which relates the number of vertices $V$, edges $E$ and faces $F$ of any tessellation of a $2-$manifold $M$:
\begin{equation}\label{eq:sec2-euler}
V-E+F=\chi\,,
\end{equation}
where $\chi$ is the Euler characteristic of $M$. If $M$ is an orientable closed surface, one can show that $\chi$ is determined uniquely by an integer $g \ge 0$, called the \emph{genus} of $M$, which represents the number of ``handles'' of $M$; namely $\chi=2(1-g)$. Two orientable closed surfaces with the same genus (thus the same Euler characteristic) are homeomorphic: they can be mapped into one another without changing their topological properties. In a surface with boundary, the Euler characteristic is given by $\chi=2(1-g)-h$, where $h$ is the number of boundaries or ``holes'' of $M$. Thus a sphere, which has no handles nor boundary ($g=h=0$) has $\chi=2$, while the embedded torus ($g=1$ and $h=0$) has $\chi=0$. A disk, on the other hand, has $\chi=1$ ($g=0$ and $h=1$) and is topologically equivalent to a sphere with one hole. In Secs. \ref{sec:3}-\ref{sec:5} we discuss ordered structures on three important topologies: the sphere, the disk and the torus. 

In the case of two-dimensional crystals with $6-$fold local order Eq. \eqref{eq:sec2-euler} can be rephrased in a form that is particularly useful in describing the presence of defects in the lowest energy state by defining a \emph{topological charge} as the departure from the ideal coordination number of a planar triangular lattice: $q_{i}=6-c_{i}$, with $c_{i}$ the coordination number of the $i$th vertex. Now, consider a tessellation in which each face is an $n-$sided polygon and let $k$ faces meet at each vertex. Since each edge is shared between two faces and links two vertices it follows that:
\[
nF = 2E = \sum_{k} k V_{k}\,,
\]
where $V_{k}$ is the number of vertices of degree $k$. For a triangulation $n=3$. From Eq. \eqref{eq:sec2-euler} it follows then:
\begin{equation}\label{eq:sec2-topological-charge}
Q = \sum_{i=1}^{V}q_{i} = 6\chi\,.
\end{equation}
In the case of a sphere, with $\chi=2$, Eq. \eqref{eq:sec2-topological-charge} implies any triangulation contains defective sites such that the total topological charge of the lattice is $Q=12$. This can be achieved, for example, by incorporating twelve $5-$fold disclinations (with $q=1$) in a network of $6-$fold coordinated sites like in a common soccer ball. These twelve disclinations, which would be suppressed in the lowest energy state of a planar crystal, result from the geometrical frustration associated with the topology of the sphere.

Eq. \eqref{eq:sec2-topological-charge} is a special case of geometrical frustration on 2-manifolds where local orientations are defined modulo $\pi/3$. More generally one can consider a $p-$atic director field for which local orientations are defined modulo $2\pi/p$. The topological charge of a disclination in this case is $q=\Delta\theta/\frac{2\pi}{p}$, where $\Delta\theta$ is the angle the director rotates in one counter-clockwise circuit of any closed contour enclosing the defect. Eq. \eqref{eq:sec2-topological-charge} becomes then:
\begin{equation}\label{eq:sec2-index-theorem}
Q = \sum_{i=1}^{N}q_{i}= p\chi
\end{equation}
where $N$ is the total number of defects. The ratio $k=q/p$ is commonly referred to as the \emph{winding number} of the disclination. In the case of a simple vector field, for instance, $p=1$ and Eq. \eqref{eq:sec2-index-theorem} corresponds to the well known Poincar\'e-Hopf theorem according to which the sum of the indices of all the isolated zeros of a vector field on a oriented differentiable manifold $M$ is equal to the Euler characteristic $\chi$ of $M$. Thus for a sphere a vector field must have at least one sink and one source, each having topological charge one, while on a torus ($\chi=0$) a vector field can be defect free. A nematic director $\bm{n}$, on the other hand, has $p=2$ (i.e. physical configurations are invariant under inversions $\bm{n}\rightarrow -\bm{n}$). Thus disclinations with $\pm 1$ topological charge correspond to configurations where the director rotates $\pm\pi$ in one circuit enclosing the defect. These elementary disclinations have semi-integer winding number $k=\pm 1/2$. As a consequence of Eq. \eqref{eq:sec2-index-theorem} the total topological charge of a nematic texture on the sphere is $Q=4$, corresponding for instance to the typical baseball texture consisting of four $q=1$ ($k=1/2$) disclinations located at the vertices of a regular tetrahedron \cite{LubenskyProst:1992,VitelliNelson:2006}.

\subsection{\label{sec:2b}Mathematical preliminaries and notation}

\begin{figure}[t]
\centering
\includegraphics[width=0.3\textwidth]{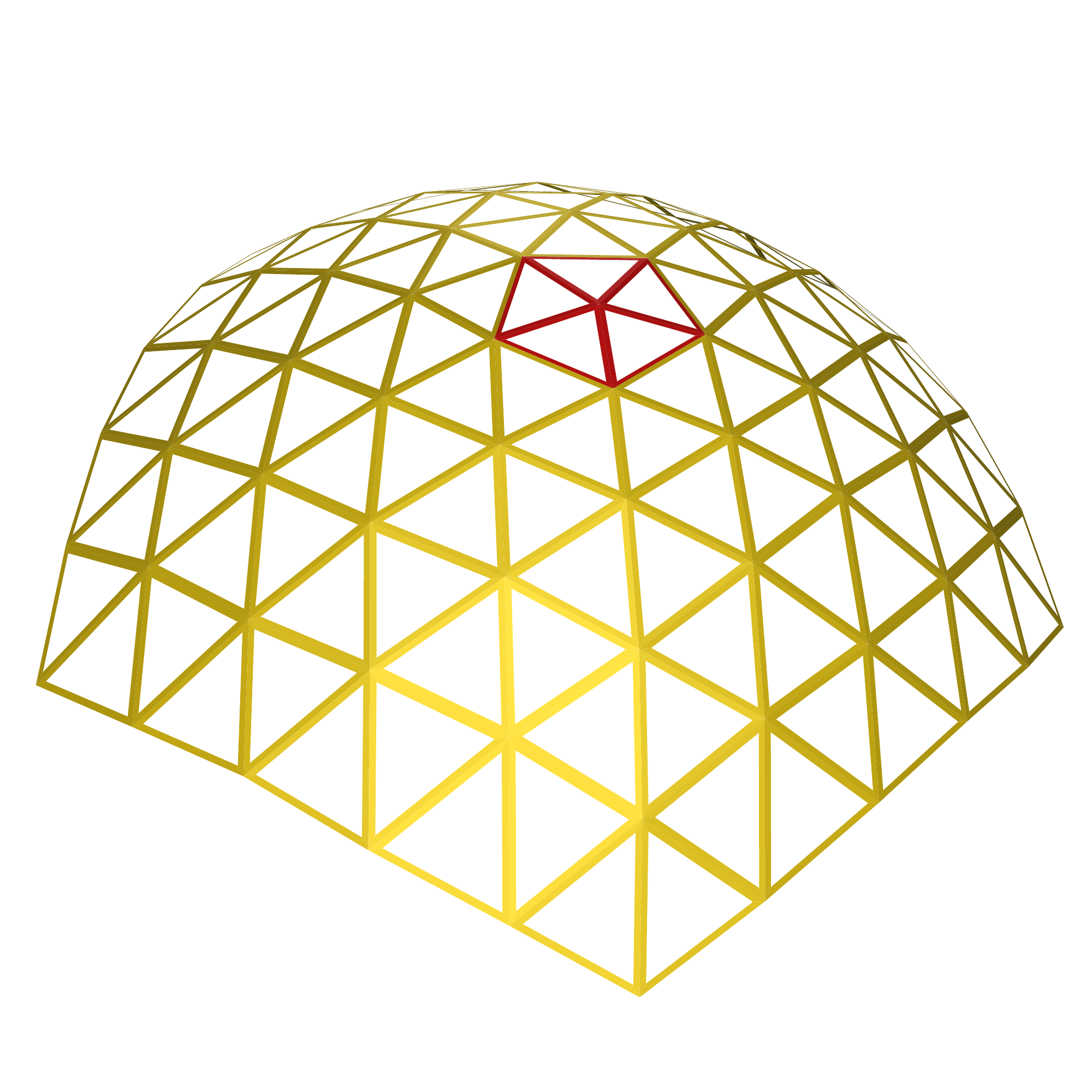}\qquad
\includegraphics[width=0.3\textwidth]{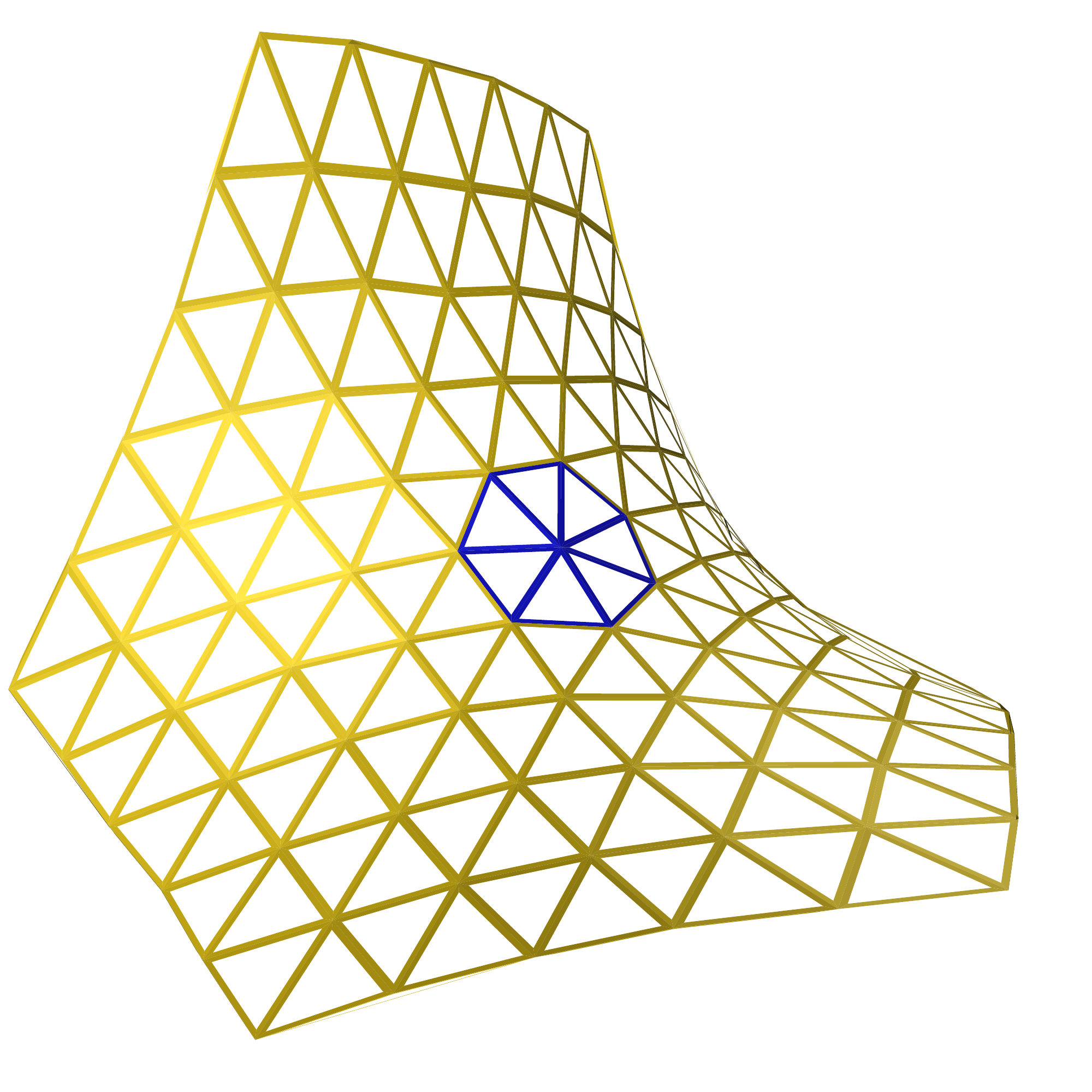}
\caption{\label{fig:sec2-disclinations}(Color online) Examples of $5-$fold (left) and $7-$fold (right) disclinations on a triangular lattice in regions of positive and negative Gaussian curvature respectively.}
\end{figure}

The equilibrium structure of two-dimensional locally ordered systems on curved substrates depends crucially on the existence and arrangement of the defects. The seminal work of Kosterlitz, Thouless, Halperin, Nelson and Young (KTHNY) on defect-mediated melting in two dimensions makes it clear that it is useful to employ a theoretical framework where the fundamental objects in the system are the defects themselves, while treating the microscopic constituents within a continuum elastic theory. This approach has the advantage of far fewer degrees of freedom than a direct treatment of the microscopic interactions and allows one to explore the origin of the emergent symmetry observed in non-Euclidean ordered structures as a result of the interplay between defects and geometry. This interplay is one of the fundamental hallmarks of order on two-dimensional manifolds and leads to universal features observed in systems as different as viral capsids and carbon macromolecules. In this section we will briefly review some concepts of differential geometry, mostly to establish notation. In the next section we will review some fundamentals of the elasticity of defects in two dimensions. The coupling mechanism between curvature and defects will be introduced in Sec. \ref{sec:2d}.

Points on a two-dimensional surface $S$ embedded in $\mathbb{R}^{3}$ are specified by a three-dimensional vector $\bm{R}(\bm{x})$ as a function of a two-dimensional coordinate $\bm{x}=(x^{1},x^{2})$. For each point of $S$ we define three vectors:
\begin{gather}
\bm{g}_{i} = \partial_{i}\bm{R}\qquad i=1,\,2\\[5pt]
\text{and}\qquad\bm{n} = \frac{\bm{g}_{1}\times\bm{g}_{2}}{|\bm{g}_{1}\times\bm{g}_{2}|}\,,
\end{gather}
where $\partial_{i}=\partial/\partial x^{i}$. The vectors $\bm{g}_{i}$ belong to $T_{\bm{R}}S$, the tangent space of $S$ at $\bm{R}$ whereas $\bm{n}$ is a normal vector. Note that while $\bm{n}$ is a unit vector, $\bm{g}_{i}$ are generally not of unit length. The metric, or \emph{first fundamental form}, of $S$ is defined as:
\begin{equation}\label{eq:sec2-metric}
ds^{2} = g_{ij} dx^{i}dx^{j}
\end{equation}
where $g_{ij}$ is the metric tensor:
\begin{equation}\label{eq:sec2-metric-tensor}
g_{ij} = \bm{g}_{i}\cdot\bm{g}_{j}
\end{equation}
The dual tensor is denoted as $g^{ij}$ and is such that:
\[
g_{ik}g^{jk} = \delta_{i}^{j}\,,
\]
with $\delta_{i}^{j}$ the Kronecker symbol. This allows us to introduce contravariant tangent-plane vectors $\bm{g}^{i}=g^{ij}\bm{g}_{j}$, satisfying $\bm{g}_{i}\cdot\bm{g}^{j}=\delta_{i}^{j}$. Any vector $\bm{v}$ on the tangent plane can be expressed as a linear combination of basis vectors $\bm{g}_{i}$ or $\bm{g}^{i}$: $\bm{v}=v^{i}\bm{g}_{i}=v_{i}\bm{g}^{i}$, where $v_{i}=g_{ij}v^{j}$. The extrinsic curvature of the surface $S$ is encoded in the \emph{second fundamental form} $b_{ij}$ (also known as the extrinsic curvature tensor):
\begin{equation}\label{eq:sec2-curvature-tensor}
b_{ij} =  - \bm{g}_{i}\cdot\partial_{j}\bm{n} = \bm{n}\cdot\partial_{j}\bm{g}_{i}\,.
\end{equation}
The eigendirections of $b_{ij}$ at a given point correspond to the principal curvature directions of $S$ at that point and the associated eigenvalues $\kappa_{1}$ and $\kappa_{2}$ are the extremal (or principal) curvatures. The mean curvature $H$ and the Gaussian curvature $K$ are defined as the sum and the product of the principal curvatures:
\begin{gather}
2H = \kappa_{1}+\kappa_{2} = g_{ij}b^{ij}\\[5pt]
K = \kappa_{1}\kappa_{2} = \tfrac{1}{2}\epsilon^{ik}\epsilon^{jl}b_{ij}b_{kl}
\end{gather}
where $\epsilon^{ij}$ is the dual of the Levi-Civita tensor, whose components are given by:
\begin{gather*}
\epsilon_{11} = \epsilon_{22} = 0\\[5pt]
\epsilon_{12} = - \epsilon_{21} = \sqrt{g}
\end{gather*}
where $g=\det g_{ij}$. The contravariant form is $\epsilon^{ij}=\epsilon_{ij}/g$ and satisfies $\epsilon_{ik}\epsilon^{jk}=\delta_{i}^{j}$. Since $\epsilon_{ij}v^{i}v^{j}=0$, where $v^{i}$ is any contravariant vector, it follows that the vector $\epsilon_{ij}v^{i}$ is perpendicular to $v^{j}$. Thus inner multiplication by $\epsilon_{ij}$ rotates a vector by $\pi/2$. The covariant derivative of a vector field $\bm{v}$ in the $i$th coordinate direction is defined as usual by:
\begin{subequations}
\begin{align}\label{eq:sec2-covariant-derivative}
\nabla_{i}v^{k}&=\partial_{i}v^{k}+\Gamma_{ij}^{k}v^{j}\\[5pt]
\nabla_{i}v_{k}&=\partial_{i}v_{k}-\Gamma_{ik}^{j}v_{j}
\end{align}
\end{subequations}
where $\Gamma_{ij}^{k}$ is the Christoffel symbol:
\begin{equation}\label{eq:sec2-christoffel}
\Gamma_{ij}^{k} = \tfrac{1}{2}g^{kl}(\partial_{j}g_{il}+\partial_{i}g_{lj}-\partial_{l}g_{ij})\,.
\end{equation}
Both the metric and the Levi-Civita tensor are invariant under parallel transport. This translates into:
\[
\nabla_{k}g_{ij} = \nabla_{k}g^{ij} = 0 \qquad
\nabla_{k}\epsilon_{ij} = \nabla_{k}\epsilon^{ij} = 0\,. 
\] 
Much of the elastic theory of defects, either in flat or curved systems, relies on the calculation of the Green function of the Laplace operator. On a generic $2-$manifold the latter obeys:
\begin{equation}
\Delta G_{L}(\bm{x},\bm{y}) = \delta(\bm{x},\bm{y})\,,
\end{equation}
where $\Delta$ is the Laplace-Beltrami operator:
\begin{equation}\label{eq:sec2-laplace-beltrami}
\Delta = \frac{1}{\sqrt{g}}\partial_{i}(\sqrt{g}\,g^{ij}\partial_{j})
\end{equation}
and $\delta$ the delta-function:
\begin{equation}
\delta(\bm{x},\bm{y}) = \frac{\delta(x^{1}-y^{1})\delta(x^{2}-y^{2})}{\sqrt{g}}\,.
\end{equation}
The Stokes theorem is frequently invoked when calculating elastic energies of defects. In covariant form it can be stated as follows: given a vector field $\bm{v}$ on a $2-$manifold $M$ with boundary $\partial M$, the following identity holds:
\begin{equation}\label{eq:sec2-stokes}
\oint_{\partial M} dx^{k}\,v_{k} = \int_{M} d^{2}x\,\epsilon^{ij}\nabla_{i}v_{j}\,.
\end{equation}
For consistency we will adopt covariant notation throughout this review. When discussing planar systems, in particular, we have: $b_{ij}=H=K=0$ and the elements of the metric tensor are given in Cartesian coordinates by $g_{xx}=g_{yy}=1$, $g_{xy}=g_{yx}=0$ and in polar coordinates by $g_{rr}=1$, $g_{\phi\phi}=r^{2}$, $g_{r\phi}=g_{\phi r}=0$.

The definition of orientational order on a surface clearly requires a non-ambiguous notion of angular distance between vectors on the same tangent plane. This is traditionally achieved by introducing a pair of orthonormal vectors $\bm{e}_{\alpha}$ ($\alpha=1,\,2$) called \emph{vielbein} (note that the canonical coordinate vectors $\bm{g}_{i}$ are generally neither orthonormal nor orthogonal) so that:
\begin{equation}\label{eq:sec2-vielbin1}
\bm{e}_{\alpha}\cdot\bm{e}_{\beta} = (e_{\alpha})_{i}(e_{\beta})^{i}=\delta_{\alpha\beta}\,,\qquad\qquad
(e_{\alpha})_{i}(e_{\alpha})_{j} = g_{ij}\,,
\end{equation}
where $\delta_{\alpha\beta}$ is the usual Kronecker symbol in the indices $\alpha$ and $\beta$. A vector field $\bm{v}=v^{i}\bm{g}_{i}$ can be expressed alternatively in the basis $\bm{e}_{\alpha}$:
\[
\bm{v} = v^{\alpha}\bm{e}_{\alpha}\qquad\qquad
v^{\alpha} = v^{i}(e_{\alpha})_{i}\,.
\]
Since the coordinates $v^{\alpha}$ are locally Cartesian, $\delta_{\alpha\beta}=\delta^{\alpha\beta}$ and there is no distinction between upper and lower Greek indices: $v_{\alpha}=v^{\alpha}$. Vielbein are constructed to be invariant under parallel transport, and thus $\nabla_{i}(e_{\alpha})_{j}=0$ and:
\[
\nabla_{i}v_{\alpha} = (e_{\alpha})_{j}\nabla_{i}v^{j} = \partial_{i}v_{\alpha}+\Omega_{i\alpha\beta}v_{\beta}\,,
\]
where $\Omega_{i\alpha\beta}$ is the so called \emph{spin connection}, and is given by:
\[
\Omega_{i\alpha\beta} 
= \bm{e}_{\alpha}\cdot\partial_{i}\bm{e}_{\beta}
= \Gamma_{ij}^{k}(e_{\alpha})^{j}(e_{\beta})_{k}-(e_{\beta})_{k}\partial_{i}(e_{\alpha})^{k}\,.
\]
Taking the derivative of the left equation in \eqref{eq:sec2-vielbin1} one immediately sees: 
\[
(e_{\alpha})^{k}\partial_{i}(e_{\beta})_{k} = -(e_{\beta})^{k}\partial_{i}(e_{\alpha})_{k}\,,
\]
from which it follows that $\Omega_{i\alpha\beta}=-\Omega_{i\beta\alpha}$. In two dimensions this permits the parametrization of the spin-connection $\Omega_{i\alpha\beta}$ by a single covariant vector $\bm{\Omega}$ (i.e. the antisymmetry under exchange of the Greek indices leaves only $d^{2}(d-1)/2=2$ independent components in $d=2$ dimensions). Thus
\begin{equation}\label{eq:sec2-spin-connection}
\Omega_{i\alpha\beta} = \epsilon_{\alpha\beta}\Omega_{i}\,,
\end{equation}
with $\epsilon_{\alpha\beta}=\delta_{\alpha}^{1}\delta_{\beta}^{2}-\delta_{\beta}^{1}\delta_{\alpha}^{2}$ the antisymmetric symbol. 

It is well known how the curvature of a manifold manifests itself when a vector is parallel transported around a closed path. Taking an infinitesimal square loop of sides $dx$ and $dy$, the parallel transported vector $\bm{v}'$ differs from the original vector $\bm{v}$ by an amount:
\begin{equation}\label{eq:sec2-parallel-transport}
v'^{k}-v^{k} = R_{lij}^{k}v^{l}dx^{i}dy^{j}\,,
\end{equation}
where $R_{lij}^{k}$ is the Riemann tensor given in two-dimensions by:
\begin{equation}\label{eq:sec2-riemann-tensor}
R_{ijk}^{l} = K \epsilon_{i}^{l}\epsilon_{jk}\,.
\end{equation}
In the orthonormal basis $\bm{e}_{\alpha}$, Eq. \eqref{eq:sec2-parallel-transport} reads:
\[
v'_{\alpha}-v_{\alpha} = R_{ij\alpha\beta}v_{\beta}dx^{i}dy^{j}\,,
\]
where $R_{ij\alpha\beta}$ is the curvature tensor associated with the spin-connection $\Omega_{i\alpha\beta}$: 
\begin{equation}\label{eq:sec2-riemann-omega}
R_{ij\alpha\beta} = \partial_{i}\Omega_{j\alpha\beta}-\partial_{j}\Omega_{i\alpha\beta}+\Omega_{i\alpha\gamma}\Omega_{j\gamma\beta}-\Omega_{j\alpha\gamma}\Omega_{i\gamma\beta}\,.
\end{equation}
In two dimensions, using Eq. \eqref{eq:sec2-riemann-tensor}, one can write:
\[
R_{ij\alpha\beta} 
= (e_{\alpha})^{k}(e_{\beta})^{l}R_{ijkl} 
= (e_{\alpha})^{k}(e_{\beta})^{l}\epsilon_{ij}\epsilon_{kl}K
= \epsilon_{\alpha\beta}\epsilon_{ij} K\,,
\]
which, combined with Eqs. \eqref{eq:sec2-riemann-omega} and \eqref{eq:sec2-spin-connection}, yields
\begin{equation}\label{eq:sec2-mixed-riemann}
R_{ij\alpha\beta} = \epsilon_{\alpha\beta}(\partial_{i}\Omega_{j}-\partial_{j}\Omega_{i}) = \epsilon_{\alpha\beta}\epsilon_{ij} K\,,
\end{equation}
which also implies:
\begin{equation}\label{eq:sec2-spin-curl}
\nabla\times\bm{\Omega} = \epsilon^{ij}\nabla_{i}\Omega_{j} = K\,.
\end{equation}

\begin{figure}
\centering
\includegraphics[width=0.8\textwidth]{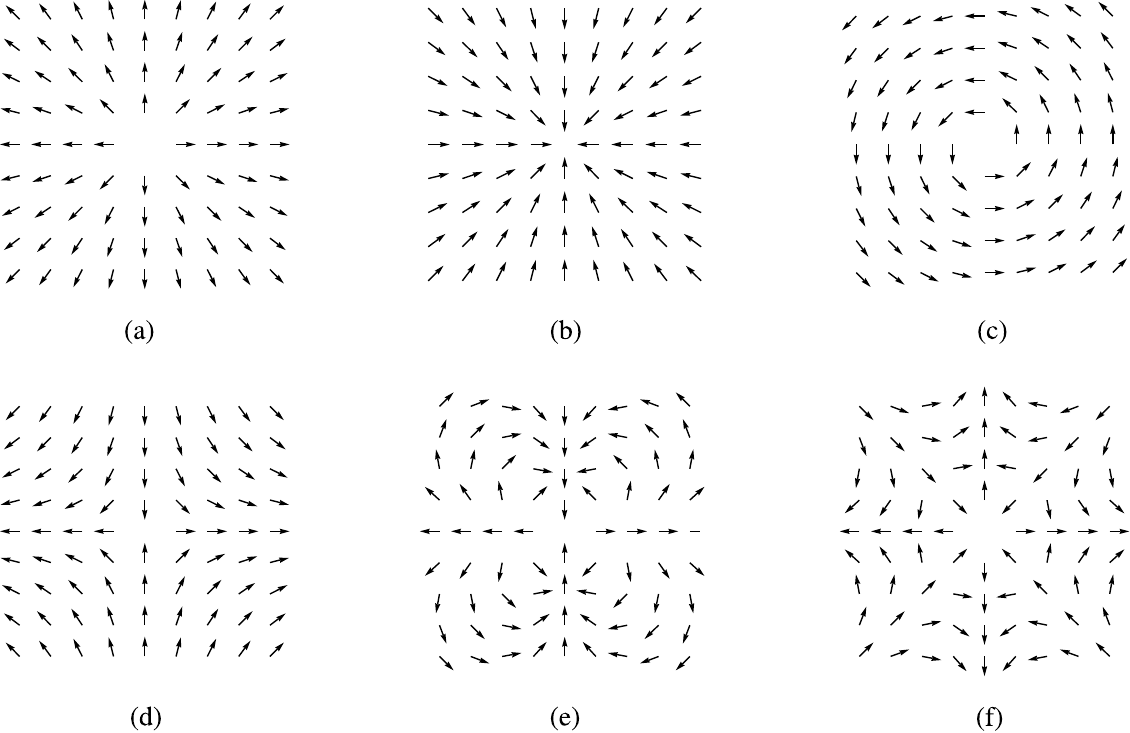}
\caption{\label{fig:sec2-defects-vector}Examples of topological defects in a vector field. (a), (b) and (c) have topological charge $q=1$, (d) $q=-1$ and (e) and (f) have charge $q=3$ and $q=-3$ respectively.}
\end{figure}

\subsection{\label{sec:2c}Elasticity of defects in the plane}

The $XY-$model is the simplest setting where particle-like objects emerge from a purely continuum theory \cite{KosterlitzThouless:1972,KosterlitzThouless:1973,Kosterlitz:1974}. The order parameter is the angular field $\theta\in[0,2\pi]$, which may represent the orientation of two-dimensional vectors $\bm{S}=S(\cos\theta,\sin\theta)$ or the phase of a complex field $\psi=|\psi|e^{i\theta}$. The interaction that tends to minimize the spatial variations of the order parameter results from the continuum free energy:
\begin{equation}\label{eq:sec2-xy-energy}
F_{el} = \frac{1}{2}K_{A}\int d^{2}x\,|\nabla\theta(\bm{x})|^{2}\,.
\end{equation}
Despite its simplicity this model successfully describes several aspects of the physics of vortices in superfluid $\,^{4}$He or thin superconducting films, where the angle $\theta$ is identified with the phase of the collective wave function. Eq. \eqref{eq:sec2-xy-energy} also describes nematic liquid crystals in the limit of equal splay and bending moduli. This approximation, however, favors $q=\pm 2$ ($k=\pm 1$) disclinations rather than the more natural $q=\pm 1$ ($k=\pm 1/2$) disclinations of nematics and is not particularly suitable to describe the ground state. At $T>0$, thermal fluctuations drive the two elastic constants to the same value at long wavelengths, so that there is a unique Kosterlitz-Thouless transition temperature \cite{NelsonPelcovits:1977}. As discussed by Deem \cite{Deem:1996}, the essential effect of unequal elastic constants is to create a distinct long-range contribution to the core energy of each defect. More generally Eq. \eqref{eq:sec2-xy-energy} can be considered as the simplest phenomenological free energy describing $p-$atic order (assuming $\theta$ defined modulo $2\pi/p$). Calling $\bm{v}=\nabla\theta$, the minimization of the free energy \eqref{eq:sec2-xy-energy} leads to the equilibrium condition:
\begin{equation}\label{eq:sec2-xy-laplace}
\Delta \theta = \nabla_{i}v^{i} = 0\,.
\end{equation}
In the presence of a number of disclinations of topological charge $q_{\alpha}$, $\theta$ changes by $(2\pi/p)\sum_{\alpha}q_{\alpha}$ in one circuit around any contour enclosing a total topological charge $\sum_{\alpha}q_{\alpha}$:
\begin{equation}\label{eq:sec2-contour-integral}
\oint d\theta = \oint dx^{i}\,v_{i} = \frac{2\pi}{p}\sum_{\alpha}q_{\alpha}\,.
\end{equation}
Using Stokes' theorem, Eq. \eqref{eq:sec2-contour-integral} can be translated into the requirement:
\begin{equation}\label{eq:sec2-patic-incompatibility}
\epsilon^{ij}\nabla_{i}v_{j}=\epsilon^{ij}\nabla_{i}\nabla_{j}\theta=\eta(\bm{x})\,,
\end{equation}
where: 
\[
\eta(\bm{x}) = \frac{2\pi}{p}\sum_{\alpha} q_{\alpha}\delta(\bm{x},\bm{x}_{\alpha})
\]
is the topological charge density. Using standard manipulations (see for example \cite{ChaikinLubensky}) a vector field $\bm{v}$ satisfying Eq. \eqref{eq:sec2-xy-laplace} with the constraint \eqref{eq:sec2-contour-integral} can be found in the form:
\begin{equation}\label{eq:sec2-v}
v_{i}(\bm{x}) = -\epsilon_{i}^{j}\nabla_{j}\int d^{2}y\,G_{L}(\bm{x},\bm{y})\eta(\bm{y}) \ ,
\end{equation}
where $G_{L}(\bm{x},\bm{y})$ is the Laplacian Green function. Using Eq. \eqref{eq:sec2-v} in Eq. \eqref{eq:sec2-xy-energy} leads to the well known expression:
\begin{equation}\label{eq:sec2-xy-coulomb}
F_{el} = -\frac{1}{2}K_{A}\int d^{2}x\,d^{2}y\,G_{L}(\bm{x},\bm{y})\eta(\bm{x})\eta(\bm{y})\,.
\end{equation}
Like charged particles, disclinations interact via a Coulomb potential, which in two dimensions is proportional to the logarithm of the distance in the plane. As noted, this framework is valid for both nematic liquid crystals in the one elastic constant approximation and superfluids. An important difference between these two systems lies in the choice of the boundary condition for the field $\theta$. Nematogens are typically forced to be normal to boundary of the substrate and this implies a constraint for $\theta$, while there is no such constraint for $\,^{4}$He films since the wave function is defined in a different space from that to which the superfluid is confined. This difference is crucial on a curved substrate (see Ref. \citep{TurnerEtAl:2008} for a detailed review of this topic).

Eq. \eqref{eq:sec2-xy-coulomb} represents the elastic energy associated with the distortion introduced by defects in the \emph{far field}, where the elastic variables change slowly in space. This expression breaks down in the neighborhood (or \emph{core}) of a defect, where the order parameter is destroyed and the actual energetic contribution depends on microscopic details. In order to describe a defective system at any length scale, Eq. \eqref{eq:sec2-xy-coulomb} must be corrected by adding a core energy $F_{c}$ representing the energetic contribution within the core of a defect, where standard elasticity breaks down. A detailed calculation of the core energy requires some microscopic model and is usually quite complicated. Nonetheless its order of magnitude can be estimated by writing:
\begin{equation}\label{eq:sec2-core-energy1}
F_{c} = \pi a^{2} f_{c}\,, 
\end{equation}
where $a$ is the core radius, corresponding to the short distance cut-off of the elastic theory, and $f_{c}$ is some unknown energy density independent of $a$. $f_{c}$ can then be estimated by minimizing the total energy of the system with respect to $a$. For a planar system with a single disclination of winding numer $k$, the total energy can be easily calculated from Eqs. \eqref{eq:sec2-xy-coulomb} and \eqref{eq:sec2-core-energy1}:
\begin{equation}\label{eq:sec2-core-energy2}
F = \pi K_{A} k^{2} \log\left(\frac{R}{a}\right) + \pi a^{2} f_{c}\,.
\end{equation}
Minimizing Eq. \eqref{eq:sec2-core-energy2} with respect to $a$, one finds $f_{c}=K_{A}k^{2}/(2a^{2})$ from which:
\begin{equation}\label{eq:sec2-core-energy3}
F_{c} = \frac{1}{2}\pi K_{A} k^{2} = \epsilon_{c}k^{2}\,.
\end{equation}
The quantity $\epsilon_{c}$ is the energy needed to increase the number of defects by one, regardless of its location. 

The elasticity of dislocations and disclinations in solids resembles in many respects that of vortex lines in the $XY-$model. In two dimensional elasticity a pure in-plane deformation is encoded in a displacement field $u^{i}$, $i=1,\,2$, which maps any point in the system $\bm{x}$ to:
\begin{equation}\label{eq:sec2-displacement}
\bm{x}' = \bm{x}+u^{i}\bm{g}_{i} \ ,
\end{equation}
where $\bm{g}_{i}$ is a suitable basis in the coordinates of the undeformed system. If there are no defects, the displacement field is a single-valued mapping of the plane into itself. Topological defects introduce an incompatibility in the displacement field, in the sense that $u^{i}$ is not a single-valued mapping anymore. The elastic stress in the region surrounding a defect appears at the macroscopic level from the Hooke's law of elasticity:
\begin{equation}\label{eq:sec2-hooke}
u_{ij} = \frac{1+\nu}{Y}\,\sigma_{ij}-\frac{\nu}{Y}\,g_{ij}\sigma_{k}^{k} \ ,
\end{equation}
where $Y$ and $\nu$ are the two-dimensional Young modulus and Poisson ratio respectively and $\sigma_{ij}$ is the stress tensor. Eq. \eqref{eq:sec2-hooke} can be obtained, for example, by minimizing the elastic energy:
\begin{equation}\label{eq:sec2-landau-energy}
F = \int d^{2}x\,\left(\tfrac{1}{2}\lambda u_{i}^{i\,2}+\mu u_{ij}u^{ij}\right) \ ,
\end{equation}
where $\lambda$ and $\mu$ are the Lam\'e coefficients in two dimensions:
\[
\lambda = \frac{Y\nu}{1-\nu^2}\,,\qquad\qquad
\mu = \frac{Y}{2(1+\nu)}\,.
\]
In the absence of body forces, the force balance equation requires $\sigma_{ij}$ to be divergence free:
\begin{equation}\label{eq:sec2-sigma-divergence}
\nabla_{j}\sigma^{ij} = 0\,.
\end{equation}
The strain tensor $u_{ij}$ represents the variation in the first fundamental form of the surface due to the deformation field \eqref{eq:sec2-displacement}, namely:
\begin{equation}\label{eq:sec2-strain}
2 u_{ij} = g_{ij}(\bm{x}+\bm{u})-g_{ij}(\bm{x}) = \nabla_{i}u_{j}+\nabla_{j}u_{i} + O(u^{2})\,.
\end{equation}
In the presence of a dislocation line $L$ the function $\bm{u}$ becomes multivalued so that, while traversing any closed counterclockwise loop $C$ containing $L$:
\begin{equation}\label{eq:sec2-dislocation}
\oint_{C} d\bm{u}= \bm{b}\,,
\end{equation}
where $\bm{b}$ is the Burgers vector representing the amount by which the image of a closed loop under the mapping \eqref{eq:sec2-displacement} fails to close in presence of a dislocation line \cite{Nabarro,HirtLothe}. If $L$ is an isolated straight dislocation with origin at the point $\bm{x}_{0}$, Eq. \eqref{eq:sec2-dislocation} implies:
\begin{equation}\label{eq:sec2-dislocation-density}
\epsilon^{ij}\nabla_{i}\nabla_{j} u_{k} = b_{k}\delta(\bm{x},\bm{x}_{0})\,.
\end{equation}

Generally speaking, maps such as that of Eq. \eqref{eq:sec2-displacement}, induce variations in both the length and the orientation of an infinitesimal distance vector $d\bm{x}$ in the deformed medium. This statement can be clarified by writing:
\[
dx_{i}'-dx_{i} = (\nabla_{j}u_{i})\,dx^{j} = (u_{ij}-\omega_{ij})\,dx^{j}
\]
where: 
\begin{equation}\label{eq:sec2-rotation1}
\omega_{ij} = \tfrac{1}{2}\,(\nabla_{i}u_{j}-\nabla_{j}u_{i})
\end{equation}
is the infinitesimal rotation tensor induced by the deformation Eq. \eqref{eq:sec2-displacement}. In two dimensions the latter can be conveniently written in the form: 
\begin{equation}\label{eq:sec2-rotation2}
\omega_{ij} = \epsilon_{ij}\,\Theta
\end{equation}
with:
\begin{equation}\label{eq:sec2-rotation3}
\Theta = \tfrac{1}{2}\,\epsilon^{ik}\nabla_{i}u_{k}\,.
\end{equation}
Since the application of $\epsilon_{ij}$ to a vector rotates the vector by $\pi/2$ clockwise, the coefficient $\Theta$ in Eq. \eqref{eq:sec2-rotation3} can be interpreted as the \emph{average rotation} of an infinitesimal line element under a distortion of the form Eq. \eqref{eq:sec2-displacement}. In presence of an isolated disclination the field $\Theta$ becomes multi-valued. If $s$ is the angular deficit associated with the disclination, integrating along a circuit enclosing the disclination core yields:
\begin{equation}\label{eq:sec2-disclination}
\oint_{C} d\Theta = s \ ,
\end{equation}
which can be rephrased as a statement about commutativity of partial derivatives:
\begin{equation}\label{eq:sec2-disclination-density}
\epsilon^{ij}\nabla_{i}\nabla_{j}\Theta = s\,\delta(\bm{x},\bm{x}_{0})\,,
\end{equation}
where $\bm{x}_{0}$ is the position of the disclination core. In a lattice with $n-$fold rotational symmetry, $s=(2\pi/n)q$, where $q$ is the topological charge introduced above. A comparison between Eq. \eqref{eq:sec2-patic-incompatibility} and Eq. \eqref{eq:sec2-disclination-density} clarifies the common identification of $\Theta$ with the \emph{bond angle field} of a crystal \cite{NelsonHalperin:1979}.

Eqs. \eqref{eq:sec2-dislocation-density} and \eqref{eq:sec2-disclination-density} provide a set of relations between the fundamental features of topological defects in crystals (i.e. the Burgers vector $\bm{b}$ and the topological charge $q$) and the partial derivatives of the displacement field $\bm{u}$. These relations can be used to derive an equation for the elastic stress arising in a system as a consequence of the defects. A simple and elegant way to achieve this task in the context of planar elasticity relies on the parametrization of the stress tensor via a single scalar field $\chi$ known as the Airy stress function (see for example \cite{GreenZerna}). Taking advantage of the commutativity of partial derivatives in Euclidean space, one can write 
\begin{equation}\label{eq:sec2-airy-stress}
\sigma^{ij}=\epsilon^{ik}\epsilon^{jl}\nabla_{k}\nabla_{l}\chi
\end{equation}
so that the force balance condition \eqref{eq:sec2-sigma-divergence} is automatically satisfied. Applying the operator $\epsilon^{ik}\epsilon^{jl}\nabla_{k}\nabla_{l}$ to both sides of Eq. \eqref{eq:sec2-hooke} and using Eq. \eqref{eq:sec2-airy-stress} one finds the following fourth order Poisson-like problem:
\begin{equation}\label{eq:sec2-biharmonic}
\Delta^{2}\chi(\bm{x}) = Y\eta(\bm{\bm{x}})\,,
\end{equation}
where $\Delta^{2}$ is the biharmonic operator and $\eta$ is the defect charge density:
\[
\eta(\bm{x}) = \frac{2\pi}{n}\sum_{\alpha}q_{\alpha}\delta(\bm{x}-\bm{x}_{\alpha})+\sum_{\beta}\epsilon^{ij}b_{i}^{\beta}\,\nabla_{j}\delta(\bm{x}-\bm{x}_{\beta})\,,
\]
where $s_{\alpha}$ denotes the topological charge of a disclination at $\bm{x}_{\alpha}$ and $\bm{b}^{\beta}$ the Burgers vector of a dislocation at $\bm{x}_{\beta}$.

Let's now turn our attention to the case of a planar crystal with local $6-$fold orientational symmetry populated with $N$ disclinations of density:
\begin{equation}\label{eq:sec2-density}
\eta(\bm{x}) = \frac{\pi}{3}\sum_{\alpha=1}^{N}q_{\alpha}\delta(\bm{x},\bm{x}_{\alpha})\,. 
\end{equation}
Assuming a free boundary (i.e. all the components of the stress tensor are zero along the boundary and thus $\chi=\nabla\chi=0$), the stress function $\chi$ can be expressed in the Green form:
\begin{equation}\label{eq:sec2-airy}
\frac{\chi(\bm{x})}{Y} = \int d^{2}y\,G_{2L}(\bm{x},\bm{y})\eta(\bm{y})\,,
\end{equation}
where $G_{2L}(\bm{x},\bm{y})$ is the biharmonic Green function. The elastic energy of the system is given by Eq. \eqref{eq:sec2-landau-energy} and \eqref{eq:sec2-airy} in the form:
\begin{equation}\label{eq:sec2-crystals-energy}
F_{el} = \frac{1}{2}Y\int d^{2}x\,d^{2}y\,G_{2L}(\bm{x},\bm{y})\eta(\bm{x})\eta(\bm{y})\, .
\end{equation}
Eqns. \eqref{eq:sec2-xy-energy} and \eqref{eq:sec2-crystals-energy} are the fundamental equations in the elastic theory of defects in two-dimensional planar $p-$atics and crystals. In the next section we will see how a non-zero Gaussian curvature in the underlying medium affects these energies by effectively screening the topological charge of the defects.

\subsection{\label{sec:2d}Coupling mechanisms between curvature and defects}

The elasticity of topological defects on curved substrates equipped with local orientational order was first considered by Nelson and Peliti in the context of hexatic membranes \cite{NelsonPeliti:1987}.  It is now standard knowledge in the statistical mechanics of membranes that long range forces appearing as a consequence of local orientational order crucially affect the behavior of membranes at finite temperature. The stiffness associated with orientational correlations leads to an enhancement of the bending rigidity that counteracts the thermal softening that occurs in fluid membranes. Furthermore, for planar membranes, the stabilizing effect induced by the orientational stiffness opposes the entropy-driven tendency to \emph{crumple}, causing a transition between a flat and crumpled phase at $T>0$. In this review we focus on the ground state properties of ordered structures on curved surfaces and we refer the reader to the literature for a discussion of finite temperature physics \cite{MembranesAndSurfaces}. 

In this section we will show how an underlying non-zero Gaussian curvature couples with the defects by \emph{screening} their topological charge. As a result, topological defects, whose existence would be energetically suppressed if the same system was lying on a flat substrate, instead proliferate. The universality of such a curvature screening mechanism is remarkable in the sense that it occurs in a conceptually identical fashion in both $p-$atics and solids, although the kernel of the elastic interactions between defects differs. As a consequence of curvature screening topological defects organize themselves on a rigid surface so as to match the Gaussian curvature of the substrate. On the other hand, if the geometry of the substrate is allowed to change (for instance by lowering the bending rigidity) the system eventually enters a regime where the substrate itself changes shape in order to accommodate some preferential in-plane order. The latter is the fundamental mechanism behind the \emph{buckling} of crystalline membranes \cite{SeungNelson:1988} and is believed to be the origin of the polyhedral geometry of large spherical viruses such as bacteriophages \cite{LidmarMirnyNelson:2003}.
 
In the case of manifolds equipped with a pure rotational degree of freedom like $p-$atics, the curvature affects the elastic energy through the connection which determine how vectors change when parallel transported. This statement can be clarified by rewriting the elastic energy \eqref{eq:sec2-xy-energy} in the form:
\begin{equation}\label{eq:sec2-xy-curved1}
F_{el} = \frac{1}{2}K_{A}\int d^{2}x\,\nabla_{i} m^{j}\nabla^{i}m_{j} \ ,
\end{equation}
where $\bm{m}$ is a $p-$atic director field which can be conveniently expressed in a local orthonormal frame $\bm{e}_{\alpha}$ ($\alpha=1\,2$):
\[
\bm{m} = \cos\theta\,\bm{e}_{1}+\sin\theta\,\bm{e}_{2}\,,
\]
with $\theta$ defined modulo $2\pi/p$. Since $\partial_{i}m_{\alpha}=-\epsilon_{\alpha\beta}m_{\beta}\partial_{i}\theta$, using the properties of vielbein outlined in Sec. \ref{sec:2b}, we have:
\begin{equation}
\nabla_{i}m_{j} 
= (e_{\alpha})_{j}(\partial_{i}m_{\alpha}+\Omega_{i\alpha\beta}m_{\beta})
= -\epsilon_{\alpha\beta}m_{\beta}(e_{\alpha})_{j}(\partial_{i}\theta-\Omega_{i})
\end{equation}
The elastic energy \eqref{eq:sec2-xy-curved1} thus becomes
\begin{equation}\label{eq:sec2-xy-curved2}
F_{el} = \frac{1}{2}K_{A}\int d^{2}x\,g^{ij}(\partial_{i}\theta-\Omega_{i})(\partial_{j}\theta-\Omega_{j}) \ .
\end{equation}
Using Eq. \eqref{eq:sec2-spin-curl}, the vector field $\bm{\Omega}$ can be expressed as:
\begin{equation}\label{eq:sec2-omega_i}
\Omega_{i} = - \epsilon_{i}^{j}\nabla_{j} \int d^{2}x\,G_{L}(\bm{x},\bm{y})K(\bm{y})\,,
\end{equation}
where $G_{L}(\bm{x},\bm{y})$ is again the Green function of the Laplace-Beltrami operator \eqref{eq:sec2-laplace-beltrami}. Substituting Eqns. \eqref{eq:sec2-omega_i} and \eqref{eq:sec2-v} in the expression for the elastic energy \eqref{eq:sec2-xy-curved2} we obtain:
\begin{equation}\label{eq:sec2-xy-curved3}
F_{el} = -\frac{1}{2}K_{A}\int d^{2}x\,d^{2}y\,G_{L}(\bm{x},\bm{y})[\eta(\bm{x})-K(\bm{x})][\eta(\bm{y})-K(\bm{y})]\,.
\end{equation}
As anticipated, the Gaussian curvature of the underlying substrate couples with the defects by screening their topological charge. Eq \eqref{eq:sec2-xy-curved3} is identical to the Coulomb energy of a multi-component plasma of charge density $\eta$ in a background of charge density $-K$. Like particles in a plasma, we can expect defects to screen each other's charge, thus forming clusters of zero net charge, and matching the charge distribution of the surrounding background. This implies that disclinations will be attracted to regions of like-sign Gaussian curvature. 

\begin{figure}
\centering
\includegraphics[width=0.3\textwidth]{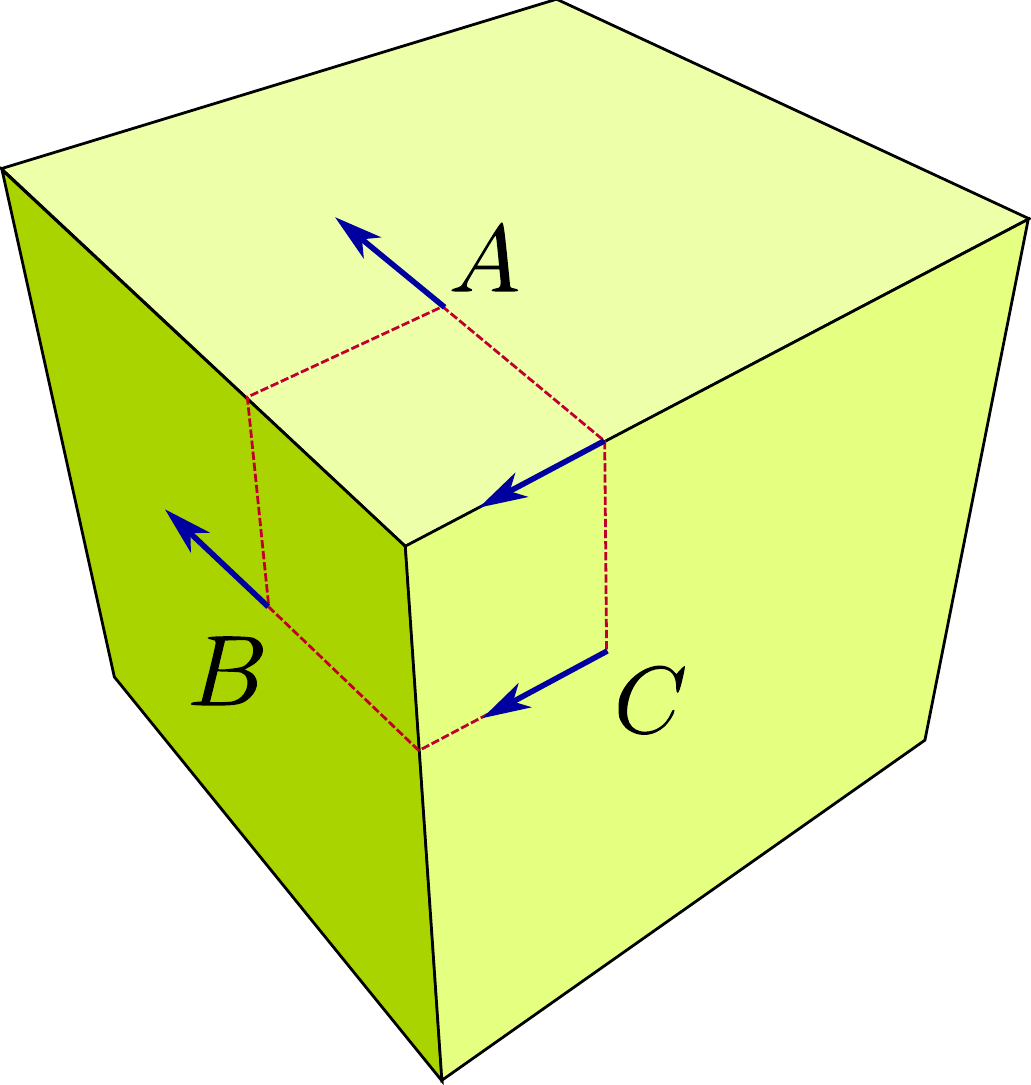}
\caption{\label{fig:sec2-parallel-transport}(Color online) Parallel transport of a vector around a corner of a cube. As a consequence of the discrete Gaussian curvature, the vector is rotated by $\pi/2$ after parallel transport.}
\end{figure}

Crystalline surfaces (i.e. non-Euclidean crystals) differ from manifolds equip\-ped with a $p-$atic director field because bonds, whose local orientation is encoded in the field $\theta$ in $p-$atics, can now be compressed and sheared to relieve part of the elastic stress due to the geometrical frustration provided by the embedding manifold. A first attempt at describing the elasticity of defects on a two-dimensional curved crystal was made by Dodgson \cite{Dodgson:1996}, who studied the ground state of the Abrikosov fulx lattice in a model thin-film superconductor on a sphere (subject to a field radiating from a magnetic monopole at the center) and found evidence for twelve $5-$fold disclinations at the vertices of an icosahedron in a otherwise $6-$fold coordinated environment (see Sec. \ref{sec:3} for a detailed review of spherical crystals). Later, Dodgson and Moore \cite{DodgsonMoore:1997} proposed adding dislocations to the ground state of a sufficiently large spherical vortex crystal to screen out the strain introduced by the twelve, topologically required, disclinations. A general framework to describe the elasticity of defects in non-Euclidean crystals was proposed by Bowick, Nelson and Travesset (BNT) in 2000 \cite{BowickNelsonTravesset:2000}. The BNT model, obtained from the covariantization of the elastic energy \eqref{eq:sec2-crystals-energy}, relies on the following expression for the elastic energy of a collection of disclinations in a triangular lattice of underlying Gaussian curvature $K$:
\begin{equation}\label{eq:sec2-bnt}
F_{el} = \frac{1}{2}Y\int d^{2}x\,d^{2}y\,G_{2L}(\bm{x},\bm{y})[\eta(\bm{x})-K(\bm{x})][\eta(\bm{y})-K(\bm{y})]\,,
\end{equation}
where $\eta(\bm{x})$ is the topological charge density \eqref{eq:sec2-density} and $G_{2L}(\bm{x},\bm{y})$ the Green function of the covariant biharmonic operator on the manifold. The origin of the coupling between topological charge and curvature, in this case, is rooted in a profound result of discrete geometry originally due to Descartes which can be considered the oldest ancestor of the Gauss-Bonnet theorem of differential geometry. Let $P$ be a convex polyhedron and define the \emph{angular deficit} of a vertex $v$ of $P$ as:
\begin{equation}\label{eq:sec2-angular-deficit}
k(v)=2\pi-\sum_{i=1}^{c(v)}\alpha_{i}(v)\,,
\end{equation}
where $\alpha_{i}(v)$ with $i\in[1,c(v)]$ are the angles formed by all the faces meeting at $v$. Descartes' theorem states that the sum of the angular deficits of a convex polyhedron is equal to $4\pi$. More generally:
\begin{equation}\label{eq:sec2-descartes-theorem}
\sum_{v\in P} k(v) = 2\pi\chi \ ,
\end{equation}
which is exactly the Gauss-Bonnet theorem for the case in which the Gaussian curvature is concentrated in a finite number of points (vertices) rather than smoothly distributed across the whole surface. The deficit angle $k(v)$ is thus the discrete analog of the Gaussian curvature. This analogy is not limited exclusively to Eq. \eqref{eq:sec2-angular-deficit}. Consider, for example, the corner of a cube and imagine parallel transporting a vector $\bm{v}$ around a closed loop encircling the corner (see Fig. \ref{fig:sec2-parallel-transport}). After parallel transport, the vector has rotated by $k(v)=\pi/2$ with respect to its original orientation. Analogously, a vector parallel transported around a closed loop on a surface rotates by an angle
\[
\delta = \int d^{2}x\,K(x) \ ,
\]
where the integral is over the portion of the surface enclosed by the loop. Now, on a triangulated surface $\alpha(v)\approx \pi/3$ and:
\[
k(v) = 2\pi-\frac{\pi}{3}\,c(v) = \frac{\pi}{3}[6-c(v)]\,.
\]
Thus the source term $\eta-K$ figuring in Eq. \eqref{eq:sec2-bnt} corresponds to the difference between the pre-existing Gaussian curvature of the manifold with local $6-$fold orientational order and the additional Gaussian curvature induced by a disclination of topological charge $q$. A more formal discussion can be found in Ref. \cite{BowickTravesset:2001}. As in the case of a purely rotational degree of freedom, disclinations in non-Euclidean crystals organize so as to match the Gaussian curvature of the underlying medium while maximizing their reciprocal distance as a consequence of the repulsive interaction between like-sign defects. In the next three sections we will see how the formalism outlined here applies to three physically relevant examples of manifolds: the sphere, the topological disk and the torus.

\section{\label{sec:3}Crystalline and nematic order on the sphere}

\subsection{\label{sec:3a}Introduction}

Experimental realizations of spherical crystals are found in emulsions of two immiscible fluids, such as oil and water, stabilized against droplet coalescence by coating droplets of one phase (say water) with small colloidal particles \cite{Pickering:1907,SacannaEtAl:2007}. The colloidal ``armor plating" of these Pickering emulsions also plays a role in colloidosomes, colloid-coated lipid bilayer vesicles used for encapsulation and delivery of flavors, fragrances and drugs \cite{DinsmoreEtAl:2002}. Identical micron-sized particles tend to crystallize under typical experimental conditions. The defect structure of crystalline ground states on a sphere is important in this context since defects influence the strength of the colloidal armor.
 
Defects play an essential role in describing crystalline particle packings on the sphere. At least twelve particles with 5-fold coordination (i.e., 12 disclinations) are required for topological reasons, and like the 5-fold rings in carbon fullerenes, one might expect that the energy would be minimized if the disclination positions approximated the vertices of a regular icosahedron. This expectation, which also plays a role in geodesic domes and in the protein capsomere configurations of spherical virus shells \cite{CasparKlug:1962,ZandiEtAl:2004}, is nevertheless \emph{violated} when the shells are sufficiently large and disclination buckling \cite{LindmarEtAl:2003} out of the spherical environment is suppressed by surface tension.

To understand the structure of spherical crystals, we must find the ground state of $M$ particles distributed on a surface $\Sigma$ and interacting with, say, pairwise repulsive potentials. For simplicity consider power law potentials $V(r)=e^2/r^{\gamma}$, where $e$ is a generalized electric charge. The energy of $M$ particles at positions $\bm{r}(\bm{\ell})$, with $\bm{\ell}=(l_1,l_2) \in \mathbb {Z} \otimes \mathbb{Z}$, becomes $2E_0=\sum_{\bm{\ell} \neq \bm{\ell}^{\prime}} e^2/|\bm{r}(\bm{\ell})-\bm{r}(\bm{\ell}^{\prime})|^{\gamma}$. This covers most of the interesting physical systems. The case $\gamma=1$ corresponds to the pure 3D Coulomb potential, appropriate for charged colloids. This is known as the Thomson problem in the physics and mathematics literature and has been widely studied for over 100 years \cite{Thomson:1904,SaffKuijlaars:1997,HardinSaff:2005}. The case $\gamma=3$ corresponds to a dipole interaction, appropriate for neutral colloids at the interface between two liquids. The different dielectric constants of the two liquids leads to an asymmetric distribution of charge on the colloids and a net dipole moment. The case $\gamma=12$ is the repulsive part of the Lennard-Jones potential and is the important piece of the interaction for driving crystallization. Finally the limit $\gamma \rightarrow \infty$ is an extreme short range interaction that selects the pair of particles with the minimum distance. The equivalent packing of spherical caps was introduced in 1930 by the biologist Tammes \cite{Tammes:1930} and connects the present problem to the rich physics and mathematics of packing \cite{AsteWeaire,ConwaySloane}.

The packing of particles on surfaces of nontrivial topology is relevant to a broad range of physical, chemical and biological systems in addition to its intrinsic mathematical interest. An almost literal realization of the Thomson problem is provided by multi-electron bubbles \cite{AlbrechtLeiderer:1992,Leiderer:1995}. Electrons trapped on the surface of liquid helium have long been used to investigate two dimensional melting \cite{GrimesAdams:1979,GlattliAndreiWilliams:1998}. Multi-electron bubbles result when a large number of electrons ($10^5{-}10^7$) at the helium interface subduct in response to an increase in the anode potential and coat the inside wall of a helium vapor sphere of radius $10{-}100 \mu\rm{m}$. Typical electron spacings, both at the interface and on the sphere, are of order 2000 $\AA$, so the physics is entirely classical, in contrast to the quantum problem of electron shells which originally motivated J.J. Thomson \cite{Thomson:1904}. Information about electron configurations on these bubbles can, in principle, be inferred from studying capillary wave excitations \cite{LenzNelson:2001,LenzNelson:2003}. Similar electron configurations should arise on the surface of liquid metal drops confined in Paul traps \cite{Davis:1997}.

A Thomson-like problem also arises in determining the arrangements of the protein subunits which comprise the shells of spherical viruses \cite{CasparKlug:1962,MarzecDay:1993,LindmarEtAl:2003,ZandiEtAl:2004,NguyenBruinsmaGelbart:2005}. Here, the ``particles" are clusters of protein subunits arranged on a shell. This problem of protein arrangements was solved in a beautiful paper by Caspar and Klug \cite{CasparKlug:1962} for intermediate values of $R/a$, where $R$ is the sphere radius and $a$ is the mean particle spacing. Caspar and Klug constructed icosadeltahedral particle packings characterized by integers $m$ and $n$, which provide regular tessellations of magic numbers $M=10(n^2+nm+m^2)+2$ particles on the sphere. The Caspar-Klug tessellations provide an excellent starting point for finding low energy particle configurations on the sphere for intermediate values of $M \approx 8\pi(R/a)^2/\sqrt{3}$. Particle numbers $M$ that are not magic numbers can be accommodated by introducing vacancies or interstitials into these regular packings.

Other realizations of Thomson-like problems include regular arrangements of colloidal particles in colloidosome cages \cite{BauschEtAl:2003,LipowskyEtAl:2005,EinertEtAl:2005,Manoharan:2006} proposed for protection of cells or drug-containing vesicles \cite{DinsmoreEtAl:2002} and fullerene patterns of spherical carbon atoms \cite{KrotoEtAl:1985}. An example with long range (logarithmic) interactions is provided by the Abrikosov lattice of vortices which would form at low temperatures in a superconducting metal shell with a large monopole (solenoid tip) at the center \cite{DodgsonMoore:1997}.

The problem of the best possible packing on spheres also has applications in the micro-patterning of spherical particles \cite{MasudaItohKoumoto:2005}, relevant for photonic crystals, and in understanding the structure of Clathrin cages, responsible for the vesicular transport of cargo in cells \cite{Alberts}. Crystalline domains covering a fraction of the sphere are also of experimental interest. In the context of lipid rafts \cite{SimonsVaz:2004}, confocal fluorescence microscopy studies have revealed the coexistence of fluid and solid domains on giant unilamellar vesicles made of lipid mixtures. The shapes of these solid domains include stripes of various widths and orientations \cite{KorlachEtAl:1999,FeigensonBuboltz:2001,ScherfeldKahyaSchwille:2003,SchneiderGompper:2005}.

\subsection{\label{sec:3b}Crystals of point particles}

Consider a collection of classical point particles constrained to a frozen (non-dynamical) two-dimensional surface $\Sigma$ embedded in three-dimensional Euclidean space. The particles interact through a general potential defined in the three dimensional embedding space or solely within the 2D curved surface itself. We focus
primarily on the potential \be\label{Part_Potential} V(r)=\frac{e^2}{r^{\gamma}} \ . \ee Here, $e$ is an ``electric charge" such that if $r$ is some quantity with dimensions of length then $e^2/r^{\gamma}$ has dimension of energy. The case $\gamma=1$ corresponds to the Coulomb potential in three dimensions. Although we do not treat this problem explicitly here, the replacement 
\be\label{Part_Potential_log} 
V(r) \rightarrow \frac{e^2}{\gamma}(r^{-\gamma}-1) 
\ee
allows us to treat the two dimensional Coulomb potential by taking the limit $\gamma \rightarrow 0$, 
\be\label{twoD_Potential_log} 
V(r) \rightarrow -e^2\log(r) \ . 
\ee
The electrostatic energy of a system of $M$ particles at positions $r({\bm{\ell}})$, interacting via Eq.(\ref{Part_Potential}), with ${\bm{\ell}}=(l_1,l_2)$, $l_1,l_2 \in {\mathbb{Z}}$, becomes
\be\label{Coul_gamma_Energy} 
2E_0=\sum_{\bm{\ell} \neq {\bm{\ell}^{\prime}}}^M \frac{e^2}{|\bm{r}(\bm{\ell})-\bm{r}(\bm{\ell}^{\prime})|^{\gamma}} \ .
\ee 
Note that with this definition the power law interaction acts across a chord of the sphere, as would be the case for electron bubbles in helium. Although we focus here on curved crystals, there are some quantities which are insensitive to the curvature of the surface, and the simpler geometry of the plane can be used to compute them. The following two subsections hence focus on planar crystalline arrays of particles interacting via the potential Eq.(\ref{Part_Potential}).

\subsubsection{\label{sec:3c}Planar Crystals}

The electrostatic energy Eq. \eqref{Coul_gamma_Energy} and the corresponding elastic tensor, from which follows the elastic constants of the system, may be explicitly computed for crystalline orderings of particles in a triangular lattice. For any non-compact surface, like the plane, the energy Eq. (\ref{Coul_gamma_Energy}) is divergent for all $\gamma \leq 2$. If $\gamma > 0$, the divergence comes exclusively from the zero mode $\bm{G}=0$ associated with the thermodynamic limit of infinite system size. This term (which would be subtracted off if a uniform background charge were present) can be isolated by setting $|\bm{G}| \equiv \varepsilon \ll 1$ for this mode. The final result for the ground state energy reads 
\bea\label{Energy_Total}
2E_0&=&-\frac{Me^2}{\Gamma(\gamma/2)}\left(\frac{\pi}{A_C}\right)^{\frac{\gamma}{2}}
\left[\frac{4}{\gamma(2-\gamma)}-\sigma(\gamma)\right]\nonumber\\&+& M e^2 \frac{\pi}{A_C}
\frac{\Gamma(1-\frac{\gamma}{2})}{\gamma/2} \lim_{{|\bm{G}|}\rightarrow \vap} \frac{2^{2-\gamma}}{|\bm{G}|^{2-\gamma}}
\nonumber\\
&=& M e^2 \theta(\gamma)
{\left(\frac{4\pi}{A_C}\right)}^{\gamma/2}+M e^2
\frac{\pi}{A_C} \frac{\Gamma(1-\frac{\gamma}{2})}{\gamma/2}
\lim_{|\bm{G}|\rightarrow \vap} \frac{2^{2-\gamma}}{|\bm{G}|^{2-\gamma}} \ . 
\eea 
$A_C$ is the area of the unit cell of the triangular Bravais lattice ($A_C=\frac{\sqrt{3}}{2} a^2$) and $\Gamma$ is the Euler Gamma function. The coefficient $\sigma$ is a
sum over Misra functions. The coefficient $\theta(\gamma)$ parameterizes the nonsingular part of the energy; its dependence on the exponent $\gamma$ is shown in Table~\ref{Tab__NumCoeff}. This negative quantity parameterizes the binding energy of the triangular lattice after the positive ``zero mode" contribution is subtracted off. For $\gamma=1$, we have a two dimensional ``jellium" model.  In the case of a sphere no neutralizing background is necessary and the energy is rendered finite by the compact nature of the sphere itself. The maximum distance between points on the sphere provides a natural infrared cut-off.

For small displacements of the particles from their equilibrium positions, one has 
\begin{equation}\label{Diff_Energy}
E-E_{0} = \frac{e^{2}}{2}\sum_{\bm{\ell}\ne\bm{\ell}^{\prime}}
\left[\frac{1}{|\bm{r}(\ell)+\bm{u}(\bm{\ell})-\bm{r}(\bm{\ell}^{\prime})-\bm{u}(\bm{\ell}^{\prime})|^{\gamma}}-
\frac{1}{|\bm{r}(\ell)-\bm{r}(\bm{\ell}^{\prime})|^{\gamma}}\right]
\end{equation}
where $\bm{u}(\bm{\ell})$ is a small displacement of the particle $\bm{\ell}$ in the plane of the surface from its equilibrium position $\bm{r}(\bm{\ell})$, and therefore a tangent vector to the surface. The elastic tensor $\Pi_{ij}(\bm{\ell},\bm{\ell}^{\prime})$ is defined as the leading term in an expansion of Eq.(\ref{Diff_Energy}),
\be\label{Phonon_Energy}
E-E_0=\frac{e^2}{2}\sum_{\bm{\ell}\ne\bm{\ell}^{\prime}} 
\Pi_{ij}(\bm{\ell},\bm{\ell}^{\prime}) u_{i}(\bm{\ell}) u_{j}(\bm{\ell}^{\prime}) \ . 
\ee 
In deriving Eq.(\ref{Diff_Energy}), we assume a constraint of fixed area per particle, enforced by a uniform background charge density or boundary conditions. This eliminates the term linear in $u_{i}(\bm{\ell})$.  The final result is 
\bea\label{Pi_Momentum} 
\Pi_{ij}(\bm{p})&=&A_C\sum_{\bm{\ell}} e^{i\bm{p} \cdot \bm{r}(\bm{\ell})}
\Pi_{ij}(\bm{\ell},\bm{0})\nonumber
\\\nonumber
&=&\frac{2^{2-\gamma}\pi}{A_C}\frac{\Gamma(1-\gamma/2)}{\Gamma(\gamma/2)}
\frac{p_{i} p_{j}}{|\bm{p}|^{2-\gamma}}+
\\\nonumber
&+&\frac{\eta(\gamma)}{A_C^{\gamma/2}} \left[|\bm{p}|^2\delta_{ij}+\rho(\gamma)(\delta_{k i}
\delta_{l j}+\delta_{k j}\delta_{l
i})p_{k}p_{l}\right]\nonumber\\
&+&{\cal O}(|\bm{p}|^4) . 
\eea 
The coefficients $\eta(\gamma)$ and $\rho(\gamma)$ depend only on the potential. In Table~\ref{Tab__NumCoeff}, some values of the coefficients for a range of potentials with $0< \gamma < 2$ are listed.

\begin{table}
\tbl{Coefficients of the response function
Eq. \eqref{Pi_Momentum} and the energy Eq. \eqref{Energy_Total}.
Results are accurate up to six digits. The coefficient $\rho$ is a
rational function of $\gamma$.}
{\begin{tabular}{llll}
\toprule
$\gamma$ & $\eta$ & $\rho$ & $-\theta$ \\
\colrule
$1.875$ & $0.699652$ & $31$ & $47.763$ \\
$1.75$ & $0.619256$ & $15$ & $22.647$ \\
$1.625$ & $0.544152$ & $87/9$ & $14.288$ \\
$1.5$ & $0.474268$ & $7$ & $10.118$ \\
$1.375$ & $0.409548$ & $27/5$ & $7.625 $ \\
$1.25$ & $0.349812$ & $13/3$ & $5.9701$ \\
$1.125$ & $0.295033$ & $25/7$ & $4.7955$ \\
$1$ & $0.245065$ & $3$ & $3.9210$ \\
$0.875$ & $0.199772$ & $23/9$ & $3.2471$ \\
$0.75$ & $0.159010$ & $11/5$ & $2.7138$ \\
$0.625$ & $0.122622$ & $21/11$ & $2.283$ \\
$0.5$ & $0.090439$ & $5/3$ & $1.9294$ \\
$0.375$ & $0.062279$ & $1.46154$ & $1.6352$ \\
$0.25$ & $0.037955$ & $9/7$ & $1.3881$ \\
$0.125$ & $0.017265$ & $43/30$ & $1.1787$ \\
\botrule
\end{tabular}}
\label{Tab__NumCoeff}
\end{table}

\subsubsection{\label{sec:3d}Continuum free energy}

When the deviations from the ground state are small, the long wavelength lattice deformations may be described by a continuous Landau elastic energy of the form \eqref{eq:sec2-landau-energy}. The elastic tensor Eq. \eqref{Phonon_Energy}, within Landau elastic theory, is then
\begin{equation}\label{Pi_fromGoldstone}
e^2 \Pi_{ij}(\bm{p})=A_C\left[\mu |\bm{p}|^2\delta_{ij}+(\lambda+\mu)p_{i}p_{j}\right]\ . 
\end{equation}
A comparison with Eq. \eqref{Pi_Momentum} immediately yields an explicit expression for the elastic constants of the crystal 
\bea\label{elastic_const}
\mu&=&\eta(\gamma)\frac{e^2}{A_C^{1+\gamma/2}} \ \ , \ \
\lambda=\infty
\\
Y&=&\frac{4 \mu (\lambda+\mu)}{2\mu+\lambda}=
4\eta(\gamma)\frac{e^2}{A_C^{1+\gamma/2}} \ , 
\eea 
where $Y$ is Young's modulus. The result $\lambda=\infty$ is equivalent to a divergent compressional sound velocity as $p\rightarrow 0$ and for $\gamma=1$ is just a statement of the incompressibility of a two-dimensional Wigner crystal. Alternatively, we can allow for wavevector-dependent elastic constants $\mu(p)$ and $\lambda(p)$ in Eq. \eqref{Pi_fromGoldstone}. In this case $\lambda(p)$ diverges as $p \rightarrow 0$: 
\begin{equation}
\lambda(p) \approx \frac{2^{2 - \gamma}\pi}{A_C} \frac{\Gamma(1 - \gamma/2)}{\Gamma(\gamma/2)} \frac{1}{p^{2 - \gamma}}\,,
\end{equation}
while $\lim_{p \rightarrow 0} \mu(p)$ is given by Eq. \eqref{elastic_const}.

\subsubsection{\label{sec:3e}Spherical Crystals}

\begin{figure}[t]
\begin{center}
\includegraphics[width=3in]{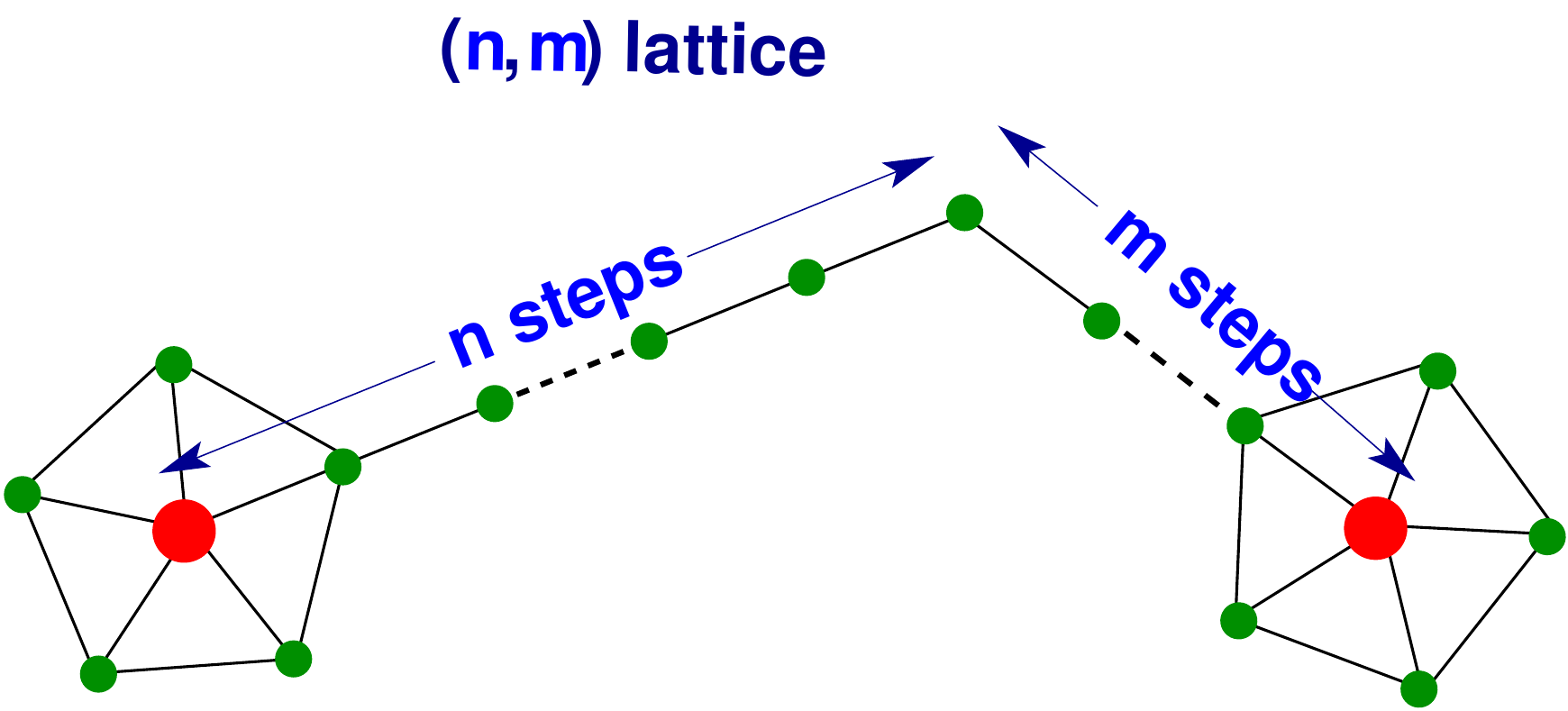}
\caption{(Color online) Construction of an $(n,m)$
icosadeltahedral lattice. The filled circles indicate two
nearest-neighbor five fold disclinations. Because these defects
sit on the vertices of an icosahedron, they are separated by a
geodesic distance $R \cos^{-1}(1/\sqrt{5})$, where $R$ is the
sphere radius.}\label{fig__cons_icos}.
\end{center}
\end{figure}

\begin{figure}[ctb]
\centering
\includegraphics[width=3.6in]{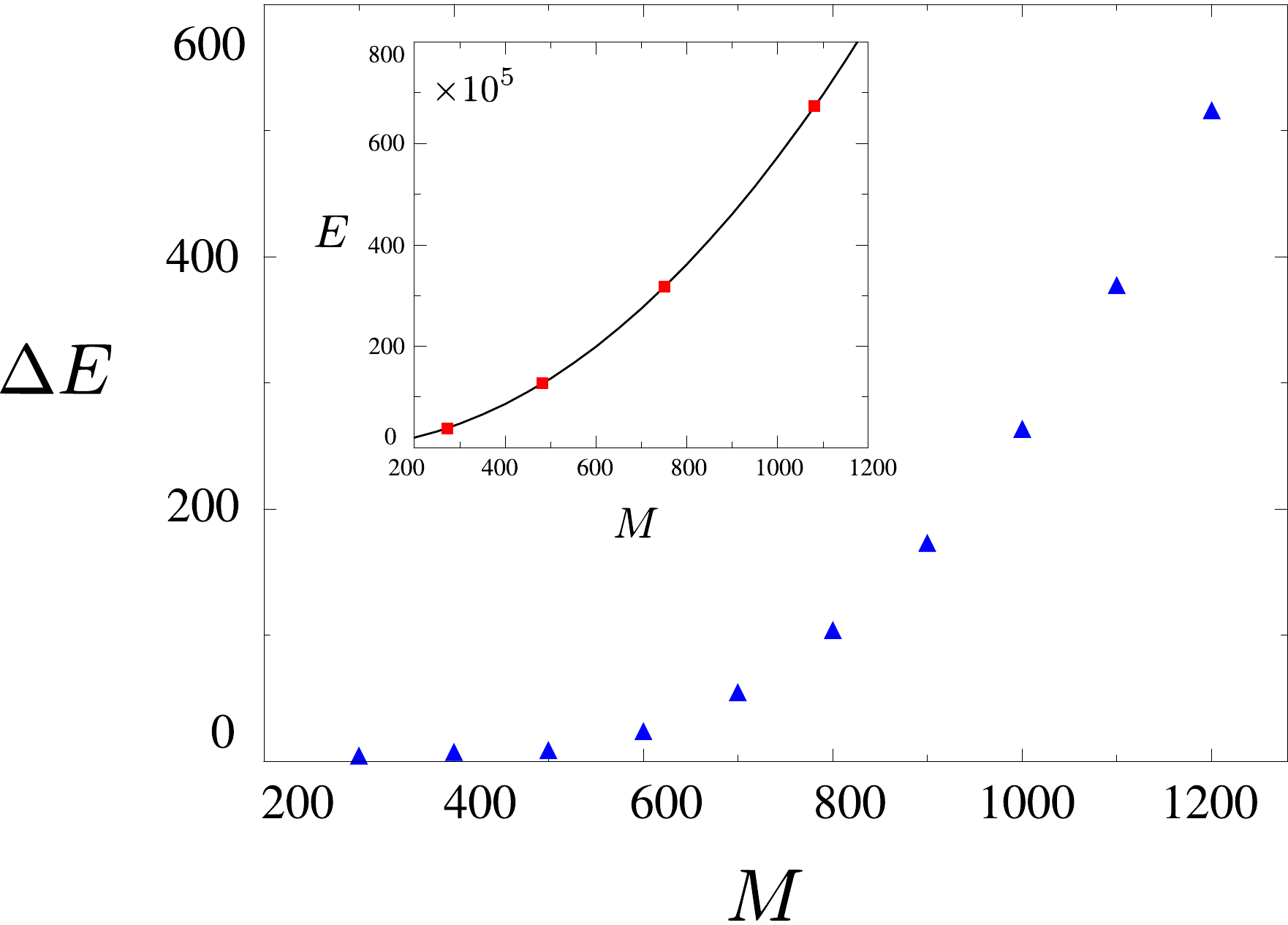}
\caption{(Color online) Difference in energy of $(n,n)$ and
$(n,0)$ configurations. Results are for a power law potential with $\gamma=1.5$ and
energies are plotted in units of $\frac{e^2}{R^{\gamma}}$.}
\label{fig__nn_versus_n0}
\end{figure}

Spherical crystals have many properties not shared by planar ones, one of the most remarkable being that there is an excess of twelve positive ($5-$fold) disclinations. These disclinations repel, and the simplest spherical crystals will be those having the minimum number of defects (i.e. twelve) located at the vertices of an icosahedron. Triangular lattices on the sphere with an icosahedral defect pattern are classified by a pair of integers $(n,m)$, as illustrated in Fig.~\ref{fig__cons_icos}. The path from one disclination to a
neighboring disclination for an $(n,m)$ icosadeltahedral lattice consists of $n$ straight steps, a subsequent $60^{\circ}$ turn, and $m$ final straight steps. The geodesic distance between nearest-neighbor disclinations on a sphere of radius $R$ is $d=R\cos^{-1}(1/\sqrt{5})$. The total number of particles $M$ on the sphere described by this $(n,m)$-lattice is $M=N_{nm}$, where
\be\label{nm_latt_num} 
N_{nm}=10(n^2+m^2+nm)+2 \ . 
\ee
Such $(n,m)$ configurations were once believed to be ground states for relatively small numbers ($M \leq 300$, say) of particles interacting through a Coulomb potential \cite{AltschulerEtAl:1994,AltschulerEtAl:1997,ErberHockney:1995,MorrisDeavenHo:1996,AltschulerGarrido:2005}. The energy of discrete particle arrays described by Eq. \eqref{Coul_gamma_Energy} can be evaluated by starting with some configuration close to an $(n,m)$ one and relaxing it to find a minimum. It is found that the $(n,m)$ configurations are always local minima. Whether these icosahedral configurations are global minima as well will be analyzed later. Results for the energy $E(M)$ are shown in the inset to Fig. \ref{fig__nn_versus_n0}.

\begin{figure}[t]
\centering
\includegraphics[width=0.3\textwidth]{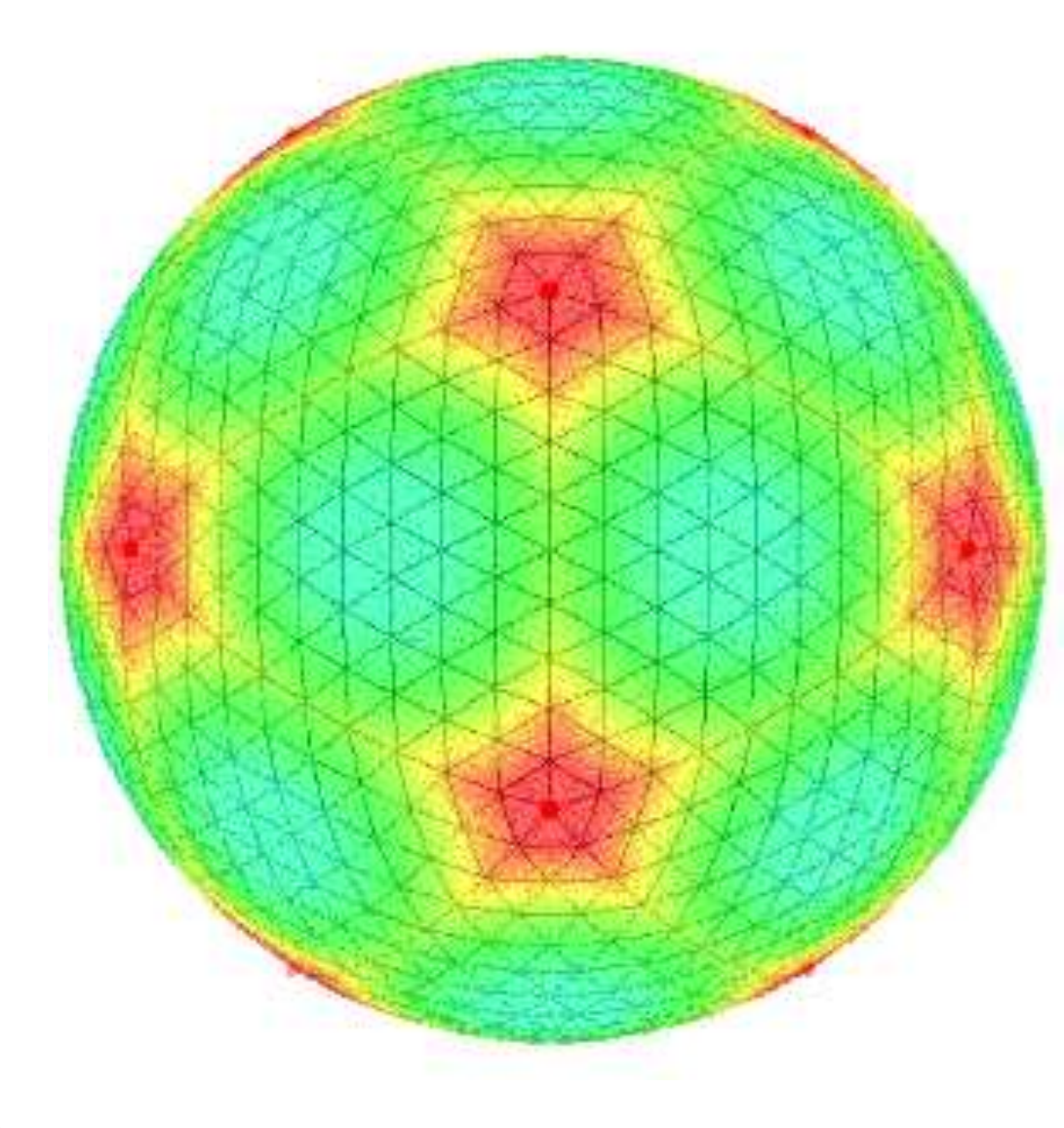}\qquad
\includegraphics[width=0.3\textwidth]{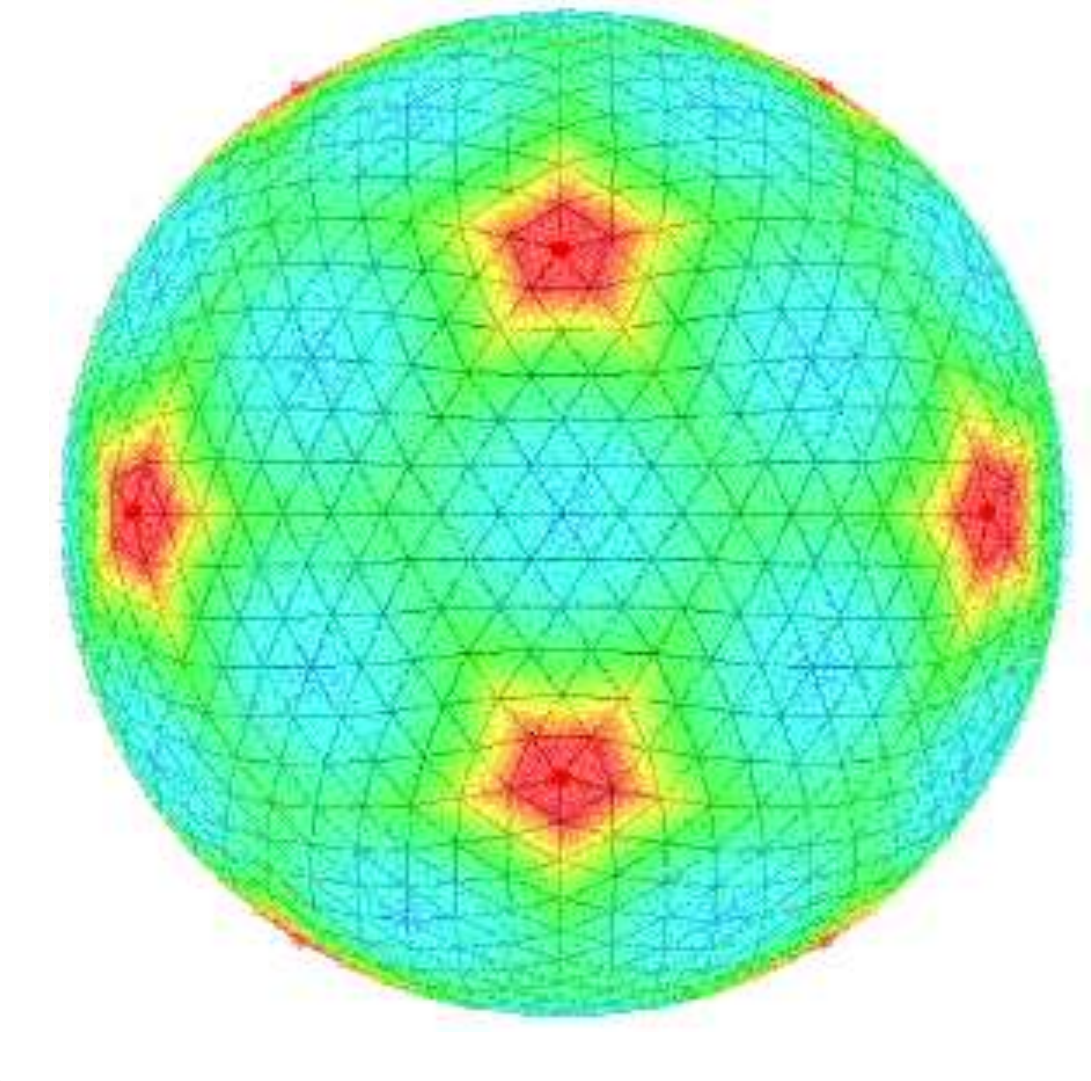}
\caption{(Color online) Potential energy distribution for a
$(n,0)$ configuration with n=10 and M=1002 (top) and a $(n,n)$
configuration with n=6 and M=1082 (bottom).} \label{fig__colornn}
\end{figure}

From Fig. \ref{fig__nn_versus_n0} it is clear that energies grow very fast for increasing volume. More interestingly, the $(n,0)$ and $(n,n)$ configurations show a growing difference in energy for increasing volume size, implying that the energy of icosahedral configurations does not tend to a universal value for large numbers of particles but rather remains sensitive to the $(n,m)$ configuration, a result also noted by other authors \cite{AltschulerEtAl:1997}. Further insight comes from investigating the distribution of energy. Plots of the local electrostatic energy, the electrostatic energy at point $\bm{x}$ on the sphere, as defined in Eq. \eqref{Coul_gamma_Energy} are shown in Fig. \ref{fig__colornn} \cite{Java}.
From Fig.~\ref{fig__colornn} it should be noted that the triangles obtained by the Voronoi-Delaunay construction, after minimization of the potential Eq. \eqref{Coul_gamma_Energy}, are very close to equilateral. The distribution of the local energies for the $(n,0)$ and $(n,n)$ configurations are very different. The $(n,0)$ configuration shows maximum energies along the paths joining the defects. The $(n,n)$ configuration, on the other hand, has its maximum energies along the directions defined by the triangles formed by three nearest neighbor disclinations. The size of these regions of differing electrostatic energy turns out to scale with system size, making
it very plausible that there might be small differences in the energy per particle for $(n,0)$ and $(n,n)$ configurations in the limit $n \rightarrow \infty$. 

\subsection{\label{sec:3f}Geometric Formalism on the Sphere}

Spherical substrates provide the simplest example of the problem of crystals on curved surfaces. The study of spherical crystals is simplified by two important properties:
there is a unique scale with dimensions of length, the radius $R$, and there is a fixed excess disclinicity of twelve following from the Euler theorem \eqref{eq:sec2-euler}
\be\label{constr_sphere} 
Q = \sum_{i=1}^N q_i=6\chi=12 \ . 
\ee
The free energy Eq. \eqref{eq:sec2-bnt}, applied to the sphere, is tractable analytically because the inverse biharmonic operator on a sphere of radius $R$ can be computed explicitly. It is shown in \cite{BowickNelsonTravesset:2000} that the Green's function for the biharmonic, in spherical coordinates $(\theta,\phi)$, has the following simple form on a unit sphere:
\be\label{bi_harm_sol} 
\chi(\theta^a,\phi^a;\theta^b,\phi^b)= 1 +
\int^{\frac{1-cos\beta}{2}}_0 dz \, \frac{\log z}{1-z} \ , 
\ee 
where $\beta$ is the geodesic distance between two disclinations located at $(\theta^a,\phi^a)$ and $(\theta^b,\phi^b)$, 
\be\label{cosbeta}
\cos \beta=\cos \theta^a\cos\theta^b+\sin\theta^a\sin\theta^b
\cos(\phi^a-\phi^b) \ . 
\ee
The total energy of a spherical crystal with an arbitrary number of disclinations follows from Eq. \eqref{eq:sec2-bnt} and Eq. \eqref{bi_harm_sol} and has the simple form \cite{BowickNelsonTravesset:2000}
\be\label{energy_cosb} 2E(Y)=E_0 +
\frac{\pi Y}{36} R^2 \sum_{i=1}^N\sum_{j=1}^{N} q_i q_j
\chi(\theta^i,\phi^i;\theta^j,\phi^j)+N \, E_{c} \ , 
\ee 
where $\{\theta_i,\phi_i\}_{i=1,\cdots, N}$ are the coordinates of $N$ defects and we restrict ourselves to $5-$fold ($q_i=+1$)and $7-$fold ($q_i=-1$) defects. The quantity $E_0$ is the zero point energy and is defined in Eq. \eqref{Energy_Total}. Although $5$ and $7$-fold disclinations will in general have different core energies \cite{Nelson}, we assume equal core energies here for simplicity. What matters for our calculations in any case is the {\em dislocation} core energy $E_d$, which we take to be $E_d =
E_5 + E_7 \equiv 2E_{c}$.

The value of the Young's modulus and the flat space ground state energy $E_0$ have been computed in Sec. \ref{sec:3c}. When the sphere radius $R$ is large compared to the particle spacing $a$, we can use flat space values of $Y$ and the flat space energy $E_0(M)$ associated with a finite number of particles $M$. To obtain the leading terms in the expansion of the ground state energy for large but finite $M$, the precise compactification of the plane employed is irrelevant {--} it may be achieved by periodic boundary conditions, for example. For a sufficiently large plane the finite size effects will be negligible. The density $\sigma$ of particles is then $M$ divided by the total
surface of the compact plane, taken to be the surface area $S$ of a sphere of radius $R$, 
\be\label{Total_density}
\sigma=1/A_C=\frac{M}{S} \ , \ S=4\pi R^2 \ . 
\ee 
From Eq. \eqref{elastic_const} the expression for the Young's modulus suitable for $M$ particles on a spherical crystal of radius $R$ with $0<\gamma<2$ is then 
\be\label{Energy_final}
Y=4\mu=\frac{4\eta(\gamma) M^{1+\gamma/2}}{(4 \pi)^{1+\gamma/2}}
\frac{e^2}{R^{2+\gamma}} \ . 
\ee 
One remaining detail is the divergent contribution to the energy $E_0$ in Eq. \eqref{Energy_Total}. Since the divergent part comes solely from the zero mode, the spatial variations in the density of the actual distribution are irrelevant. It may therefore be computed for a uniform density of charges. The divergent part is identical
to the energy of a constant continuum of charges as described by the density Eq. \eqref{Total_density}. We now evaluate this divergent part of the energy on a sphere, instead of a plane.
\bea\label{Dens_Tot} E_D&\equiv&Me^2 \frac{\pi}{A_C}
\frac{\Gamma(1-\frac{\gamma}{2})}{\gamma/2}
\lim_{|\bm{G}|\rightarrow \vap}
\frac{2^{2-\gamma}}{|\bm{G}|^{2-\gamma}}
\nonumber\\
&\rightarrow& e^{2}\int d^{2}x\,d^{2}y\,\frac{\rho(\bm{x})\rho(\bm{y})}{|\bm{x}-\bm{y}|^{\gamma}}
\\\nonumber
&=&\frac{M^2}{2^{\gamma-1}(2-\gamma)} \frac{e^2}{R^{\gamma}} \ .
\eea 
The divergent part has thus been regularized, and the energy is finite and well-defined for all $M < \infty$. Note that for the case $\gamma<2$,  $E_D \sim M^{2-\gamma/2}(M/S)^{\gamma/2}$. Hence $E_D$ is not simply a function of the particle density $M/S$, as one would expect for a short range interaction.

\subsubsection{\label{sec:3g}The energy of spherical crystals}

Upon substituting the elastic constant of Eq. \eqref{Energy_final} into Eq. \eqref{energy_cosb}, one arrives at 
\bea\label{Energy_Y}
2E&=&E_0 + \frac{\pi Y}{36}R^2\sum_{i=1}^N \sum_{j=1}^N q_i q_j
\chi(\theta^i,\phi^i;\theta^j,\phi^j)+NE_{c}
\nonumber\\ &=&E_0+\frac{4
\eta(\gamma)}{(4\pi)^{1+\gamma/2}}\frac{\pi}{36} C(i_1\cdots i_N)
M^{1+\gamma/2} \frac{e^2}{R^{\gamma}}+NE_{c} \ ,
\eea 
where $E_0$ is defined in Eq. \eqref{Energy_Total} and the function $C(i_1 \cdots i_N)$ depends on the position $i_1=(\theta_1,\phi_1)$ etc. of the $N$ disclination charges and is universal with respect to the potential. The total energy of a spherical crystal, including the contributions to $E_0$ is then
\begin{align}\label{Energy}
2E_{\mathrm{tot}} 
&= \frac{e^{2}}{R^{\gamma}}\left\{\frac{M^{2}}{2^{\gamma-1}(2-\gamma)}\right.\nonumber\\
&+ \left.\left[\frac{\theta(\gamma)}{(4\pi)^{\gamma/2}}+\frac{\pi}{6}\frac{\eta(\gamma)}{(4\pi)^{\gamma/2+1}}\,C(i_{1},\ldots,i_{N})\right]M^{\gamma/2+1}\right\}+NE_{c}
\end{align}
Note that the leading correction to the zero mode energy proportional to $M^2$ varies as $M^{1 + \gamma/2}$, and depends both on the flat space function $\theta(\gamma)$ and on the $C$-coefficient
\be\label{C_coefficient}
C(i_1,\cdots,i_N)=\sum_{j=1}^N\sum_{i=1}^N q_i q_j
\chi(\theta^i,\psi^i;\theta^j,\psi^j) \ , 
\ee 
associated with a particular configuration of disclinations. Note that the core energies contribute to the second sub-leading coefficient. For short-range potentials, such as $\gamma > 2$, the ground energy is extensive, and the leading term varies as $M^{1 + \gamma/2}$. The extensive nature of the $M^{1 + \gamma/2}$ term becomes clear upon noting that 
\be\label{dim_inverse_Lapl} 
M^{1+\gamma/2}\frac{e^2}{R^{\gamma}} \propto R^2 \times
\frac{e^2}{a^{\gamma+2}} \ , 
\ee 
where $a$ is the particle spacing. Comparison with Eq. \eqref{energy_cosb} shows that the dimension of Young's modulus $Y$ arises solely from the lattice constant $a$ and
the electric charge $e$, consistent with elastic constants arising from physics on the scale of the lattice constant in an essentially flat geometry. This observation is now generalized to the rest of the couplings discussed in the previous section.

For a fluid membrane not on a frozen topography, Helfrich terms arising from the mean curvature $H$ as well as the Gaussian curvature can be important. Their scaling behavior is given by
\begin{equation}\label{dim_curv} 
\kappa_{g}\int d^{2}x\,K \sim \kappa \int d^{2}x\,H^{2} \sim \kappa \sim \frac{e^{2}}{R^{\gamma}}\,M^{\gamma/2}  \ .
\end{equation}
Both terms would therefore contribute to the same order in the $M$ expansion, although the last term is purely topological. For crystals embedded in a frozen topography we expect an expansion along the lines of Eq. \eqref{Energy}, 
\be\label{General_form_serie}
2E_{\mathrm{tot}}(M)=\left(a_0M^2 - \sum_{i=1} a_i
M^{\gamma/2+1-i}\right)\frac{e^2}{R^{\gamma}} \ . 
\ee 
The nonextensive term $a_0M^2$ arises from the long range interactions. The next extensive contribution comes from the interaction between Gaussian curvature and defects as well as the extensive energy per particle in flat space. Bending rigidity contributions are higher order in $1/M$ and can be absorbed into a redefinition of the disclination core energy. Core energies also depend on non-universal details of the short-distance physics. 

The results presented so far are strictly for systems at zero temperature. In systems with short range interactions, the elastic constants can be strongly temperature-dependent. An extreme example is hard disks of radius $a_0$, which may be viewed as a limiting case of a power law potential of the form $V(r) \simeq \epsilon_0
(a_0/r)^{\gamma}$, with $\gamma \rightarrow \infty$. In this case, the elastic constants are strictly proportional to temperature. It is straightforward, however, to adapt the techniques presented here to the simpler problem of short range pair potentials.

\begin{figure}[ctb]
\centering
\includegraphics[width=0.3\textwidth]{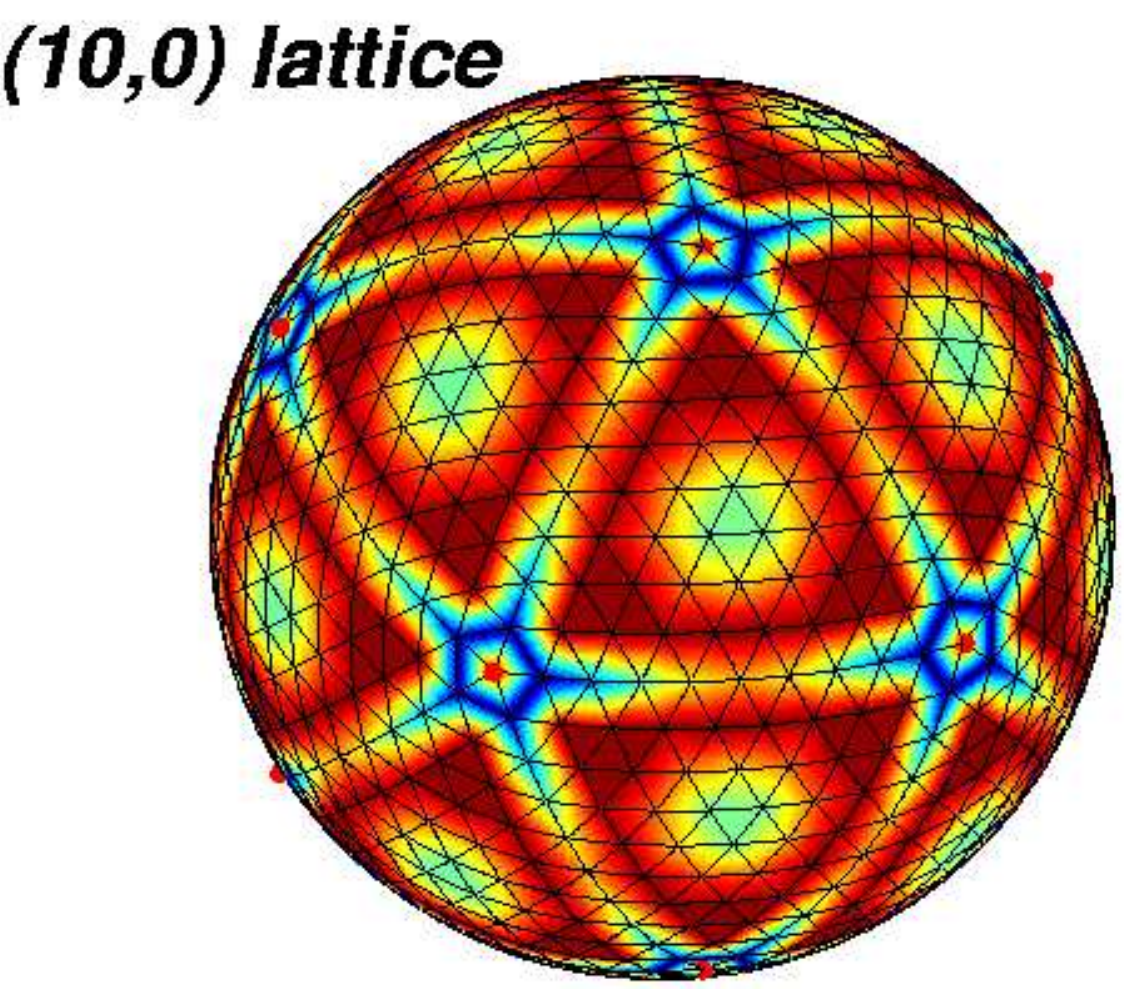}\qquad
\includegraphics[width=0.3\textwidth]{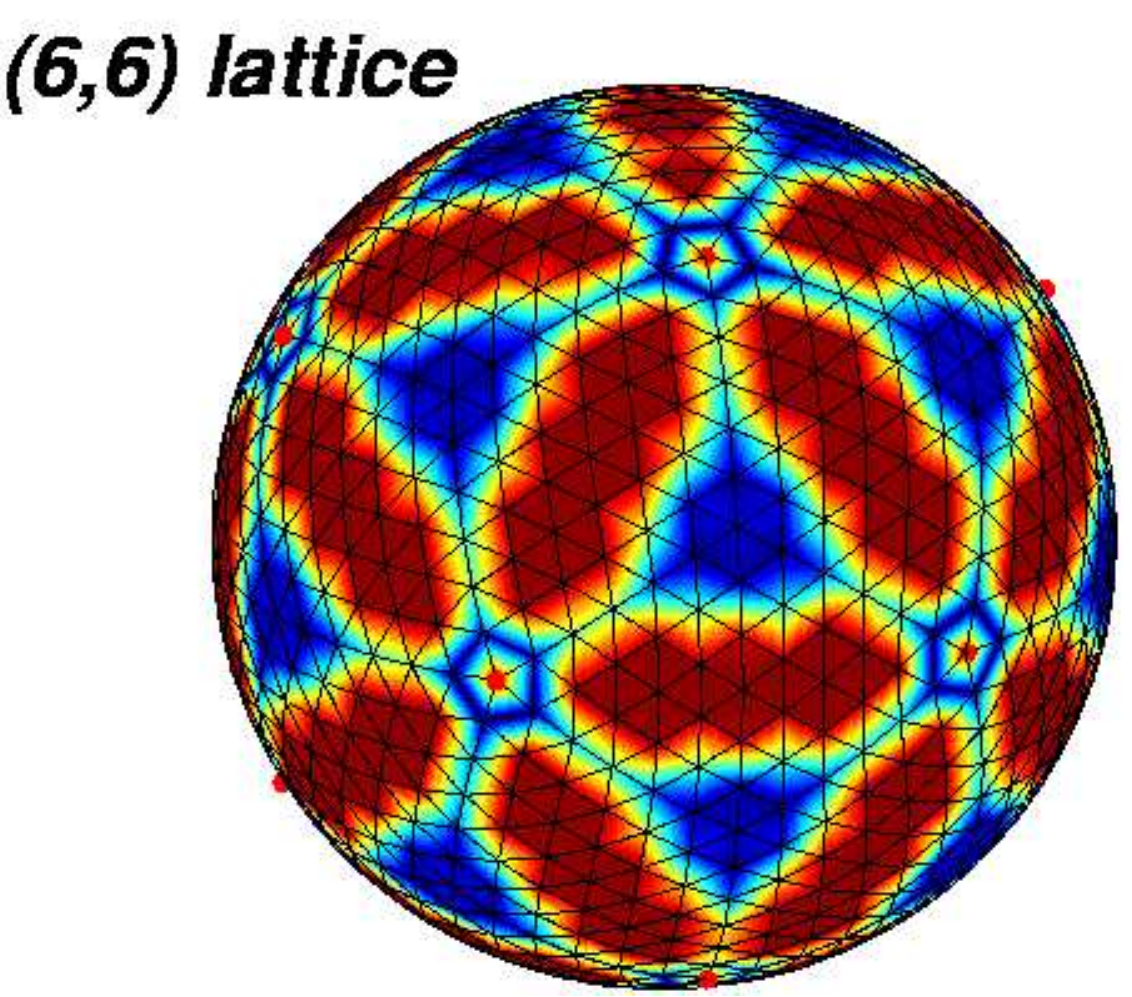}
\caption{(Color online) Illustration of the calculation discussed in the
text. The energy of the discrete $(n,n)$ configuration on the left
is extrapolated for large $M$ and compared to the energy computed
with the continuum model on the right. While in the continuum model
only twelve degrees of freedom (the 12 disclinations) need to be
considered, the direct calculation of a family of discrete models
requires the consideration of the full lattice and a careful
extrapolation of the energies to large $M$.} \label{fig__comparison}
\end{figure}

\subsubsection{\label{sec:3h}Energies of Icosahedral configurations}

The configuration on a sphere with the minimum number of charge $\pm 1$ defects is twelve $+1$ ($5-$valent) disclinations, which minimize their energy by sitting at the vertices of an icosahedron ${\cal Y}$. The energies of such configurations will be computed for the discrete spherical tessellations described in Sec. \ref{sec:3e} and compared with the predictions of continuum elastic theory, as illustrated in Fig. \ref{fig__comparison}. It is well established that for sufficiently large values of $M$ configurations with more than 12 disclinations (i.e., those with ``grain boundary scars") have lower energies \cite{GarridoDodgsonMoore:1997,BowickNelsonTravesset:2000}. It is of interest, however, to study simple icosahedral configurations for large $M$, as metastable states with a well defined energy.

\begin{figure}[ctb]
\includegraphics[width=0.9\textwidth]{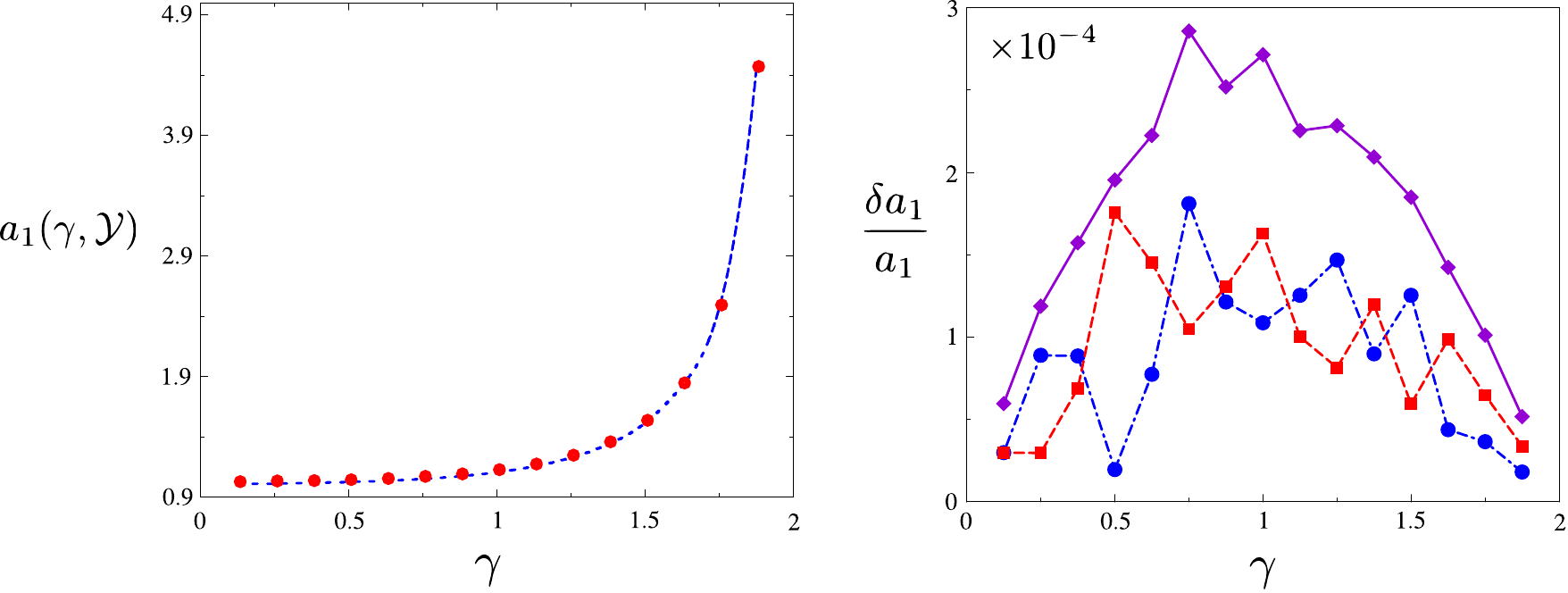} 
\caption{(Color online) Energy coefficient $a_1$ as a function of gamma (solid
line) and from the numerical results with $(n,m)$ configurations
(filled circles), for the icosahedral configurations. Plot of
$a_1(\gamma,{\cal Y})$ - $a_1(\gamma,{\cal Y})^{(n,n)}$ (circles),
$a_1(\gamma,{\cal Y})$ - $a_1(\gamma,{\cal Y})^{(n,0)}$
(diamonds), $a_1(\gamma,{\cal Y})^{(n,n)}$ - $a_1(\gamma,{\cal
Y})^{(n,0)}$ (squares).} \label{fig__gamma}
\end{figure}

\begin{figure}[ctb]
\centering
\includegraphics[width=3in]{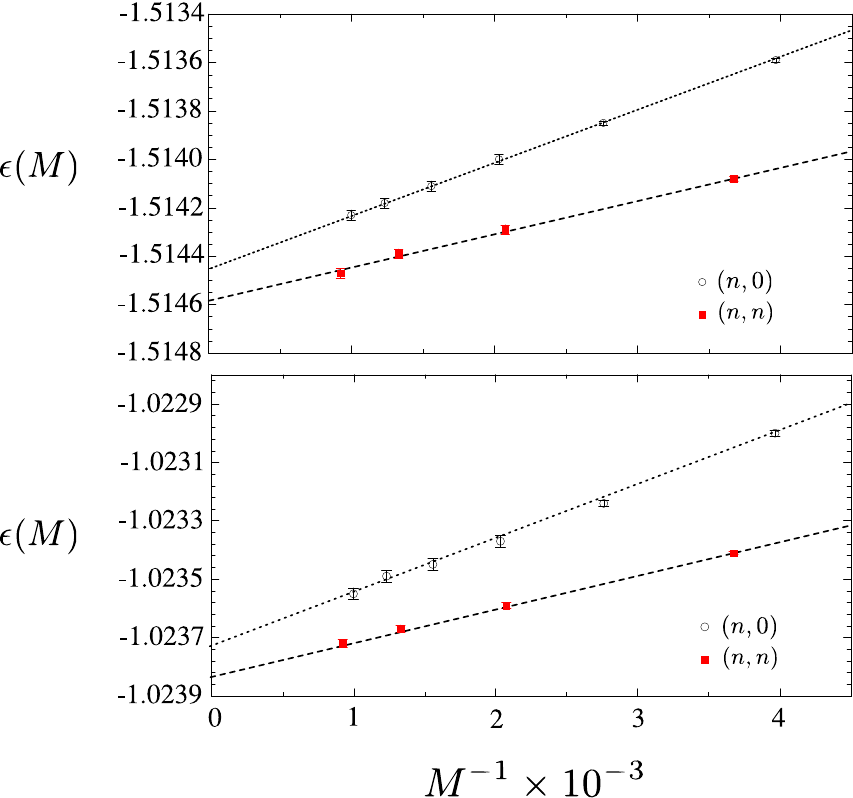}
\caption{(Color online) Numerical estimate of $\epsilon (M)$
as a function of $1/M$ for $(n,0)$ and $(n,n)$ icosadeltahedral
lattices with $\gamma =(1.5,0.5)$.} \label{Both}
\end{figure}

Within the continuum elastic theory it can be shown that twelve disclinations at the vertices of an icosahedron minimize the energy \cite{BowickNelsonTravesset:2000} when no further defects are allowed. The $C$-coefficient of Eq. \eqref{Energy} for this configuration of defects has been computed in \cite{BowickNelsonTravesset:2000}  
\be\label{C_icos} 
C({\cal
Y})=0.6043 \ . 
\ee 
${\cal Y}$ here stands for a particle configuration with 12 defects at the vertices of an icosahedron. Using the energy of Eq. \eqref{Energy}, the coefficient $a_1(\gamma,{\cal Y})$ appearing in the expansion of Eq. \eqref{General_form_serie} may be computed, with the results shown in Fig. \ref{fig__gamma} and Table \ref{Tab__Thomson}.
From the results described in Sec. \ref{sec:3e}, the $a_1$ coefficient may be extrapolated to very large numbers of particles using the expansion derived from Eq. \eqref{General_form_serie}. Indeed, as shown in Fig. \ref{Both}, plots of 
\be\label{enex} \epsilon(M) \equiv \frac{2R^{\gamma}E_{\mathrm{tot}}(M)/e^2 -
a_0(\gamma)M^2}{M^{1 + \gamma/2}} 
\ee 
vs $1/M$ are linear, with a slope that determines $a_1(\gamma)$ and an intercept related to the higher order core energy-like contribution. The results of these extrapolations are shown in Table~\ref{Tab__Thomson}. The agreement between the continuum elastic theory and the explicit computation for the $(n,n)$ configuration is remarkable, holding
to almost five significant figures. For the $(n,0)$ lattice there is agreement to four significant figures. This agreement is even more striking when it is recalled that the $a_1$-coefficient is obtained after subtraction of the term $a_0(\gamma)M^2$, as illustrated in Fig. \ref{fig__nn_versus_n0}. Furthermore, in the range from $\gamma=0.125$ to $\gamma=1.875$, all the significant digits vary and yet the accuracy of the calculation is virtually independent of $\gamma$.

\begin{table}[cbt]
\tbl{Numerical values of the coefficient $a_1(\gamma,{\cal Y})$ (twelve disclinations on the vertices of an icosahedron) using the $C$-coefficient from Eq.(\ref{C_icos}). The same coefficients from the $(n,n)$ and $(n,0)$ lattices}
{\begin{tabular}{lccc}
\toprule
$\gamma$ & $a_{1}(\gamma,{\cal Y})$ & $(n,n)$ & $(n,0)$ \\
\colrule
$1.875$ & $4.45118$ & $4.45110(4)$ & $4.45095(4)$ \\
$1.75$ & $2.47175$ & $2.47166(3)$ & $2.47150(3)$ \\
$1.625$ & $1.82629$ & $1.82621(2)$ & $1.82603(2)$ \\
$1.5$ & $1.51473$ & $1.51454(2)$ & $1.51445(2)$ \\
$1.375$ & $1.33695$ & $1.33683(4)$ & $1.33667(4)$ \\
$1.25$ & $1.22617$ & $1.22599(7)$ & $1.22589(7)$ \\
$1.125$ & $1.15366$ & $1.1535(2)$ & $1.15340(2)$ \\
$1.0$ & $1.10494$ & $1.10482(3)$ & $1.10464(3)$ \\
$0.875$ & $1.07187$ & $1.07174(3)$ & $1.07160(3)$ \\
$0.75$ & $1.04940$ & $1.04921(6)$ & $1.04910(6)$ \\
$0.625$ & $1.03421$ & $1.03413(5)$ & $1.03398(5)$ \\
$0.5$ & $1.02392$ & $1.02390(4)$ & $1.02372(4)$ \\
$0.375$ & $1.01672$ & $1.01663(6)$ & $1.01656(6)$ \\
$0.25$ & $1.01115$ & $1.01106(3)$ & $1.01103(3)$ \\
$0.125$ & $1.00595$ & $1.00592(2)$ & $1.00589(2)$ \\
\botrule 
\end{tabular}}
\label{Tab__Thomson}
\end{table}

\subsubsection{\label{sec:3i}The Energy difference of the $(n,m)$ lattices}

The $a_1$-coefficient computed within the continuum elastic approach above does not depend on the icosadeltahedral class $(n,m)$. Results from the direct minimization of particles do, however, show a weak dependence (in the $4th$ significant digit) on the particular $(n,m)$ configuration, as is apparent from Fig. \ref{fig__nn_versus_n0} and Table \ref{Tab__Thomson}. It should be noted that the discrepancy from the continuum result has a well defined sign, and is therefore reasonably attributed to a term not present in the energy expansion.

\subsection{\label{sec:3j}Thomson problem with a continuous distribution of dislocations}

When the number of particles is extremely large, the minimum energy configurations can be approximated by a closed analytical form, upon assuming a continuous distribution of defects.  

The formal elimination of the geometric frustration introduced by the Gaussian curvature may be formulated as a concrete set of equations in the case of the sphere. We shall use the identity
\bea\label{discl_delta} s(\bm{x})&=&\frac{\pi}{3\sqrt{g}}
\sum_{i=1}^N q_i \delta({\bm{x}},\bm{x}_i)
\\\nonumber
&=&\frac{1}{R^2}+\frac{\pi}{3 R^2}\sum_{l=1}^{\infty} \sum_{m=-l}^l
Y^{l\ast}_m(\theta,\phi) \sum_{i=1}^N q_i Y^{l
\ast}_m(\theta_i,\phi_i)
\\\nonumber
&=& K(\bm{x})+\frac{\pi}{3 R^2}\sum_{l=1}^{\infty} \sum_{m=-l}^l
Y^{l\ast}_m(\theta,\phi) \sum_{i=1}^N q_i Y^l_m(\theta_i,\phi_i) ,
\eea 
which follows from the topological constraint Eq. \eqref{constr_sphere}. Provided a disclination configuration exists such that 
\be\label{zero_screen} \sum_{i=1}^N q_i
Y^{l}_m(\theta_i,\phi_i)=0 \ , 
\ee 
for each $(l \geq 1 ,m)$, the disclination density completely screens the Gaussian curvature. A configuration of defects satisfying Eq.(\ref{zero_screen}) is an absolute minimum of the elastic energy, a result easily understood by writing the energy in the form 
\be\label{energy_harm}
E=E_0+\frac{\pi^2Y}{9} R^2 \sum_{l=1}^{\infty} \sum_{m=-l}^{l}
\frac{\left| \sum_{i=1}^N q_i Y^{l}_m(\theta_i,\phi_i) \right|^2}
{l^2(l+1)^2} +N\, E_{c}\ , 
\ee 
where the zero point energy $E_0$ in Eq. \eqref{Energy_Total} is kept. A configuration satisfying Eq. \eqref{zero_screen}) will be denoted by ${\cal G}$. For this hypothetical configuration, the $C$-coefficient in Eq. \ref{Energy} vanishes, although there is now a large contribution (linear in $R$) from the dislocation core energies
represented by the last term of Eq. \eqref{energy_harm}.

\begin{table}[h]
\tbl{Value of the $a_1$ coefficients for the ${\cal G}$ configuration Eq. \eqref{zero_screen}.}
{\begin{tabular}{ll}
\toprule
$\gamma$ & $a_{1}(\gamma,{\cal G})$ \\
\colrule
$1.875$ & $4.45227$ \\
$1.75$ & $2.47289$ \\
$1.625$ & $1.82746$ \\
$1.5$ & $1.51592$ \\
$1.375$ & $1.33815$ \\
$1.25$ & $1.22737$ \\
$1.125$ & $1.15485$ \\
$1$ & $1.10610$ \\
$0.875$ & $1.07297$ \\
$0.75$ & $1.05044$ \\
$0.625$ & $1.03515$ \\
$0.5$ & $1.02473$ \\
$0.375$ & $1.01737$ \\
$0.25$ & $1.01161$ \\
$0.125$ & $1.00620$ \\
\botrule
\end{tabular}}
\label{Tab__infty}
\end{table}

The ${\cal G}$ configuration may be characterized more explicitly. It consists of a density of dislocations that converges to the local Gaussian curvature. It can be shown that upon approximating the dislocations (each regarded as a disclination dipole with spacing $a$) as a continuum distribution, this dislocation density for a sphere becomes  \be\label{LC_right} 
\bm{b}(\theta,\phi)=\frac{1}{6R}\sum_{k=1}^6 \cot
[\alpha_k(\theta,\phi)] e^k_{\phi} \ . 
\ee 
The summation here runs over the six coordinates of the northern hemisphere of an icosahedron, $(0,0)$ and $(\theta_{\cal Y},2 \pi k/5)$, where $\theta_{\cal Y}=\arccos(1/\sqrt{5})$ and $\alpha_k$ is the angle $\theta$ relative to a coordinate system with the north pole located at $(\theta_{\cal Y},2 \pi k/5)$ for $k=1,\cdots\,5$. This angle is given implicitly by
\be\label{costheta}
\cos[\alpha_k(\theta,\phi)]=\cos(\theta)\cos(\theta_{\cal
Y})-\sin(\theta) \sin(\theta_{\cal Y})\cos\left(\frac{2\pi}{5}k+\phi\right)
\ . 
\ee
The implicit form of Eq. \eqref{LC_right} can be further simplified
\bea\label{vec_vec} \bm{e}_{\phi}^{k}&=&f^k(\theta,\phi)\left\{-\sin(\theta_{\cal Y})
\sin\left(\frac{2\pi}{5}k+\phi\right)\bm{g}_{\theta}
\right. \\
&+&\left. [\cos(\theta_{\cal Y}) \sin \theta+\sin(\theta_{\cal
Y})]\cos \theta\cos\left(\phi+ \frac{2\pi}{5}k\right)\bm{g}_{\phi}\right\}  \nonumber 
\eea 
where $f^k(\theta,\phi)=1/\sin [\alpha_k(\theta,\phi)]$. Close to one of the 12 disclinations with charge $s=2\pi/6$ Eq. \eqref{LC_right} predicts a singularity in the dislocation density \cite{BowickTravesset:2001} 
\be\label{gen_res} b \approx \frac{s}{2
\pi R a} \ . \ee 
For small angles, close to each disclination, there is a short-distance singularity 
\be\label{disl_dens_ok} b(\theta) =
\frac{\pi}{3 R \theta} + \cdots \ , 
\ee 
in agreement with known results in flat space. Eq. \eqref{LC_right} represents a continuous distribution of dislocations, and neglects both dislocation discreteness and their
mutual interactions. It represents six families of dislocations with azimuthal Burgers vectors associated with antipodal pairs of the 12 original disclinations in the icosahedron. When discreteness and interactions are taken in account, we expect these dislocations to condense into grain boundary arms, containing with quantized Burgers vectors and variable spacing in the radial direction \cite{BowickNelsonTravesset:2000,Travesset:2003,Travesset:2005}. The total number of discrete arms remains, therefore, the variable that needs to be determined for a discrete solution of the Thomson problem.

\subsubsection{\label{sec:3k}The intermediate regime}

Within the continuum elastic approach, the dominant configurations for a small number of particles are 12 defects with an icosahedral symmetry \cite{BowickNelsonTravesset:2000}. We have just seen, however, that adding a continuous distribution of dislocations, as might be appropriate when the particle number is large, can more efficiently screen the Gaussian curvature on a sphere. The natural problem then becomes to determine the precise structure of the defect arrays for intermediate numbers of particles when the discreteness of interacting dislocations is taken into account.
\begin{figure}[ctb]
\begin{center}
\centering
\includegraphics[width=0.4\textwidth]{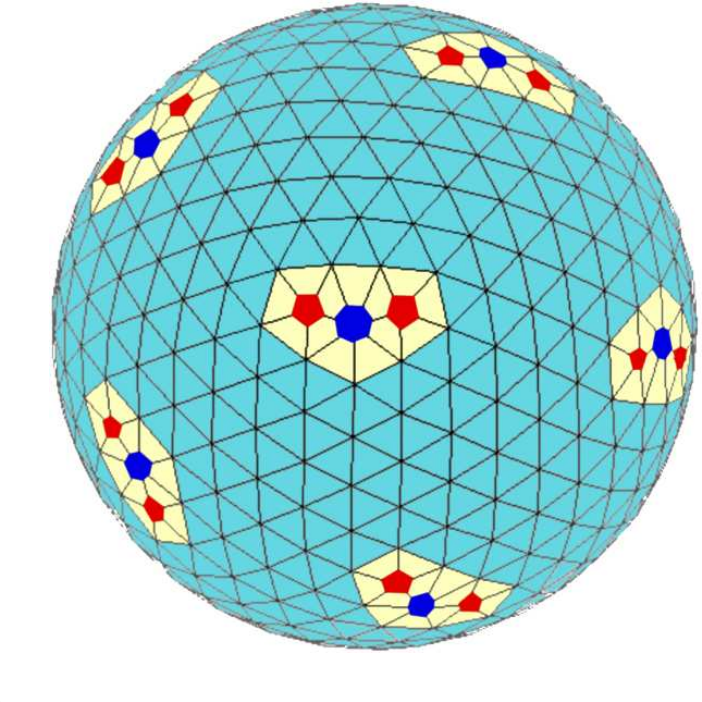} \caption{(Color online) Results
of a minimization of 500 particles interacting with a Coulomb
potential, showing the appearance of scars.
}\label{fig__scars_thomson}
\end{center}
\end{figure}

\begin{figure}[ctb]
\centering
\includegraphics[width=0.3\textwidth]{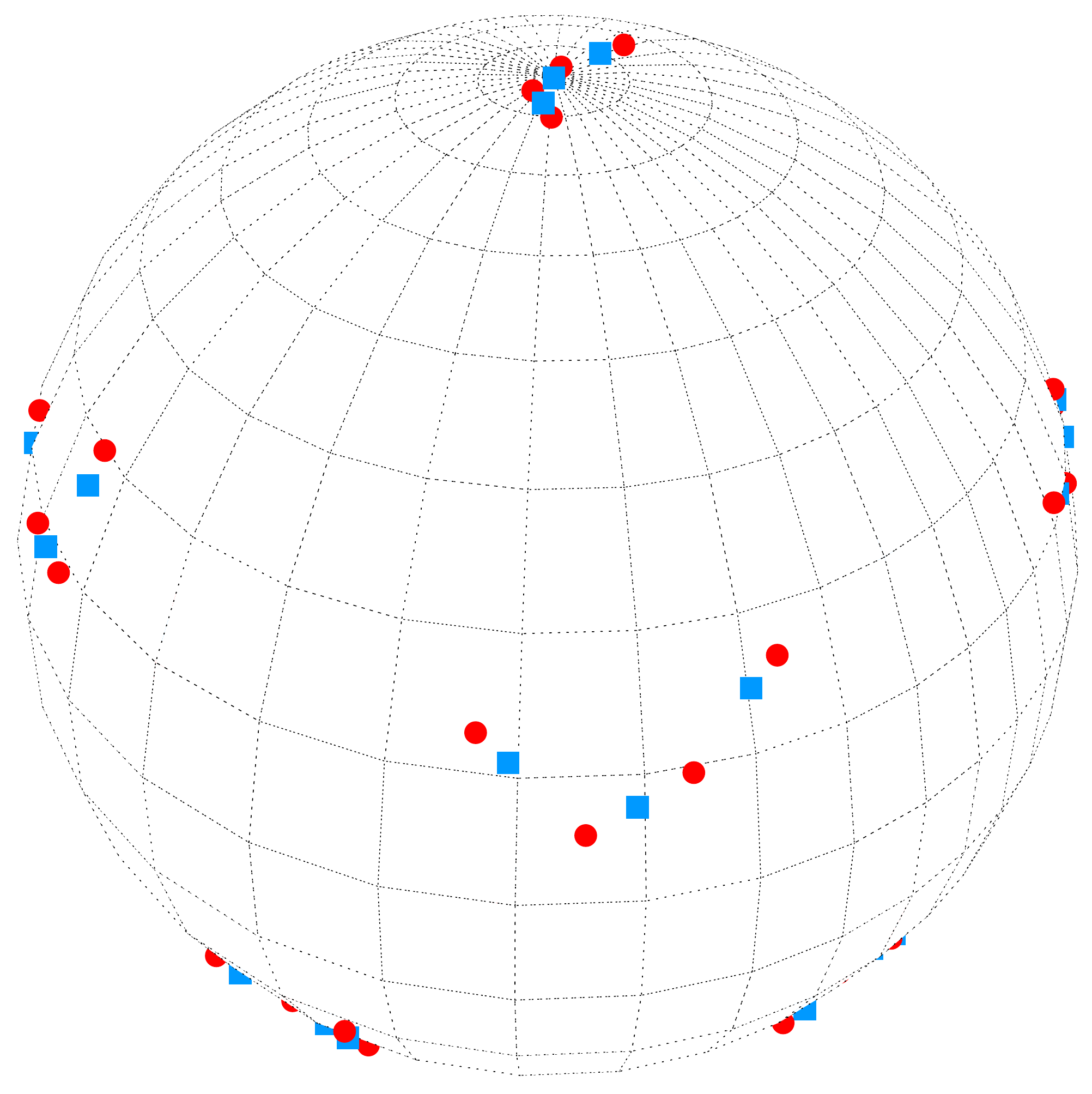}
\includegraphics[width=0.3\textwidth]{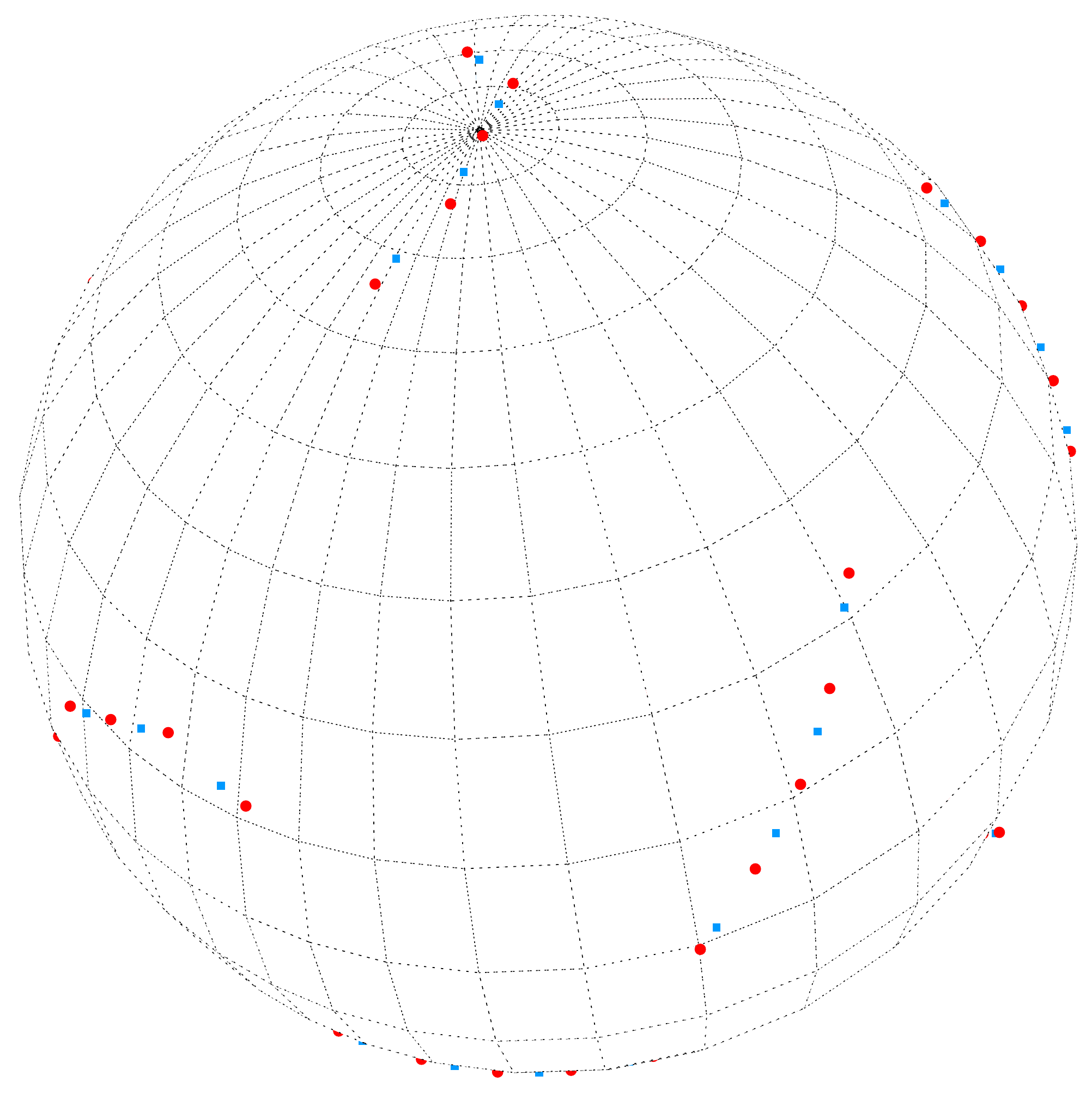}\\
\includegraphics[width=0.3\textwidth]{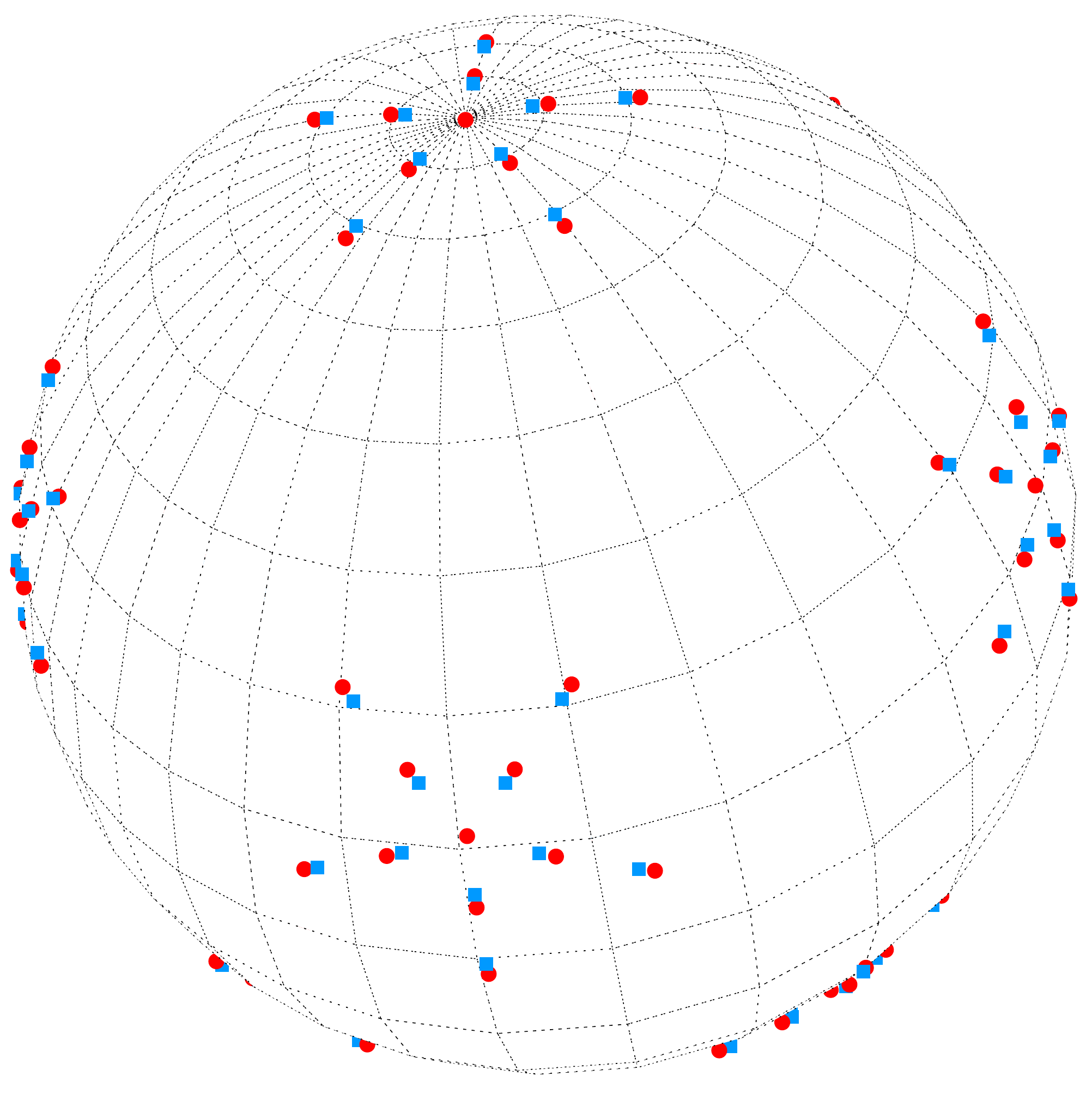}
\includegraphics[width=0.3\textwidth]{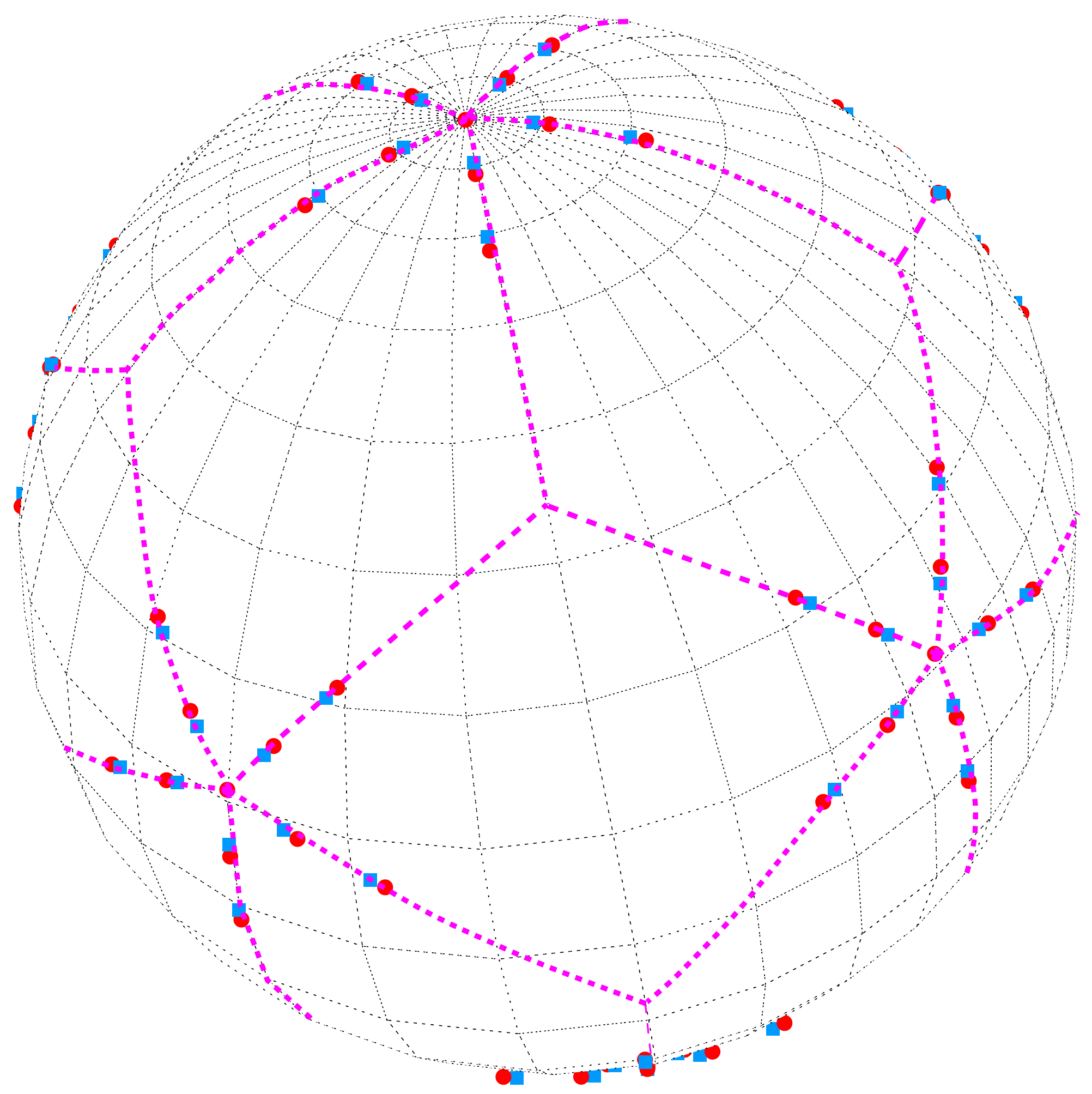}
\caption{(Color online) Ground state configurations for $M\approx2000$ particles
obtained from the continuum elastic formalism. In the top figure one
finds scars ($m=2$) and in the bottom pentagonal buttons ($m=5$)
forming a rhombic tricontahedron.} \label{fig__defects}
\end{figure}

We note first that the particular arrangement of defects dominating in this regime will not be fully universal. The particular array structure favored can vary from system to system with fixed particle number, depending, e.g., on details such as the dislocation core energy. This result may be traced back to the $M$-expansion of Eq. \eqref{General_form_serie}, in which the sub-leading terms which depend on non-universal properties influence the dominant terms for finite values of $M$. Some typical defect configurations obtained by direct minimization of particles on the sphere are shown in Fig. \ref{fig__scars_thomson} and show incipient scars, already at 500 particles (in \cite{BowickNelsonTravesset:2000} the minimum number of particles where scars are systematically found is predicted to be around 400). By using the geometrical model described here, where the energy is parameterized just by a Young's modulus and a dislocation core energy \cite{BowickNelsonTravesset:2000,BauschEtAl:2003} one can simulate larger particle numbers and one obtains results as in Fig. \ref{fig__defects}. Note the occurrence of low energy configurations with scars ($m=2$) in one instance and pentagonal buttons ($m=5$) in another. The dislocation spacing decreases the further a dislocation is from the central disclination.

\begin{figure}[ctb]
\centering
\includegraphics[width=0.7\textwidth]{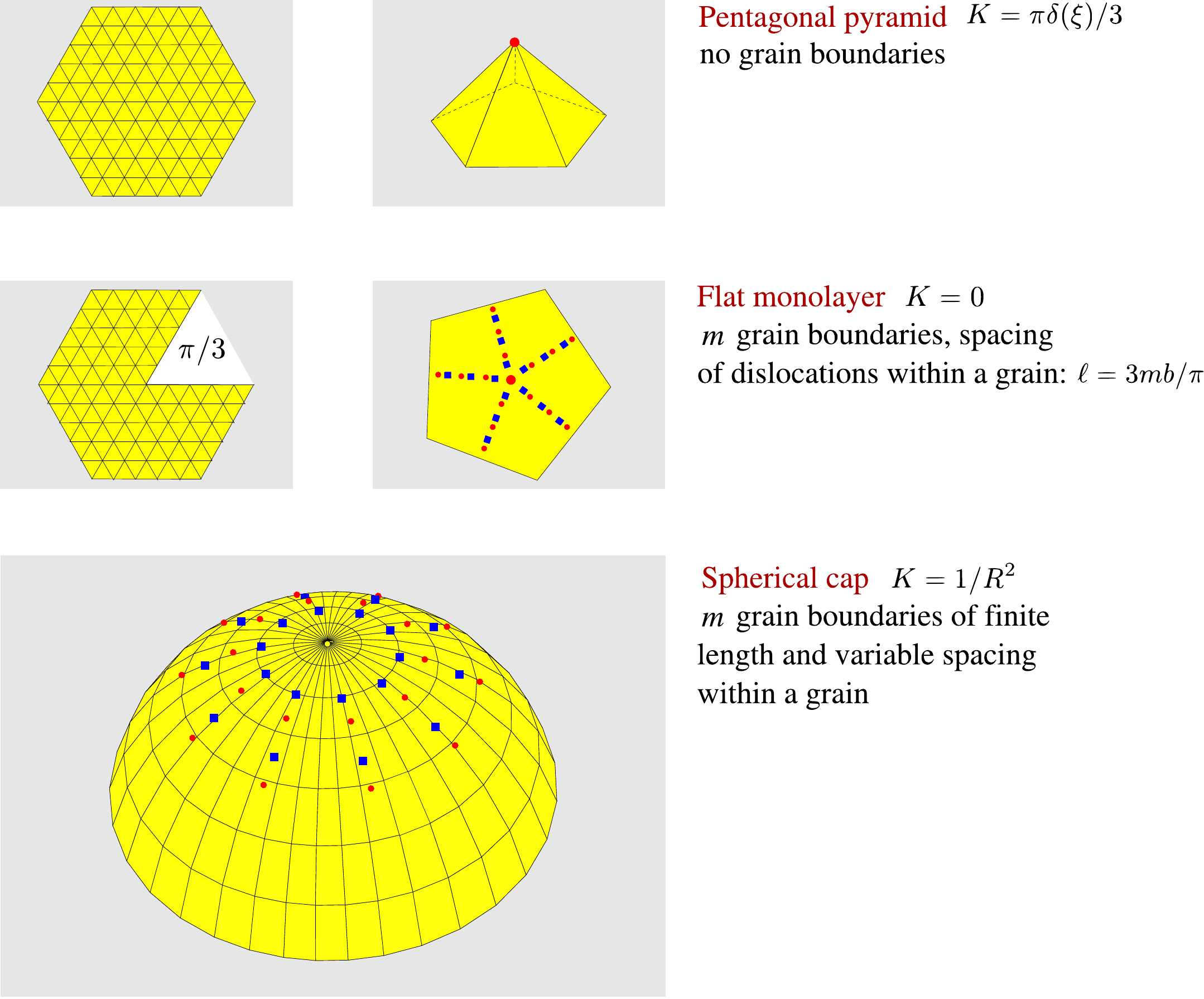}
\caption{(Color online) Schematic illustrating the genesis of grain boundary scars.
A disclination is first constructed from a perfect lattice. If this
disclination is placed on a tip of a cone, with a delta function of
Gaussian curvature balancing the defect charge, then no additional
defects form. If the crystal is forced into a monolayer, grain
boundaries radiating out of the disclination radiate all the way to
the boundary. In the intermediate regime of constant non-zero
Gaussian curvature, $m$ grain boundaries of finite length and
variable spacing of dislocations form.}\label{fig__c_argument}
\end{figure}

An overview of results involving grain boundary scars is presented in Fig. \ref{fig__c_argument}. If a disclination is placed on a perfect crystal, no additional defects will appear if the disclination is located on the tip of a cone with total Gaussian curvature equal to the disclination charge. If a disclination is forced into a flat monolayer, then $m$ low angle grain boundaries, with constant spacing between dislocations as shown in Fig. \ref{fig__c_argument} and grains going all the way to the boundary, will be favored (see \cite{Travesset:2003} for a detailed discussion). In the intermediate situation where a finite Gaussian curvature is spread over a finite area, as in the case of
a spherical cap, a disclination arises at the center of the cap and finite length grain boundaries stretched out over an area of $(\pi/3)R^2$ with variable spacing dominate, again as illustrated in Fig. \ref{fig__c_argument}.  

Additional results may be obtained for the number of arms within the grain boundary, the actual variable spacing between dislocations within the grain and the length of the grains as a function of the number of particles.  

When grain boundary scars appear, one can estimate the number of excess dislocations which decorate each of the 12 curvature-induced disclinations on the sphere using ideas from Ref. \cite{BowickNelsonTravesset:2000}. This estimate is in reasonable agreement with experiments probing equilibrated assemblies of polystyrene beads on water droplets \cite{BauschEtAl:2003}. Consider the region surrounding one of the 12 excess disclinations, with charge $s=2\pi/6$, centered on the north pole. As discussed in Ref. \cite{BowickNelsonTravesset:2000}, one expects the stresses and strains at a fixed geodesic distance $r$ from the pole on a sphere of radius $R$ to be controlled by an {\em effective} disclination charge 
\bea\label{deficit} 
s_{eff}(r) 
&=& s - \int_{0}^{2\pi} d\phi \int_{0}^{r} dr'\,\sqrt{g}\,K \nonumber \\
&=& \frac{\pi}{3} - 4\pi \sin^{2}\left(\frac{r}{2R}\right) \, . 
\eea 
Here the Gaussian curvature is $K=1/R^2$ and the metric tensor associated with spherical polar coordinates $(r,\phi)$, with distance element $ds^2 = d^2r + R^2\sin^2(r/R)d^2\phi$, gives $\sqrt{g}=R\sin(r/R)$. Suppose $m$ grain boundaries radiate from the disclination at the north pole. Then, in an approximation which neglects interactions between the individual arms, the spacing between the dislocations in these grains is \cite{BowickNelsonTravesset:2000} 
\be 
l(r) =
\frac{am}{s_{eff}(r)} \ , 
\ee 
which implies an effective dislocation density 
\bea 
n_d(r) = \frac{1}{l(r)} = \frac{1}{ma}\bigg[\frac{\pi}{3} - 4\pi\sin^{2}\left(\frac{r}{2R}\right)\bigg] = \frac{2\pi}{ma}\bigg[\cos\left(\frac{r}{R}\right) - \frac{5}{6}\bigg] \ . 
\eea 
This density vanishes when $r \rightarrow r_c$, where 
\be 
r_c = R\arccos \left(\frac{5}{6}\right) \approx 33.56^{\circ}R \ , 
\ee 
which is the distance at which the $m$ grain boundaries terminate. The total number of dislocations residing within this radius is thus 
\bea N_d &=&
m\int^{r_c}_0 dr\,n_d(r) \nonumber \\
&=& \frac{\pi}{3a}r_{c} -
\frac{4\pi}{a}\int^{r_c}_0 dr\,\sin^2\left(\frac{r}{2R}\right) \nonumber \\
&=& \frac{\pi}{3}\bigg[\sqrt{11} -5\arccos\left(\frac{5}{6}\right)\bigg]\left(\frac{R}{a}\right)
\nonumber
\\
&\approx& 0.408\,\left(\frac{R}{a}\right)\ . 
\eea 
As discussed in Ref. \cite{BowickEtAl:2002}, it is also of interest to consider $2\pi$ disclination defects (appropriate to crystals of tilted molecules \cite{DierkerPindakMeyer:1986}) on the sphere. The icosahedral configuration of 12 $s=2\pi/6$ disclinations is now replaced by just two $s=2\pi$ disclinations at the north and south poles. Using the approximation discussed above, it is straightforward to show that the density of dislocations in each of $m$ (noninteracting) grain boundary arms now reads 
\be 
n_d(r) = \frac{2\pi}{ma}\cos\left(\frac{r}{R}\right) \
. 
\ee 
This density vanishes at $r_c = (\pi/2)R$, corresponding to a hemisphere of area on the sphere for each cluster of arms.

It is of considerable interest to repeat the above calculation for a square lattice, as found for example in the protein surface layers (s-layers) of some bacteria \cite{PumMessnerSleytr:1991,SleytrEtAl:2001}. In this case the basic disclination has $s=2\pi/4$. The effective dislocation density becomes 
\bea 
n_d(r) 
= \frac{1}{l(r)} 
= \frac{1}{ma}\left[\frac{\pi}{2} - 4\pi\sin^2\left(\frac{r}{2R}\right)\right]
= \frac{2\pi}{ma}\left[\cos\left(\frac{r}{R}\right) - \frac{3}{4}\right] \ . 
\eea 
This density vanishes when $r \rightarrow r_c$, where 
\be 
r_c = R\arccos\left(\frac{3}{4}\right) \approx 41.4^{\circ}R \ , 
\ee 
which is the distance at which the $m$ grain boundaries terminate. The longer angular length of square lattice scars reflects the larger initial disclination charge ($90^{\circ}$) that must be screened. The total number of dislocations residing within this radius is thus
\bea 
N_d &=&
m\int^{r_c}_0 dr\,n_d(r) \nonumber \\
&=& \frac{\pi}{2a}\,r_{c} -
\frac{4\pi}{a}\int^{r_c}_0 dr\,\sin^{2}\left(\frac{r}{2R}\right) \nonumber \\
&=& \frac{\pi}{2}\left[\sqrt{7} -3\arccos\left(\frac{3}{4}\right)\right]\left(\frac{R}{a}\right)
\nonumber
\\
&\approx& 0.75\left(\frac{R}{a}\right) \ . 
\eea 
Thus the angular length of scars and the total number of excess dislocations is a measure of the underlying topology of the lattice tiling the sphere.

\subsection{\label{sec:3l}Interstitial fractionalization on the sphere}

\subsubsection{\label{sec:3m}Introduction}

Consistent with theoretical expectations, experiments on particle-coated water droplets in oil~\cite{BauschEtAl:2003} reveal that the twelve excess disclinations sprout grain boundary ``scars" for sufficiently large $R/a$. When triangulations of microscopic particle packings are used to reveal the local coordination number, these grain boundaries appear as additional dislocations, i.e., 5-7 pairs, arrayed around an unpaired 5, in a pattern such as 5-7--5-7--5--7-5--7-5. Although the critical value of  $R/a$ above which grain boundaries appear depends on microscopic details, both theoretical estimates and experiments indicate that this instability arises as soon as $R/a \gtrsim 5-6$, i.e., when the total number of particles exceeds several hundred. Thus, unlike crystals in flat space, dislocations arrayed in grain boundaries are an intrinsic part of the ground state.   These grain boundaries can, moreover, stop and start freely on the sphere, unlike their flat space counterparts. Such terminations occur naturally (and with low energy cost) because crystalline grains rotate under parallel transport due to the nonzero Gaussian curvature of the sphere.

If disclinations and dislocations are crucial for understanding spherical crystallography, what can we say about \emph{vacancies and interstitials}, which are well known to play a key role in conventional crystals \cite{AshcroftMermin,BowickNelsonShin:2007}. It is natural to introduce vacancies and interstitials in an attempt to understand spherical particle packings that deviate from the sequence of ``magic numbers" $N_{nm}=10(n^{2}+nm+m^{2})+2$. It is tempting to ignore the instability to grain boundary scars for large $N_{nm}$ and regard the commensurate $(n,m)$ lattice as an interesting metastable state. It would then be natural to introduce vacancies and interstitials to describe candidate ground state packings for $N_{nm} \pm t$ particles, where $t=1, 2,...$ and much less than the distance to the next magic number.   There is, however, another surprise in store: in contrast to flat space, where vacancy and interstitial defects are stable and well defined, one finds that interstitials and vacancies are typically ripped apart into dislocation fragments by the strain fields of nearby 5-fold disclinations. The dislocations then combine with some of the excess 5's to form defect clusters such 5-7-5. Thus, vacancies and interstitials lose their integrity via fragmentation in spherical crystals, and mediate formation of small grain boundary scars.    

 \begin{figure}[!h]
\centering
\includegraphics[scale=0.4]{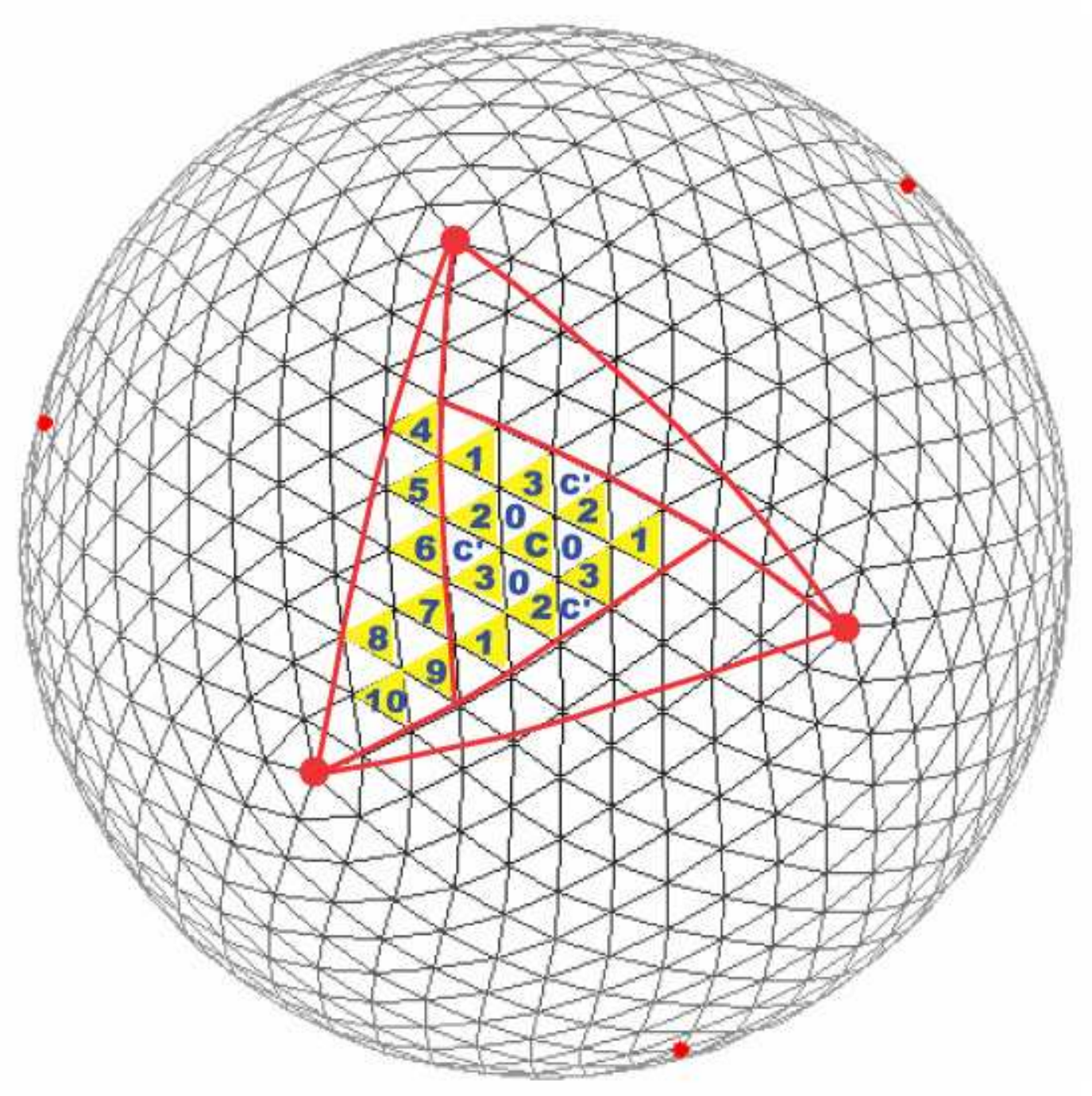}
\caption{(Color online) Initial icosahedral configuration of an (8,3) spherical
crystal lattice with 12 disclination defects.  This lattice is
chiral, and is arranged such that 8 steps along a Bragg row,
followed by 3 steps to the right along a Bragg row at a
$120^{\circ}$ angle, connect neighboring 5-fold disclinations.
Distinct initial locations for interstitials are shown as triangular
plaquettes labeled by characters and numbers.} \label{initial}
\end{figure}

\subsubsection{\label{sec:3n}Interstitial fractionalization}

We are interested in studying the effect of inserting an interstitial or a vacancy into a regular $(n,m)$ icosahedral lattice by adding or removing a single particle, giving rise to particle numbers $N_{nm}\pm1$ not falling in the classification of icosadeltahedral lattices and in studying the relation of interstitials to the extra dislocation defects (scars) found in spherical crystals above a critical system size \cite{BowickNelsonTravesset:2000,BauschEtAl:2003}. Refs. \cite{CockayneElser:1991,JainNelson:2000, FisherHalperinMorf:1979,FreyNelsonFisher:1994,PertsinidisLing:2001a, PertsinidisLing:2001b,PertsinidisLing:2005,LibalEtAl:2007} study configurations and energies of interstitial and vacancy defects and their energetics in triangular lattices in flat space.

\begin{figure}[t]
\centering
\includegraphics[scale=0.5]{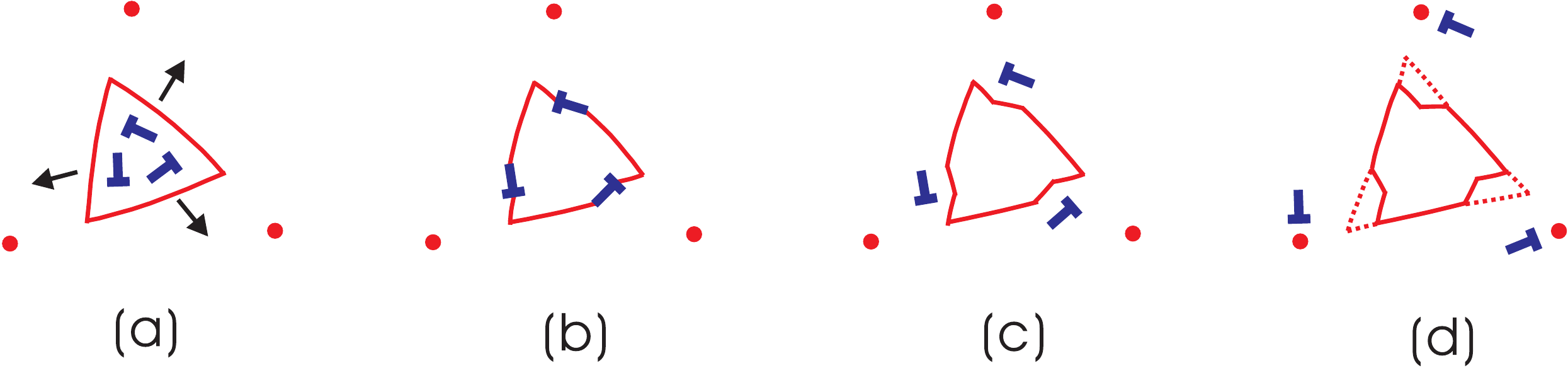}
\caption{(Color online) A schematic of interstitial fractionization. The $\dashv$
symbols are alternative ways of representing dislocations depicted
elsewhere as 5-7 pairs.} \label{fracschematic}
\end{figure}

\begin{figure}[!h]
\centering
\includegraphics[scale=0.7]{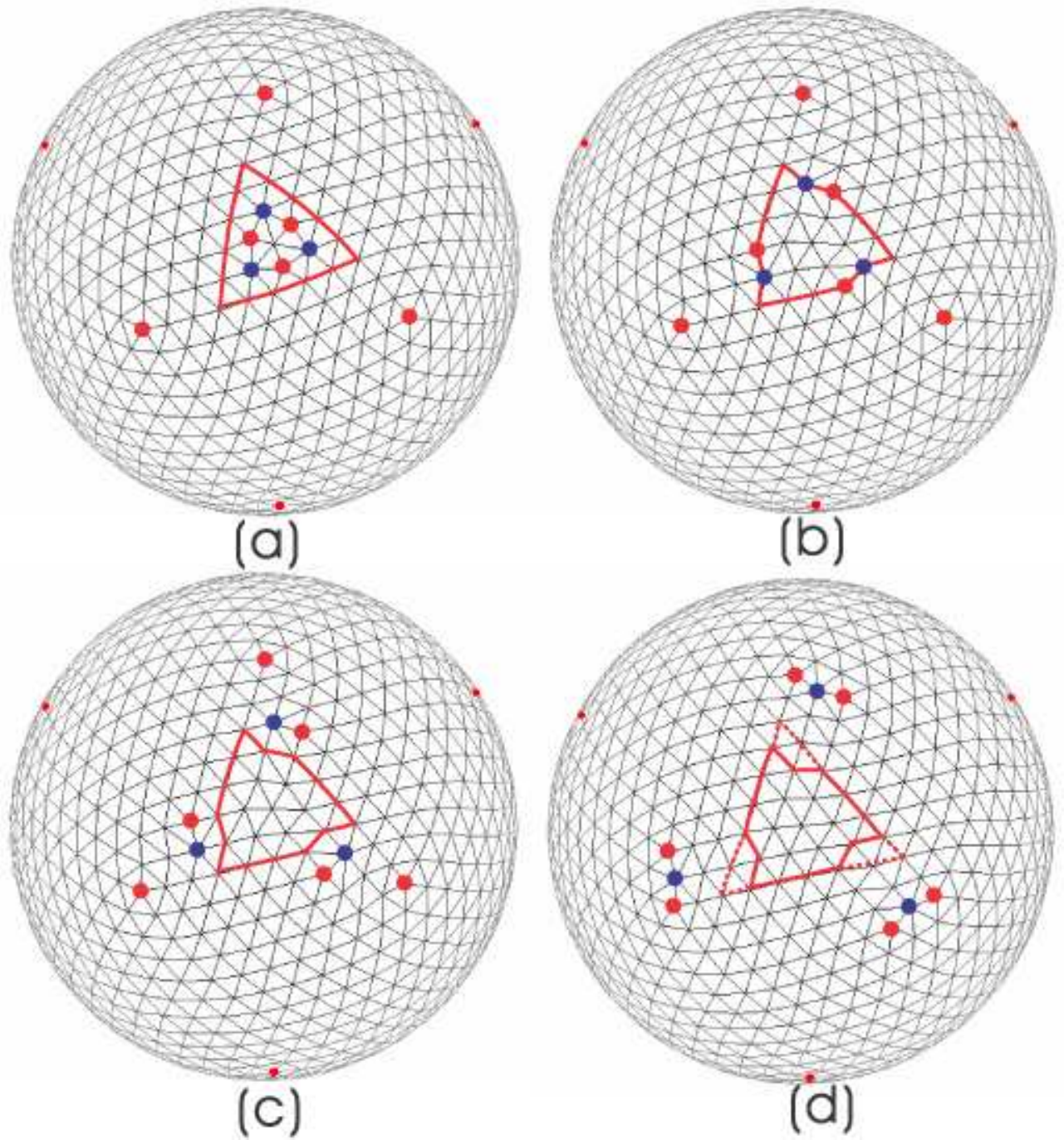}
\caption{(Color online) The fractionalization of an interstitial defect inserted at
the center of three neighboring disclinations in an (8,3) spherical
tessellation: (a) the initial interstitial defect configuration
($I_3$) shown surrounded by a triangular reference contour; (b) the
bound defect unbinds to form three separate dislocations (5-7)-
pairs; (c) each dislocation glides (i.e., moves parallel to its
Burgers vector) towards the nearest disclination; (d) three
mini-grain boundary scars are formed. We also keep track of the
evolution of the initial defect in (a) by illustrating the
deformation of the triangular contour around the initial defect for
(b), (c), and (d) induced by the passage of the dislocation.}
\label{frac}
\end{figure}

The presence of excess dislocation defects in the ground state of spherical crystals is dramatically illustrated by the following numerical experiment: we start with a regular icosadeltahedral tessellation of the sphere {--} say an $(8,3)$, corresponding to $N_{83}=972$ (Fig.~\ref{initial}). This may be done with the applet located at \cite{Java} using the Construct $(m,n)$ algorithm. Although the true ground state for $972$ particles on the sphere with most pair potentials has additional dislocation defects (i.e.
tightly bound pairs of 5- and 7- coordinated particles) arrayed in grain boundary scars~\cite{BowickNelsonTravesset:2000}, the regular icosadeltahedral lattice is a local minimum from which it is difficult to escape without the addition of thermal noise. In fact it is a major challenge to find fast and reliable algorithms to locate the true ground state (global minimum) in this problem with its complex energy landscape. Now add a single particle to the lattice at the center of mass of a spherical triangle whose vertices are 3 nearest-neighbor 5-fold disclinations (using commands shift + click). The {\em self-interstitial} so formed is then relaxed by a standard relaxation algorithm, with sufficient thermal noise to allow dislocation glide over the Peierls potential \cite{LipowskyEtAl:2005}. One immediately finds that an interstitial is structurally unstable. In a few time steps it morphs into a complex of dislocations with zero net Burgers vector. The most common structure observed is a set of three dislocations, with Burgers vectors perpendicular to a line joining each 5-7 pair, inclined at $120^{\circ}$ angles to each other. Eventually the interstitial complex is ripped apart entirely, as illustrated schematically in Fig.~\ref{fracschematic} (see also Fig.~\ref{frac}). Intermediate configurations and final states as the dislocations glide apart will be classified later. Most often three separate dislocations are formed which each glide toward a $5$-fold disclination. The end result is the formation of a ``mini-scar" (a $5-7-5$ grain boundary) at each of the vertex $5$s. Subsequent removal of a particle to restore the particle number to the original $972$ and relaxing still leaves scars with total energy lower than the starting configuration with $12$ isolated $5$'s. This observation confirms that scars are definitely lower energy states and not simply artifacts of the relaxation algorithm.

The above phenomenon of low-temperature ($T \gtrsim 0$) {\em unbinding} of dislocations by spatial curvature is a curved space analog of melting at finite temperature. The extended nature of fractionated interstitials (each separating dislocation component involves an extra row of particles) means that they cannot be treated as small perturbations from the initial spherical crystal with particle number $N_{nm}$.

Let's return to the specific case of the $N=973$ particle configuration generated by an interstitial inserted into one triangular plaquette of a regular $(8,3)$ lattice of $N_{nm}=972$ particles with the requisite $12$ disclination defects (5-fold coordinated particles) at the vertices of a regular icosahedron, as shown in Fig. \ref{initial}. The spherical crystal is distorted by the additional particle {--} the local configuration adopted by the interstitial changes as the crystal relaxes toward a lower energy state. As in the case of planar lattices, one also finds here that various interstitial defect configurations appear, such as the twofold symmetric interstitial $I_2$, the threefold symmetric interstitial $I_3$, and the fourfold symmetric interstitial $I_4$ (see Figs.~\ref{frac},~\ref{rot2}, and ~\ref{rot_I_4_2}). The most common intermediate complex formed by the interstitial is threefold symmetric in the rough form of a triangular loop composed of three dislocations with radially oriented Burgers vectors. All of the configurations adopted by an interstitial prior to unbinding can be described as a set of dislocations with zero net Burgers vector.

In marked contrast to interstitial defects in a planar crystal, interstitial defect configurations in a spherical crystal are metastable states with characteristic decay processes. As we shall see, the instability is caused by interactions with the inevitable disclinations associated with the nonzero Gaussian curvature of the
sphere. A representative evolution of an interstitial inserted at the center of three neighboring disclinations in an $(8,3)$ icosadeltahedron is shown in Fig. \ref{frac}. After some local relaxation the interstitial configuration denoted $I_3$ is formed, as shown in Fig.~\ref{frac} (a). We also show there the construction of a triangular contour surrounding the original interstitial defect. The presence of an interstitial follows because the contour encloses $7$ particles instead of $6$, the number appropriate to a perfect triangular lattice. During the annealed relaxation illustrated in Figs.~\ref{frac}(b) through (d), the dislocations which are bound together in the initial interstitial unbind into individual dislocations and subsequently glide towards nearby disclinations, eventually forming minimal 5-7-5 grain boundary scars. If one thinks of the initial dislocations as internal degrees of freedom within the interstitial, one could say that one-third of an interstitial is present in each mini-scar in the final state and thus that the interstitial demonstrates $1/3$ fractionalization. For other initial conditions the fractionation is into two dislocations, each representing $1/2$ of the original interstitial. The instability of interstitial defects in curved crystals may be studied via continuum elasticity theory by calculating the interaction energies between defects at each stage of the relaxation process of Fig. \ref{frac} \cite{BowickNelsonShin:2007}. We note that the triangular plaquette around the initial defect has been deformed such that it conforms as closely as possible to the regular triangular lattice during the relaxation process. Its deformation reveals the passage of the escaping dislocations.

\subsubsection{\label{sec:3o}Interstitial defect energies}

In this section, we discuss the energy of an interstitial defect in a spherical crystal. Consider $N$ point particles constrained to lie on the two-dimensional surface of a unit sphere. The energy of $N$ particles interacting through a generalized Coulomb potential within the curved surface is given by 
\be
\label{general}
E=\sum_{i<j}^{1,N} \frac{1}{|\bm{x}_i-\bm{x}_j|^\gamma} \ , 
\ee 
where $\bm{x}$ is the position of the particle in three dimensions and $\gamma$ is an integer. For a flat triangular lattice with periodic boundary conditions, the interstitial defect energy at constant density was defined in \cite{FisherHalperinMorf:1979,FreyNelsonFisher:1994} as 
\be
E_I = E_{relaxed} - E_{per} \ , 
\ee 
where $E_{relaxed}$ is the relaxed energy of the rearranged lattice of $N$ particles with the interstitial defect in the area $A$, and $E_{per}$ is the energy of the perfect crystal at the same areal density $N/A$. In curved space, however, the definition of a ``perfect crystal" is more subtle, since disclination defects resulting from the Gaussian
curvature and the topology are inevitable. We will take as a reference crystal the $(n,m)$ icosadeltahedral configurations corresponding to triangular tessellations of a magic number of particles $N_{nm}= 10 (n^2 + m^2 + nm) + 2$. Once an interstitial or vacancy is added to such a $(n,m)$ configuration, we are no longer at a magic number of particles since these are quite sparsely distributed. We thus need to define the energy of the perfect crystal. Here we define the energy of the interstitial (vacancy) defect at constant density in the spherical crystal as 
\be \label{E_I} 
E_I = E_{local} - E^*_{annealed} \ , 
\ee 
where $E_{local}$ measures the energy of the relaxed interstitial while the constituent dislocations are still bound and $E^*_{annealed}$ is the minimum energy of all possible final states attained after \emph{annealed relaxation} leading to interstitial {\em fractionalization}. This definition will be more explicitly discussed in the following section (see Table~\ref{anneal}). We note that both $E_{local}$ and $E^*_{annealed}$ are measured at the same areal density $(N_{nm} \pm 1)/A$, where $\pm 1$ correspond to an interstitial (vacancy) respectively. The lowest relaxed energy $E^*_{annealed}$ plays the role of the energy of the perfect lattice in the planar case at the density of $(N_{nm} \pm 1)/A$.

Numerical measurements of $E_{local}$ and $E^*_{annealed}$ for the power-law potentials with $\gamma = 1$, $3$, $6$ and $12$ have been performed by adding one interstitial at the center of a spherical triangle formed by three nearest-neighbor disclinations in the $(8,3)$ lattice (the location represented by $C$ in Fig. \ref{initial}) \cite{BowickNelsonShin:2007}. $E_{local}$ is measured by quenching the system at the moment just prior to the fractionation of the interstitial into individual dislocations [(a) in Fig. \ref{frac}]. The results are reported in Table~\ref{s}.

\begin{table}[h!]
\tbl{The lowest local and annealed relaxed energy with the central interstitials created by putting a particle at $C$ in Fig. \ref{initial} of the $(8,3)$ lattice, for $\gamma=1$, $3$, $6$, and $12$. The differences between two relaxed energies are calculated. Because the particles are embedded in a sphere of unit radius with our conventions, near-neighbor particle spacings are of order $a \sim N^{-1/2} \ll 1$, leading to a strong $\gamma$ dependence in the total energy given by Eq. \eqref{general}.}
{\begin{tabular}{ccccc}
\toprule
$\gamma$ & 1 & 3 & 6 & 12 \\
\colrule
Local  & 456601.99 & 2840600.7 & 9.62182 $\times 10^8$ & 3.03015 $\times 10^{14}$ \\
Annealed & 456600.91 & 2840025.5 & 9.60570 $\times 10^8$ & 3.00313 $\times 10^{14}$ \\
$E_{local}-E^*_{annealed}$ & 1.08 & 575.2 & 0.01612 $\times 10^8$ & 0.02702 $\times 10^{14}$ \\
\botrule
\end{tabular}}
\label{s}
\end{table}

\subsubsection{\label{sec:3p}Position dependence of interstitial defect energies}

By adding a particle in different plaquettes within the large spherical triangle of Fig. \ref{initial}, one can explore the dependence of the final state on the initial interstitial location. In contrast to the case for planar crystals, both the {\em location} and {\em orientation} of the interstitial defect relative to nearby disclinations influences the resultant configuration and its corresponding evolution, leading to distinct relaxed configurations. The insertion of a particle at the center of the large spherical triangle leads to an $I_3$-type initial configuration, whereas adding the interstitial to the plaquette along an edge results in an $I_2$-type initial configuration.

During the relaxation process, one also finds that the dislocation complex representing an interstitial can rotate so as to reorient the Burgers vectors so that constituent dislocations can glide towards a nearby disclination and bind to it. This phenomenon is especially noticeable if one places one extra particle slightly off the absolute center $C$, such as at the locations $C'$ or $0$ in Fig.~\ref{initial}.

\begin{figure}[b!]
\centering
\includegraphics[scale=0.5]{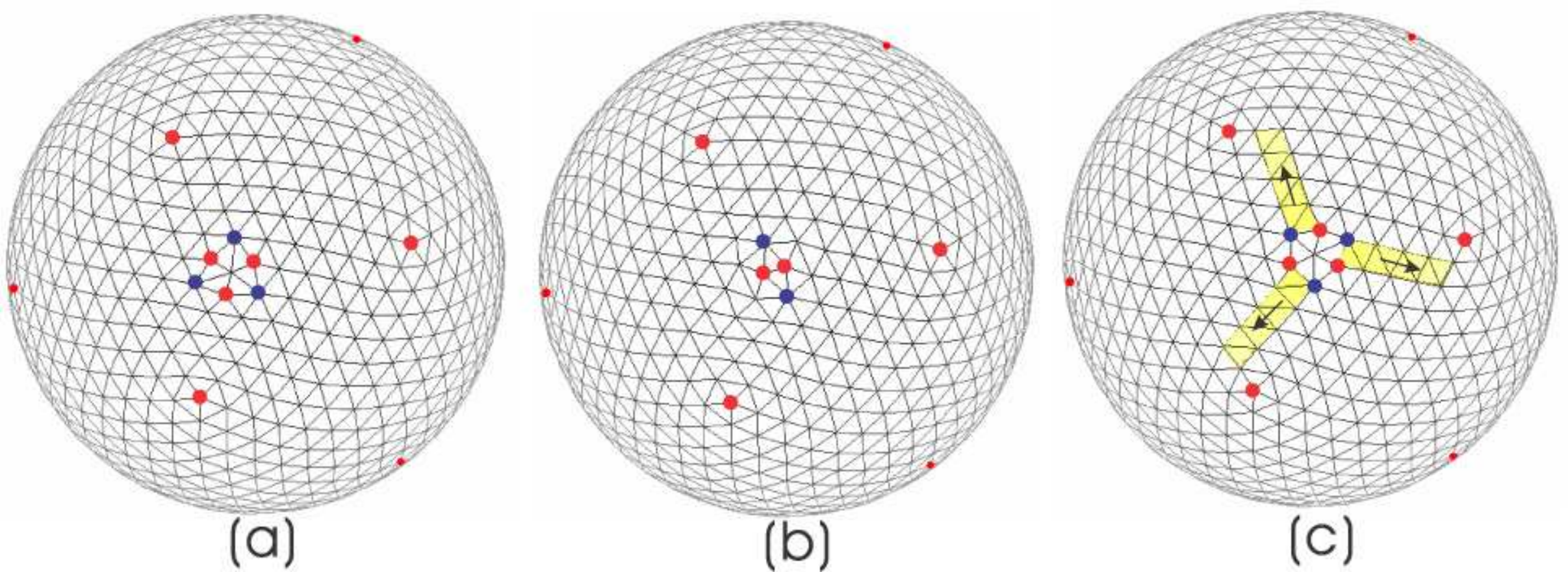}
\caption{(Color online) The rotational motion of an interstitial configuration
(created with the initial location $C'$ in Fig.~\ref{initial})
mediated by the transition: $I_3 \rightarrow I_2 \rightarrow I_3$.}
\label{rot2}
\end{figure}

\begin{figure}[t]
\centering
\includegraphics[scale=0.5]{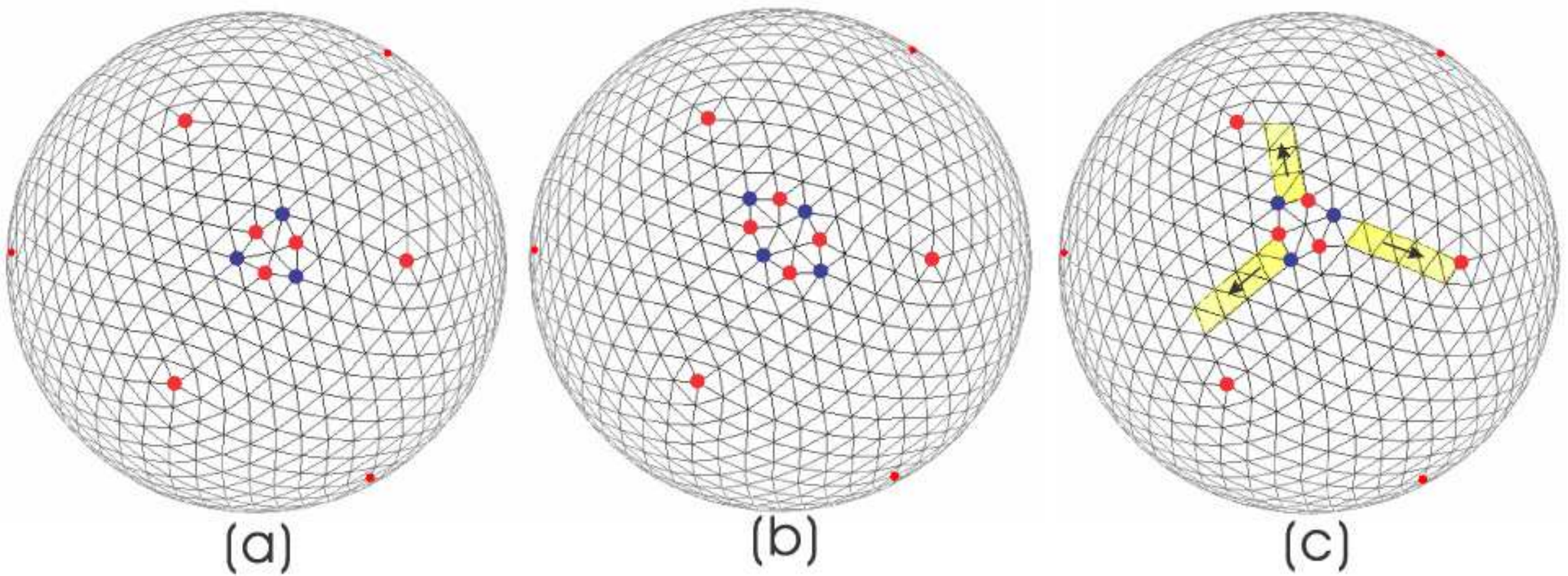}
\caption{(Color online) The rotational motion of an interstitial configuration
(created with the initial location $0$ in Fig.~\ref{initial})
mediated by the transition: $I_3 \rightarrow I_4 \rightarrow I_3$.}
\label{rot_I_4_2}
\end{figure}

In Figs.~\ref{rot2} and Fig.~\ref{rot_I_4_2} this phenomenon is illustrated explicitly. In Fig.~\ref{rot2}(a), an interstitial initially placed at the position $C'$ in Fig.~\ref{initial} morphs quickly to an $I_3$ configuration. In (a), however, the orientations of the dislocations within $I_3$ are not appropriate for dislocation glide to the surrounding disclinations since the $5$ end of the dislocations point towards rather than away from these 5-fold disclinations, causing them to be repelled. Glide-induced
fractionalization is therefore prohibited in this orientation. Remarkably, though, the entire complex of dislocations can change its orientation by a transition through an intermediate $I_2$ configuration (shown in Fig.~\ref{rot2}(b)) and subsequently to a second $I_3$ configuration [Fig.~\ref{rot2}(c)]. The final $I_3$ configuration is rotated by $60^{\circ}$ with respect to the first $I_3$ and can now fractionate analogously to an interstitial with initial position $C$ in Fig. \ref{initial}.

One also find rotational reorientation of an interstitial defect with an $I_4$-type intermediate state, as shown in Fig.~\ref{rot_I_4_2}. This particular relaxation process reveals an interesting feature of dislocation dynamics on a curved surface. The $I_3$ configuration generated after the intermediate [see Fig.~\ref{rot_I_4_2}(c)] now has one dislocation with its glide plane such that it can  glide head-on into a vertex disclination. This disclination absorbs the dislocation but hops over one lattice spacing to accommodate the curved space Burgers vector. The other two dislocations end up bound in the form of mini-scars. In this case then the interstitial has fractionated into 2 rather than 3 parts and one say that there has been $1/2$ fractionization of the interstitial. Absorption and emission of dislocations by 5-fold disclinations are somewhat analogous to absorption and emission of vacancies and interstitials by dislocations (allowing dislocations to climb), a phenomenon well-known in flat space.

What is the dependence of the final state on the initial location of the interstitial?  The distinct initial conditions shown in Fig.~\ref{initial} lead to three different final annealed states, as summarized in Table~\ref{anneal}.  The final state with three mini-scars of the form 5-7-5 has the lowest energy of all final states and provides a measure of $E^*_{annealed}$.

\begin{table}[t!]
\tbl{The three classes of final annealed state depending on the initial interstitial location. The relaxed energies are measured for the power law potential $\gamma = 6$. Here $E_{annealed}$ with 3 scars corresponds to $E^*_{annealed}$.}
{\begin{tabular}{ccc}
\toprule
Initial Location & $E_{annealed}$ & Annealed State \\
\colrule
$C$, $C'$, 7, 8, 10 & 9.60570 $\times 10^8$ & 3 scars (5-7-5) \\
0, 2, 3, 4, 6, 9  & 9.61011 $\times 10^8$ & 2 scars (5-7-5) \\
1, 5 &  9.61062 $\times 10^8$ & 2 scars (5-7-5-7-5) \\
\botrule
\end{tabular}}
\label{anneal}
\end{table}

\begin{table}[t!]
\tbl{The energy of interstitial defects created at different initial positions within the spherical crystal. The energies shown are for a power law potential with $\gamma=6$.}
{\begin{tabular}{cccc}
\toprule
$n$ & Transition & $E_{local}$ of $I_3$ & $E_{local}-E^{*}_{annealed}$ \\
\colrule
$C$  & $I_3$ & 9.62182 $\times 10^8$ &0.01612 $\times 10^8$ \\
$C'$ & $I_3 \rightarrow I_2 \rightarrow I_3$ & 9.62195 $\times 10^8$ & 0.01655 $\times 10^8$\\
0 & $I_3 \rightarrow I_4 \rightarrow I_3$ & 9.62274 $\times 10^8$ & 0.01708 $\times 10^8$\\
1 & $I_3 \rightarrow I_2$ &  9.62391 $\times 10^8$ &  0.01820 $\times 10^8$\\
2 & $I_3 \rightarrow I_2$ &  9.62296 $\times 10^8$ &  0.01725 $\times 10^8$\\
3 & $I_3$                 &  9.62269 $\times 10^8$ &  0.01698 $\times 10^8$\\
4 & $I_3 \rightarrow I_2$ &  9.62595 $\times 10^8$ &  0.02025 $\times 10^8$   \\
5 & $I_3 \rightarrow I_2$ &  9.62479 $\times 10^8$ &  0.01909 $\times 10^8$     \\
6 & $I_3 \rightarrow I_2$ &  9.62420 $\times 10^8$ &  0.01849 $\times 10^8$   \\
7 & $I_3 \rightarrow I_2$ &  9.62350 $\times 10^8$ &  0.01780 $\times 10^8$\\
8 & $I_3 \rightarrow I_4$ &  9.62494 $\times 10^8$ &  0.01924 $\times 10^8$    \\
9 & $I_3 \rightarrow I_2$ &  9.62526 $\times 10^8$ &  0.01956 $\times 10^8$    \\
10& $I_3 \rightarrow I_4$ &  9.62608 $\times 10^8$ &  0.02037 $\times 10^8$   \\
\botrule
\end{tabular}}
\label{EI}
\end{table}

It is also informative to track the position dependence of the interstitial energy after it relaxes. Numerical measurements of local relaxed energy for the interstitial as a function of initial location  are presented in Table ~\ref{EI}. In most cases the initial $I_3$ complexion undergoes transitions to more stable interstitial configurations, except the configuration that starts from the very center location $C$ in Fig.~\ref{initial}. For the initial conditions, $C$, $C'$, $0$, and $3$, $I_3$ is the most stable. Note that the interstitial created at the exact center $C$ has the lowest defect energy, while the one nearest to the disclination from the location $10$ requires the largest energy for interstitial defect formation.

\subsection{\label{sec:3q}Spherical Nematics}

\begin{figure}[t]
\begin{center}
\includegraphics[width=0.4\textwidth]{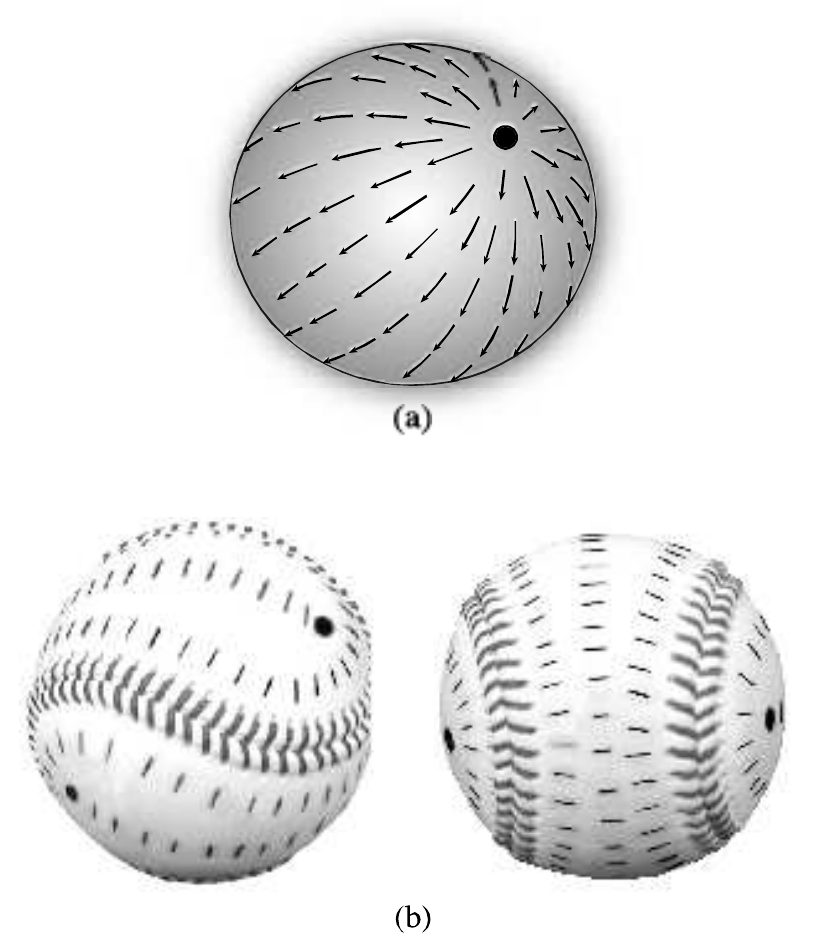}
\caption{(a) Vector order on the sphere with its 2 vortex defects;
(b) Nematic order on the sphere with its 4 type $1/2$ disclination
defects. Taken from \cite{Nelson:2002}.}\label{fig__vectororder}
\end{center}
\end{figure}

The case of nematic order on the sphere is also of great interest from both a fundamental and materials science viewpoint \cite{Nelson:2002,VitelliNelson:2006}. The free energy for nematogens on the sphere is given by 
\be 
F = - \frac{\pi K}{8} \sum_{i \neq
j}n_in_j\ln(1 - \beta_{ij}) + E(R)\sum_j{n_j}^2 \ , 
\ee 
where $\beta_{ij}$ is the geodesic angle between defects $i$ and $j$ with winding numbers $n_i$ and $n_j$, and $E(R)$ is the defect self-energy. Nematic spheres might be made by coating a droplet with gemini lipids, ABA triblock copolymers or nanorods. Microfluidic techniques for creating a thin spherical shell of liquid crystal in a double emulsion have been explored in \cite{Fernandez-Nieves:2007,LopezNieves:2009}. Although the tetrahedral array of four type-$1/2$ nematic disclinations might be unstable to Boojum formation (type-$1$ defects) for spheres immersed in a nematic bulk fluid \cite{Fernandez-Nieves:2004}, Huber and Stark \cite{HuberStark:2004} have shown that nematic wetting for spheres immersed in an isotropic fluid is a promising alternative route to a spherical nematic.

An important motivation for exploring nematic order on spherical topologies is the search for superatoms of low valency. We have seen that spherical crystals have 12 distinguished regions, each region being marked by the presence of either an isolated 5-disclination or an extended defect array like a scar. By functionalizing these regions one could hope to engineer 12-valent supermolecules. In particular one might be able to take advantage of the unusual presence of 5-7 pairs (dislocations) in scars to convert scars to sticky zones of the spherical superatom. Twelve is very likely too high a coordination number, however, to produce viable molecules or bulk materials. How can we change the coordination number of spherical superatoms? The solution lies in changing the local order on the sphere. An ordered state on the sphere with $p$-fold symmetry will have $2p$ net topological defects by Eq. \eqref{eq:sec2-index-theorem}. The crystalline case discussed so far corresponds to $p=6$. Vector-like order on the sphere corresponds to $p=1$ and the defects correspond to index-one vortices {--} one can also visualize the source and sink of fluid flow on a sphere [see Fig.~\ref{fig__vectororder}(a)] or imagine the base and crown bald spots of hair combed on a sphere. Thus a 2-sphere decorated with magnetic (or any other vector) field lines is a candidate for a valence-2 superatom. Consider now {\em nematic} order: $p=2$. In this case the vortex configurations above each break up into 2 type-$1/2$ nematic disclinations, leading to a total of 4 defects located at the corners of a tetrahedron (in the one Frank constant approximation) \cite{LubenskyProst:1992} (see Fig.~\ref{fig__vectororder}(b)). This raises the possibility of making tetravalent superatoms with $sp^{3}$ type bonding \cite{Nelson:2002}.

The precise geometry of the defect structure is important for determining what kind of chemical linkers could be attached to the defects to facilitate the creation of large scale structures like a diamond-type lattice of linked tetravalent colloids. Diamond lattices are desirable for photonic crystals (optical analogs of semiconductors) since they are known to possess a large photonic band gap \cite{HoChangSoukoulis:1990}.  

Although the elementary type-$1/2$ disclinations favor a tetrahedral geometry if splay and bend deformations are degenerate in energy this changes when elastic anisotropy is present. In the limit of pure splay or pure bend the energy of a $+1$ defect becomes degenerate with two $1/2$ defects.  Thus two $+1$ defects sitting at the north and south pole can be separated be cutting the sphere on a great circle joining the two defects and rotating by an arbitrary angle $\alpha$. This gives rise to a set of four $1/2$ disclinations lying on the same great circle and therefore coplanar in the embedding space \cite{ShinBowickXing:2008}. Thus the tetrahedral configuration crosses over to a {\em great-circle} configuration as one increases the anisotropy of the elastic constants. This explicitly demonstrates that defect positions can be controlled by varying the elastic anisotropy. Thus the structure of directional bonding for the associated supermolecules is controllable.

\begin{figure}[h]
\centering
\includegraphics[width=0.7\textwidth]{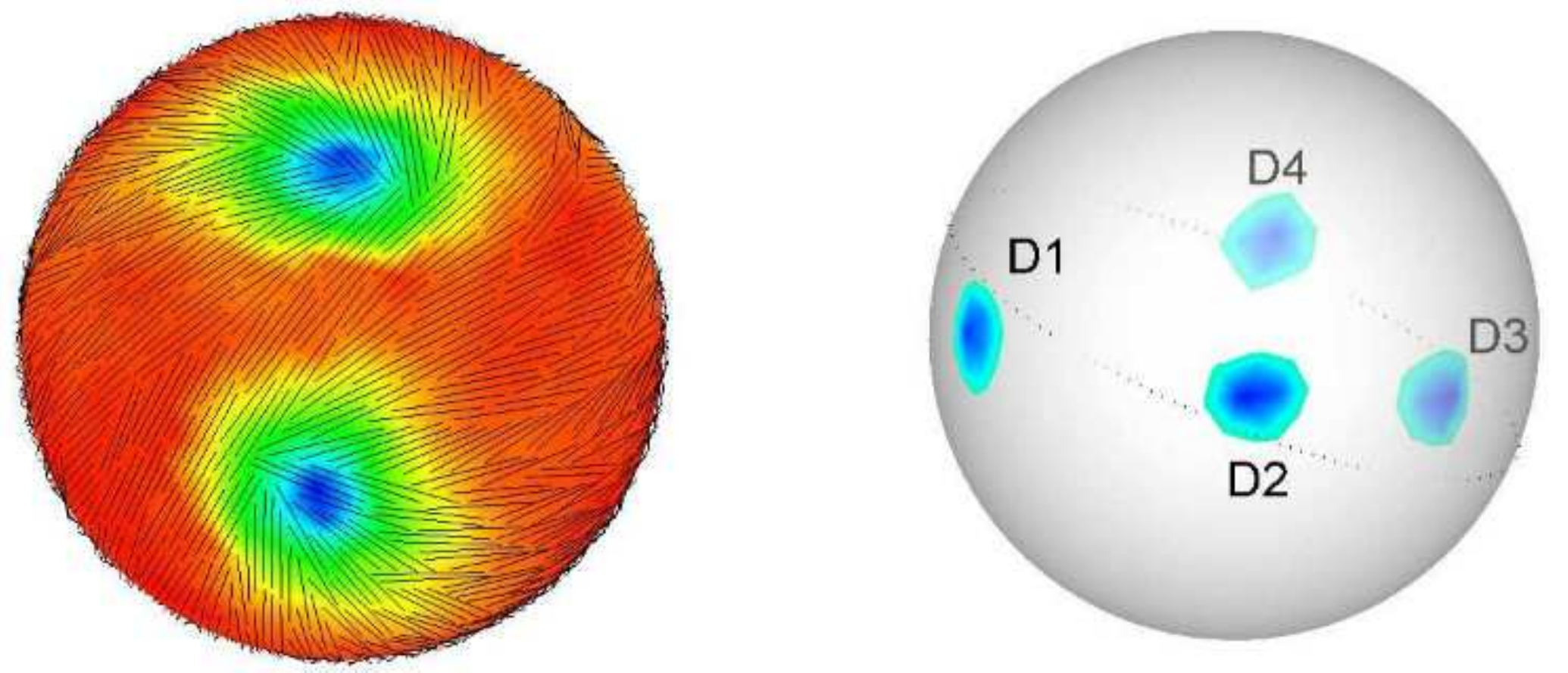}
\caption{\label{fig:sec3-spherical-nematics}(Color online) (Left) the ground state configuration of 1082 nematic rods of aspect ratio 15 on a unit sphere. (Right) the exact location of the four disclinations lying near one great circle. From \cite{ShinBowickXing:2008}.}
\end{figure}

\section{\label{sec:4}Crystalline order on surfaces with variable Gaussian curvature and boundary}

\subsection{\label{sec:4a}Introduction}

Under specific experimental conditions amphiphilic molecules in solution, such as lipids or amphiphilic block copolymers, self-assemble in a spectacular variety of shapes including spherical and cylindrical micelles, vesicles and lamellae, together with more complex geometries such as uni- and multilamellar vesicles, onion vesicles, toroidal and cage-shaped micelles. The exact shape and size of these structures has been observed to depend on both molecular details, such as molecular size, the hydrophilic/hydrophobic ratio and molecular stiffness, and collective parameters, such as the concentration or the diffusivity of the molecules. 

As noted previously, amphiphilic membranes can exist in different thermodynamic states depending on the degree of orientational and positional order and their precise molecular constituents. The $L_{\beta}'$ and $P_{\beta}$ phases, which occur in lipid membranes featuring the phosphatidylcholine (PC) group, have been found, in particular, to exhibit in-plane orientational correlations extending over 200 $\AA$, one order of magnitude larger than the typical spacing between PC groups \cite{SmithEtAl:1988,RaghunathanKatsaras:1996}. Because of thermal fluctuations, such membranes are typically corrugated, possessing spatially inhomogeneous Gaussian and mean curvature. Furthermore it is likely for lipid membranes to have pores providing a passage from the exterior to the interior of a vesicle. It is therefore natural to ask how variable Gaussian curvature and the presence of one or more boundaries affects the  phenomenology reviewed thus far for the sphere.

In this section we discuss crystalline order for two important surfaces with variable Gaussian curvature and (possibly) boundary, namely the bumpy surface that is obtained by revolving the graph of a Gaussian function around its symmetry axis (i.e. a \emph{Gaussian bump}) and the paraboloid of revolution. The former can be thought as a gentle deformation of a plane and thus can serve as a playground to analyze the onset of behavior not found in planar systems; the latter is possibly the simplest two-dimensional Riemannian surface having variable Gaussian curvature and boundary and provides a setting that is simple enough to carry out a full analytical treatment and analyze also large curvature regimes. Furthermore, since paraboloidal shapes naturally occur across the air/liquid interface of a fluid placed in a rotating cylindrical vessel, a direct physical realization of paraboloidal crystals can be constructed by assembling monodisperse objects on the surface of a rotating liquid. In Sec. \ref{sec:4e} we review a simple experiment by Bowick \emph{et al} \cite{BowickEtAl:2008} in which such a macroscopic model for a paraboloidal crystal is constructed by assembling a two-dimensional soap bubble ``raft'' on the air/liquid interface of a water-soap solution, thus extending the classic work of Bragg and Nye on planar bubble rafts. Purely orientational order on the Gaussian bump has been recently reviewed by Turner \emph{et al} \cite{TurnerEtAl:2008} and won't be discussed here.

\subsection{\label{sec:4b}Surfaces of revolution and conformal mapping}

Before analyzing the ground state structure of a crystalline paraboloid and Gaussian bump, it is useful to review the geometry of surfaces of revolution. The notion of conformal mapping of Riemannian surfaces will also be used in the following and will be briefly reviewed in this section with special attention to its application to the calculation of Green functions on simply connected $2-$manifolds. 

A surface of revolution $M$ is a surface obtained by revolving a two-dimensional curve around an axis. The resulting surface therefore always has azimuthal symmetry. The standard parametrization of a surface of revolution is:
\begin{equation}\label{eq:sec3-surface-revolution} 
\left\{
\begin{array}{l}
x = \xi(r)\cos\phi\\
y = \xi(r)\sin\phi\\
z = \eta(r)
\end{array}
\right.
\end{equation}
where $r\in[0,R]$ (with $R$ possibly infinite) and $\phi=[0,2\pi)$. The metric of the surface \eqref{eq:sec3-surface-revolution} is given by:
\begin{equation}\label{eq:sec3-metric-revolution}
ds^{2}=[(\xi')^{2}+(\eta')^{2}]dr^{2} + \xi^{2}d\phi^{2}\,,
\end{equation}
where the prime indicate a partial derivative with respect to $r$. The Gaussian curvature is a function of $r$ only and is given by:
\begin{equation}
K = \frac{\eta'(\xi'\eta''-\xi''\eta')}{\xi(\xi'^{2}+\eta'^{2})^{2}} \ .
\end{equation}\label{eq:sec3-euler-boundary}%
In the presence of a boundary $\partial M$ the condition \eqref{eq:sec2-topological-charge} for the total topological charge of any triangulation on $M$ reads:
\begin{equation}
Q = \sum_{i=1}^{V_{\partial M}}(4-c_{i})+\sum_{i=1}^{V_{M}}(6-c_{i}) = 6\chi
\end{equation}
where $V_{\partial M}$ and $V_{M}$ are the number of vertices on the boundary and the interior of the manifold respectively (with $V=V_{\partial M}+V_{M}$ the total number of vertices). $q_{i,\partial M}=4-c_{i}$ is the topological charge of a vertex of coordination number $c_{i}$ located on the boundary, where the coordination number of a perfect triangular lattice is four rather than six. A surface of revolution with boundary $\partial M = \{r=R\}\times [0,2\pi]$ is homeomorphic to a disk ($g=0$ and $h=1$) and has therefore $\chi=1$ and total topological charge $Q=6$. This topological constraint can be satisfied, for instance, by placing six isolated $3-$fold disclinations along the boundary and keeping the interior of the surface defect-free or by placing a $5-$fold disclination in the interior and the remaining five $3-$fold disclinations along the boundary. 

Let us consider now a generic Riemannian surface $M$ and two curves on $S$ intersecting at some point $\bm{x}_{0}$. The angle between the two intersecting curves is, by definition, the angle between the tangents to these curves at $\bm{x}_{0}$. A mapping of a portion $S$ of a surface onto a portion $S^{*}$ is called \emph{conformal} (or angle-preserving) if the angle of intersection of every arbitrary pair of intersecting arcs on $S^{*}$ is the same as that of the corresponding inverse images on $S$ at the corresponding point (see for example \cite{Kreyszig}). It is not difficult to prove that a mapping from a portion $S$ of a surface onto a portion $S^{*}$ is conformal if and only if, when on $S$ and $S^{*}$ the same coordinate systems have been introduced, the coefficients $g_{ij}^{*}$ and $g_{ij}$ of the metric tensor of $S^{*}$ and $S$ are related by:
\begin{equation}\label{eq:sec3-conformal-mapping1}
g^{*}_{ij} = w(\bm{x})g_{ij}\,,
\end{equation}
with $w$ a positive function of the coordinates $\bm{x}=(x^{1},x^{2})$. Eq. \eqref{eq:sec3-conformal-mapping1} implies indeed that the angle between any pair of intersecting curves is the same in $S^{*}$ and $S$. \emph{Isometries} are a special case of conformal mappings where $w=1$ and the mapping is both distance and angle-preserving. 

A special type of conformal mapping is that of a portion $S$ of a surface into a plane. This may be accomplished by introducing a set of coordinates $\bm{u}=(u^{1},u^{2})$ such that:
\begin{equation}\label{eq:sec3-conformal-mapping2}
ds^{2} = w(\bm{u})[(du^{1})^{2}+(du^{2})^{2}]\,.
\end{equation}
Coordinates $(u^{1},u^{2})$ satisfying Eq. \eqref{eq:sec3-conformal-mapping2} are called \emph{isothermal} (or conformal). In general, any simply connected Riemannian manifold with a $C^{\infty}-$smooth metric $ds^{2}$ can be equipped with a set of local isothermal coordinates. This important result can be stated by saying that any simply connected Riemannian manifold is locally conformally equivalent to a planar domain in two-dimensions. In conformal coordinates, the Gaussian curvature reads:
\begin{equation}
K = -\frac{2\Delta[\log w(\bm{u})]}{w(\bm{u})} \ .
\end{equation}
Conformal mapping is the fundamental tool behind the celebrated uniformization theorem according to which, \emph{every simply connected Riemannian surface is conformally equivalent to the unit disk, the complex plane or the Riemann sphere}. This theorem, first proved by Koebe and Poincar\'e independently in 1907, extends the Riemann mapping theorem for simply connected domains in the complex plane to all simply connected Riemannian surfaces and provides an insightful classification scheme. Its formidable power lies in the fact that the mapping that allows one to transform a generic surface into a simpler ``irreducible'' one is not an arbitrary homeomorphism but is conformal, and thus preserves part of the geometrical structure of the original manifold. In the following we will see how this feature has important consequences in the elastic theory of defects on curved surfaces. The uniformization of Riemannian surfaces is historically the first example of geometrization. In the case of manifolds of higher Hausdorff dimension, and 3-manifolds in particular, the latter program has become, following Thurston, one of the most challenging and fascinating chapters of modern geometry.

A bounded Gaussian bump and a paraboloid of revolution are both conformally equivalent to the unit disk $\mathbb{D}$ of the complex plane. Calling $z=\varrho e^{i\phi}$, the new metric will be:
\begin{equation}\label{eq:sec3-conformal-metric}
ds^{2} = w(z)(d\varrho^{2}+\varrho^{2}d\phi^{2}) \ .
\end{equation}
The conformal factor $w$ can be found by equating the metrics \eqref{eq:sec3-metric-revolution} and \eqref{eq:sec3-conformal-metric}. This yields:
\begin{equation}
w(\rho) = \left[\frac{\xi(r)}{\varrho}\right]^{2} \ ,
\end{equation}
with $\varrho$ and $r$ related by the differential equation
\begin{equation}
\frac{d\varrho}{dr}\pm\sqrt{\frac{(\xi')^{2}+(\eta')^{2}}{\xi^{2}}}\,\varrho = 0 \ ,
\end{equation}
whose solution is given by:
\begin{equation}\label{eq:sec3-conformal-distance}
\varrho = \exp\Biggl\{\pm\int dr\,\sqrt{\frac{(\xi')^{2}+(\eta')^{2}}{\xi^{2}}}\Biggr\} \ .
\end{equation}
The sign of the exponent and the integration constant in Eq. \eqref{eq:sec3-conformal-distance} can be tuned to obtain the desired scale and direction of the conformal map. 

The calculation of the Green function of the Laplace and biharmonic operator is considerably simplified once a surface of revolution has been endowed with a local system of isothermal coordinates. In this case it is easy to show the Laplace-Beltrami operator $\Delta_{g}$ takes the form:
\begin{equation}
\Delta_{g} = w^{-1}\Delta \ ,
\end{equation}
where $\Delta$ is now the Laplacian in the Euclidean metric tensor:
\begin{equation}
\gamma_{\varrho\varrho} = 1,\qquad \gamma_{\varrho\phi} = 0, \qquad \gamma_{\phi\phi} = \varrho^{2},
\end{equation}
with determinant $\gamma$. Since the determinant of the metric tensor is rescaled ($\sqrt{g}\rightarrow w\,\sqrt{\gamma}$) under conformal mapping, the Laplace and biharmonic equation for the Green function become:
\begin{subequations}
\begin{gather}
\Delta G_{L}(z,\zeta) = \delta(z,\zeta)\label{eq:sec3-conformal-laplace-green} \\[5pt]
\Delta w^{-1} \Delta G_{2L}(z,\zeta) = \delta(z,\zeta) \label{eq:sec3-conformal-biharmonic-green}
\end{gather}
\end{subequations}
where $\delta(z,\zeta)$ is the standard delta function at the point $z=\zeta$ of the unit disk. Eq. \eqref{eq:sec3-conformal-laplace-green} is now the standard Laplace-Green equation. Its associated Dirichlet problem has the familiar solution:
\begin{equation}\label{eq:sec3-disk-green}
G_{L}(z,\zeta) = \frac{1}{2\pi}\log\left|\frac{z-\zeta}{1-z\overline{\zeta}}\right| \ .
\end{equation}
Eq. \eqref{eq:sec3-conformal-biharmonic-green} is known as the \emph{weighted} biharmonic Green equation. The uniqueness of its solution requires imposing both Dirichlet and Neumann boundary conditions:
\begin{equation}
\left\{
\begin{array}{lr}
G_{2L}(z,\zeta) = 0 & z\in\partial\mathbb{D} \\[5pt]
\partial_{n(z)}G_{2L}(z,\zeta) = 0 & z\in\partial\mathbb{D} 
\end{array}
\right.
\end{equation}
where $\partial_{n(z)}$ denotes the derivative with respect to the variable $z$ along the normal direction at $\partial\mathbb{D}$. Its solution can be expressed in integral form as
\begin{equation}
G_{2L}(z,\zeta) = \int d^{2}\sigma\,G_{L}(z,\sigma)[G_{L}(\sigma,\zeta)-H(\sigma,\zeta)] \ ,
\end{equation}
where $H(\sigma,\zeta)$ is a harmonic kernel that enforces the Neumann condition. Such a function depends on the form of the conformal weight $w$. For radial weights $w=w_{0}(|z|^{2})$, such as those obtained by conformally mapping a surface of revolution, $H(\sigma,\zeta)$ has been calculated explicitly by Shimorin \cite{Shimorin:1998}:
\begin{equation}\label{eq:sec3-harmonic-kernel}
H(\sigma,\zeta) = -2\int_{|\zeta|}^{1}\frac{dt}{t}\int_{0}^{t^{2}}ds\,w_{0}(s)k\left(\frac{s}{t^{2}}\,\zeta\,\overline{\sigma}\right)
\end{equation}
where:
\begin{equation}
k(z\,\overline{\zeta}) = \sum_{n\ge 0}\frac{(z\,\overline{\zeta})^{n}}{c_{n}}+\sum_{n<0}\frac{(\overline{z}\,\zeta)^{|n|}}{c_{|n|}}
\end{equation}
and the coefficients $c_{n}$ are given by:
\begin{equation}
c_{n} = 2\int_{0}^{1}dt\,t^{n}w_{0}(t)\,.
\end{equation}

\subsection{\label{sec:4c}Crystalline order on the Gaussian bump}

The most natural way of introducing a non-vanishing Gaussian curvature on an initially flat medium is to gently deform the medium at one point in such a way that the curvature introduced by this deformation dies off at infinity. If the initial planar domain is the entire Euclidean plane $\mathbb{R}^{2}$, the Euler characteristic is zero and, independently of the value of the Gaussian curvature, an ordered phase embedded on it is not topologically required to contain defects. Such a construction is clearly ideal to detect the onset of structural behavior not occurring in the ground state of a planar system such as the appearance of dislocations and disclinations. Vitelli \emph{et al} \cite{VitelliLucksNelson:2006,VitelliNelson:2004} analyzed crystalline and $p-$atic order on the bumpy surface obtained by revolving the graph of a Gaussian function about its symmetry axis. The resulting \emph{Gaussian bump} has parametrization:
\begin{equation}\label{eq:sec3-gaussian-bump}
\left\{
\begin{array}{l}
x = r\cos\phi \\
y = r\sin\phi \\
z = h\exp\left(-\frac{r^{2}}{2r_{0}^{2}}\right)
\end{array}
\right.
\end{equation}
with $r\in[0,R]$ and $\phi\in[0,2\pi)$. In this parametrization the metric tensor $g_{ij}$ (with determinant $g$) and the Gaussian curvature are given by:
\begin{subequations}
\begin{gather}
g_{rr} = \ell(r),\qquad g_{r\phi}=0, \qquad g_{\phi\phi} = r^{2},\\[7pt]
K = \frac{\alpha^{2}e^{-\frac{r^{2}}{r_{0}^{2}}}}{r_{0}^{2}\,\ell^{2}(r)}\left(1-\frac{r^{2}}{r_{0}^{2}}\right)\,,
\end{gather}
\end{subequations}
where $\alpha=h/r_{0}$ is the aspect ratio of the bump and $\ell(r)$ is given by
\begin{equation}
\ell(r) = 1+\frac{\alpha^{2}r^{2}}{r_{0}^{2}}\,e^{-\frac{r^{2}}{r_{0}^{2}}} \ .
\end{equation}
It is instructive to verify that the Euler characteristic $\chi$ vanishes when the boundary radius $R$ is set to infinity. Employing the Gauss-Bonnet theorem one has
\begin{equation}\label{eq:sec3-gauss-bonnet}
\chi = \int_{0}^{R} dr\,\sqrt{g}\,K+\frac{1}{2\pi}\oint_{C_{R}} ds\,\kappa_{g} \ ,
\end{equation}
where $\kappa_{g}$, the geodesic curvature of the circular boundary $C_{R}$ of radius $R$, is given by
\begin{equation}
\kappa_{g} = \frac{1}{R\sqrt{\ell(R)}} \ .
\end{equation}
If the bump is unbounded the second term in Eq. \eqref{eq:sec3-gauss-bonnet} disappears and, in the limit $R\rightarrow\infty$, one has
\[
\int_{0}^{\infty} dr\,\sqrt{g}\,K = 0
\] 
which implies $\chi=0$. For finite values of $R$, on the other hand, the first integral in Eq. \eqref{eq:sec3-gauss-bonnet} gives:
\[
\int_{0}^{R} dr\,\sqrt{g}\,K = 1-\frac{1}{\sqrt{\ell(R)}} \ ,
\]
so that the terms proportional to $1/\sqrt{\ell(R)}$ cancel each other, yielding $\chi=1$. Because the infinite bump has $\chi=0$ disclinations must appear in pairs and dislocations must have total Burgers vector zero:
\[
\sum_{i}q_{i} = \sum_{i}\bm{b}_{i} = 0 \ .
\]
Vitelli \emph{et al} showed that topological defects appear in the ground state when the aspect ratio $\alpha$ exceeds a critical value $\alpha_{c}$. Because of the topological constraint, defects appear initially in the form of a pair of unbound dislocations, roughly located in the region where $K=0$. Upon increasing the aspect ratio, more dislocations appear. This mechanism clearly resembles the defect proliferation that occurs in two-dimensional melting and suggests an interpretation of the curvature as a local effective temperature.

At the onset of defect proliferation, inter-defect interactions are negligible compared to the interactions with the ``smeared out'' topological charge associated with the curvature of the underlying medium. The dislocation unbinding mechanism is then described by a non-local function of the Gaussian curvature representing the defect-curvature interaction part of the elastic energy of Eq. \eqref{eq:sec2-xy-curved3}:
\begin{equation}\label{eq:sec3-defect-curvature} 
F_{int} 
= Y\int d^{2}x\,\eta(\bm{x})\int d^{2}y\,G_{2L}(\bm{x},\bm{y})K(\bm{y})
= Y \int d^{2}x\,\eta(\bm{x})\varphi(\bm{x})\,,
\end{equation}
where $F_{int}$ is the curvature-defect interaction part of the elastic energy and $\varphi(\bm{x})$ can be interpreted as a \emph{geometric potential} associated with the Gaussian curvature of the embedding manifold:
\begin{equation}\label{eq:sec3-gp-biharmonic}
\Delta^{2}\varphi(\bm{x}) = K(\bm{x})\,.
\end{equation}

\begin{figure}
\centering
\includegraphics[width=0.6\textwidth]{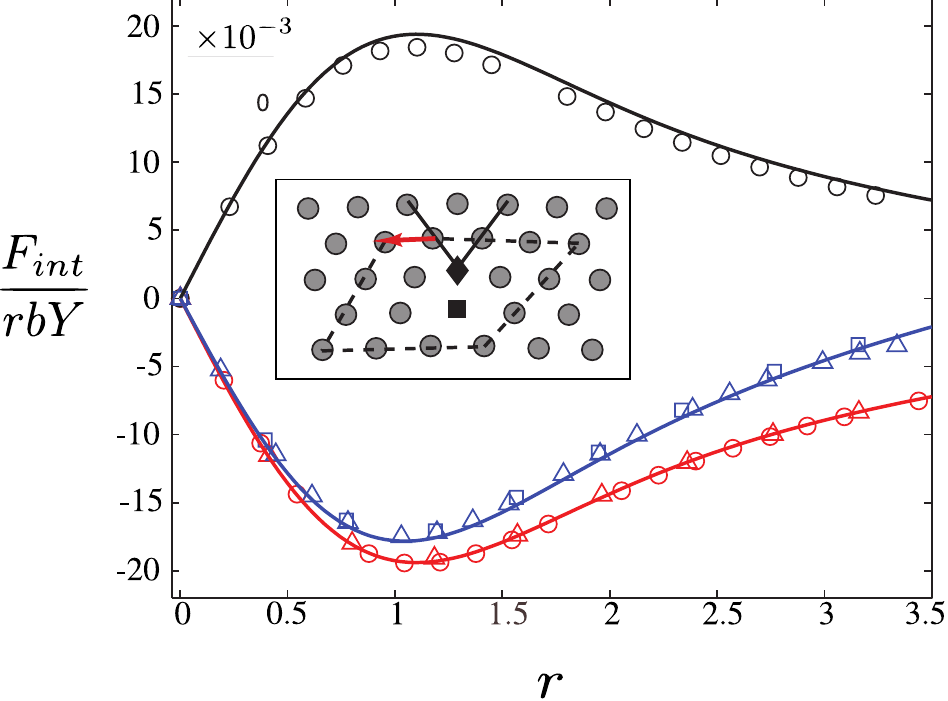}
\caption{\label{fig:sec3-dislocation-potential}(Color online) Curvature-defect interaction energy of an isolated disclination for a Gaussian bump of aspect ratio $\alpha=0.5$. Open symbols represent the data from a numerical minimization of a fixed connectivity harmonic model. The lower and upper branch are obtained from Eq. \eqref{eq:sec3-gp-dislocation} by setting $\theta=\pm\pi/2$ and letting $\Lambda$ equal $4$ (blue curve) and $8$ (red curve). [Courtesy of V. Vitelli, University of Pennsylvania, Philadelphia, PA].}
\end{figure}

Thus
\begin{equation}\label{eq:sec3-gp}
\varphi(\bm{x}) = \int d^{2}z\,d^{2}y\,G_{L}(\bm{x},\bm{y})[G_{L}(\bm{y},\bm{z})K(\bm{z})+U(\bm{y})]\,,
\end{equation}
where $U(\bm{y})$ is a harmonic function which enforces the boundary conditions. The Laplacian Green function has the form \eqref{eq:sec3-disk-green} with the conformal distance $\rho=|z|$ given here by
\begin{equation}
\varrho = \frac{r}{R}\exp\left\{-\int_{r}^{R}dr'\,\frac{\sqrt{\ell(r')}-1}{r'}\right\} \ .
\end{equation}
For a single dislocation of Burgers vector $\bm{b}$, $F_{int}$ has the form \cite{VitelliLucksNelson:2006}:
\begin{equation}\label{eq:sec3-gp-dislocation}
F_{int} = \frac{1}{8}hb\alpha^{2}Y\sin\phi\left(\frac{e^{-\lambda^{2}}-1}{\lambda}+\frac{\lambda}{\Lambda^{2}}\right)\,,
\end{equation}
where $\lambda=r/r_{0}$ and $\Lambda=R/r_{0}$. The first term in Eq. \eqref{eq:sec3-gp-dislocation} corresponds to the $R\rightarrow\infty$ geometric potential, while the second is a finite size correction arising from a circular boundary of radius $R$. A plot of the function \eqref{eq:sec3-gp-dislocation} is shown in Fig. \ref{fig:sec3-dislocation-potential}. The profile of the function $F_{int}$ can be undestood by regarding a dislocation as bound pair of disclinations of opposite topological charge. Each disclination interacts with a potential of the form \eqref{eq:sec3-gp-biharmonic}. For small $r$, positive (negative) disclinations are attracted (repelled) by the center of the bump. As a consequence disclinations experience a force which increases linearly with $r$. Thus if the positive disclination in the dipole is closer to the top, it will experience a force that is opposite and slightly less than that acting on the negative disclination that is further from the top. As a result an effective ``tidal'' force will push the dislocation downhill. For large $r$, however, the geometric potential saturates and the attractive force exerted on the positive disclination takes over and drags the dislocation toward the center of the bump. The minimum of the elastic energy corresponding to the equilibrium between these two competing forces is obtained for $\lambda \approx 1.1$. The origin of these forces is the Peach-Koehler force $f_{k}=\epsilon_{kj}b_{i}\sigma_{ij}$ acting on a dislocation of Burgers vector $b_{i}$ in an external stress field $\sigma_{ij}$. In the case of a two-dimensional crystal on a substrate with preexisting Gaussian curvature, dislocations couple with the internal stress due to the curvature of the medium, with similar effects.

\begin{figure}[t]
\centering
\includegraphics[width=0.9\textwidth]{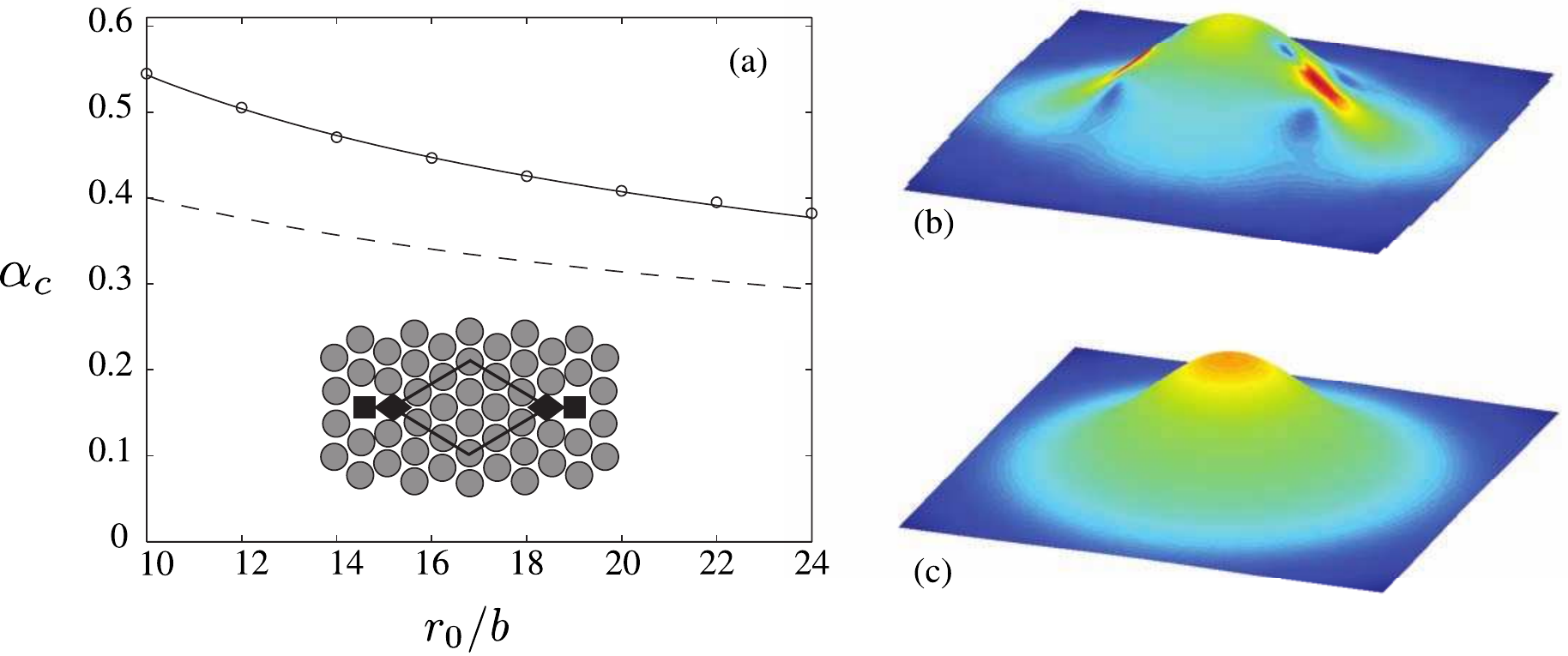}
\caption{\label{fig:sec3-dislocation-unbind}(Color online) Dislocation unbinding. (a) Critical aspect ratio $\alpha_{c}$ as a function of $r_{0}/b$. The theoretical estimate \eqref{eq:sec3-critical-alpha} is plotted versus $r_{0}/b$ for core energies $E_{d}=0$ (dashed line) and $E_{c}=0.1 B^{2}Y$ (solid line). Circles are obtained from a numerical minimization of a fixed connectivity harmonic model. On right the logarithm of the numerically calculated strain energy for a (c) defect-free configuration on a bump of $r_{0}=10b$ and $\alpha=0.7>\alpha_{c}$ and (d) the defective configuration shown in the inset. [Courtesy of V. Vitelli, University of Pennsylvania, Philadelphia, PA].}
\end{figure}

Unbound dislocations occur in the ground state of a crystalline Gaussian bump when the aspect ratio $\alpha$ exceeds a critical value. For nearly flat landscapes the energetic cost of a pair of unbound dislocations is larger than that arising from the distortion of the medium and the system is favored to be defect free. Upon increasing the aspect ratio, however, the resulting elastic strain can be partially relieved by introducing a pair of dislocations with equal and opposite Burgers vector. The transition occurs when the energy gain from placing each dislocation in the minimum of the potential energy $F_{int}$ outweights the total work needed to tear them apart plus the core energies $2E_{d}$. The resulting critical aspect ratio is given by:
\begin{equation}\label{eq:sec3-critical-alpha}
\alpha_{c}^{2} \approx \frac{b}{2r_{0}}\log\left(\frac{r_{0}}{b'}\right)\,,
\end{equation}
where $b'=(b/2)e^{-8\pi E_{c}/(Yb^{2})}$ . The number of unbound dislocations increases with increasing aspect ratio and elementary $5-7$ dislocations cluster in more complicated structures with zero net Burgers vector. Fig. \ref{fig:sec3-dislocation-unbind} shows a plot of the critical aspect ratio $\alpha_{c}$ as the function of the dimensionless parameter $r_{0}/b$ from Ref. \cite{VitelliLucksNelson:2006}, as well as density plots of the strain energy corresponding to a defect-free bump (c) and a representative configuration featuring two unbound dislocations (b).

\begin{figure}[t]
\centering
\includegraphics[width=0.9\textwidth]{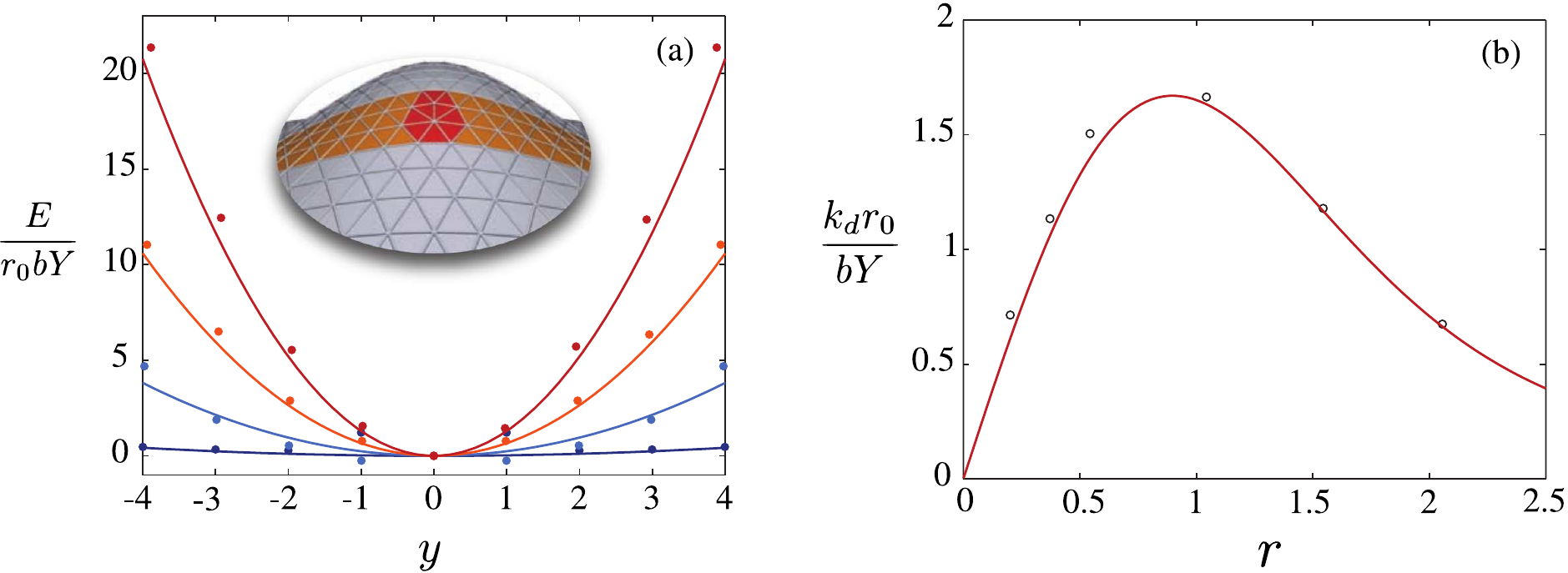}
\caption{\label{fig:sec3-dislocation-glide}(Color online) Dislocation glide. (a) Recalling potential $\frac{1}{2}k_{d}y^{2}$ acting on dislocations gliding in the direction $y$ (expressed in units of the lattice spacing $a$) for $r_{0}/a=10$, $\lambda=0.5$ (at $y=0$), $\Lambda=8$ and $\alpha=0.1$ (dark blue), $0.3$ (blue), $0.5$ (orange) and $0.7$ (red). The energy is scaled by $10^{-4}$. Effective spring constant for $r_{0}=10a$ and $\alpha=0.5$. The ordinate axis is scaled by $10^{-2}$. Filled and empty circles represent numerical data obtained from a fixed connectivity harmonic model. [Courtesy of V. Vitelli, University of Pennsylvania, Philadelphia, PA].}
\end{figure}

The interaction between defects and curvature also has some remarkable stabilizing effects on defect dynamics. Dislocation dynamics consists of two distinct processes: glide and climb. Glide is motion along the direction of the Burgers vector. It requires only local rearrangement of atoms and thus has a very low activation energy, making it dominant at low temperatures. Climb consists of motion in the direction perpendicular to the Burgers vector. It requires diffusion of vacancies and interstitials and is usually suppressed therefore relative to glide. As a consequence of the underlying curvature, however, a gliding dislocation experiences a ``recalling'' force $f_{recall}\approx k_{d}|y|$, where $y$ is the transverse displacement and $k_{d}$ is a position-dependent effective spring constant. At leading order in $y$ and $\alpha$, the latter is given by:
\begin{equation}\label{eq:sec3-spring-constant}
k_{d}(r) = \frac{b\alpha^{2}Y}{4r_{0}}\left[\frac{1-(1+\lambda^{2})e^{-\lambda^{2}}}{\lambda^{3}}\right]\,.
\end{equation}
This effective recalling force is not due to any external field nor to the interaction of dislocations with other defects, but exclusively to the coupling of the gliding dislocation with the curvature of the substrate. As expected, the effective spring constant \eqref{eq:sec3-spring-constant} vanishes for planar crystals (i.e. $\alpha\rightarrow 0$). Since $Yb^{2}$ can be hundreds of $k_{B}T$, at finite temperature the harmonic potential $\frac{1}{2}k_{d}y^{2}$ associated with the recalling force raises the activation energy of thermally induced dislocation glide. The harmonic recalling potential and the effective spring constant \eqref{eq:sec3-spring-constant} are plotted in Fig. \ref{fig:sec3-dislocation-glide}, together with numerical data obtained from a fixed connectivity harmonic model \cite{VitelliLucksNelson:2006}.

Spatial curvature also provides an effective potential in the thermal diffusion of interstitials and vacancies. These can be constructed by grouping three dislocation dipoles. The elastic energy associated with a single interstitial/vacancy can be derived from Eq. \eqref{eq:sec3-defect-curvature} in the form
\begin{equation}
F_{int}(\bm{x})\approx \frac{1}{2}Y\Omega V(\bm{x})\,, 
\end{equation}
where $V=\Delta \varphi$ and $\Omega$ is the area excess or deficit associated with the defects. As a result, interstitials tend to climb to the top of the bump while vacancies are pushed into the flat regions. Like a disclination of positive topological charge, an interstitial is attracted to regions of positive Gaussian curvature while a vacancy is attracted to regions of negative Gaussian curvature. 

\begin{figure}[h]
\centering
\includegraphics[width=0.6\textwidth]{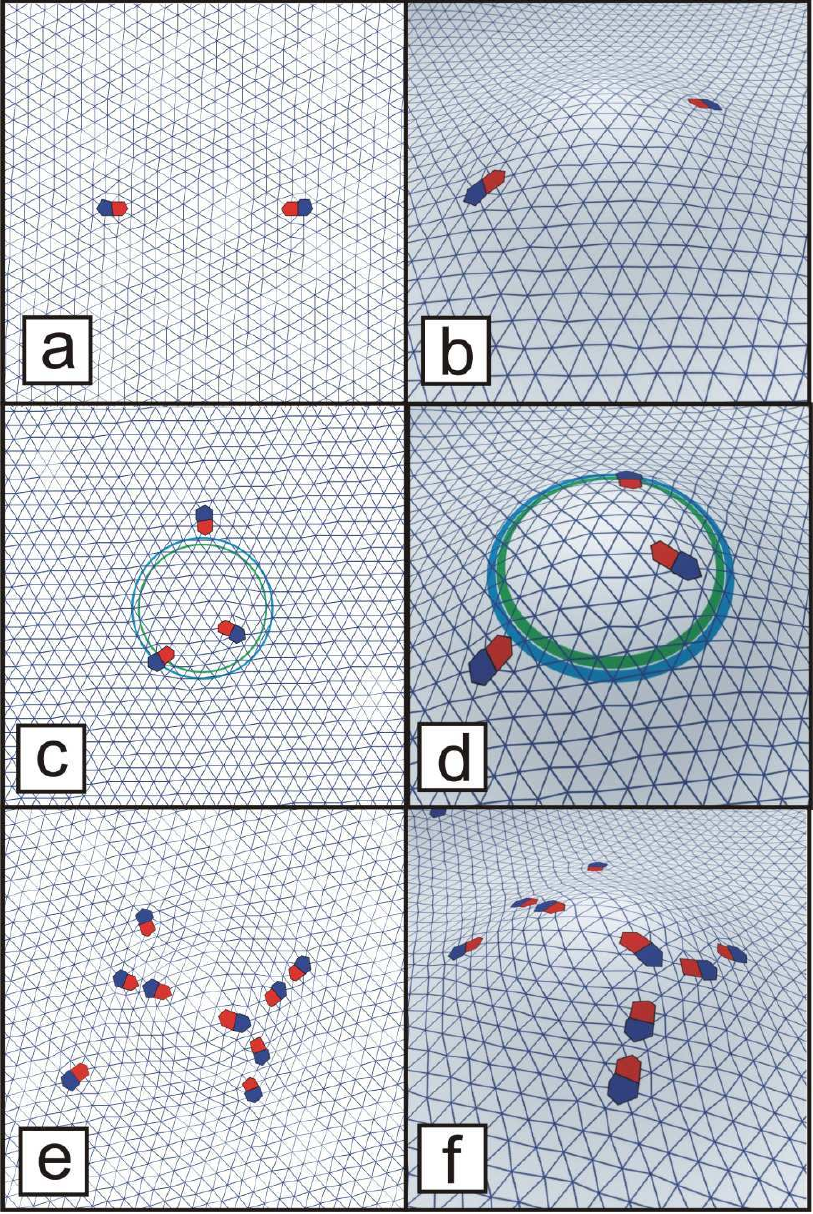}
\caption{\label{fig:sec3-bump-numerics}(Color online) Delaunay triangulation of 780 particles on a Gaussian bump. (a) and (b) Top view of the lattice containing two dislocations of opposite Burgers vector on a bump of aspect ratio $\alpha=0.82$. (c) and (d) Three dislocations arranged around a bump of aspect ratio $\alpha=0.95$. (e) and (f) A more complex arrangements of dislocations on a bump of aspect ratio $\alpha=1.58$. [Courtesy of Alexander Hexemer, UC Santa Barbara, Santa Barbara, CA. Currently at Lawrence Berkeley National Lab., Berkeley, CA].}
\end{figure}

The crystalline structure arising on a Gaussian bump has been investigated numerically by Hexemer \emph{et al} \cite{HexemerEtAl:2007}. By means of smart Monte Carlo (SMC) simulations, the authors analyzed the ground state configuration of a system of $N$ point-like particles (with $N$ up to 780) interacting on a Gaussian bump of variable aspect ratio with a Yukawa potential of the form $U(r)=\exp(-\kappa r)/r$, with $r$ the Euclidean distance between two particles in $\mathbb{R}^{3}$. This potential approximately describes the interaction between charged particles in solution with counter-ions, with $\kappa^{-1}$ being the Deybe-H\"uckel screening length. Starting from an initially defect-free lattice and relaxing it with order $10^{5}$ SMC iterations, Hexemer \emph{et al} observed the appearance of defects for increasing aspect ratio. Fig. \ref{fig:sec3-bump-numerics} shows the arrangement of 780 particles for three different aspect ratios. As predicted by the elastic theory, the proliferation of defects starts with the appearance of a pair of isolated dislocations (Figs. \ref{fig:sec3-bump-numerics}a and b). They are sitting at $\lambda=1.46$ from the center and are rotated by $180^{\circ}$, giving rise to a configuration with total Burgers vector close to zero. Upon increasing the aspect ratio, a third dislocation is observed (Fig. \ref{fig:sec3-bump-numerics}c and d). The three dislocations are rotated by $120^{\circ}$ with respect to each other. The green and blue circles in Fig. \ref{fig:sec3-bump-numerics}c and d mark the region of zero Gaussian curvature and the minimum of the geometric potential at $1.1\,r_{0}$. As shown in the figure, for $\alpha=0.95$ two of the three dislocation are not sitting at $1.1\,r_{0}$ as one might expect. One is deep inside the area of positive Gaussian curvature while the other is in the region of negative Gaussian curvature. Hexemer \emph{et al}  attributed this discrepancy to finite-size effects. 

\begin{figure}[t]
\centering
\includegraphics[width=0.5\textwidth]{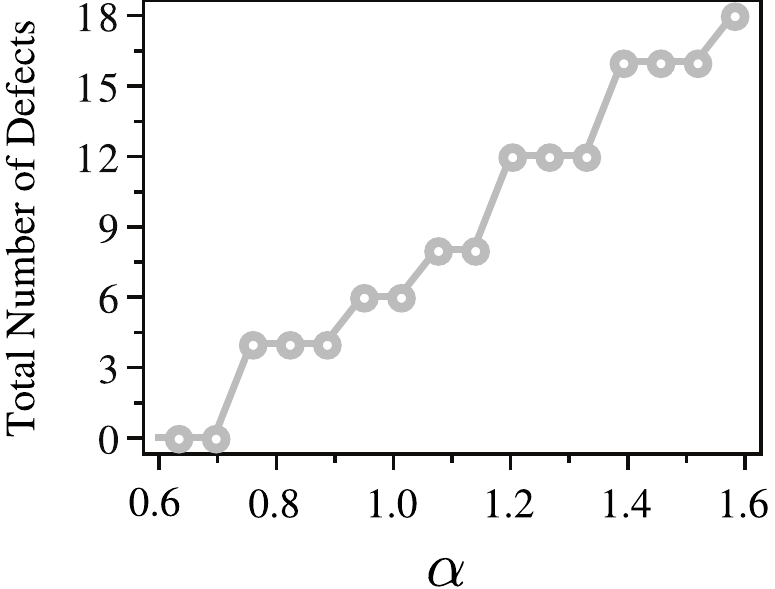}
\caption{\label{fig:sec3-dislocation-number} Total number of disclinations in the ground state as a function of the aspect ratio $\alpha$. [Courtesy of Alexander Hexemer, UC Santa Barbara, Santa Barbara, CA (currently at Lawrence Berkeley National Lab., Berkeley, CA).]}
\end{figure}

Fig. \ref{fig:sec3-bump-numerics}e and f show the lowest energy state on a bump of aspect ratio $\alpha=1.58$. In this situation dislocations appear arranged in the form of grain boundaries. This phenomenon has a simple interpretation. The total elastic energy of a collection of defects on a curved surface consists of three parts: the curvature-defect interaction discussed above, which tends to localize the defects where the geometric potential is minimal, a defect-defect interaction which is repulsive for like-sign defects and attractive for defects of opposite sign and an additional contribution due to the curvature alone. For a small number of dislocations, inter-defect interactions are negligible compared with curvature-defect interactions and dislocations tend to gather where $F_{int}$ is minimal. As shown from the data in Fig. \ref{fig:sec3-dislocation-number}, however, the total number of dislocations in the bump is linearly proportional to the aspect ratio. When the total number of dislocations on the bump exceeds some threshold value the repulsive interaction takes over and dislocations start spreading. Dislocations with parallel (antiparallel) Burgers vector are organized in such a way as to maximize (minimize) their reciprocal distance while still taking advantage of the Gaussian curvature screening. The competition leads to the appearance of the $Y-$shaped grain boundaries shown in Fig. \ref{fig:sec3-bump-numerics}. The branching allows the dislocations to lower their potential energy by keeping a large number of dislocations close to the $1.1\,r_{0}$ circle.  

\subsection{\label{sec:4d}Paraboloidal crystals}

A paraboloid of revolution is a simple surface with both variable Gaussian curvature and boundary. Its standard parametrization is given by
\begin{equation}\label{eq:sec3-paraboloid}
\left\{
\begin{array}{l}
x = r\cos\phi \\
y = r\sin\phi \\
z = \frac{h}{R^{2}}\,r^{2}
\end{array}
\right.
\end{equation}
with $r\in [0,R]$ and $\phi=[0,2\pi]$. Here $h$ is the height of the paraboloid and $R$ the maximum radius. In the following we will call $\kappa=2h/R^{2}$ the normal curvature of the paraboloid at the origin. The metric tensor $g_{ij}$ and the Gaussian curvature are given respectively by:
\begin{subequations}
\begin{gather}
g_{rr} = 1+\kappa^{2}r^{2},\qquad g_{r\phi}=0,\qquad g_{\phi\phi} = r^{2}\,,\\[7pt]
K = \frac{\kappa^{2}}{(1+\kappa^{2}r^{2})^{2}} \ .
\end{gather}
\end{subequations}
The problem of finding the optimal arrangement of disclinations in the ground state of paraboloidal crystals has been considered by the authors of this review article \cite{GiomiBowick:2007a,GiomiBowick:2007b}. As in the case of a Gaussian bump, when the maximal Gaussian curvature exceeds some critical value (depending on $\kappa$ and $R$) defects proliferate in an initially defect-free configuration. Unlike the Gaussian bump, however, the Gaussian curvature on a paraboloid is strictly positive and only vanishes at infinity. Positive isolated disclinations are thus energetically preferred to dislocation dipoles and the defect proliferation mechanism consists of a ``migration'' of one of the six topologically required $+1$-disclinations from the boundary to the origin of the paraboloid. Dislocations, on the other hand, appear in the system clustered in the form of grain boundary scars in the high density regime as in spherical crystals. Since the Gaussian curvature is not constant throughout the manifold, however, this transition is preceded by a regime in which isolated disclinations and scars coexist in the crystal.

\begin{figure}
\centering
\includegraphics[width=0.9\textwidth]{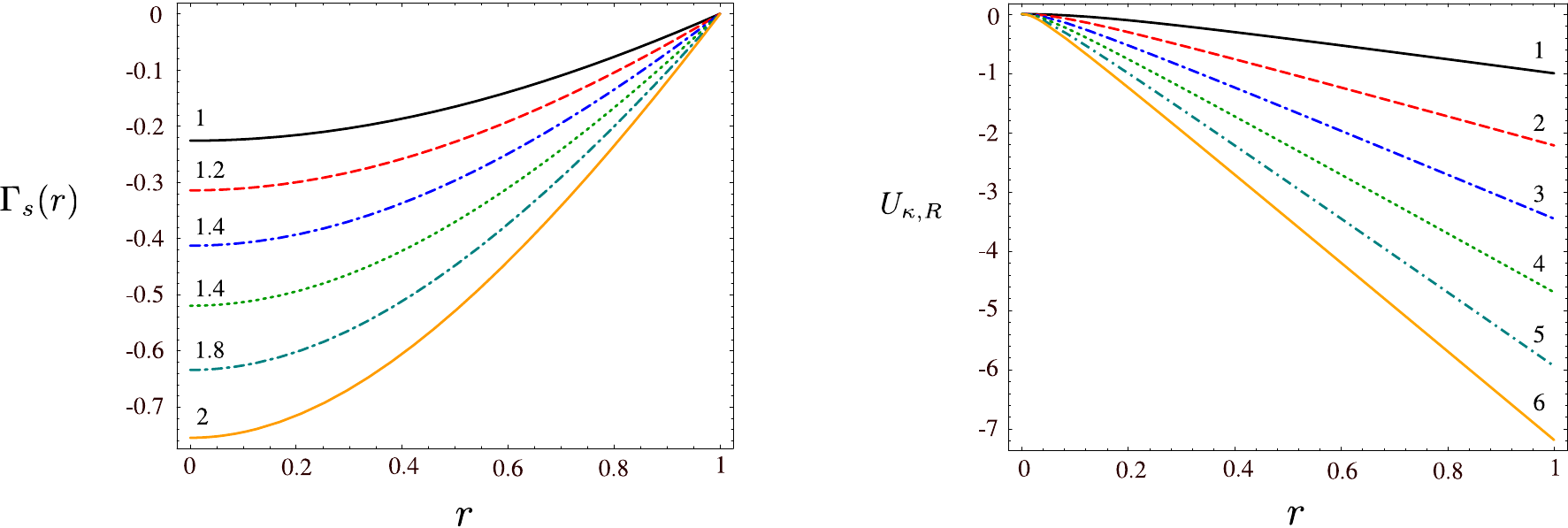}
\caption{\label{fig:sec3-paraboloid-gamma}(Color online) The function $\Gamma_{s}(r)$ and $U_{\kappa,R}$ for different values of $\kappa$.}
\end{figure}

Let $\rho(\bm{x})$ be the effective topological charge density of a system of $N$ disclinations on a background of Gaussian curvature $K(\bm{x})$:
\begin{equation}
\rho(\bm{x}) = \frac{\pi}{3}\sum_{i=1}^{N}q_{i}\delta(\bm{x},\bm{x}_{i})-K(\bm{x}) \ .
\end{equation}
If free boundary conditions are choosen, the elastic energy \eqref{eq:sec2-bnt} can be written as
\begin{equation}\label{eq:sec3-paraboloid-energy}
F_{el}=\frac{1}{2Y}\int d^{2}x\,\Gamma^{2}(\bm{x}) \ ,
\end{equation}
where $\Gamma(\bm{x})=\Delta\chi(\bm{x})$, and $\chi(\bm{x})$ satisfies the inhomogeneous biharmonic equation
\begin{equation}
\Delta^{2}\chi(\bm{x}) = Y\rho(\bm{x}) \ ,
\end{equation}
with boundary conditions
\begin{equation}\label{eq:sec3-paraboloid-bc}
\left\{
\begin{array}{lr}
\chi(\bm{x}) = 0 & \bm{x}\in\partial M \\[5pt]
\nu^{i}\nabla_{i}\chi(\bm{x}) & \bm{x}\in\partial M
\end{array}
\right.
\end{equation}
where $\nu^{i}$ is the $i$th component of the tangent vector $\nu$ perpendicular to the boundary. In the parametrization \eqref{eq:sec3-paraboloid}, the normal vector $\bm{\nu}$ is simply given by $\bm{g}_{r}/|\bm{g}_{r}|$, with $\bm{g}_{r}$ the basis vector associated with the coordinate $r$. The stress function $\Gamma(\bm{x})$ can thus be expressed as
\begin{equation}
\frac{\Gamma(\bm{x})}{Y} = \int d^{2}y\,G_{L}(\bm{x},\bm{y})\rho(\bm{y})+U(\bm{x})\,.
\end{equation}
$G_{L}(\bm{x},\bm{y})$ is given in \eqref{eq:sec3-disk-green} and $U(\bm{x})$ is a harmonic function on the paraboloid that enforces the Neumann boundary conditions in Eq. \eqref{eq:sec3-paraboloid-bc}. The conformal distance on the paraboloid is given by
\begin{equation}\label{eq:sec3-paraboloid-conformal-distance}
\varrho(r) = \lambda\frac{r e^{\sqrt{1+\kappa^{2}r^{2}}}}{1+\sqrt{1+\kappa^{2}r^{2}}}\,.
\end{equation}
Using Eqns. \eqref{eq:sec3-paraboloid-conformal-distance} and \eqref{eq:sec3-disk-green} in Eq. \eqref{eq:sec3-paraboloid-energy}, the elastic energy $F_{el}$ of a collection of $N$ disclinations in a paraboloidal crystal can be expressed as:
\begin{equation}\label{eq:sec3-paraboloid-gamma}
\frac{\Gamma(\bm{x})}{Y}
= \frac{\pi}{3}\sum_{i=1}^{N}q_{i}G_{L}(\bm{x},\bm{x}_{i})-\Gamma_{s}(|\bm{x}|)+U(\bm{x})\,,
\end{equation}
where the first term represents the bare contribution of the defects to the energy density and the second corresponds to the screening effect of the Gaussian curvature. Explicitly:
\begin{equation}\label{eq:sec3-gamma_screening}
\Gamma_{s}(|\bm{x}|) = \log\left(\frac{\alpha e^{\sqrt{1+\kappa^{2}r^{2}}}}{1+\sqrt{1+\kappa^{2}r^{2}}}\right),
\end{equation}
where $r=|\bm{x}|$ and
\begin{equation}\label{eq:sec3-alpha}
\alpha = \frac{1+\sqrt{1+\kappa^{2}R^{2}}}{\exp\left(\sqrt{1+\kappa^{2}R^{2}}\right)}
\end{equation}
is a normalization constant depending on boundary radius $R$ and the ratio $\kappa$. Fig. \ref{fig:sec3-paraboloid-gamma}a shows a plot of the screening function $\Gamma_{s}(r)$ for different values of $\kappa\in[1,2]$.  As expected, the contribution due to Gaussian curvature is maximal at the origin of the paraboloid and drops to zero at the boundary.

The calculation of the harmonic function $U(\bm{x})$ requires a little more effort. If the crystal was defect-free (or populated by a perfectly isotropic distribution of defects) the function $U(\bm{x})$ would be azimuthally symmetric and  constant on the boundary. By the maximum principle of harmonic functions, $U(\bm{x})$ would then be constant on the whole manifold and depend only on $\kappa$ and the radius $R$: $U(\bm{x})=U_{\kappa,R}$. This constant can be determined by integrating $\Delta\chi(\bm{x}) = \Gamma(\bm{x})$ and imposing the second boundary condition in Eq. \eqref{eq:sec3-paraboloid-bc}. This gives:
\begin{equation}
U_{\kappa,R} = \frac{2\pi}{A}\int_{0}^{R} dr\,\sqrt{g}\,\Gamma_{s}(r),
\end{equation}
where $A$ is the area of the paraboloid:
\begin{equation}\label{eq:sec3-area}
A = \frac{2\pi}{3\kappa^{2}}\left[\left(1+\kappa^{2}R^{2}\right)^{\frac{3}{2}}-1\right].
\end{equation}
As shown in Figure \ref{fig:sec3-paraboloid-gamma}b, the value of $U_{\kappa,R}$ quickly approaches the linear regime as the size of the radius increases:
\begin{equation}\label{eq:sec3-harmonic-constant}
U_{\kappa,R} \approx -\frac{1}{4}\,\kappa R +\frac{1}{3}\cdot
\end{equation}
Then, for a defect-free configuration, the contribution of the boundary to the energy density is a constant offset that persists even for large radii. In the presence of disclinations, on the other hand, the function $\chi(\bm{x})$ is no longer expected to be azimuthally symmetric and the harmonic function $U(\bm{x})$ will not be constant throughout the paraboloid. In this case $U(\bm{x})$ can be expressed in the integral form
\begin{equation}
U(\bm{x}) = -\int d^{2}y\,H(\bm{x},\bm{y})\rho(\bm{y}),
\end{equation}
where $H(\bm{x},\bm{y})$ is the harmonic kernel \eqref{eq:sec3-harmonic-kernel}. In the regime of large core energies $F_{c}\gg F_{el}$, the creation of defects is strongly penalized and the lattice necessarily has the minimum number of disclinations allowed by the topology of the paraboloidal substrate. From symmetry considerations, we might expect the optimal distribution of defects to consist of $b$ $+1-$disclinations arranged along the boundary at the base vertices of a $b$-gonal pyramid and a $b$-fold apex (of topological charge $q_{0}=6-b$) at the origin. The homogeneous boundary conditions adopted require the first term in Eq. \eqref{eq:sec3-paraboloid-gamma} to vanish at the boundary. In the minimal energy configuration then, the system has the freedom to tune the total number of defects along the boundary to minimize the elastic energy Eq. \eqref{eq:sec3-paraboloid-energy} for any given value of the ratio $\kappa$. This behavior is exclusive to manifolds with boundary and doesn't have any counterpart in crystals on compact surfaces like the sphere and the torus.

\begin{figure}[h]
\centering
\includegraphics[width=5cm]{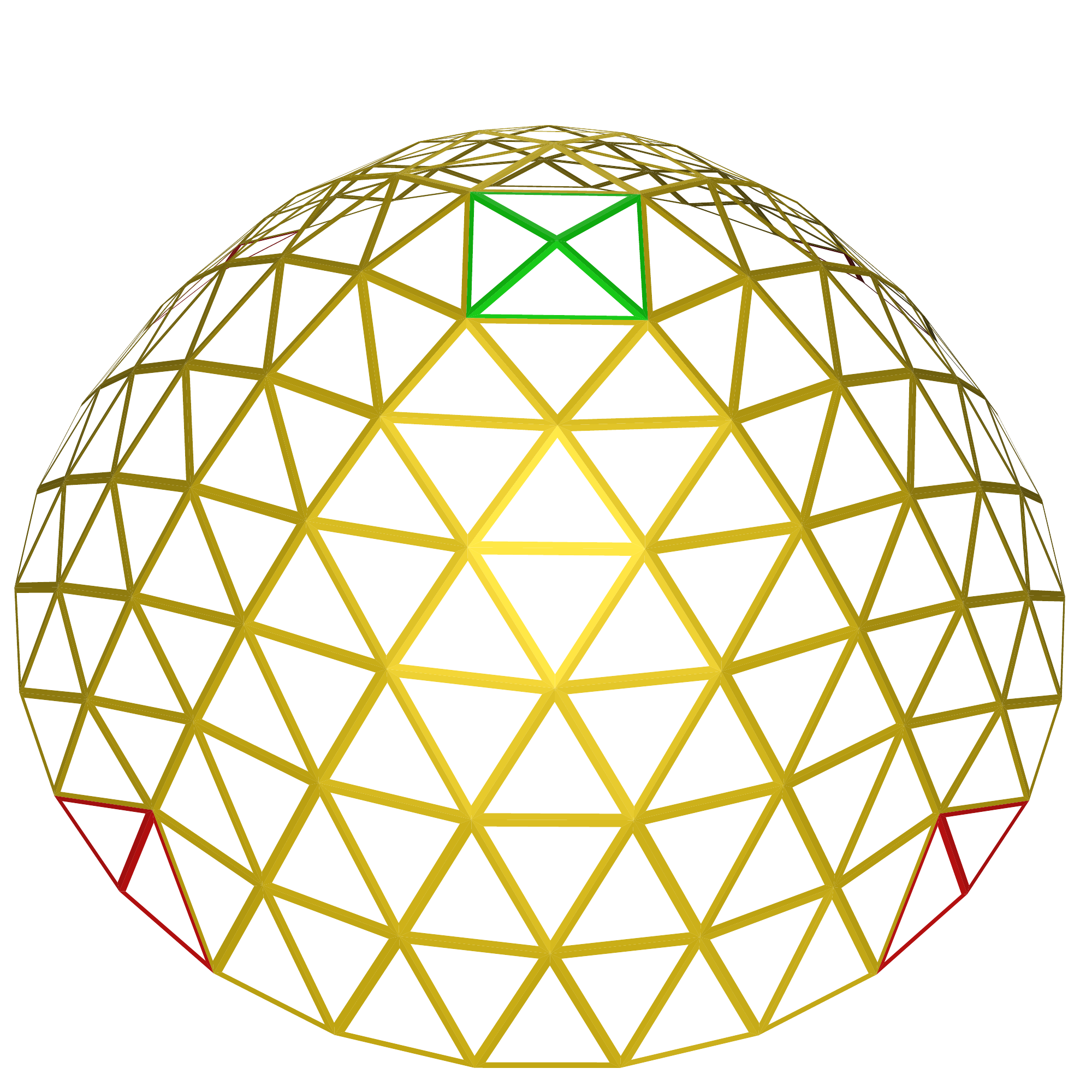}\\
\includegraphics[width=5cm]{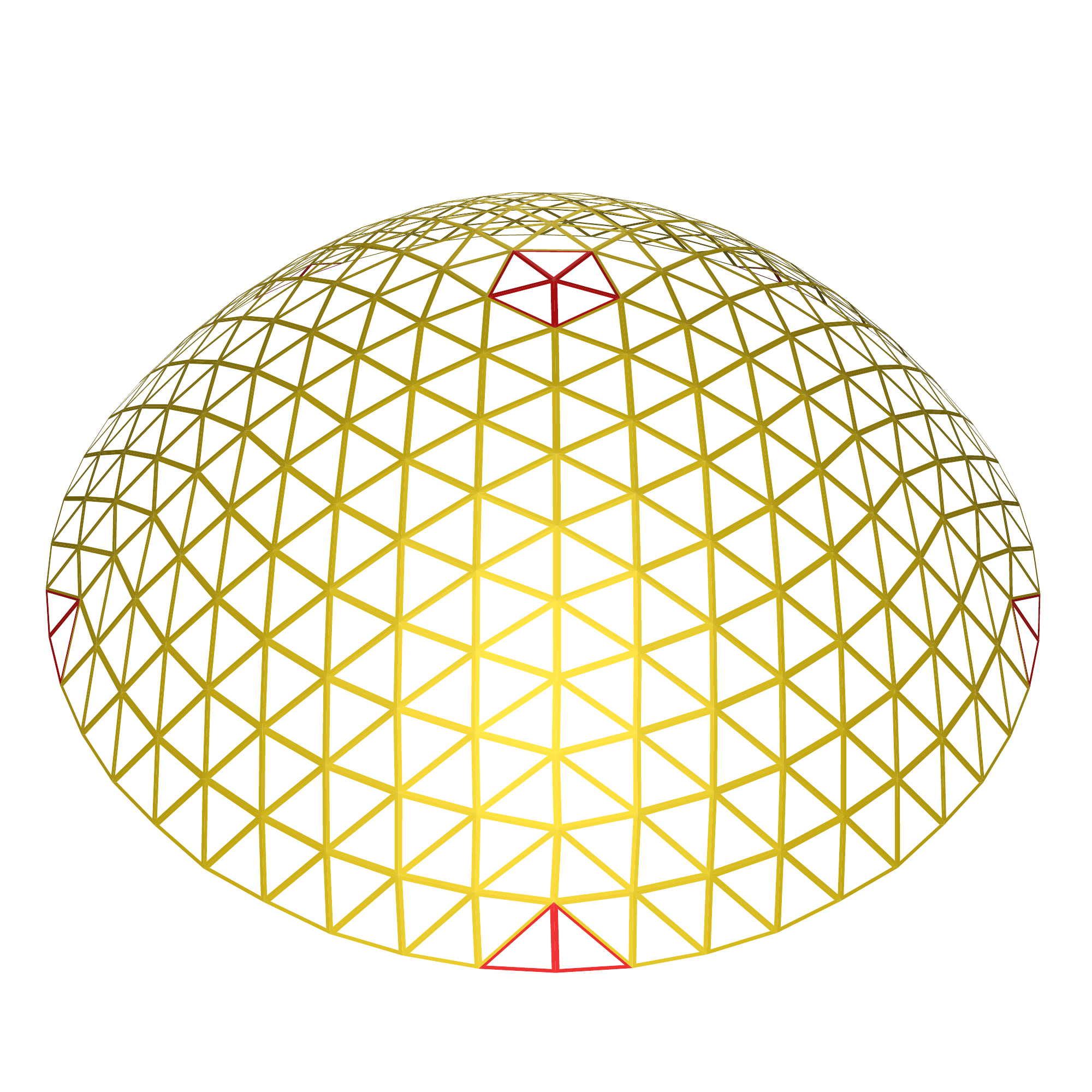}
\caption{\label{fig:sec3-pyramihedra}(Color online) Two examples of $Y_{b,n}$
triangulations of the paraboloid ($Y_{4,7}$ on the top and $Y_{5,10}$ on 
the bottom). Plaquettes with disclinations are highlighted in red, for 
$+1-$disclinations, and green for $+2-$disclinations.}
\end{figure}

\begin{figure}
\centering
\includegraphics[scale=0.65]{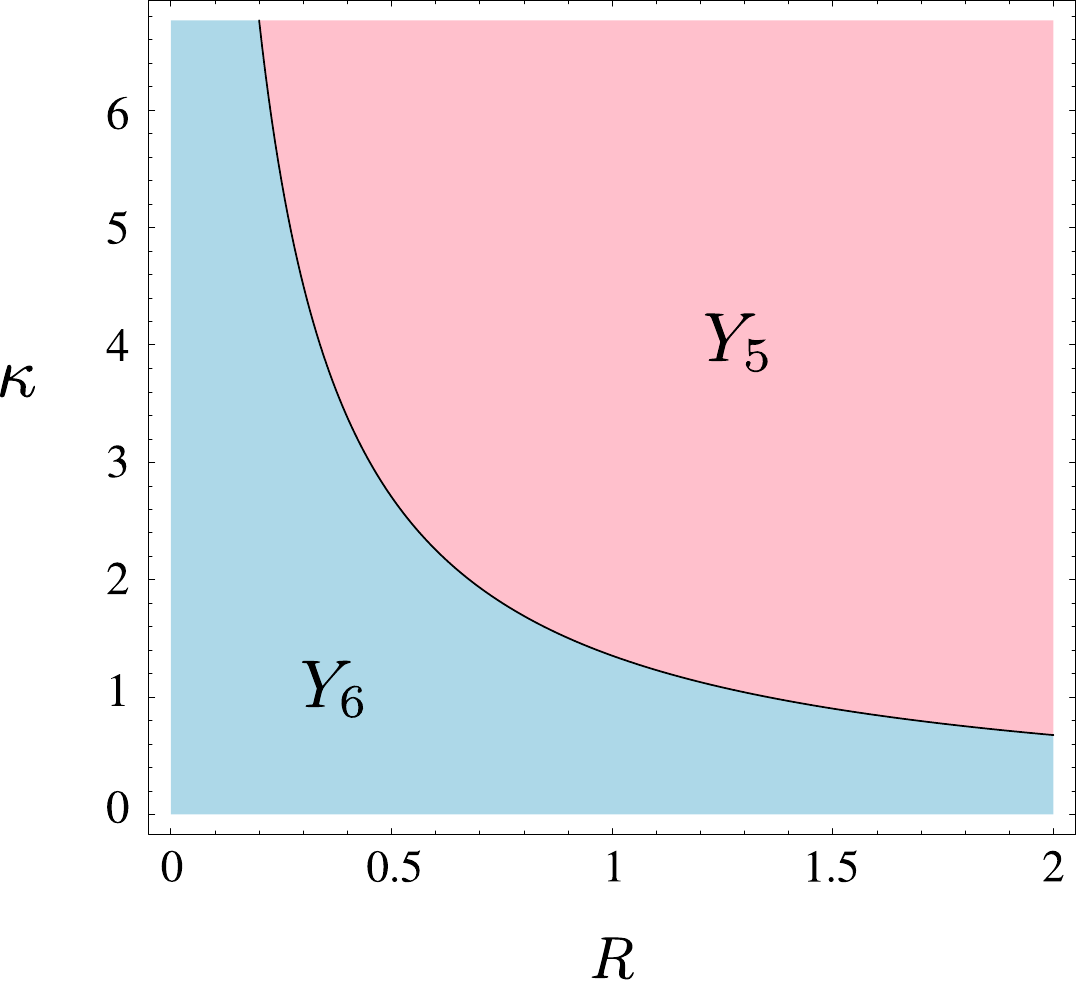}
\caption{\label{fig:sec3-pyramid_phase_diagram}(Color online) Phase diagram 
in the large core energy regime. For small $\kappa$ the lattice preserves 
the $6-$fold rotational symmetry of the flat case. As the curvature at the 
origin increases the system undergoes a transition to the $Y_{5}$ phase.}
\end{figure}

We will label a pyramidal configuration by $Y_{b}$, where $b$ denotes the number of base $+1-$disclinations. The coordinates $(r,\phi)$ of the vertices are given by
\begin{equation}\label{eq:sec3-pyramid_vertices}
Y_{b}: \qquad \left\{(0,\text{any}),\,\left(R,\frac{2\pi k}{b}\right)_{1\le k \le b}\right\}.
\end{equation}
Using the Euler theorem one can show that it is possible to construct infinite families of polyhedra with the symmetry group $C_{bv}$ from the pyramidal backbone $Y_{b}$. The number of vertices is given by:
\begin{equation}\label{eq:pyramihedra}
V = \tfrac{1}{2}bn(n+1)+1,
\end{equation}
where $n$ is a positive integer which represents the number of edges (not necessarily of the same length) of the polyhedron which separates two neighboring disclinations. In the following we will refer to these polyhedra with the symbol $Y_{b,n}$. Fig. \ref{fig:sec3-pyramihedra} illustrates two $Y_{b,n}$ lattices for the cases $b=4$ and $n=7$ (with $V=113$), and $b=5$ and $n=10$ ($V=276$). 
By a numerical minimization of the energy Eq. \eqref{eq:sec3-paraboloid-energy} one can establish that the $Y_{b}$ are indeed equilibrium configurations for $b\in[3,5]$, for   some range of the parameters $\kappa$ and $R$. The cases of $b=5,6$ are particularly significant because they are characterized by an equal number of defects ($N=6$) of the same topological charge ($q=1$). The two configurations will be associated therefore with the same core energy $F_{c}$ and this introduces the possibility of a structural transition between $Y_{5}$ and $Y_{6}$ governed by the curvature ratio $\kappa$ and the boundary radius $R$. For fixed $R$ and small values of $\kappa$, the $6-$fold symmetric configuration $Y_{6}$ is the global minimum of the free energy Eq. \eqref{eq:sec3-paraboloid-energy}. For $\kappa$ larger than some critical value $\kappa_{c}(R)$, however, the $Y_{6}$ crystal becomes unstable with respect to the $5-$fold symmetric configuration $Y_{5}$. A numerical calculation of the intersection point between the elastic energies of $Y_{5}$ and $Y_{6}$ for different values of $\kappa$ and $R$ allow us to construct the phase diagram shown in Fig. \ref{fig:sec3-pyramid_phase_diagram}. The word ``phase'' in this context refers to the symmetry of the ground state configuration as a function of the geometrical system parameters $\kappa$ and $R$. In principle, if we keep increasing the curvature we might expect the crystal to undergo a further transition to the $Y_{4}$ phase.  In this case, however, the core energy will also increase by a factor $4/3$ and so this is not generally possible in the regime in which $F_{c} \gg F_{el}$. For intermediate regimes (i.e. $F_{c} \sim F_{el}$), $Y_{5} \rightarrow Y_{4}$ and $Y_{4} \rightarrow Y_{3}$ transitions are also possible. The critical value of the parameters $\kappa$ and $R$, however, is not universal and will depend on the precise values of the core energy and the Young modulus.

When the core energy $F_{c}$ is small, the elastic energy Eq. \eqref{eq:sec3-paraboloid-energy} can be lowered by creating additional defects. Let us assume that a fivefold disclination is sitting at the point $\bm{x}_{0}=(r_{0},\phi_{0})$. We can introduce a notion of distance on the paraboloid by setting up a system of geodesic polar coordinates $(s,\varphi)$ with origin at $\bm{x}_{0}$. We expect that the stress introduced by the defect is controlled by an effective disclination charge inside a circular domain $C_{L}$ of geodesic radius $L$:
\begin{equation}\label{eq:sec3-effective_charge}
q_{eff}=q-\int_{0}^{2\pi} d\varphi\,\int_{0}^{L} ds\,\sqrt{g}\,K(s,\varphi),
\end{equation}
where $q=\pi/3$ is the charge of the isolated defect and the integral measures the screening due to the total Gaussian curvature within the domain. The metric tensor and the Gaussian curvature of a generic Riemannian manifold can be expressed in geodesic polar coordinates in the form (see for example Do Carmo \cite{DoCarmo}):
\begin{subequations}
\begin{gather}
g_{ss} = 1,\qquad g_{s\varphi}=0,\qquad g_{\varphi\varphi}=G,\\[7pt]
K(s,\varphi) = - \frac{\partial_{s}^{2}\sqrt{G}}{\sqrt{G}},
\end{gather}
\end{subequations}
where $G=\bm{g}_{\varphi}\cdot\bm{g}_{\varphi}$. Furthermore, an expansion of the metric around the origin $(0,\varphi)$ yields:
\[
\sqrt{G} = s- \tfrac{1}{6} K_{0} s^{3} + o(s^{5}).
\]
For small distance from the origin, Eq. \eqref{eq:sec3-effective_charge} becomes:
\begin{align}
q_{eff}
&= q+\int_{0}^{2\pi} d\varphi\,\int_{0}^{L} ds\,\partial_{s}^{2}\sqrt{G}\\[7pt]
&= q-\pi K_{0} L^{2} + o(L^{4}). \label{eq:sec3-charge_expansion}
\end{align}
The right hand side of Eq. \eqref{eq:sec3-charge_expansion} is a very general expression for the effective disclination charge at small distance and doesn't depend on the embedding manifold. If a grain boundary is radiating from the original disclination, we expect the spacing between consecutive dislocations to scale like $a/q_{eff}$, with $a$ the lattice spacing \cite{BowickNelsonTravesset:2000}. When $q_{eff} \rightarrow 0^{+}$ the dislocation spacing diverges and the grain boundary terminates. Since the Gaussian curvature is not constant, the choice of the origin (i.e. the position of the central disclination along the grain boundary) affects the evaluation of $q_{eff}$. One can identify upper and lower bounds by observing that:
\begin{subequations}
\begin{gather}
\max_{r} K(r) = K(0) = \kappa^{2},\\
\min_{r} K(r) = K(R) = \frac{\kappa^{2}}{(1+\kappa^{2}R^{2})^{2}}\cdot
\end{gather}
\end{subequations}
Unlike the case of surfaces of constant Gaussian curvature, the phase diagram for paraboloidal crystals consists of three regions separated by the curves:
\begin{equation}\label{eq:sec3-critical_length}
K_{0}L^{2} = \frac{1}{3}
\qquad
K_{0} = K_{\min},\,K_{\max}.
\end{equation}
When $L-L(K_{\min})\rightarrow 0^{+}$, the effective disclination charge goes to zero and the distance between two consecutive dislocations diverges at any point. On the other hand if $L-L(K_{\max})\rightarrow 0^{-}$, the disclination charge will prefer to be delocalized in the form of grain boundary scars. For $L(K_{\max})<L<L(K_{\min})$ the paraboloid will be equipped with regions where the Gaussian curvature is high enough to support the existence of isolated disclinations as well as regions where the screening due to the curvature is no longer sufficient and the proliferation of grain boundary scars is energetically favored. This leads to a three region phase diagram in which the regime of isolated disclinations is separated from the delocalized regime of scars by a novel phase in which both isolated disclinations and scars coexist in different parts of the paraboloid according to the magnitude of the Gaussian curvature.
\begin{figure}[t]
\centering
\includegraphics[scale=.65]{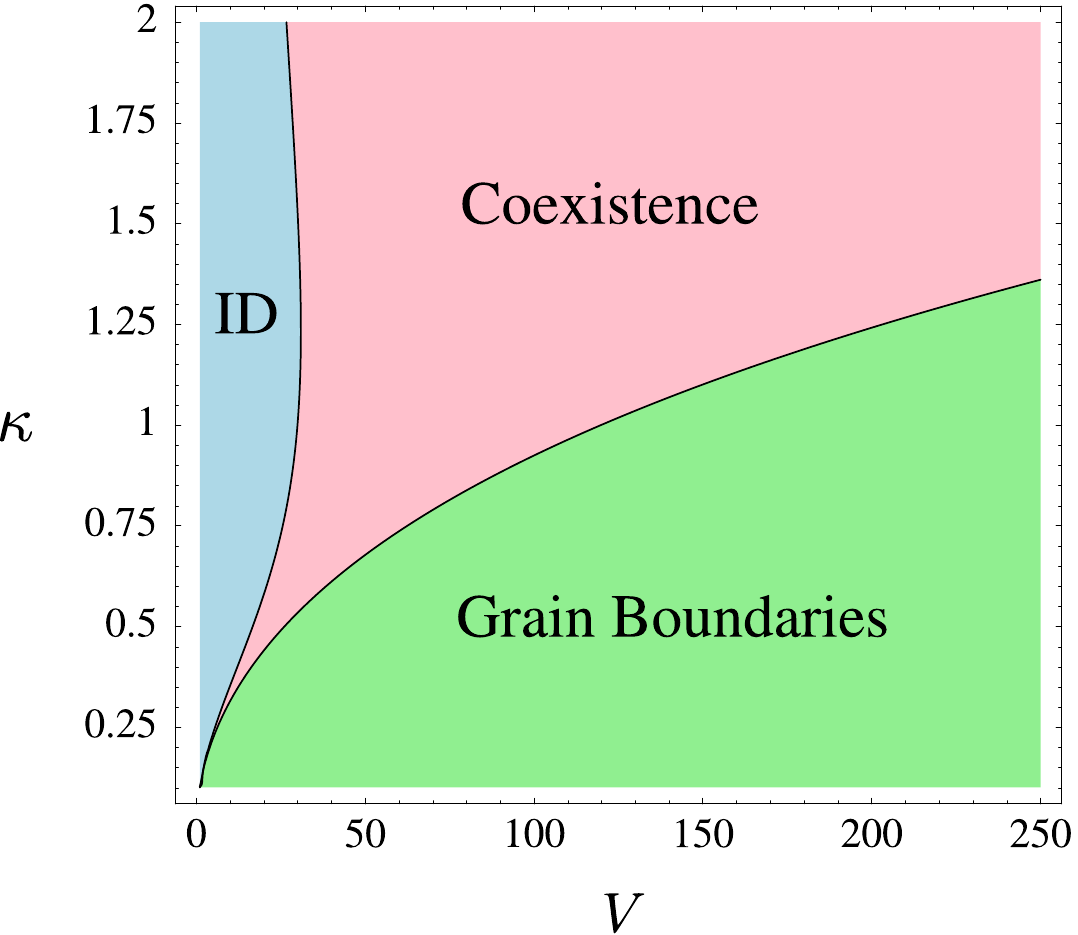}
\caption{\label{fig:sec3-paraboloid-phase-diagram}(Color online) Defect phase diagram 
for a paraboloidal crystal of radius $R=1$. The two phase boundaries 
that separate the isolated disclinations (ID) regime from the coexistence
regime and the coexistence regime from the scar phase correspond to
the solutions of Eq. \eqref{eq:sec3-critical_lines} for
$K_{0}=K_{\min}$ and $K_{0}=K_{\max}$, respectively.}
\end{figure}

It is useful to measure the distance $L_{c}$ in terms of the lattice spacing $a$ and rephrase Eq. \eqref{eq:sec3-critical_length} as a condition on $a$ (or equivalently on the number of vertices $V$). To do this we note that in order for the domain $C_{L}$ to completely screen the topological charge of the shortest scar possible (i.e. $5-7-5$), the geodesic radius $L$ has to be large enough to enclose the entire length of the scar. Calling $\ell$ the geodesic distance associated with a single lattice spacing $a$, we will then approximate $L\sim 3\ell$. This leads to the following expression for the lattice spacing $a$ at the onset of scar formation:
\begin{equation}\label{eq:sec3-critical_lines}
a^{2}\approx\frac{2}{K_{0}}\left(1-\cos\frac{1}{3\sqrt{3}}\right).
\end{equation}
The lattice spacing $a$ can be approximately expressed as a function of the number of vertices of the crystal by dividing the area $A$ of the paraboloid by the area of a hexagonal Voronoi cell of radius $a/2$ with $a^{2}\approx A/\frac{\sqrt{3}}{2}\,V$. The phase diagram arising from the solution of Eq. \eqref{eq:sec3-critical_lines} is sketched in Fig. \ref{fig:sec3-paraboloid-phase-diagram} for the case $R=1$. The two phase boundaries that separate the isolated defects (ID in the plot) regime from the coexistence regime and this one from the grain boundaries phase correspond to the solutions for $K_{0}=K_{\min}$ and $K_{0}=K_{\max}$ respectively. 

\begin{figure}
\centering
\includegraphics[scale=0.7]{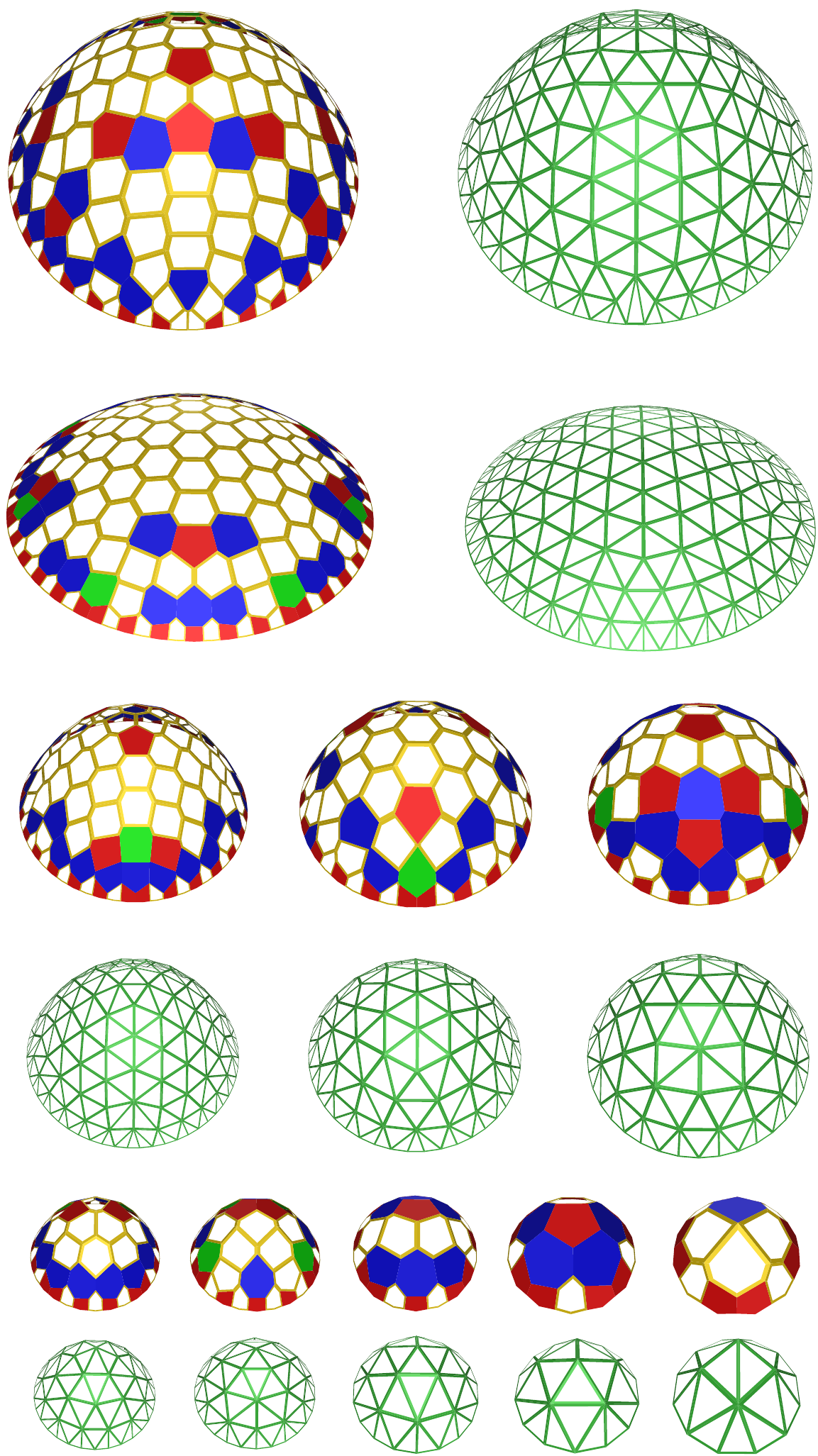}
\caption{\label{fig:sec3-paraboloid-numerics}(Color online) Voronoi lattice and Delaunay triangulations for ten selected systems from numerical simulations with $R=1$. The first row corresponds to $V=200$ and $\kappa=1.6$, while the second row is for $V=200$ and $\kappa=0.8$. In the bottom four rows $V=150\,,100\,,80\,,60\,,50\,,30\,,20\,,16$ and $\kappa=1.6$. From \cite{GiomiBowick:2007b}.} 
\end{figure}

Fig. \ref{fig:sec3-paraboloid-numerics} shows the Voronoi lattice and the Delaunay triangulations obtained from a numerical minimization of a system of $V$ classical particles (up to $V=200$) constrained to lie on the surface of a paraboloid and interacting with a Coulomb potential of the form $U(r)=1/r$, with $r$ the Euclidean distance between two particles in $\mathbb{R}^{3}$. Simulations are carried out using a parallel implementation of Differential Evolution (DE) \cite{StornPrice:1997,GiomiBowick:2007a}. In all the systems observed disclinations always appear clustered in either grain boundary scars or dislocations with the exception of isolated $+1-$disclinations which appear in the bulk as expected from the curvature screening argument discussed above. The complex aggregation of defects along the boundary together with the presence of negatively charged clusters indicates that the effect of the boundary, in the case of relatively small systems like the ones simulated, is more drastic than predicted by the homogeneous boundary conditions Eq. \eqref{eq:sec3-paraboloid-bc}. Even in the computationally expensive case of $V=100$, the distance between the origin and the boundary of the paraboloid is only four lattice spacings. In this situation we expect the distribution of particles along the boundary to play a major role in driving the order in the bulk.

For larger systems, such as $V=200$ (top of Fig. \ref{fig:sec3-paraboloid-numerics}), the behavior of the particles in the bulk is less affected by the boundary and the crystalline order reflects more closely the free-boundary problem previously presented. A comparison of the lattices in the first two rows of Fig. \ref{fig:sec3-paraboloid-numerics}, in particular, reveals substantial agreement with the scenario summarized in Fig. \ref{fig:sec3-paraboloid-phase-diagram}. For $\kappa=0.8$ and $V=200$, the defects are all localized along the boundary with the exception of one length-$3$ scar in the bulk at distance $r \approx 0.63$ from the center. For $\kappa=1.6$, the pattern of defects in the bulk is characterized by the coexistence of an isolated $+1-$disclination at the origin and a length-$5$ $W-$shaped scar displaced along a parallel one lattice spacing away from the central disclination. Apart from the evident difficulty in comparing the structures of small systems with those predicted from continuum elasticity theory, this behavior is consistent with the simple picture sketched in the phase-diagram of Fig. \ref{fig:sec3-paraboloid-phase-diagram}. The local $5-$fold symmetry at the origin of the $\kappa=1.6$ configuration, compared with $6-$fold symmetry for $\kappa=0.8$, suggests, as in the case of spherical crystals \cite{EinertEtAl:2005}, that the complicated structure of defect clusters appearing in large systems is the result of the instability of the simpler $Y_{b,n}$ configurations from which they partially inherit their overall symmetry. A more accurate numerical verification of our theory remains a challenge for the future.

The symmetry of the configurations presented in Fig. \ref{fig:sec3-paraboloid-numerics} deserves special attention. As for any surface of revolution, the circular paraboloid possesses the symmetry group $O(2)$ of all rotations about a fixed point and reflections in any axis through that fixed point. Any given triangulation of the paraboloid may destroy the full rotational symmetry completely or just partially, leaving the system in one of the following two subgroups: the pyramidal group $C_{nv}$ or the reflection symmetry group $C_{s}$. In general we found the latter symmetry group for system sizes up to $V=200$ particles. The symmetry for larger system sizes is under investigation.

\subsection{\label{sec:4e}Experimental realization of paraboloidal crystals}

Some sixty years ago Bragg and Nye used bubble rafts to model metallic crystalline structures \cite{BraggNye:1947}. A carefully made assemblage of bubbles, floating on the surface of a soap solution and held together by capillary forces, forms an excellent two-dimensional replica of a crystalline solid, in which the regular triangular arrangement of bubbles is analogous to the close packed structure of atoms in a metal. Feynman considered this technique to be important enough that the famous Feynman lectures in physics include a reproduction of the original Bragg-Nye paper in its entirety \cite{Feynman}. Bubble rafts can be made easily and inexpensively, equilibrate quickly, exhibit topological defects such as disclinations, dislocations and grain boundaries, and provide vivid images of the structure of defects. Bubble raft models have been used to study two-dimensional polycrystalline and amorphous arrays \cite{SimpsonHodkinson:1972}, nanoindentation of an initially defect-free crystal \cite{GouldstoneEtAl:2001}, and the dynamic behavior of crystals under shear \cite{WangKrishanDennin:2006}. In this section we review the experimental realization of a bubble-raft model for a paraboloidal crystal done by Bowick \emph{et al} \cite{BowickEtAl:2008} by assembling a single layer of millimeter-sized soap bubbles on the surface of a rotating liquid, thus extending the classic work of Bragg and Nye on planar soap bubble rafts. 

As we mentioned in the introduction, paraboloidal shapes naturally occur across the air/liquid interface of a fluid placed in a rotating cylindrical vessel. It is simple to verify that the height $z$ of such an interface above the $xy-$plane of a system of Cartesian coordinates is given by
\begin{equation}
z = \frac{\omega^{2}}{2g}\,r^{2}+C\,,
\end{equation}
where $\omega$ is the angular velocity and $g$ the gravitational acceleration and $C = D - \omega^{2}R^{2}/4g$ is an integration constant following from the requirement that the volume of the liquid during rotation be equal to the volume of the liquid at rest ($D$ is its height) \cite{Spiegel}. Thus the normal curvature at the origin is given by $\kappa=\omega^{2}/g$.

\begin{figure}[t]
\centering
\includegraphics[scale=0.4]{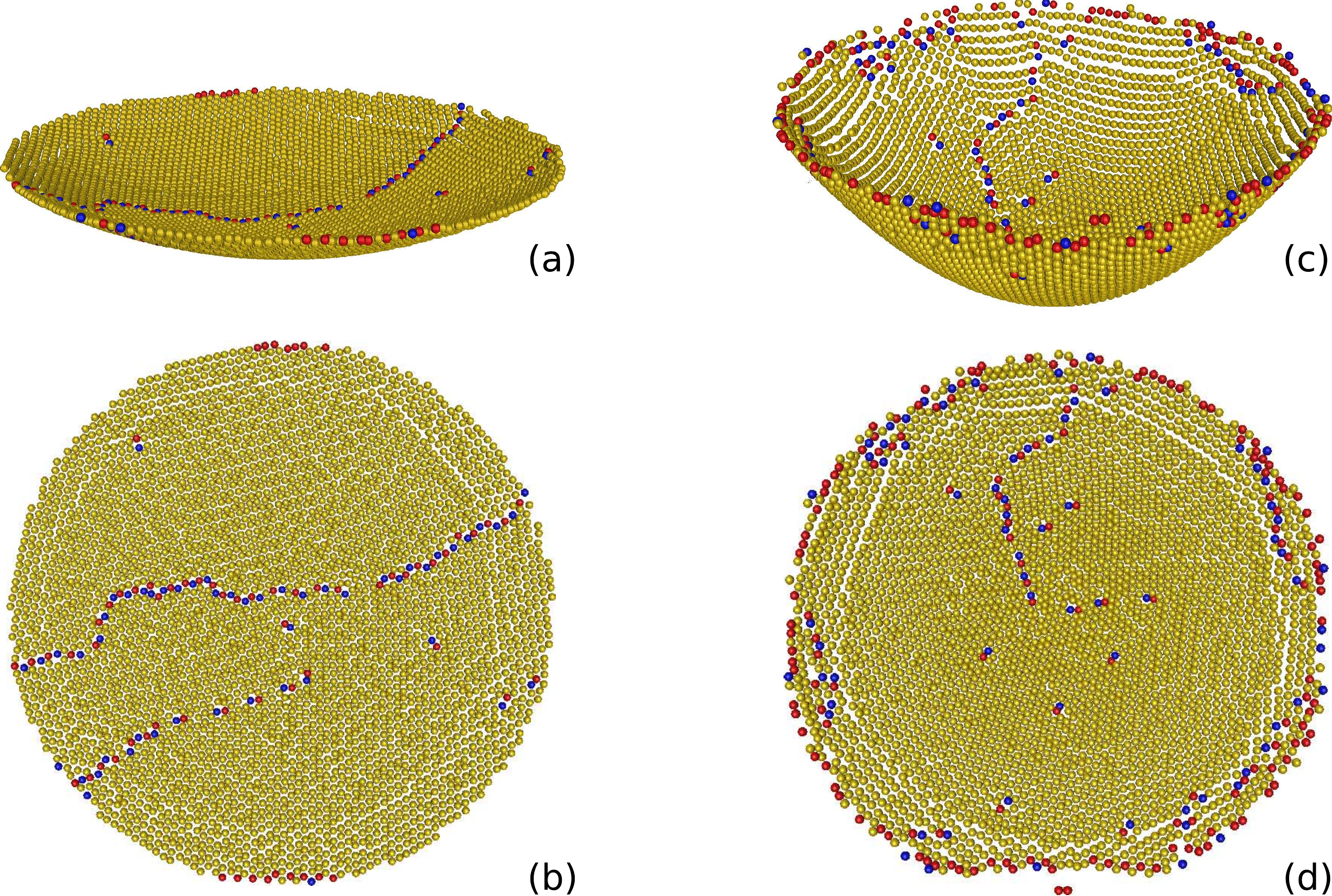}
\caption{\label{fig:sec3-bubble-raft} (Color online)
Lateral and top view of a computer reconstruction of two paraboloidal rafts
with $\kappa_{1} \approx 0.15$ cm$^{-1}$ (a, b) and $\kappa_{2} \approx 0.32$
cm$^{-1}$ (c, d). The number of bubbles is $N_{1}=3813$ and $N_{2}=3299$
respectively. The color scheme highlights the $5-$fold (red) and $7-$fold
(blue) disclinations over $6-$fold coordinated bubbles (yellow). From Ref. \cite{BowickEtAl:2008}.}
\end{figure}

\begin{figure}[t]
\centering
\includegraphics[scale=0.5]{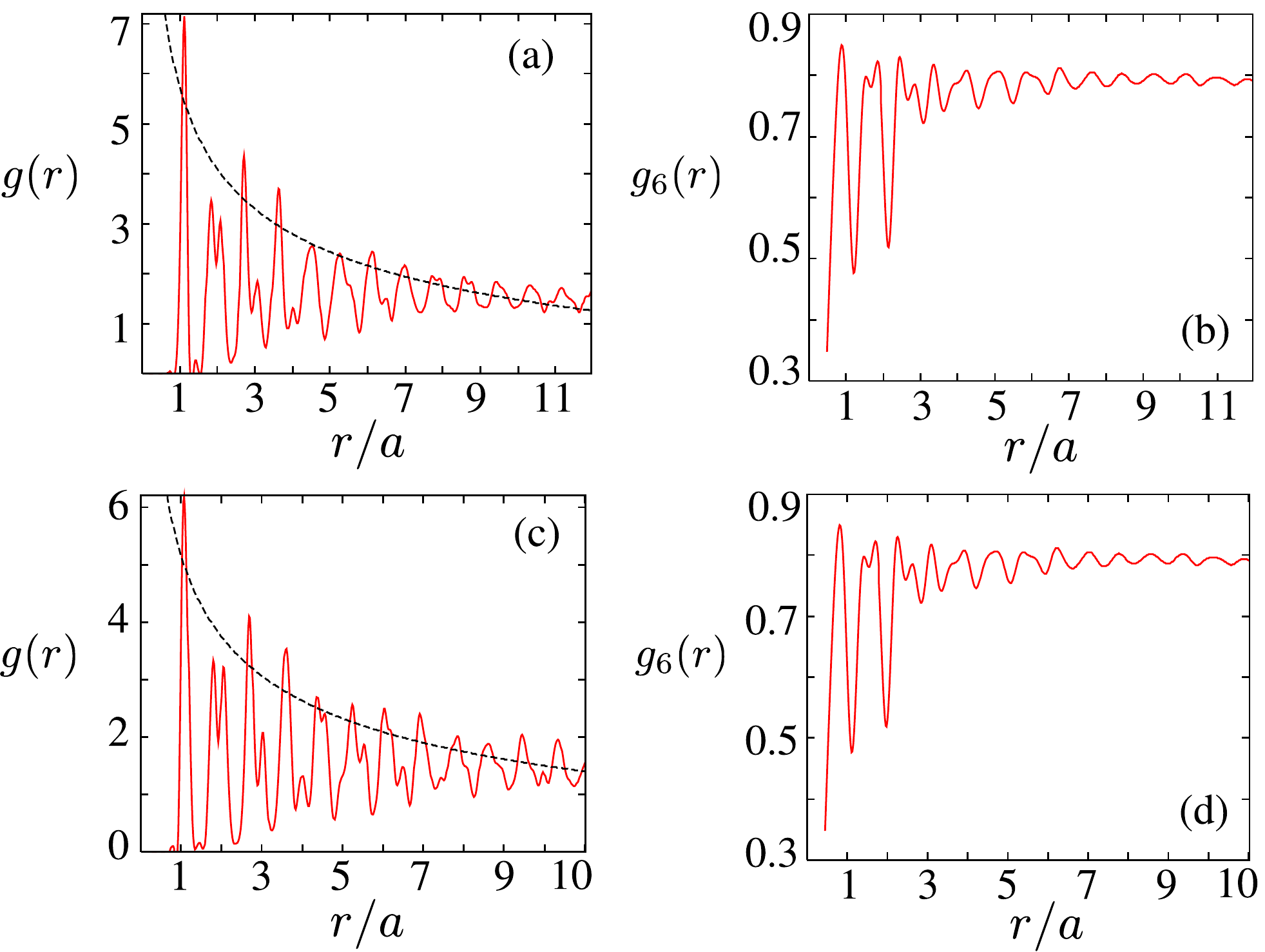}
\caption{\label{fig:sec3-correlation_functions} (Color online) Translational and
orientational correlation functions ($g$ and $g_6$, respectively) for rafts
with (a,b) $\kappa \approx 0.32$ cm$^{-1}$, $a=(0.8410 \pm 0.0025)$ mm, and
(c,d) $\kappa \approx 0.15$ cm$^{-1}$, $a=(0.9071 \pm 0.0037)$ mm. All the
curves are plotted as functions of $r/a$, where $r$ is the planar distance
from the center and $a$ is the bubble radius. The envelope for the crystalline
solid decays algebraically (dashed line), while the orientational correlation
function approaches the constant value 0.8. From Ref. \cite{BowickEtAl:2008}.}
\end{figure}

Fig. \ref{fig:sec3-bubble-raft} shows a computer reconstruction from Ref. \cite{BowickEtAl:2008} of two bubble rafts with $\kappa = 0.15$ cm$^{-1}$ and $\kappa=0.32$ cm$^{1}$ respectively. Soap bubbles have diameter $a=0.84(1)$ mm with monodispersity $\Delta a/a\approx 0.003$. The images shown in Fig. \ref{fig:sec3-bubble-raft} are taken with a CCD digital camera rotating along with a cylindrical vessel of radius $R=5$ cm and Delaunay triangulated to determine the lattice topology. The degree of order of the crystalline raft can be characterized by the translational and orientational correlation functions $g(r)$ and $g_{6}(r)$ \cite{Nelson}. The former gives the probability of finding a particle at distance $r$ from a fixed particle at the origin. The function is normalized with the density of an equivalent homogeneous system in order to ensure $g(r)=1$ for a system with no structure. Interactions between particles build up correlations in their position and $g(r)$ exhibits decaying oscillations, asymptotically approaching one. For a two-dimensional solid with a triangular lattice structure the radial correlation function is expected to exhibit sharp peaks at $r/a=\sqrt{n^{2}+nm+m^{2}}=1,\,\sqrt{3},\,2,\,2\sqrt{3}\ldots$, while the amplitude of the peaks decays algebraically as $r^{-\eta}$ with $\eta=1/3$ (dashed line in Fig.~\ref{fig:sec3-correlation_functions}). Within the precision of the data, the positional order of the paraboloidal crystals assembled with the bubble raft model reflects this behavior, although with more accurate measurements one might see dependence of the exponent $\eta$ on the curvature (because of the proliferation of defects with curvature).

The orientational correlation function $g_{6}(r)$ is calculated as the average of the product $\langle \psi(0)\psi^{*}(r) \rangle$ of the hexatic order parameter over the whole sample. For each bubble (labeled $j$) that has two or more neighbors, $\psi_{j}(r)=(1/Z_{j})\sum_{k=1}^{Z_{j}}\exp(6i\theta_{jk})$, where $Z_{j}$ is the number of neighbors of $i$ and $\theta_{jk}$ is the angle between the $j-k$ bonds and a reference axis. One expects $g_{6}(r)$ to decay exponentially in a disordered phase, algebraically in a hexatic phase and to approach a finite asymptote in the case of a crystalline solid. In the systems studied $g_{6}(r)$ approaches $0.8$ within $5$--$6$ lattice spacings.

Of particular interest is the structure of the grain boundaries appearing in the paraboloidal lattice for different values of the curvature parameter $\kappa$. Grain boundaries form in the bubble array  during the growing process as a consequence of geometrical frustration. As noted, any triangular lattice confined in a simply connected region with the topology of the disk is required to have a net disclination charge $Q=6$. In the absence of curvature, however, the elastic stress due to an isolated disclination is extremely high and defects are energetically favored to cluster in the form of a grain boundary consisting of one-dimensional arrays of tightly bound $(5,7)-$fold disclination pairs. In a planar confined system, grain boundaries typically span the entire length of the crystal. In the curved case, however, they can appear in the form of scars carrying a net $+1$ topological charge and terminating in the bulk of the crystal.

\begin{figure}[t]
\centering
\includegraphics[scale=0.5]{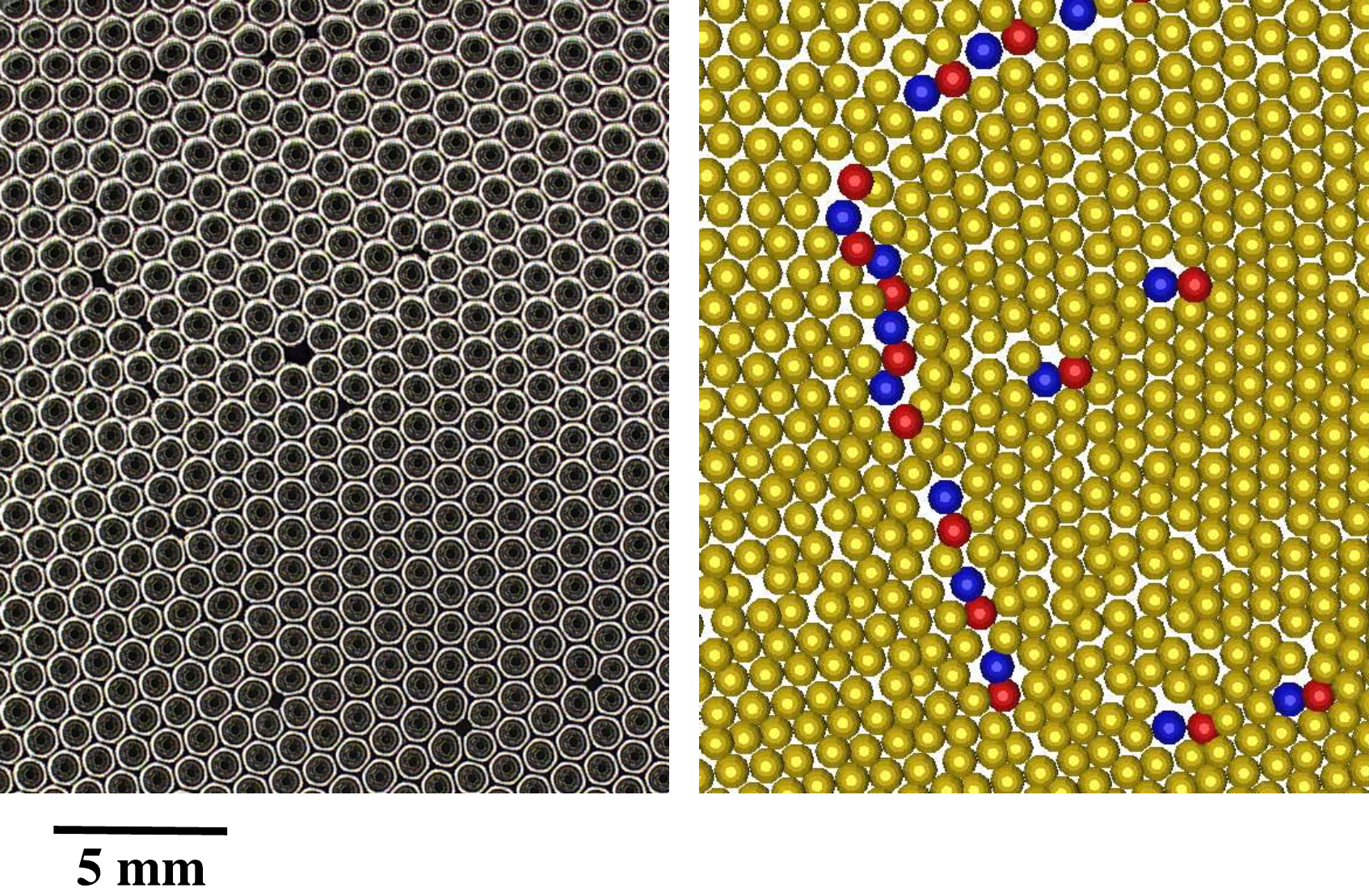}
\caption{\label{fig:sec3-scars} (Color online) An enlarged view of the
terminating grain boundary scar shown in Fig \ref{fig:sec3-bubble-raft}d for a
system with large Gaussian curvature. The scar starts from the circular
perimeter of the vessel and terminates roughly in the center
carrying a net $+1$ topological charge.  The image of the
bubbles (left) shows that they may deform slightly to better
fill space, whereas the computer reconstruction of the lattice (right)
uses perfect spheres of uniform size. From Ref. \cite{BowickEtAl:2008}.}
\end{figure}

Prominent examples of grain boundaries are visible in the two lattices shown in Fig.~\ref{fig:sec3-bubble-raft}. For a gently curved paraboloid (with $\kappa\approx 0.15$ cm$^{-1}$), grain boundaries form long (possibly branched) chains running from one side to the other and passing through the center. As the curvature of the paraboloid is increased, however, this long grain boundary is observed to terminate in the center (see Fig.~\ref{fig:sec3-bubble-raft}d; a close-up version of this image is seen in Fig.~\ref{fig:sec3-scars}). For $R=5$ cm, the elastic theory of defects predicts a structural transition at $\kappa_{c}=0.27$ cm$^{-1}$  in the limit of large core energies. In this limit the creation of defects is strongly penalized and the lattice has the minimum number of disclinations required by the topology of the embedding surface. In a low curvature paraboloid ($\kappa<\kappa_{c}$) these disclinations are preferentially located along the boundary to reduce the elastic energy of the system. When the aspect ratio of the paraboloid exceeds a critical value $\kappa_{c}(R)$, however, the curvature at the origin is enough to support the existence of a $5-$fold disclination and the system undergoes a structural transition. In the limit of large core energies, when only six disclinations are present, such a transition implies a change from the $C_{6v}$ to the $C_{5v}$ rotational symmetry group. 

Together with the theoretical argument reviewed in the previous section, these experimental observations point to the following mechanism for scar nucleation in a paraboloidal crystal. In the regime in which the creation of defects is energetically inexpensive, geometrical frustration due to the confinement of the lattice in a simply connected region is responsible for the formation of a long side-to-side grain boundary. When the curvature of the paraboloid exceeds a critical value, dependent on the radius of the circular boundary, the existence of a +1 disclination near the center is energetically favored. Such a disclination serves as a nucleation site for 5-7 dislocations and the side-to-side grain boundary is replaced by a terminating center-to-side scar.
\section{\label{sec:5}Crystalline and $p-$atic order in toroidal geometries}

\subsection{\label{sec:5a}Introduction}

Circa twenty years afer their first observation in partially polymerized dia\-ce\-tyl\-e\-nic phospholipid membranes by Mutz and Bensimon \cite{MutzBensimon:1991}, self-assembled toroidal aggregates are now considered the progenitors of a magnificent cornucopia of complex structures, also featuring branched network and micellar surfaces of high genus. The existence of such complex structures has become an experimental fact thanks to the enormous work in the past decade on the study of self-assembly of amphiphilic compounds such as lipids, surfactants and amphiphilic block copolymers. After the work of Eisenberg and coworkers \cite{ZhangEisenberg:1995,YuZhangEisenberg:1996,BurkeEisenberg:2000} it became clear that block copolymers in particular afford access to a variety of complex structures spanning an unexpectedly vast range of topologies and geometries. More recently, several experimental studies have been performed on di- and triblock copolymers with the intent of unraveling the origin of morphological complexity in copolymer surfactants and some possible pathways for micelle and vesicle formation have been proposed. 

\begin{figure}[t]
\centering
\includegraphics[scale=0.25]{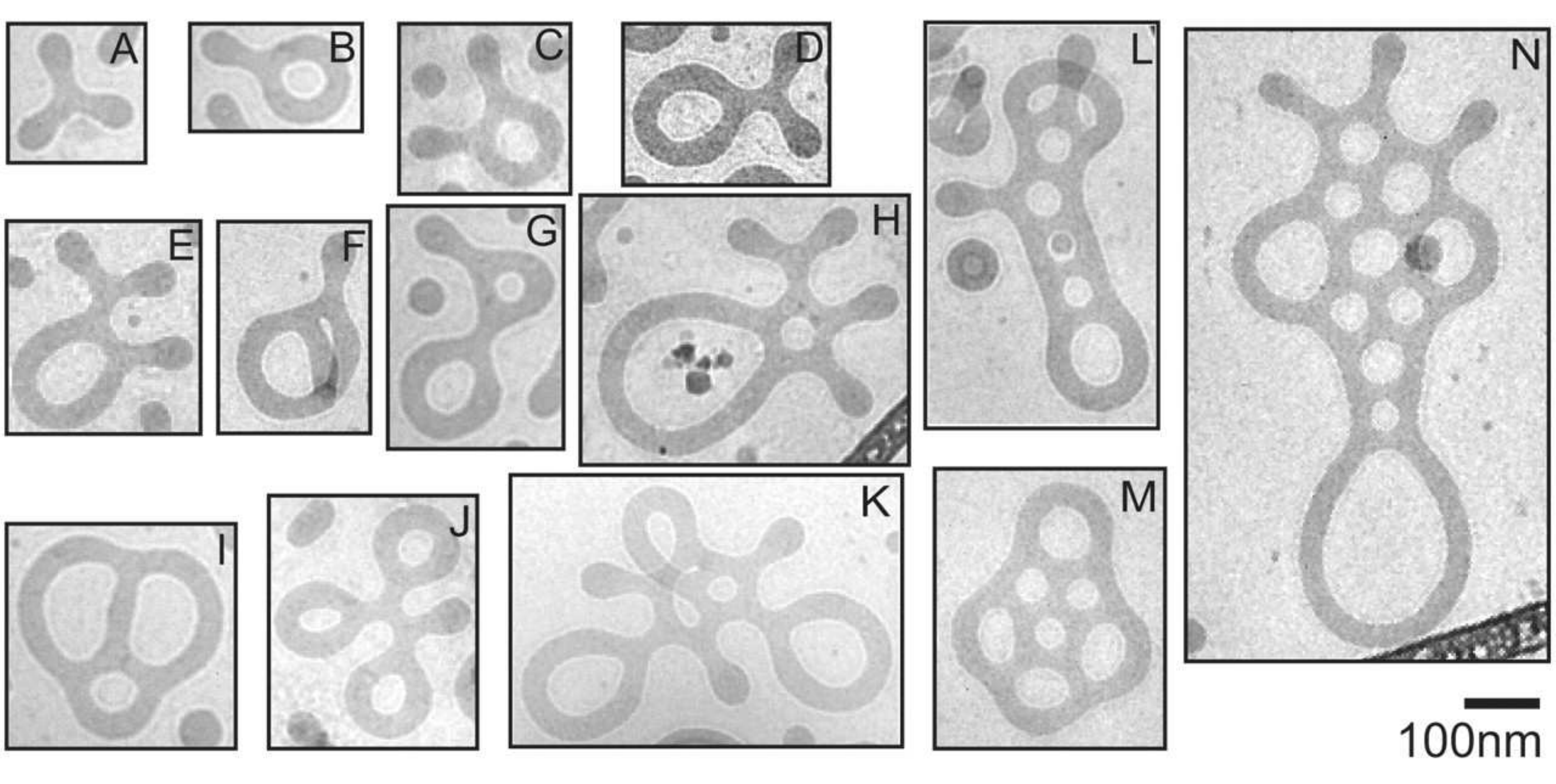}
\caption{\label{fig:sec5-cornucopia}Complex non-simply connected structures from self-assembly of diblock copolymer poly(1,2-butadiene-b-ethylene oxide) (PB-PEO). From Ref.~\cite{JainBates:2003}.}
\end{figure}

Jain and Bates, for instance, reported the formation of several non-simply-connected micellar structures from the self-assembly of diblock copolymer poly(1,2-butadiene-$b$-ethylene oxide) (PB-PEO) \cite{JainBates:2003} (see Fig. \ref{fig:sec5-cornucopia}). These polymers self-assemble in $Y-$shaped junctions and these form the building blocks for more complex micellar structures via re-assembly. Later, Pochan \emph{et al}. \cite{PochanEtAl:2004} found that almost all of the microstructures assembled from poly(acrylic acid-$b$-methyl acrylate-$b$-styrene) (PAA99-PMA73-PS66) triblock copolymers are ringlike or toroidal micelles. The route to toroidal micelles in block copolymers, however, is believed to be different from that to more complex network structures suggested by Jain and Bates, since residual $Y-$shaped aggregates are very rarely found in the sample. According to the authors of Ref. \cite{PochanEtAl:2004}, the formation of toroidal micelles has to be attributed primarly to the collapse of cylindrical micelles. On the other hand, mere end-to-end connection of cylindrical micelles doesn't appear to be the exclusive ring-forming mechanism in triblock systems since the average circumference of the self-assembled tori appears smaller than the contour length of the average cylindrical micelles. It seems to be a more complicated process, possibly involving interactions and exchange of matter between neighboring cylinders.

Toroidal micelles have also been observed in recent experiments by Kim \emph{et al} \cite{KimEtAl:2006} from the self-assembly of amphiphilic dumbbell molecules based on a aromatic rod segment that is grafted by hydrophilic polyether dendrons at one end and hydrophobic branches at the other end. Molecular dumbbells dissolved in a selective solvent self-assemble in an aggregate structure due to their amphiphilic character. This process has been observed to yield coexisting spherical and open-ended cylindrical micelles. These structures, however, change slowly over the course of a week to toroidal micelles which thus appear more stable. The formation of ring-like structures was explained in this case as result of the coalescence of spherical micelles occurring to reduce the contact between hydrophobic segments and water molecules.

An alternative pathway for the self-assembly of toroidal structures in copolymer solutions has been proposed and numerically tested by He and Schimd \cite{HeSchmid:2008}. In this pathway, the micelles do not coalesce, but simply grow by attracting copolymers from the solution. Once a critical micelle size is exceeded, copolymers start to flip-flop in such a way that the micelle core becomes itself solvent-philic (semi-vesicle state). Finally the solvent diffuses inside the core and the semi-vesicle swells into a vesicle.  

The existence of toroidal aggregates in amphiphilic compounds was suggested theoretically for smectic-C (SmC) membranes, based on the argument that orientational order is not frustrated on a surface with zero Euler characteristic and thus a toroidal topology may be energetically favoured in $p-$atic membranes with large order parameter coupling compared with the bending rigidity \cite{LubenskyProst:1992,Evans:1995}. This hypothesis was tested by Evans in 1995 with the result that a toroidal topology is indeed prefered over the spherical one for wide range of geometrical and mechanical parameters in defect-free $p-$atic tori. The role of disclinations in $p-$atic tori was investigated systematically by Bowick, Nelson and Travesset nine years after the original work of Evans with the conclusion that, even if not required by topological constraints, unbound disclinations can be energetically convenient even in very dense systems \cite{BowickNelsonTravesset:2004}. The precise number of unbound disclinations is controlled primarly by the aspect ratio of the torus. The existence of defects in the ground state of an ordered phase on the torus is crucial for toroidal crystals where it leads to some unique structural features as well as a spectacular example of a curvature-driven transition to a disordered, liquid-like, state in the limit the aspect ratio approaches to one \cite{GiomiBowick:2008a,GiomiBowick:2008b}. In the remainder of this section we will review these three examples of order on embedded tori.

\subsection{\label{sec:5b}Geometry of the torus}

\begin{figure}[t]
\centering
\includegraphics[width=0.5\textwidth]{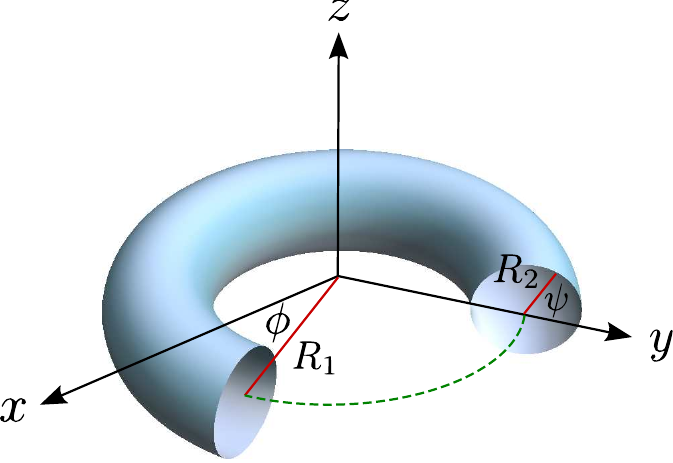}
\caption{\label{fig:torus-section}(Color online) The standard parametrization of 
a circular torus of radii $R_{1}$ and $R_{2}$.}
\end{figure}

The standard two-dimensional axisymmetric torus embedded in $\mathbb{R}^{3}$ is obtained by revolving a circle of radius $R_{2}$ about a coplanar axis located at distance $R_{1}\ge R_{2}$ from its center. Choosing the symmetry axis as the $z$ direction of a Cartesian frame, a convenient parametrization can be found in terms of the polar angle $\psi$ along the revolving circle and the azimuthal angle $\phi$ on the $xy-$plane:
\begin{equation}\label{eq:sec5-parametrization}
\left\{
\begin{array}{l}
x = (R_{1}+R_{2}\cos\psi)\cos\phi\\
y = (R_{1}+R_{2}\cos\psi)\sin\phi\\
z = R_{2}\sin\psi
\end{array}
\right.\,,
\end{equation}
where $\psi\,,\phi\in[-\pi,\pi)$. These coordinates satisfy the Cartesian equation
\begin{equation}
\left(R_{1}-\sqrt{x^{2}+y^{2}}\right)^{2}+z^{2}=R_{2}^{2}
\end{equation}
In this parametrization the metric tensor $g_{ij}$ and Gaussian curvature $K$ are given by
\begin{gather}
g_{\psi\psi}=R_{2}^{2}\,,\qquad 
g_{\psi\phi}=0\,,\qquad
g_{\phi\phi}=(R_{1}+R_{2}\cos\psi)^{2}\,,\\[7pt]
K = \frac{\cos\psi}{R_{2}(R_{1}+R_{2}\cos\psi)}\,.
\end{gather}
The Gaussian curvature is therefore positive on the outside of the torus, negative on the inside and zero along the two circles of radius $R_{1}$ at $\psi=\pm\pi/2$ (see Fig. \ref{fig:torus-section}). Moreover $K$ is maximally positive along the external equator at $\psi=0$ and maximally negative on the internal equator at $\psi=\pm\pi$. The Gauss-Bonnet theorem requires the total topological charge and the integrated Gaussian curvature to be zero on the torus:
\begin{equation}
\sum_{k=1}^{N} q_{k} = \int d^{2}x\,K(\bm{x}) = 0 \ .
\end{equation}
A global measure of the curvature of the embedded torus is provided by the aspect ratio $r=R_{1}/R_{2}$. Our discussion will be limited to the case $r\ge 1$. A ``fat'' torus with $r=1$ is obtained by taking $R_{1}=R_{2}$ and is characterized by a singularity in the Gaussian curvature along the internal equator at $\psi=\pm\pi$. The case $r<0$ corresponds to a self-intersecting torus whose symmetry axis lies in the interior of the revolving circle. The ``skinny'' torus limit, $r\rightarrow\infty$, can be obtained either by taking $R_{2}\rightarrow 0$ with finite $R_{1}$ or by letting $R_{2}$ stay finite and taking $R_{1}\rightarrow\infty$. The Gaussian curvature diverges in the first case and goes to zero in the second. Neither case, however, reflects a real physical situation since the area $A=(2\pi)^{2}R_{1}R_{2}$ of the torus becomes zero or infinite respectively.

Non-axisymmetric tori can be generated in various ways by deforming the surface of revolution \eqref{eq:sec5-parametrization}. A special and physically relevant transformation consists of an inversion wrt the unit sphere, a translation about a vector $\bm{\beta}$ and a second inversion. Thus a point $\bm{R}$ on the surface is mapped onto:
\begin{equation}\label{eq:sec5-non-axisymmetric}
\bm{R}'=\frac{\frac{\bm{R}}{R^{2}}+\bm{\beta}}{|\frac{\bm{R}}{R^{2}}+\bm{\beta}|^{2}}
\end{equation}
This map is conformal and transforms an a\-xi\-sym\-met\-ric shape to a non-a\-xi\-sym\-met\-ric one if the vector $\bm{\beta}$ is not parallel to the symmetry axis \cite{Seifert:1991}. As $\bm{\beta}$ is increased in magnitude, the asymmetry increases until, in the limit, the torus becomes a perfect sphere with an infinitesimal handle. The physical importance of mapping \eqref{eq:sec5-non-axisymmetric} relies on the fact that it leaves the bending energy $\kappa\int d^{2}x\,H^{2}$ invariant with significant consequences for the equilibrium shape and stability of fluid toroidal membranes. 	

As in any closed manifold the traditional Green-Laplace equation doesn't have a solution on the torus. An alternative Green function, can be obtained from the modified equation:
\begin{equation}
\Delta_{g}G_{L}(\bm{x},\bm{y}) = \delta_{g}(\bm{x},\bm{y})-A^{-1}
\end{equation}
As usual the calculation of the Green function can be simplified considerably by conformally mapping the torus to a domain of the Euclidean plane via a 
suitable system of isothermal coordinates. Intuitively the torus is conformally equivalent to a rectangular domain described by a system of Cartesian coordinates. To make this explicit, one can equate the metric of the torus in the coordinates $(\psi,\phi)$ to a conformally Euclidean metric in the coordinates $(\xi,\eta)$:
\[
ds^{2}
= R_{2}^{2}d\psi^{2}+(R_{1}+R_{2}\cos\psi)^{2}d\phi^{2}
= w\,(d\xi^{2}+d\eta^{2})\,,
\]
where $w$ is a positive conformal factor. Taking $\eta=\phi$ and $w=(R_{1}+R_{2}\cos\psi)^{2}$, the coordinate $\xi$ is determined by the differential equation:
\begin{equation}\label{eq:conformal_ode}
\frac{d\xi}{d\psi}
= \pm \frac{1}{r+\cos\psi}\,,
\end{equation}
where $r=R_{1}/R_{2}$, the aspect ratio of the torus, may be taken greater than or equal to one without loss of generality. Choosing the plus sign and integrating both sides of Eq.~\eqref{eq:conformal_ode} we find:
\begin{equation}\label{eq:xi}
\xi = \int_{0}^{\psi}\frac{d\psi'}{r+\cos\psi'}\,.
\end{equation}
Taking $\psi\in[-\pi,\pi)$, the integral \eqref{eq:xi} yields:
\[
\xi = \kappa\atan\left(\omega\tan\frac{\psi}{2}\right)\,,
\]
where
\begin{equation}\label{eq:kappa-omega}
\kappa = \frac{2}{\sqrt{r^{2}-1}}\,,
\qquad\qquad
\omega = \sqrt{\frac{r-1}{r+1}}\,.
\end{equation}
In the transformed coordinate system $(\xi,\eta)$ the modified Green-Laplace equation reads:
\begin{equation}\label{eq:green3}
\Delta G_{L}(\bm{x},\bm{y}) = \delta(\bm{x},\bm{y})-\frac{w}{A}\,,
\end{equation}
where $\Delta$ and $\delta$ are now the Euclidean Laplacian and delta function. The function $G_{L}(\bm{x},\bm{y})$ can be expressed in the form:
\[
G_{L}(\bm{x},\bm{y})
= G_{0}(\bm{x},\bm{y})
-\langle G_{0}(\bm{x},\cdot\,)\rangle 
-\langle G_{0}(\cdot\,,\bm{y})\rangle 
+\langle G_{0}(\cdot\,,\cdot\,) \rangle\,,
\]
where $G_{0}(\bm{x},\bm{y})$ is the Laplacian Green function on a periodic rectangle and the angular brackets stand for the normalized integral of the 
function $G_{0}(\bm{x},\bm{y})$ with respect to the dotted variable:
\begin{equation}
\langle G_{0}(\bm{x},\cdot\,)\rangle = \int \frac{d^{2}y}{A}\,G_{0}(\bm{x},\bm{y})\,. 
\end{equation}
Analogously the function $\langle G_{0}(\cdot\,,\cdot\,) \rangle$ is given by
\[
\langle G_{0}(\cdot\,,\cdot\,) \rangle = \int \frac{d^{2}x\,d^{2}y}{A^{2}}\,G_{0}(\bm{x},\bm{y}) 
\]
and ensures the neutrality property:
\begin{equation}\label{eq:green_neutrality}
\int d^{2}x\,G_{L}(\bm{x},\bm{y}) = \int d^{2}y\,G_{L}(\bm{x},\bm{y}) = 0\,.
\end{equation}
The modified Laplacian Green function on a periodic rectangle of edges $p_{1}$ and $p_{2}$ can be conveniently calculated in the form:
\begin{equation}
G_{0}(\bm{x},\bm{y}) = \sum_{\lambda \ne 0} \frac{u_{\lambda}(\bm{x})\overline{u}_{\lambda}(\bm{y})}{\lambda}\,,
\end{equation}
where $u_{\lambda}$ is the eigenfunction of the Laplace operator with periodic boundary conditions:
\begin{equation}
\Delta u_{\lambda}(\bm{x}) = \lambda u_{\lambda}(\bm{x})\,,
\end{equation}
such that:
\[
\left\{
\begin{array}{c}
u_{\lambda}(0,\eta) = u_{\lambda}(p_{1},\eta) \\[7pt]
u_{\lambda}(\xi,0) = u_{\lambda}(\xi,p_{2})
\end{array}
\right.\,.
\]
In Cartesian coordinates the eigenfunctions are simple plane waves of the form:
\begin{equation}
u_{\lambda}(\xi,\eta) = \frac{e^{i(\lambda_{n}\xi+\mu_{m}\eta)}}{\sqrt{p_{1}p_{2}}}\,,
\end{equation}
where $\lambda_{n}$ and $\mu_{m}$ are given by:
\[
\lambda_{n} = \frac{2\pi n}{p_{1}} \qquad
\mu_{m} = \frac{2\pi m}{p_{2}} \qquad
n,\,m = 0,\,\pm 1,\,\pm 2\ldots
\]
and the eigenvalue $\lambda$ is given by:
\begin{equation}
\lambda = - \lambda_{n}^{2} - \mu_{m}^{2}\,.
\end{equation}
Calling for simplicity $\bm{x}=(x,y)$ and $\bm{y}=(\xi,\eta)$, the function 
$G_{0}$ is given by:
\begin{equation}\label{eq:g0_1}
G_{0}(\bm{x},\bm{y}) 
= -\frac{1}{p_{1}p_{2}}\sum_{(n,m)\ne(0,0)} \frac{e^{i\lambda_{n}(x-\xi)}e^{i\mu_{m}(y-\eta)}}{\lambda_{n}^{2}+\mu_{m}^{2}}\,.
\end{equation}
Summing this series eventually leads to \cite{GiomiBowick:2008b}:
\begin{equation}\label{eq:final_green_function}
G_{0}(\bm{x},\bm{y})
= \frac{\log 2}{6\pi}
-\frac{1}{2\,p_{1}p_{2}}|y-\eta|^{2}
+\frac{1}{2\pi}\log\left|\frac{\vartheta_{1}(\frac{z-\zeta}{p_{1}/\pi}|\frac{ip_{2}}{p_{1}})}
 {\vartheta_{1}'^{\frac{1}{3}}(0|\frac{ip_{2}}{p_{1}})}\right|\,,
\end{equation}
where $\vartheta_{1}(u|\tau)$ is the Jacobi theta function \cite{Mumford,Polchinski} defined as:
\begin{equation}
\vartheta_{1}(u|\tau)
= 2q^{\frac{1}{4}}\sin u \prod_{n=1}^{\infty}\left(1-2q^{2n}\cos2u+q^{4n}\right)\left(1-q^{2n}\right)\,,
\end{equation}
with $q=\exp(i\pi\tau)$, $z=x+iy$ and $\zeta=\xi+i\eta$.

\subsection{\label{sec:5c}Defect-free $p-$atic textures and \emph{genera} transition}

There has been much effort in recent years to shed light on possible pathways in the self-assembly of toroidal micelles and vesicles from block copolymer solutions and other amphiphilic compounds. In particular, the spontaneous formation of structures of genus $g \ge 1$ from preexisting spherical objects, a process appearing in several different scenarios proposed in the literature, has attracted the most attention and debate because of its exotic character. The simplest question one can ask in this context is whether a vesicle can be energetically favored to change its topology from spherical to toroidal once the total surface area, and therefore the number of constituent molecules in the vesicle, is specified. Evans addressed this problem in 1995 \cite{Evans:1995} and showed how such a transition between \emph{genera} is indeed possible in fluid membranes and is even enhanced if the membrane is endowed with in-plane orientational order. Although oversimplified if compared to the great complexity of the real self-assembly mechanism, Evans' calculation provides insight into the delicate problem of stability of toroidal vesicles as well as a good starting point for our discussion of order on the torus. 

Let $\theta$ be the local orientation of a $p-$atic director field on a surface as defined in Sec. 2. A standard orientational order parameter is given by scalar field:
\begin{equation}
\psi(\bm{x}) = \langle e^{ip\theta(\bm{x})} \rangle\,,
\end{equation}
where $\langle\cdot\rangle$ deontes a thermal average. The total elastic energy of the vesicle consists of a pure bending term $F_{b}$, describing the elasticity of the membrane in the liquid state, and a term $F_{p}$ associated with the internal $p-$atic order. Thus $F=F_{p}+F_{b}$, with:
\begin{subequations}
\begin{gather}
F_{p} = \int d^{2}x\,\left(\tau|\psi|^{2}+\frac{u}{2}|\psi|^{4}+C|(\nabla-ip\bm{\Omega})\psi|^{2}\right)\label{eq:sec5-patic-field}\\
F_{b} = \int d^{2}x\,\left(2\kappa H^{2}+\kappa_{g}K\right)\label{eq:sec5-patic-bending}
\end{gather}
\end{subequations}
where $\kappa$ and $\kappa_{g}$ are the bending and Gaussian rigidity respectively\footnote{The expression of the bending energy used in Eq. \eqref{eq:sec5-patic-bending} is that originally gave by Helfrich for a membrane with zero pre-existing curvature \cite{Helfrich:1973}. Often the equivalent expression $\int d^{2}x\,(\frac{1}{2}\kappa H^{2}+\kappa_{g}K)$ is found in the literature. In this case, however, the mean curvature $H$ is defined as the sum of the principal curvatures rather than their average.} and the operator $\nabla-ip\bm{\Omega}$ is obtained from the covariant derivative of a $p-$atic tensor order parameter expressed in a local orthonormal frame as described in Sec. \ref{sec:2d}, with $\bm{\Omega}$ the covariant vector of Eq. \eqref{eq:sec2-omega_i} that parametrizes the spin-connection. The free energy \eqref{eq:sec5-patic-field} is invariant under rotations of the local reference frame by an arbitrary angle $\chi$: 
\begin{subequations}
\begin{gather}
\psi \rightarrow \psi\,e^{ip\chi} \\[5pt]
\Omega_{i} \rightarrow \Omega_{i}-\partial_{i}\chi\,,
\end{gather}
\end{subequations}
which is a typical gauge transformation. The gradient term in Eq. \eqref{eq:sec5-patic-field} is the same as in Eq. \ref{eq:sec2-xy-curved2}, upon identifying $K_{A}=p^{2}C$. As observed by Park \emph{et al}. \cite{ParkLubenskyMacKintosh:1992}, Eqns. \eqref{eq:sec5-patic-field} and \eqref{eq:sec5-patic-bending} closely resemble the Landau-Ginzburg Hamiltonian of a superconductor in an external magnetic field:
\begin{equation}
\mathcal{H} 
= \int d^{3}x\,\left\{
 \left|\left(\nabla-\frac{2ie}{\hbar c}\bm{A}\right)\psi\right|^{2}
+\tau|\psi|^{2}+\frac{u}{2}|\psi|^{4}+\frac{1}{8\pi}|\nabla\times\bm{A}-\bm{H}|^{2}\right\}\,.
\end{equation}
When exposed to an external magnetic field, a super-conducting material can undergo a second order mean-field transition from a metal to a super-conductor characterized by an Abrikosov lattice of vortices whose density is determined by the temperature and the applied magnetic field $\bm{H}$. The magnetic field, in particular, is conjugate to the vortex number $N_{v}$ since:
\begin{equation}
\int d^{3}x\,\nabla\times\bm{A} = \phi_{0}LN_{v}\,,
\end{equation}
where $L$ is the length of the sample in the direction of $\bm{H}$ and $\phi_{0}=hc/2e$ is the flux quantum. In this context the vector potential associated with the applied magnetic field is replaced by the covariant vector $\bm{\Omega}$ whose curl is the Gaussian curvature. Morover, on a closed surface:
\[
\int d^{2}x\,\nabla\times\bm{\Omega} = \int d^{2}x\,K = \frac{2\pi}{p}\sum_{i=1}^{N}q_{i}\,.
\]
Thus a $p-$atic phase on a closed $2-$manifold is analogous to an Abrikosov phase with a fixed total vorticity rather than a fixed applied magnetic field. 

The transition to an ordered phase in controlled by the parameter $\tau$ in Eq. \eqref{eq:sec5-patic-field}. For $\tau$ above a critical value $\tau_{c}$ ($\tau_{c}=0$ on a flat surface), $\psi=0$ and the system is in the isotropic phase. In this regime the only contribution to the energy is determined by the shape of the vesicle as expressed by the bending term \eqref{eq:sec5-patic-bending}. For $\tau<\tau_{c}$, on the other hand, the gauge symmetry is spontaneously broken and $\psi\ne 0$, with the exception of a number of isolated defective points. Within a meanfield approach, thermal fluctuations can be neglected and the preferred configuration of the system corresponds to the minimum of the free energy. A ground state configuration for the $p-$atic order parameter can be found by rewriting the gradient energy term in Eq. \eqref{eq:sec5-patic-field} as
\begin{align*}
\int d^{2}x\,|(\nabla-ip\bm{\Omega})\psi|^{2}
&= \int dx^{1}\,dx^{2} \sqrt{g}\,D^{k}\psi\, D_{k}^{*}\psi^{*}\\ 
&=-\int dx^{1}\,dx^{2}\,\psi^{*} \left(D^{k}\sqrt{g}\,D_{k}\right)\psi\,,
\end{align*}
with $D_{k}=\partial_{k}-ip\Omega_{k}$, and expressing $\psi$ in the basis of eigenfunctions of the operator in parentheses:
\begin{equation}\label{eq:sec5-linear-combination}
\psi = \sum_{n}a_{n}\varphi_{n}
\end{equation}
with $\varphi_{n}$ satisfying the Hermitean equation
\begin{equation}\label{eq:sec5-eigenvalues}
-\frac{1}{\sqrt{g}}\,D^{k}\left(\sqrt{g}\,D_{k}\right)\varphi_{n} = \frac{4\pi\lambda}{A}\,\varphi_{n}
\end{equation}
with the normalization condition
\begin{equation}
\int d^{2}x\,\varphi_{n}^{*}\varphi_{m} = \frac{A}{4\pi}\,\delta_{nm} \ .
\end{equation}
The complex coefficients $a_{k}$ are then determined by minimization of the free energy $F$. The eigenfunctions are sometimes referred to as ``Landau levels'' and are typically degenerate. If $\psi$ is expanded in this complete set of eigenfunctions then, in meanfield, the partition function is dominated by configurations involving only Landau levels with the lowest eigenvalue $\lambda_{0}$. Taking lowest levels only, the $p-$atic free energy \eqref{eq:sec5-patic-field} becomes
\begin{equation}\label{eq:sec5-lowest-landau}
F_{p} = \int d^{2}x\,\left[\left(\tau+\frac{4\pi\lambda_{0}C}{A}\right)|\psi|^{2}+\frac{u}{2}|\psi|^{4}\right] \ .
\end{equation} 
To find the lowest energy configuration of the field $\psi$ one now has to solve the eigenvalue problem \eqref{eq:sec5-eigenvalues} to determine the lowest eigenvalue $\lambda_{0}$ and minimize the free energy with respect to the parameters of the linear combination \eqref{eq:sec5-linear-combination}. Notice that the mean field transition has been lowered from $\tau_{c}=0$ to $\tau_{c}=-4\pi\lambda_{0}C/A$. Although $\lambda_{0}=0$ on a flat surface, it is finite and positive-definite on a surface with finite Gaussian curvature \cite{Evans:1995}.

For a sphere of unit radius parametrized in standard spherical coordinates $(\theta,\phi)$, the eigenvalue problem \eqref{eq:sec5-eigenvalues} is solved by $\lambda_{0}=p$ and $\varphi_{n}$ of the form \cite{Evans:1996}
\begin{equation}\label{eq:sec5-sphere-eigenfunctions}
\varphi_{n} = \sqrt{\frac{2p+1}{4\pi(p+n)!(p-n)!}}\,\sin^{p+n}\left(\frac{\theta}{2}\right)\cos^{p-n}\left(\frac{\theta}{2}\right)\,e^{in\phi}
\end{equation}
for integer values of $n$ between $-p$ and $p$. These functions have $2p$ zeros, corresponding to topological defects of unit charge and winding number $1/p$. Because of the lowest Landau level approximation used to derive Eq. \eqref{eq:sec5-lowest-landau}, eigenfunctions \eqref{eq:sec5-sphere-eigenfunctions} are degenerate to quadratic order in $\psi$. This implies the freedom to place defects anywhere on the sphere with no additional energy cost of order $\psi^{2}$. The $\psi^{4}$ term, on the other hand, lifts this degeneracy and make the defects repel.

\begin{figure}[t]
\centering
\includegraphics[scale=0.5]{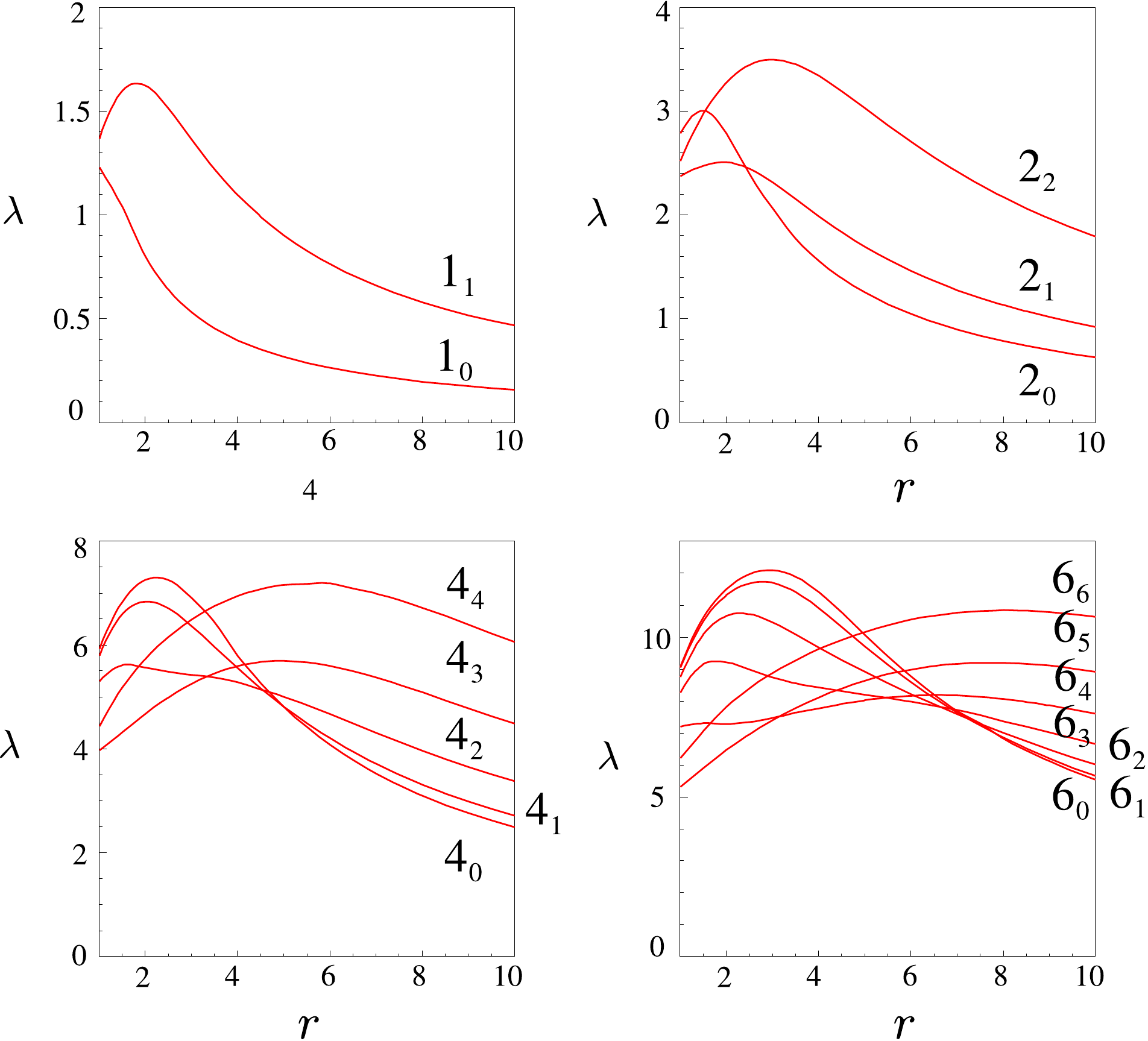}
\caption{\label{fig:sec5-evans-lambda}(Color online) Eigenvalues $\lambda$ versus the aspect ratio $r$ for varius $p-$atic configurations (labelled $p_{n}$) on the axisymmetric torus. The number of disclination pairs for each configuration is $2n$. Data are taken from \cite{Evans:1995}.}
\end{figure}

For axisymmetric tori the eigenvalue problem \eqref{eq:sec5-eigenvalues} was solved numerically by Evans with the result shown in Fig. \ref{fig:sec5-evans-lambda}. Eigenvalues are plotted as a function of the aspect ratio $r$ of the torus for various $p-$atic textures labelled $p_{n}$ with $0 \le n \le p$. Eigenfunctions $\varphi_{n}$ have $4n$ zeros corresponding to $2n$ disclinations of topological charge $q=1$ (winding number $1/n$) and $2n$ disclinations of topological charge $q=-1$ (winding number $-1/n$). Except for vector order ($p=1$), defective configurations are energetically favored for small aspect ratios and the number of disclination pairs increases with $r$ as a consequence of the larger (in magnitude) Gaussian curvature. 

Carrying out the integrals in Eq. \eqref{eq:sec5-patic-field} and \eqref{eq:sec5-patic-bending} in the lowest Landau level approximation, the total elastic free energy of spherical and toroidal vesicles can be written as
\begin{subequations}
\begin{gather}
F_{\mathrm{sphere}}
= 4\pi(2\kappa+\kappa_{g})
- 2\pi\left(\frac{C^{2}}{uA}\right)\frac{\left[\frac{\tau A}{4\pi C}+p\right]^{2}}{J_{\mathrm{sphere}}(p)}\label{eq:sec5-patic-sphere}\\[5pt]
F_{\mathrm{torus}} 
= \kappa\frac{2\pi^{2}r^{2}}{\sqrt{r^{2}-1}}
- 2\pi\left(\frac{C^{2}}{uA}\right)\frac{\left[\frac{\tau A}{4\pi C}+\lambda_{0}(r)\right]^{2}}{J_{\mathrm{torus}}(p,r)}\label{eq:sec5-patic-torus}
\end{gather}
\end{subequations}
where $J$ is the minimal value of the integral of $|\psi_{\mathrm{MF}}|^{4}$ and
\[
\psi_{\mathrm{MF}} = 
\left\{
\begin{array}{lc}
\sum_{n=-p}^{p}a_{n}\varphi_{n} & \text{sphere} \\[7pt]
a_{+}\varphi_{p}+a_{-}\varphi_{-n} & \text{torus}
\end{array}
\right.
\]
is the linear combination of eigenfuctions that are degenerate to quadratic order in $\psi$. The second term in Eqs. \eqref{eq:sec5-patic-sphere} and \eqref{eq:sec5-patic-torus} disappears in the isotropic phase when $\tau>-4\pi\lambda_{0}C/A$ due to the intrinsic orientational order on the manifold. Even in this case, Eqs. \eqref{eq:sec5-patic-sphere} and \eqref{eq:sec5-patic-torus} reveal the existence of a transition line between spherical and toroidal shape. In the absence of in-plane orientational order the bending energy term in Eq. \eqref{eq:sec5-patic-torus} is minimized by the so called Clifford torus with $r=\sqrt{2}$. In this case $F_{\mathrm{torus}}=4\pi^{2}\kappa$, which is smaller than $F_{\mathrm{sphere}}=4\pi(2\kappa+\kappa_{g})$ when $\kappa_{g}/\kappa>\pi-2$. Toroidal vesicles with $r=\sqrt{2}$ are therefore energetically favored in the fluid phase for large values of $\kappa_{g}/\kappa$. This argument cannot be invoked to explain the complex self-assembly of toroidal structures from homogeneous solutions mentioned in the introduction, but does provide a simple (but non-trivial) example of a toroidal shape being energetically preferred to a spherical one. 

\begin{figure}[t]
\centering
\includegraphics[scale=0.55]{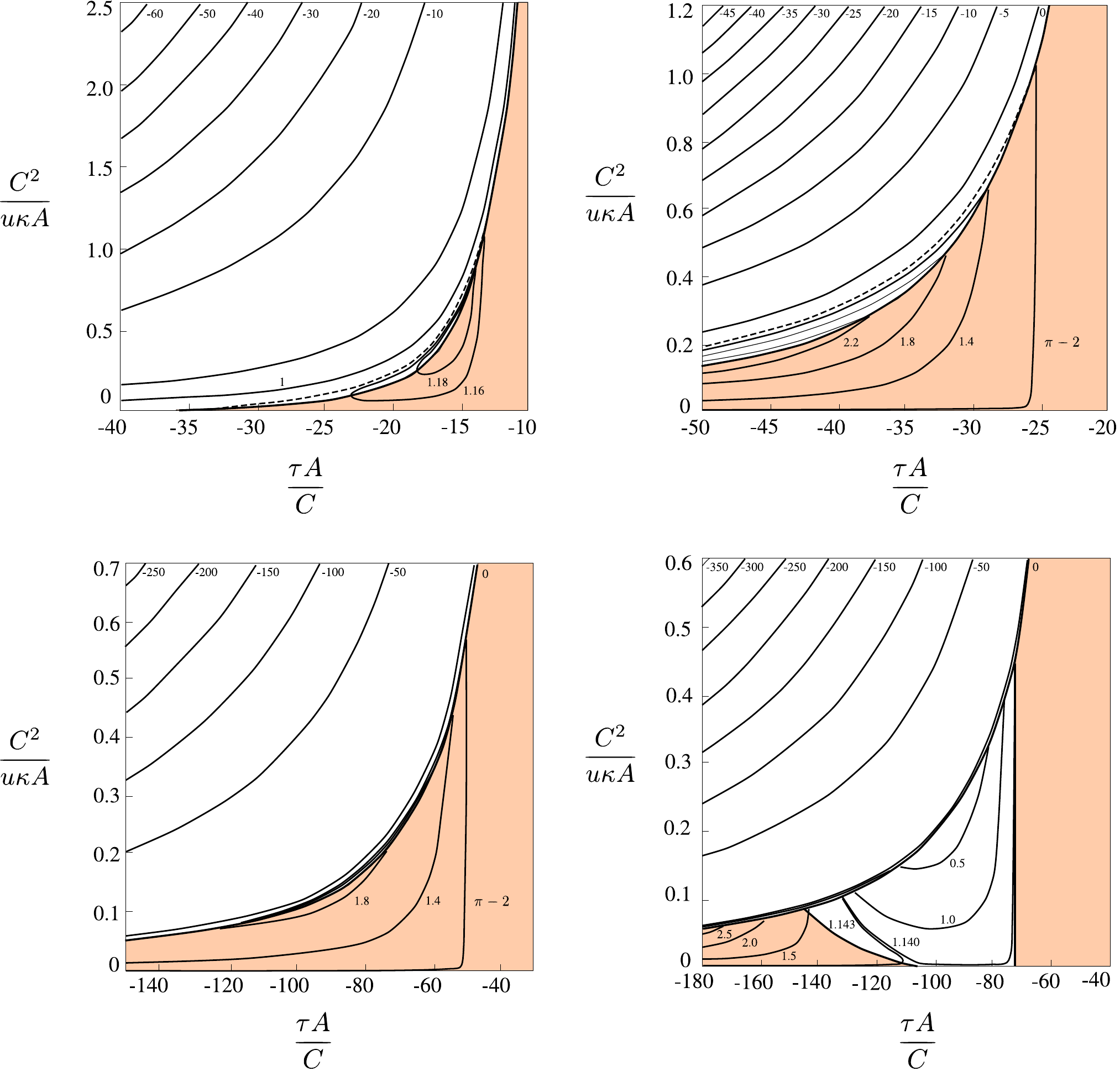}
\caption{\label{fig:sec5-evans-phase-diagram}(Color online) Phase diagram for vesicles of constant area of genus zero and one with intrinsic vector (top left), nematic (top right), tetradic (bottom left) and hexatic (bottom right) order. The position of the transition line depends on the value of $\kappa_{g}/\kappa$. Toroidal vesicles that exist in the shaded region are non-axisymmetric. Data are taken from \cite{Evans:1995}.}
\end{figure}

The critical value of $\kappa_{g}/\kappa$ of the genera transition is lowered for vesicles possessing in-plane $p-$atic order. This is displayed in the phase diagrams of Fig. \ref{fig:sec5-evans-phase-diagram} for the case of vector ($p=1$), nematic ($p=2$), tetradic ($p=4$) and hexatic ($p=6$) order. The lines, whose position depends on $\kappa_{g}/\kappa$, separate spheres (above the line) from tori (below the line). Even for simple vector order, stable toroidal vesicles exist as well for $\kappa_{g}/\kappa<\pi-2$. For $\kappa_{g}/\kappa>\pi-2$, furthermore, transition lines are closed, with spheres on the inside and tori on the outside. In the shaded regions non-axisymmetric tori are favored over symmetric ones. The hexatic phase diagram (bottom right corner of Fig. \ref{fig:sec5-evans-phase-diagram}) deserves special attention. The shaded region is split in two parts. That on the right contains spheres if $\kappa_g/\kappa<\pi-2$ and non-axisymmetric tori otherwise, while that on the left contains spheres above the transition line and non-axisymmetric tori below. The white region of the phase diagram is also divided in two. The top left region of the diagram behaves as described above, while the other part is characterized by closed lines separating spheres (outside) and small tori (inside). These last class of tori exhibit ten pairs of $q=\pm 1$ disclinations with positive disclinations distributed on the external equator of the torus and negative disclinations along the internal equator. This last feature is an important property sheared by toroidal objects with in-plane order and will be clarified in the following sections. The results reviewed here are valid within the mean-field approach and the lowest Landau level approximation, which both hold deep in the ordered phase that we are most interested in. At higher temperatures Evans proved the approximations are still valid away from the transition line and for $C/A\gg \kappa/k_{B}T \gg 1$ \cite{Evans:1995}. The latter condition is generally fulfilled by several systems (i.e. $\kappa/k_{B}T=1-10$ for lipid bilayers).

\subsection{\label{sec:5d}Defective ground states in hexatics}

Evans' analysis, summarized in the previous section, indicates that disclinations, even if not required by topological constraints, can nonetheless appear in the ground state of $p-$atic tori as a consequence of the coupling between in-plane orientational order and spatial curvature. The occurrence of defects in the ground state of toroidal hexatic vesicles was systematically investigated by Bowick, Nelson and Travesset based on the formalism outlined in Sec. 2 \cite{BowickNelsonTravesset:2004}. Before embarking on a detailed analysis of the elasticity of defects in hexatic tori, it is instructive to obtain a rough estimate of how many pairs of $\pm 1$-disclinations would be required to achieve perfect screening of the background topological charge associated with the Gaussian curvature of the torus. Consider a wedge of angular width $\Delta\phi$ on the outside wall of positive Gaussian curvature. The net curvature charge associated with this region is
\begin{equation}
s_{eff} = \int_{\phi_{0}-\frac{\Delta\phi}{2}}^{\phi_{0}+\frac{\Delta\phi}{2}} d\phi \int_{-\frac{\pi}{2}}^{\frac{\pi}{2}}d\psi\,\sqrt{g}\,K = 2\Delta\phi\,.
\end{equation}
Upon equating $s_{eff}$ to $2\pi/6$, the charge of a single disclination, one finds that $\Delta\phi=2\pi/12$, independently of $R_{1}$ and $R_{2}$. Thus, $2\pi/\Delta\phi=12$ positive disclinations would be required to completely compensate the negative curvature of the inner wall. This simple argument neglects core energies and interactions between disclinations, effects which will cause the preferred number of defect pairs to be less than twelve.

The total elastic energy of a toroidal hexatic vesicle containing $N$ disclinations of topological charge $q_{i}$ ($i=1\,\ldots N$) takes the form
\begin{multline}\label{eq:sec5-bnt-energy}
 \frac{F}{K_{A}}
= \frac{\pi^{2}}{9} \sum_{i < j}^{1,N} q_{i}q_{j}\mathcal{Q}(\bm{x}_{i},\bm{x}_{j}) 
- \frac{\pi}{3}\sum_{i=1}^{N} q_{i}\mathcal{L}(\bm{x}_{i})\\
+ \frac{2\pi^{2}}{r+\sqrt{r^{2}-1}}
+ \frac{\kappa}{K_{A}}\frac{2\pi^{2}r^{2}}{\sqrt{r^{2}-1}}
+ \frac{\epsilon_{c}}{K_{A}}\sum_{i=1}^{N}q_{i}^{2}
\end{multline}
where the first two terms represent the contributions due to the pair interaction between defects and the interaction of defects with the topological charge of the substrate associated with the Gaussian curvature. The third term in Eq. \eqref{eq:sec5-bnt-energy} is the spin-wave part of the frustrated hexatic energy while the last two terms represents the bending energy and the defect core energy respectively. The defect-defect interaction potential has been calculated in Ref. \cite{BowickNelsonTravesset:2004}:
\begin{align}
\mathcal{Q}(\bm{x}_{i},\bm{x}_{j})
= &-\frac{1}{4\pi}\log\frac{\left|\vartheta_{1}\left(\frac{\phi_{i}-\phi_{j}}{2\pi}+\frac{i}{2\pi}\frac{\alpha_{i}-\alpha_{j}}{\sqrt{r^{2}-1}}\Big|\frac{i}{\sqrt{r^{2}-1}}\right)\right|^{2}}{4\pi^{2}\left|\eta\left(\frac{i}{\sqrt{r^{2}-1}}\right)\right|^{6}}\notag\\[5pt]
&+\frac{1}{2\sqrt{r^{2}-1}}\left(\frac{\alpha_{i}-\alpha_{j}}{2\pi}\right)^{2} \ ,
\end{align}
where $\alpha$ is related to the polar angle $\psi$ on a cross-section of the torus by:
\begin{equation}
\cos\psi = \frac{r\cos\alpha-1}{r-\cos\alpha}
\end{equation}
and $\eta$ is the Dedekind eta function:
\begin{equation}
\eta(\tau) = e^{\frac{2\pi i \tau}{24}}\prod_{n=1}^{\infty}(1-e^{2\pi in\tau})\,.
\end{equation}
The one-body interaction between defects and curvature, on the other hand, is expressed via the potential energy
\begin{equation}
\mathcal{L}(\bm{x}_{i}) = \log\left(\frac{1}{r-\cos\alpha_{i}}\right)\,.
\end{equation}

For opposite sign disclinations, the defect-defect interaction, determined by the function $\mathcal{Q}(\bm{x}_{i},\bm{x}_{j})$, is attractive for all separations at constant $\phi$. If only this term were present, the attraction would bring both charges as close as possible, binding all disclinations into dipoles which have a higher energy than a defect-free configuration. Thus if no other terms were present, the ground state configuration would be defect free. The defect-curvature interaction term $\mathcal{L}(\bm{x}_{i})$, however, favors the appearance of additional defects. This term acts like an electric field pulling the positive (negative) disclinations into regions of positive (negative) Gaussian curvature. 

\begin{figure}
\centering
\includegraphics[width=0.9\textwidth]{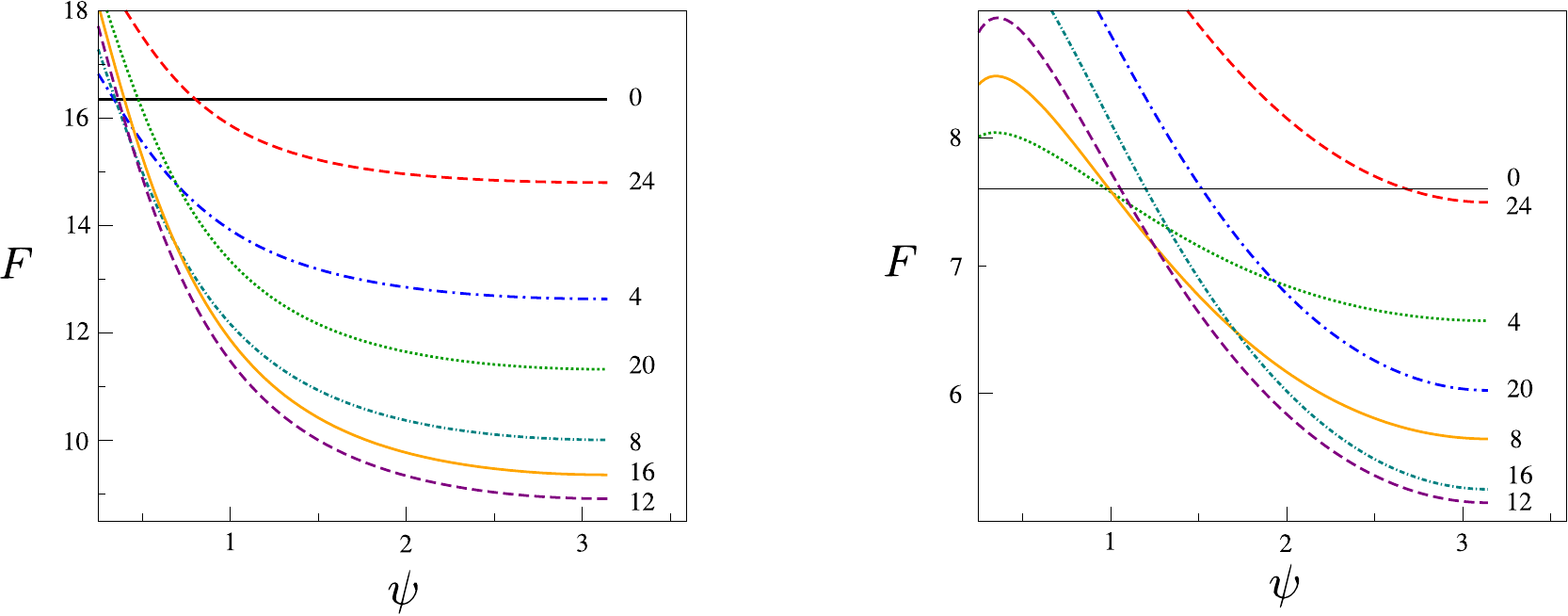}
\caption{\label{fig:sec5-bnt-energies}(Color online) The total energy (in units of $K_{A}/2$ of hexatic tori of aspect ratio $r=\sqrt{2}$ (left) and $r=2.6926$) for varying number of defects. The bending energy at fixed $r$ is subtracted off. The disclination core energy is set to $0.1K_{A}$, which is $0.2$ in the above units.}
\end{figure}

The elastic energy \eqref{eq:sec5-bnt-energy} was analyzed numerically in Ref. \cite{BowickNelsonTravesset:2004} for various aspect ratios and defect core energies. A plot of the energy of configurations obtained by placing a ring of $(N/2)$ $(+1)$-disclinations equally spaced along the same parallel on the outside of the torus and a second ring of $(N/2)$ $(-1)$-disclinations on the inside, is shown in Fig. \ref{fig:sec5-bnt-energies} for various $N$ and variable angular separation $\Delta\psi$ between the rings. The disclination core energy is taken to be $\epsilon_{c}=cK_{A}$, with $c$ a numerical constant taken equal to 0.1 in the plots. The energy at the maximum separation ($\Delta\psi=\pi$) first decreases and then increases with $N$. The optimal number of defect pairs $N/2$ is $6$ and $7$ (the latter curve is not shown in the plot) for $r=\sqrt{2}$ and $r=2.6926$ respectively.

The existence of defects in the ground state of a toroidal vesicle with hexatic order is also affected by the total number of molecules $\mathcal{N}$ forming the hexatic phase. In Ref. \cite{BowickNelsonTravesset:2004} it was found that defects disappear in the limit $\mathcal{N}\rightarrow\infty$, but are present for numbers of molecules as large as $\mathcal{N}=10^{10}$. This number is several orders of magnitude larger than the typical number of molecules of biological vesicles such as red-blood cells (i.e. $\mathcal{N}\sim 10^{8}$). To make this estimate, one considers a pair of opposite sign disclinations that have been pulled apart from the circle of zero Gaussian curvature to the equators in the regions of like sign Gaussian curvature; thus $\psi_{+}=0$ and $\psi_{-}=\pi$. Upon approximating the defect-pair energy by its flat space value, the total energy reads:
\begin{equation}\label{eq:sec5-bnt-dipole}
F = -\frac{\pi}{3}K_{A}\log\left(\frac{r+1}{r-1}\right)+\frac{\pi}{18}K_{A}\log\left(\frac{R_{2}}{a_{0}}\right)+2\epsilon_{c}\,,
\end{equation}
where $a_{0}$ is the lattice spacing. Eq. \eqref{eq:sec5-bnt-dipole} changes sign when
\begin{equation}
\frac{R_{2}}{a_{0}} = e^{-\frac{36\epsilon_{c}}{K_{A}}}\left(\frac{r+1}{r-1}\right)^{6}\,.
\end{equation}
Taking
\[
\mathcal{N}\approx \frac{A}{\frac{\sqrt{3}}{2}a_{0}^{2}}=\frac{8\pi^{2}}{\sqrt{3}\,r\left(\frac{R_{2}}{a_{0}}\right)^{2}}
\]
leads to the conclusion that defects are favored for:
\begin{equation}\label{eq:sec5-critical-size}
\mathcal{N}<\frac{8\pi^{2}}{\sqrt{3}}e^{-\frac{72\epsilon_{c}}{\pi K_{A}}}\left[r\left(\frac{r+1}{r-1}\right)^{12}\right]
\approx 4.6\,r\left(\frac{r+1}{r-1}\right)^{12}
\end{equation}
where the last identity has been obtained by taking $\epsilon_{c}/K_{A}=0.1$. This result establishes that defects are present in the ground state of a hexatic tours for any fixed number of molecules provided the torus is sufficiently fat. For the energetically favored Clifford torus with $r=\sqrt{2}$, Eq. \eqref{eq:sec5-critical-size} predicts a critical number of molecules to be order $\mathcal{N}\sim 10^{10}$.

In absence of defects the ratio between the hexatic stiffness $K_{A}$ and the bending rigidity $\kappa$ dictates the optimal shape of the toroidal vesicle. Taking $q_{i}=0$ in Eq. \eqref{eq:sec5-bnt-energy} one obtains in this case:
\begin{equation}\label{eq:sec5-bnt-defect-free}
F =\frac{2\pi^{2}K_{A}}{r+\sqrt{r^{2}-1}}+\kappa\frac{2\pi^{2}r^{2}}{r^{2}-1}\,.
\end{equation}
The first term represents the energetic cost associated with the distortion of the hexatic director field due to the Gaussian curvature alone. In the limit of large and small hexatic stiffness Eq. \eqref{eq:sec5-bnt-defect-free} is minimized by:
\[
\begin{array}{cc}
r = \sqrt{2} &\qquad K_{A}\ll\kappa \\[7pt]
r = \sqrt{\frac{K_{A}}{2\kappa}} &\qquad K_{A} \gg \kappa 
\end{array}
\]
which provides a compelling example of the interplay between order and geometry: if the stiffness associated with the intrinsic hexatic order is much smaller than the bending rigidity, the Clifford torus is the optimal geometry; if on the other hand, the hexatic stiffness dominates, then a thin torus, similar to a bicycle tire, is optimal. 

\subsection{\label{sec:5e}Toroidal crystals}

Crystalline assemblages of identical sub-units packed together and elastically bent in the form of a torus have been found in the past ten years in a variety of systems of surprisingly different nature, such as viral capsids, self-assembled monolayers and carbon nanomaterials. In the introduction we mentioned the self-assembly of toroidal micelles and vesicles from homogeneous solutions of amphiphilic molecules such as oligomers of aromatic compounds or block copolymers. Toroidal geometries also occur in microbiology in the viral capsid of the coronavirus \emph{torovirus} \cite{SnijderHorzinek}. The torovirus is an RNA viral package of maximal diameter between $120$ and $140$ nm and is surrounded, as other coronaviridae, by a double wreath/ring of cladding proteins.

Carbon nanotori form another fascinating and technologically promising class of toroidal crystals \cite{LiuEtAl:1997} with remarkable magnetic and electronic properties. The interplay between the ballistic motion of the $\pi$ electrons and the geometry of the embedding torus leads to a rich variety of quantum mechanical properties including Pauli paramagnetism \cite{MeunierEtAl:1998} and Aharonov-Bohm oscillations in the magnetization \cite{LiuChenDing:2008}. Ring closure of carbon nanotubes by chemical methods \cite{SanoEtAl:2001} suggest that nanotubes may be more flexible than at first thought and provides another technique of constructing carbon tori. In this section we review some recent developments in the study of the geometry and the elasticity of toroidal crystals. Additional details can be found in Refs. \cite{GiomiBowick:2008a,GiomiBowick:2008b}.

\subsubsection{\label{sec:5f}Geometry of toroidal polyhedra}

Toroidal crystals with local $6-$fold order can be described as toroidal polyhedra with all triangular faces.  After the discovery of carbon nanotubes in 1991 and the subsequent theoretical construction (later followed by the experimental observation) of graphitic tori, this class of polyhedra has drawn considerable attention and many possible tessellations of the circular torus have been proposed by the community \cite{Dunlap:1992,Dunlap:1994a,Dunlap:1994b,ItohEtAl:1993a,ItohEtAl:1993b,ItohEtAl:1993c,Kirby:1994,LaszloEtAl:2001a,LaszloEtAl:2001b,DiudeaEtAl:2001,Diudea:2002}. Here we review the construction of a defect-free triangulated torus and we show how the most symmetric defective triangulations can be generally grouped into two fundamental classes corresponding to symmetry groups $D_{nh}$ and $D_{nd}$ respectively. 

\begin{figure}
\centering
\includegraphics[width=.4\columnwidth]{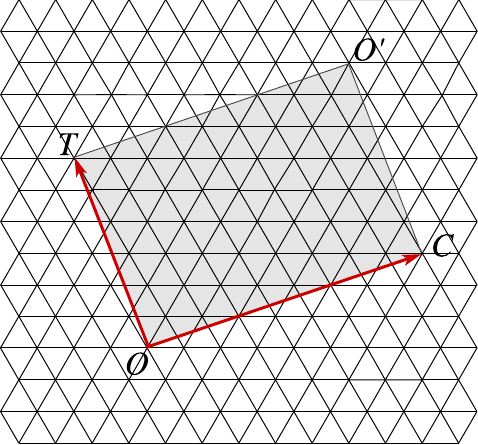}\qquad
\includegraphics[width=.4\columnwidth]{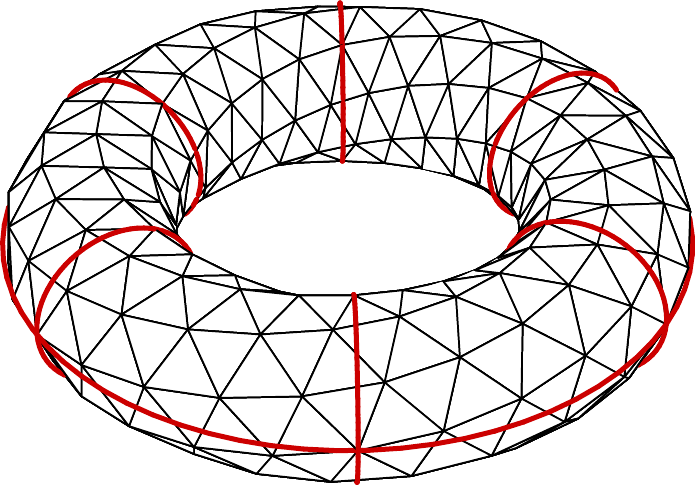}
\caption{\label{fig:triangulation}(Color online) Construction of a defect-free triangulation of 
the torus. On top planar map of the triangulated torus corresponding to the 
$(n,m,l)$ configuration $(6,3,1)$. On the bottom $(6,3,6)$ chiral torus. The edges 
of each one of the six tubular segments has been highlighted in red.}
\end{figure}

The structure of a triangulated cylinder can be specified by a pair of triangular lattice vectors $\bm{c}$ and $\bm{t}$, called the \emph{chiral} and \emph{translation} vector respectively, which together define how the planar lattice is rolled up. In the canonical basis $\bm{a_{1}}=(1,0)$ and $\bm{a}_{2}=(\frac{1}{2},\frac{\sqrt{3}}{2})$, the vector $\bm{c}$ has the form
\begin{equation}\label{eq:chiral_vector}
\bm{c} = n\bm{a}_{1}+m\bm{a}_{2}\qquad n,\,m\in\mathbb{Z}\,.
\end{equation}
The translation vector $\bm{t}$, on the other hand, can be expressed as an integer multiple 
\begin{equation}\label{eq:translation_vector}
\bm{t}=l\bm{e}_{t}\qquad l\in\mathbb{Z}
\end{equation}
of the shortest lattice vector $\bm{e}_{t}$ perpendicular to $\bm{c}$. The vector $\bm{e}_{t}$ is readily found to be of the form
\[
\bm{e}_{t} = \frac{(n+2m)\bm{a}_{1}-(2n+m)\bm{a}_{2}}{(n+2m:2n+m)}\,,
\]
where $(a:b)$ denotes the greatest common divisor of $a$ and $b$ and enforces the minimal length. The three-dimensional structure of the torus is obtained by connecting the edge $\overline{OT}$ of the rectangle in Fig. \ref{fig:triangulation} (left) to $\overline{O'C}$ and $\overline{OC}$ to $\overline{O'T}$. The edge $\overline{OT}$ is then mapped to the external equator of the torus while the edge $\overline{OC}$ to the $\phi=0$ meridian. The resultant toroidal lattice has characteristic chirality related to the initial choice of the vector $\bm{c}$. In the nanotubes literature \emph{armchair} refers to the lattice obtained by choosing $n=m$, \emph{zigzag} to that obtained for $m=0$ and \emph{chiral} to all other lattices. An example of an $(n,m,l)$ chiral torus is shown in Fig.~\ref{fig:triangulation} (right) for the case $n=6$, $m=3$ and $l=6$. The chirality is extremely important in graphitic carbon nanotube or nanotori, where it determines whether the electronic behavior of the system is metallic or semiconducting.

By Euler's theorem one can prove that the number of triangular faces $F$ and the number of vertices $V$ of a triangular toroidal lattice is given by
\[
V = \tfrac{1}{2}F\,.
\]
Denoting $A_{R}$ the area of the rectangle with edges $\bm{c}$ and $\bm{t}$ and $A_{T}$ the area of a fundamental equilateral triangle, the number of vertices of a defect-free toroidal triangulation is then:
\begin{equation}\label{eq:vertices}
V 
= \frac{A_{R}}{2A_{T}}
= \frac{2l\,(n^{2}+nm+m^{2})}{(n+2m:2n+m)}\,.
\end{equation}
The planar construction reviewed above allows only lattices with an even number of vertices. Defect-free toroidal deltahedra with an odd number of vertices are also possible and their construction is generally achieved by assembling congruent octahedral building blocks. We refer the reader to Ref. \cite{Webber:1997} for an comprehensive review
of the topic.

\begin{figure}
\centering
\includegraphics[width=.6\columnwidth]{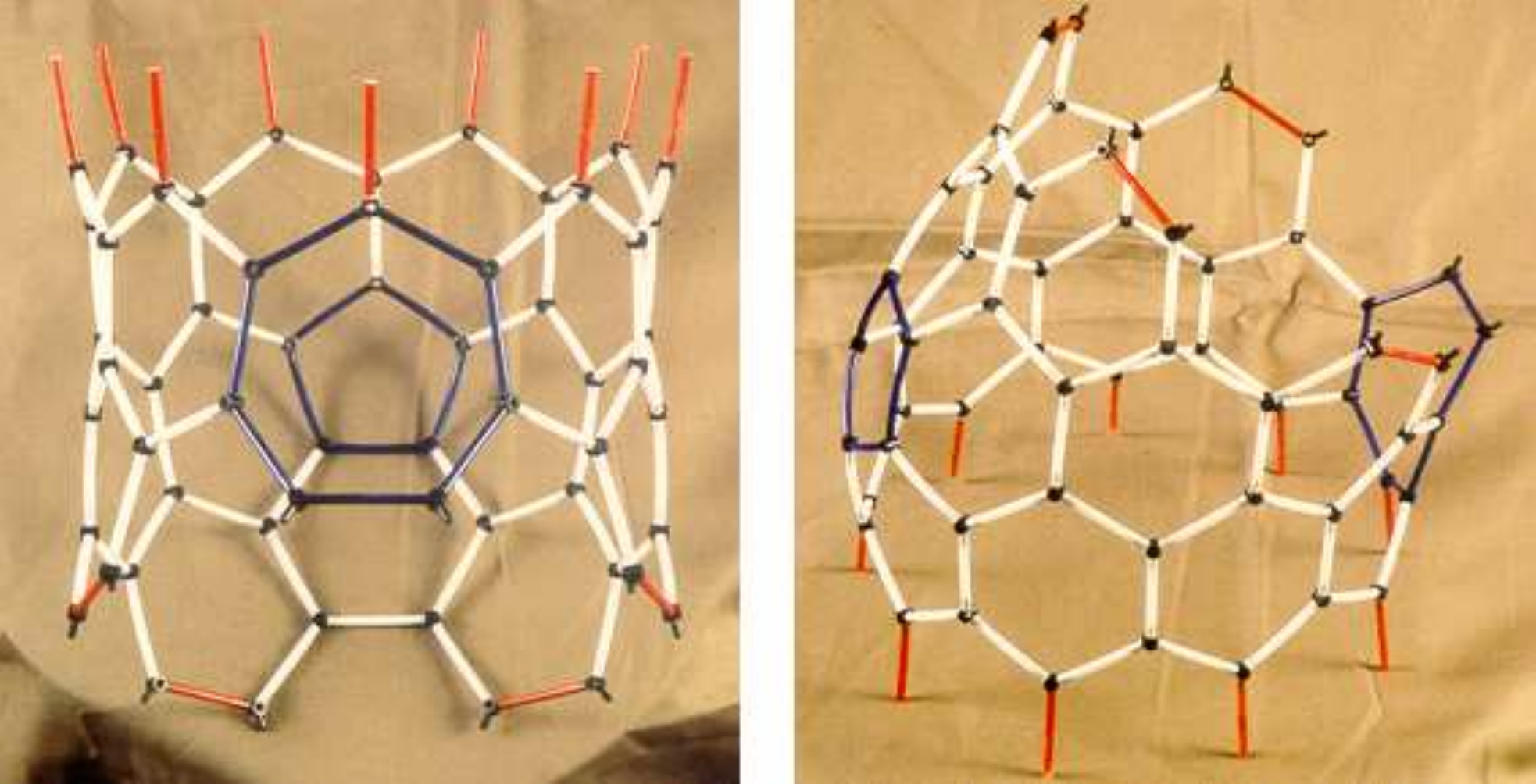}
\caption{\label{fig:knees}(Color online) Dunlap knees obtained by joining two straight tubular segments with $(n,0)$ and $(m,m)$ chirality. [Courtesy of A. A. Lucas and A. Fonseca, Facult\'es Universitaries Notre-Dame de la Paix, Namur, Belgium].}
\end{figure}

\begin{figure}[h]
\centering
\includegraphics[width=.5\columnwidth]{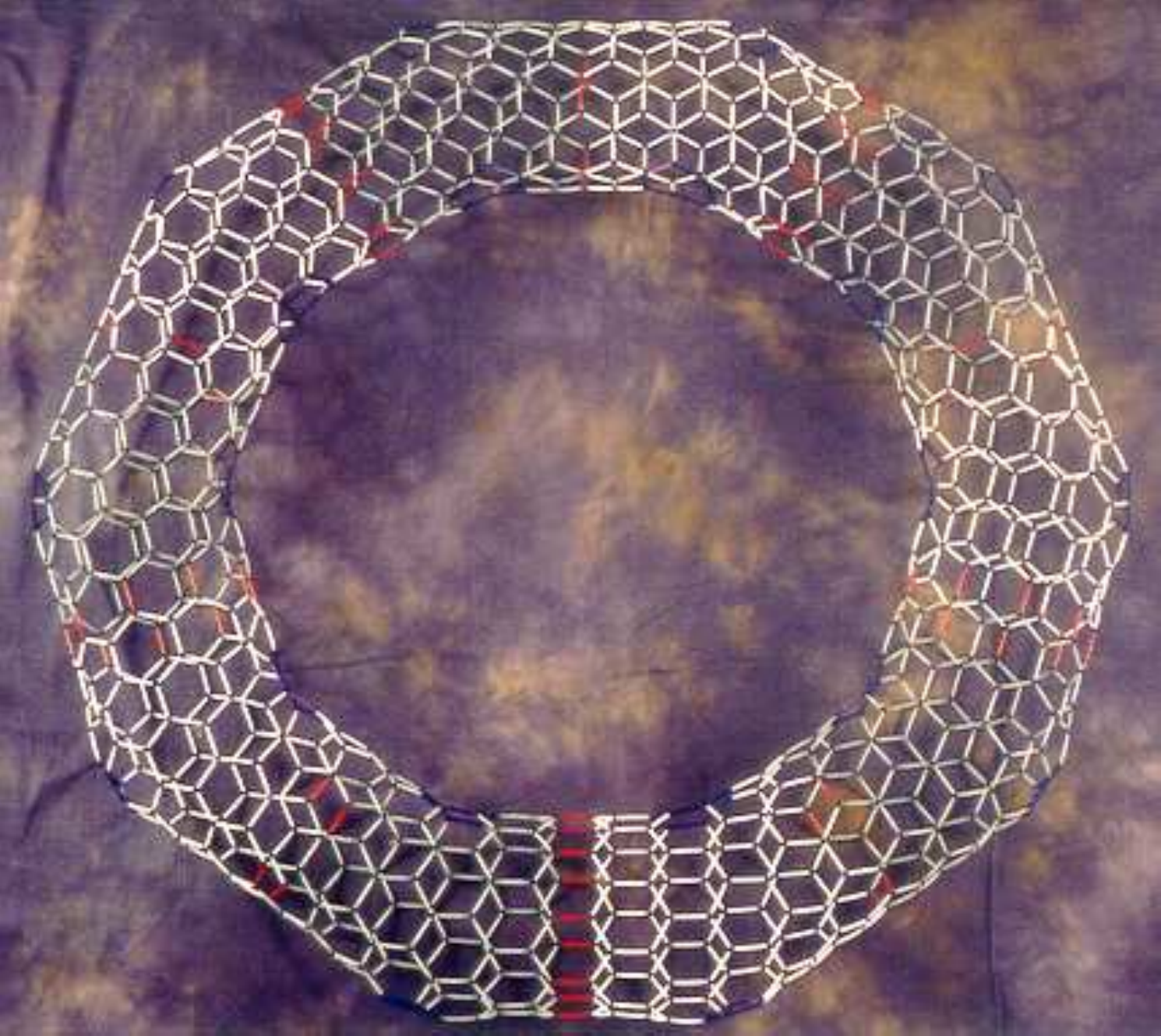}
\caption{\label{fig:lucas-torus}(Color online) Five-fold polygonal torus obtained by joining $(5,5)$ and $(9,0)$ tubular segments via ten pairs of $5-7$ rings. This structure was originally proposed by the authors of Ref.~\cite{LambinEtAl:1995,FonsecaEtAl:1996} as a possible low-strain configuration for carbon nanotori. [Courtesy of A. A. Lucas and A. Fonseca, Facult\'es Universitaries Notre-Dame de la Paix, Namur, Belgium].}
\end{figure}

The embedding of an equal number of pentagonal and heptagonal disclinations in the hexagonal network was first proposed by Dunlap in 1992 as a possible 
way to incorporate positive and negative Gaussian curvature into the cylindrical geometry of carbon tubules \cite{Dunlap:1992}. According to the Dunlap 
construction the necessary curvature is incorporated by the insertion of ``knees'' (straight cylindrical sections of the same diameter joined with a kink) in correspondence with each pentagon-heptagon pair arising from the junction of tubular segments of different chirality (see Fig.~\ref{fig:knees}).   
\begin{figure}[h]
\centering
\includegraphics[width=.45\columnwidth]{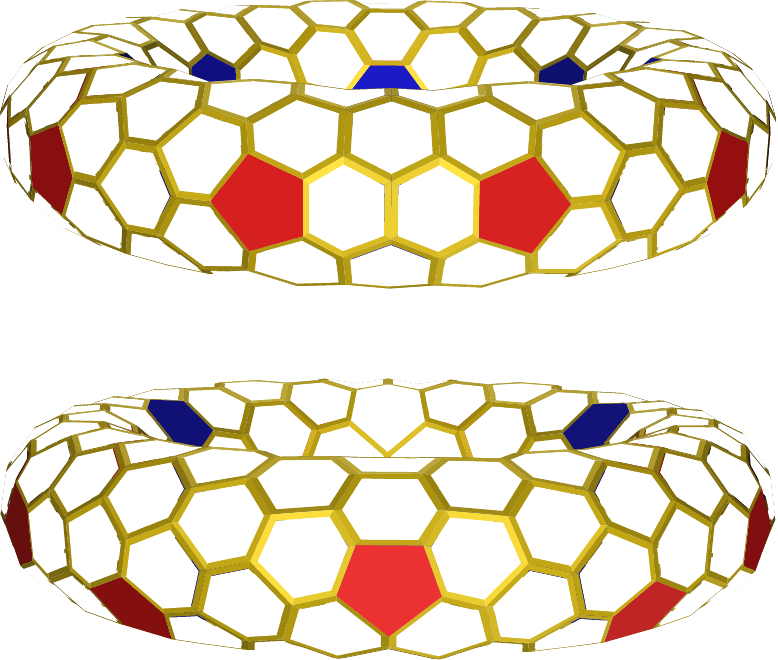}
\caption{\label{fig:prism}(Color online) Voronoi lattices of a TP$n$ prismatic (top) and 
TA$n$ antiprismatic (bottom) toroids with $R_{1}=1$ and $R_{2}=0.3$.}
\end{figure}
In particular, a junction between a $(n,0)$ and a $(m,m)$ tube can be obtained by placing a $7-$fold disclination along the internal equator of the torus and a $5-$fold disclination along the external equator. Since the radii of the two segments of a junction are different by construction, the values of $n$ and $m$ are commonly chosen to minimize the ratio $|\bm{c}_{(n,0)}|/|\bm{c}_{(m,m)}|=n/\sqrt{3}m$. By repeating the $5-7$ construction periodically it is possible to construct an infinite number of toroidal lattices with an even number of disclinations pairs and dihedral symmetry group $D_{nh}$ (where $2n$ is the total number of $5-7$ pairs, Fig. \ref{fig:prism} top and Fig. \ref{fig:lucas-torus}). The structure of the lattice is described by the alternation of two motifs with crystalline axes mutually rotated by $30^{\circ}$ as a consequence of the connecting disclination. One of the fundamental aspects of Dunlap's construction is that all the disclinations are aligned along the two equators of the torus where the like-sign Gaussian curvature is maximal. As we will see below, this feature makes these arrangements optimal in releasing the elastic stress due to curvature. Other defective triangulations with $D_{nh}$ symmetry group have been proposed in Ref. \cite{GiomiBowick:2008b} and won't be discussed here.

Another class of crystalline tori with dihedral antiprismatic symmetry $D_{nd}$ was initially proposed by Itoh \emph{et al} \cite{ItohEtAl:1993a,ItohEtAl:1993b,ItohEtAl:1993c} shortly after Dunlap. Aimed at reproducing a structure similar to the C$_{60}$ fullerene, Itoh's original construction implied ten disclination pairs and the point group $D_{5d}$. In contrast to Dunlap tori, disclinations are never aligned  along the equators in antiprismatic tori, instead being staggered at some angular distance $\delta\psi$ from the equatorial plane. Hereafter we will use the symbols TP$n$ and TA$n$ to refer to toroidal lattices with $2n$ disclination pairs and symmetry group $D_{nh}$ and $D_{nd}$ respectively.

\begin{figure}[h]
\centering
\includegraphics[scale=0.4]{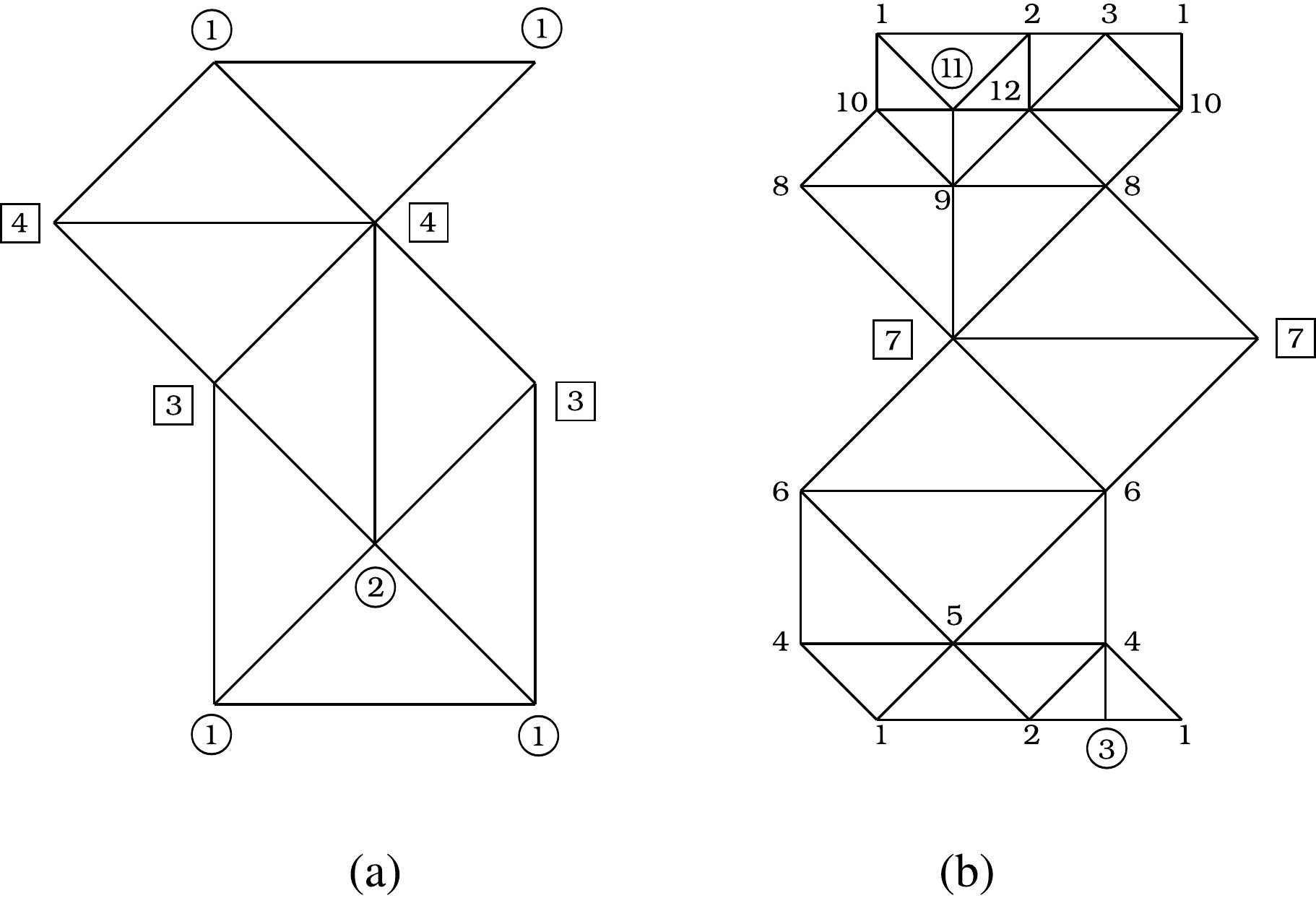}
\caption{\label{fig:graph1}(a) Unit cell for toroidal antiprisms. $5-$fold vertices are circled and $7-$fold vertices are boxed. (b) Unit cell of a $D_{nd}$ torus in the Berger-Avron construction. The graph consists of four generation of tiles and the internal equator of the torus is mapped into the horizontal line passing to the mid-point between the 6th and the 7th vertex.}
\end{figure}

\begin{figure}[t]
\centering
\includegraphics[width=.6\columnwidth]{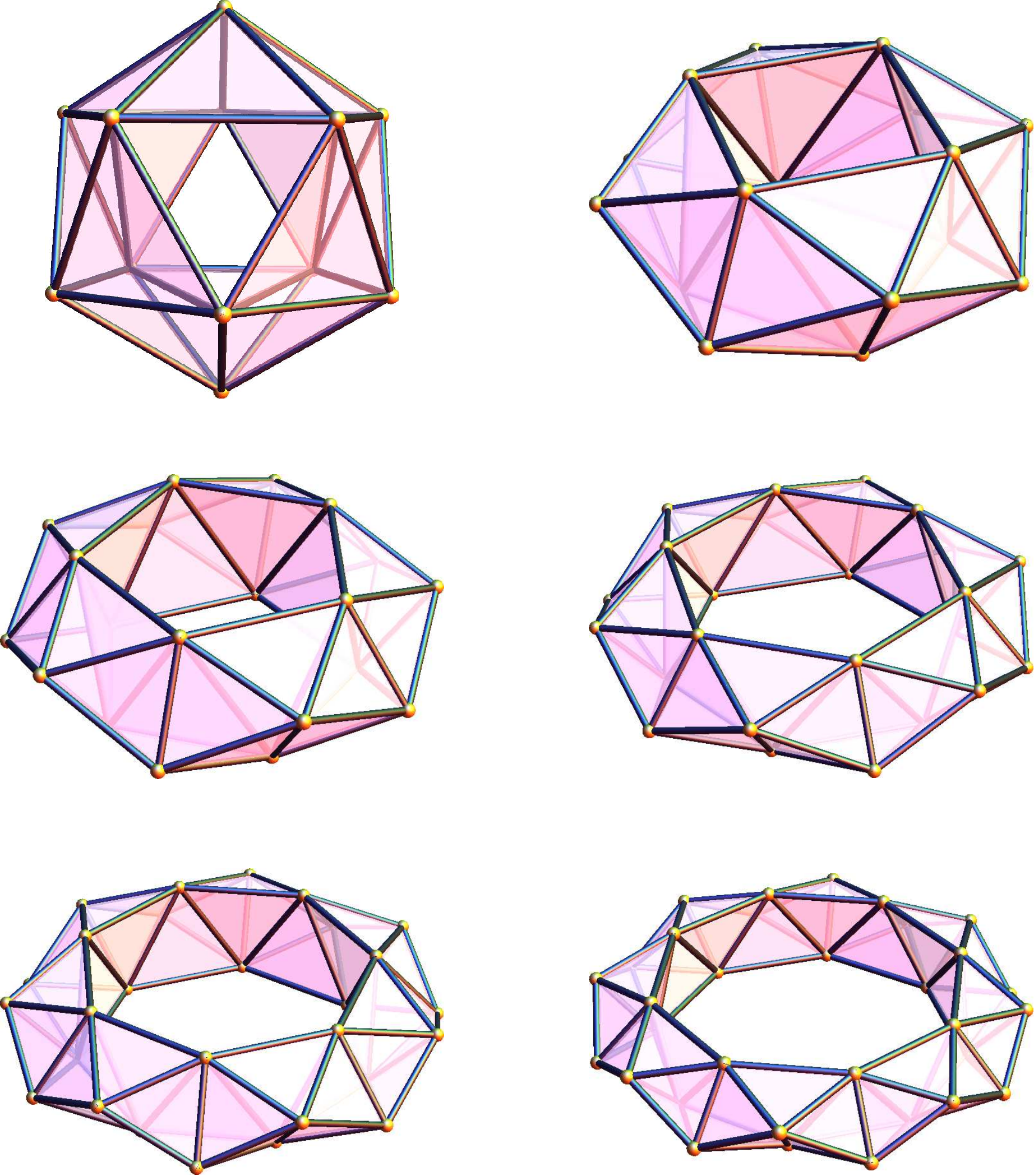}
\caption{\label{fig:toroidal-antiprisms}(Color online) First six toroidal antiprisms obtained
by repeating the unit cell of Fig.~\ref{fig:graph1}. The first polyhedron on 
the left is the ``drilled icosahedron''.}
\end{figure}

A systematic construction of defected triangulations of the torus can be achieved in the context of planar graphs \cite{Lavrenchenko:1990,BergerAvron:1995a,BergerAvron:1995b}. A topological embedding of a graph in a two-dimensional manifold corresponds to a triangulation of the manifold if each region of the graph is bounded by exactly three vertices and three edges, and any two regions have either one common vertex or one common edge or no common elements of the graph. The simplest example of toroidal polyhedra with $D_{nd}$ symmetry group, featuring only $5-$fold and $7-$fold vertices, can be constructed by repeating $n$ times the unit cell of Fig.~\ref{fig:graph1}a. These \emph{toroidal antiprisms} have $V=4n$ vertices and can be obtained equivalently from the edge skeleton of a $n-$fold antiprism by attaching at each of the base edges a pentagonal pyramid and by closing the upper part of the polyhedron with $n$ additional triangles. By counting the faces one finds $F=5n+2n+n=8n$ from which $V=4n$. The simplest polyhedron of this family has $V=12$ and $D_{3d}$ symmetry group (see top left of Fig. \ref{fig:toroidal-antiprisms}) and corresponds to the ``drilled icosahedron'' obtained by removing two parallel faces of an icosahedron and connecting the corresponding edges with the six lateral faces of an antiprism with triangular base (i.e. a prolate octahedron). Starting from this family of toroidal antiprisms a number of associated triangulations having the same defect structure can be obtained by geometrical transformations such as the Goldberg inclusion \cite{Goldberg:1937,CasparKlug:1962,VirusMacromolecules}. Such transformations, popularized by Caspar and Klug for the construction of the icosadeltahedral structure of spherical viruses \cite{CasparKlug:1962}, consist in partitioning each triangular face of the original graph into smaller triangular faces in such a way that old vertices preserve their valence and new vertices have valence six. The partition is obtained by specifying two integer numbers $(L,M)$ which define how the original vertices of each triangle are connected by the new edges so that the total number of vertices is increased by a factor $T = L^{2}+LM+M^{2}$. 

A general classification scheme for $D_{nd}$ symmetric tori was provided by Berger and Avron \cite{BergerAvron:1995a,BergerAvron:1995b} in 1995. Their scheme is based on the construction of unit graphs comprising triangular tiles of different \emph{generations}. In each generation, tiles are scaled in length by a factor $1/\sqrt{2}$ with respect to the previous generation. This rescaling approximates the non-uniformity of the metric of a circular torus. 

In the past few years, alternative constructions of triangulated tori have been proposed as well as novel geometrical and graph-theoretical methods to express the coordinates of their three-dimensional structures (see for example Kirby \cite{Kirby:1994}, L\'{a}szl\'{o} \emph{at al} \cite{LaszloEtAl:2001a,LaszloEtAl:2001b}, Diudea \emph{et al} \cite{DiudeaEtAl:2001,Diudea:2002}). Here we choose to focus on the defect structure associated with the two most important class TP$n$ and TA$n$ with groups $D_{nh}$ and $D_{nd}$.

\subsubsection{\label{sec:5g}Elasticity of defects on the torus}

The total free energy of a toroidal crystal with $N$ disclinations can be expressed as usual as
\begin{equation}\label{eq:sec5-gb-free-energy}
F = \frac{1}{2Y}\int d^{2}x\,\Gamma^{2}(\bm{x})+\epsilon_{c}\sum_{i=1}^{N}q_{i}^{2}+F_{0}
\end{equation}
where $Y$ is the two dimensional Young modulus and
\begin{equation}\label{eq:gamma_total}
\Gamma(\bm{x})=\frac{\pi}{3}\sum_{k=1}^{N}q_{k}\Gamma_{d}(\bm{x},\bm{x}_{k})-\Gamma_{s}(\bm{x})\, .
\end{equation}
The defect part of the stress function $\Gamma(\bm{x})$ can be found by integrating the Green function \eqref{eq:final_green_function} and takes the form
\begin{multline}\label{eq:gamma_defects}
\frac{\Gamma_{d}(\bm{x},\bm{x}_{k})}{Y}
= \frac{\kappa}{16\pi^{2}}\left(\psi_{k}
-\frac{2}{\kappa}\,\xi_{k}\right)^{2}
-\frac{1}{4\pi^{2}\kappa}(\phi-\phi_{k})^{2}
+\frac{1}{4\pi^{2}r}\log(r+\cos\psi_{k})\\[5pt]
-\frac{\kappa}{4\pi^{2}}\Real\{\Li(\alpha e^{i\psi_{k}})\}
+\frac{1}{2\pi}\log\left|\vartheta_{1}\left(\frac{z-z_{k}}{\kappa}\bigg|\frac{2i}{\kappa}\right)\right|\,,
\end{multline}
where $\Li$ is the usual Eulerian dilogarithm and
\begin{equation}
\alpha = \sqrt{r^{2}-1}-r\,.
\end{equation}
The function $\Gamma_{s}(\bm{x})$ representing the stress field due to the Gaussian curvature of the torus, on the other hand, is given by
\begin{equation}\label{eq:gamma_screening}
\frac{\Gamma_{s}(\bm{x})}{Y}
= \log\left[\frac{r+\sqrt{r^{2}-1}}{2(r+\cos\psi)}\right]+\frac{r-\sqrt{r^{2}-1}}{r}\,.
\end{equation}
A derivation of the functions $\Gamma_{d}(\bm{x},\bm{x}_{k})$ and $\Gamma_{s}(\bm{x})$ is reported in Ref. \cite{GiomiBowick:2008b} and won't be repeated here. 

\begin{figure}[t]
\centering
\includegraphics[width=0.65\columnwidth]{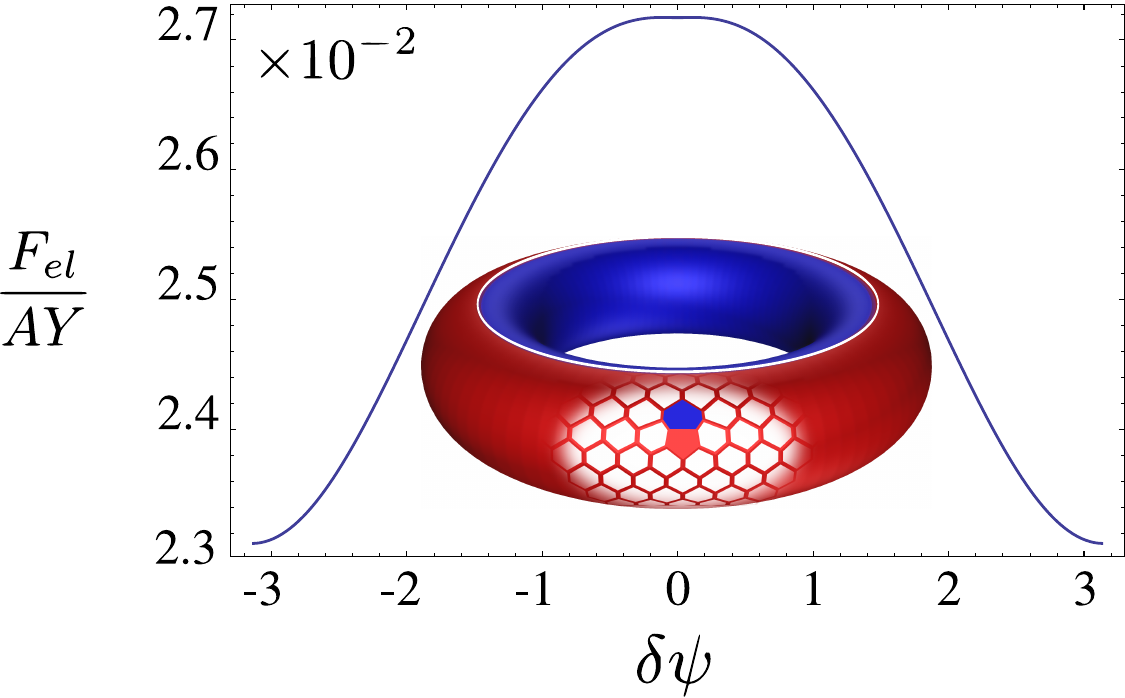}
\caption{\label{fig:redbluetorus}(Color online) Elastic energy of a $5-7$ disclination dipole constrained to lie on the same meridian, as a function of the angular separation. In the inset, illustration of a circular torus of radii $R_{1}>R_{2}$. Regions of positive and negative Gaussian curvature have been shaded in red and blue respectively.}
\end{figure}

\begin{figure}[t]
\centering
\includegraphics[width=0.65\columnwidth]{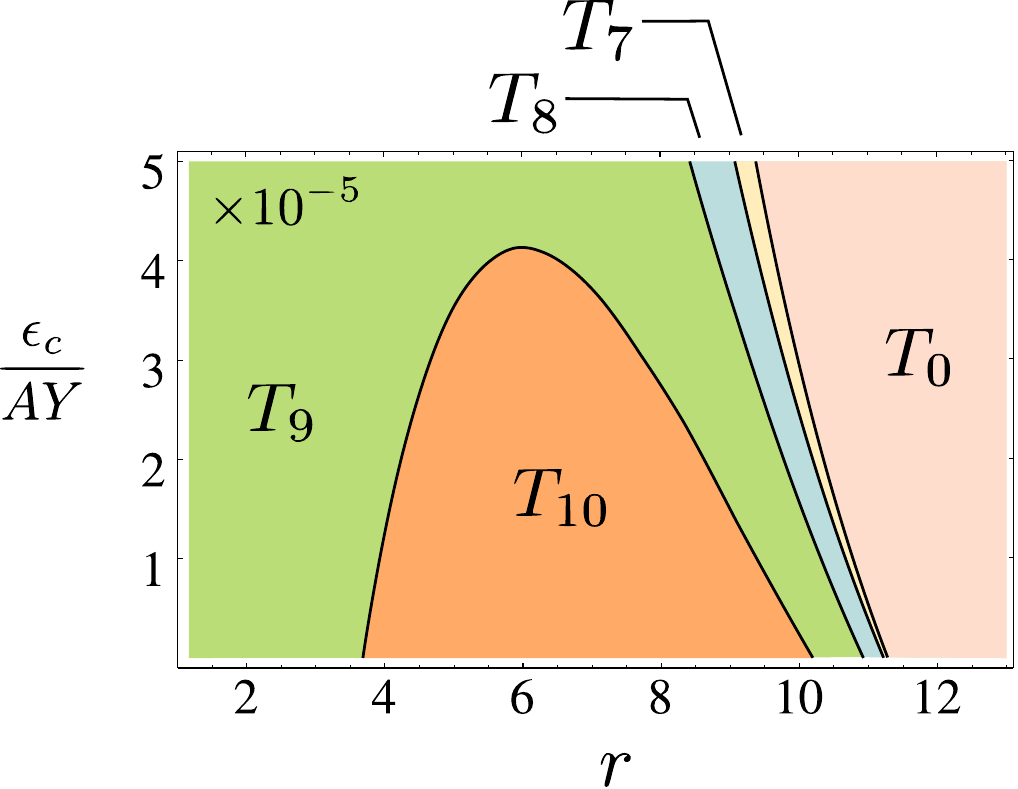}
\caption{\label{fig:phase_diagram1}(Color online) Phase diagram for $T_{p}$ configurations in the plane $(r,\epsilon_{c}/AY)$. For $r\in[3.68,\,10.12]$ and 
$\epsilon_{c}\sim 0$ the structure is given by a $T_{10}$ configuration with symmetry group $D_{5h}$.}
\end{figure}

To analyze the elastic free energy \eqref{eq:sec5-gb-free-energy} we start by considering the energies of two opposite sign disclinations constrained to lie on the same meridian. The elastic free energy of this system is shown in Fig.~\ref{fig:redbluetorus} as a function of the angular separation between the two disclinations. The energy is minimized for the positive ($5-$fold) disclination on the external equator (maximally positive Gaussian curvature) and the negative ($7-$fold) disclination on the internal equator (maximally negative Gaussian curvature). The picture emerging from this simple test case suggests that a good \emph{ansatz} for an optimal defect pattern is a certain number $p$ of equally spaced $+1$ disclinations on the external equator matched by the same
number of equally spaced $-1$ disclinations on the internal equator.
We name this configuration with the symbol $T_{p}$, where $p$ stands for the 
total number of disclination pairs.
\begin{equation}\label{eq:s_n}
T_{p}:\quad
\left\{
\left( 0, \frac{2\pi k}{p}\right)_{1 \le k \le p};\,
\left(\pi,\frac{2\pi k}{p}\right)_{1 \le k \le p}
\right\}\,,
\end{equation}
where the two pairs of numbers specify the $(\psi,\phi)$ coordinates of the positive and negative disclinations respectively. A comparison of the energy 
of different $T_{p}$ configurations, as a function of aspect ratio and disclination core energy, is summarized in the phase diagram of Fig. \ref{fig:phase_diagram1}. We stress here that only $T_{p}$ configurations with $p$ even have an embedding on the torus corresponding to lattices of the TP$\frac{p}{2}$ class. Nevertheless a comparison with $p-$odd configurations can provide additional information on the stability of $p-$even lattices. For small core energies, moreover, thermally excited configurations with a large number of defects and similar $p-$polar distributions of topological charge are expected to exhibit an elastic energy comparable in magnitude with that of these minimal constructions.      
The defect core energy has been expressed here in the form
\begin{equation}\label{eq:core_energy}
F_{c} 
= \epsilon_{c}\sum_{i=1}^{2p} q_{i}^{2} 
= 2p\epsilon_{c} \ .
\end{equation}
The core energy $\epsilon_{c}$ of a single disclination depends on the details of the crystal-forming material and the corresponding microscopic interactions. A simple phenomenological argument (see for example Ref.~\cite{KlemanLavrentovich}) gives
\[
\frac{\epsilon_{c}}{Y} \sim \frac{a^{2}}{32\pi} \ ,
\]
where $a$ the lattice spacing. Taking $a^{2}=A/\frac{\sqrt{3}}{2}V$, with $A$ the area of the torus, yields
\begin{equation}\label{eq:core_magnitude}
\frac{\epsilon_{c}}{AY} \sim \frac{1}{16\sqrt{3}\,\pi V} \sim \frac{10^{-2}}{V}\,.
\end{equation}
For a system of order $V=10^{3}$ subunits, then, the dimensionless core energy on the left hand side of Eq.~\eqref{eq:core_magnitude} is of order $10^{-5}$. This estimate motivates our choice of the scale for $\epsilon_{c}/(AY)$ in Fig. \ref{fig:phase_diagram1}. 

For dimensionless core energies below $4\cdot 10^{-5}$ and aspect ratios $r$ between 3.68 and 10.12 the ground state structure is the TP5 lattice corresponding to a double ring of $+1$ and $-1$ disclinations distributed on the external and internal equators of the torus as the vertices of a regular 
decagon (the $T_{10}$ configuration). The TP5 lattice has dihedral symmetry group $D_{5h}$. That this structure might represent a stable configuration for polygonal carbon toroids has been conjectured by the authors of Ref. \cite{LambinEtAl:1995}, based on the argument that the 36$^{\circ}$ angle  arising from the insertion of ten pentagonal-heptagonal pairs into the lattice would optimize the geometry of a nanotorus consistently with the structure of the $sp^{2}$ bonds of the carbon network (unlike the 30$^{\circ}$ angle of the $6-$fold symmetric configuration originally proposed by Dunlap). In later molecular dynamics simulations, Han \cite{Han:1997,Han:1998} found that a $5-$fold symmetric lattice, such as the one obtained from a (9,0)/(5,5) junction (see Fig.~\ref{fig:lucas-torus}), is in fact stable for toroids with aspect ratio less then $r\sim 10$. The stability, in this case, results from the strain energy per atom being smaller than the binding energy of carbon atoms. Irrespective of the direct experimental observation of such disclinated toroidal crystals, which is still open, we show here how continuum elasticity predicts that a $5-$fold symmetric lattice indeed constitutes a minimum of the elastic energy for a broad range of aspect ratios and defect core energies.

\begin{figure}[t]
\centering
\includegraphics[width=0.65\columnwidth]{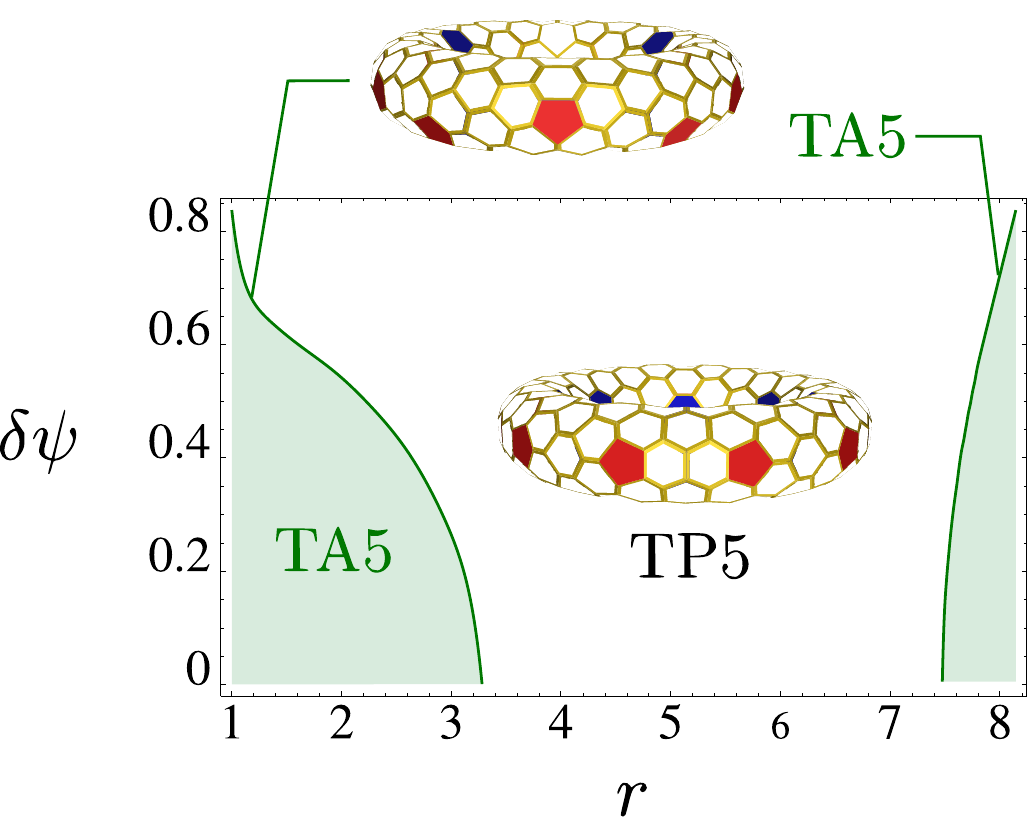}
\caption{\label{fig:phase_diagram2}(Color online) Phase diagram of a $5-$fold symmetric lattice in the plane $(r,\delta\psi)$. For small $\delta\psi$ and $r$ in the range $[3.3,\,7.5]$ the prismatic TP5 configuration is energetically favored. For $r<3.3$ the system undergoes a structural transition to the antiprismatic phase TA5.}
\end{figure}

For small aspect ratios the $5-$fold symmetric configuration becomes unstable and is replaced by the $9-$fold symmetric phase $T_{9}$. As we mentioned, however, this configuration doesn't correspond to a possible triangulation of the torus. It is likely that the ground sate in this regime consists of ten skew disclination pairs as in the antiprismatic TA$n$ lattice. The latter can be described by introducing a further degree of freedom $\delta\psi$ representing the angular displacement of defects from the equatorial plane:
\begin{equation}
\mathrm{TA}n:\quad
\left\{
\left( (-1)^{2k}\delta\psi, \frac{2\pi k}{n}\right)_{1 \le k \le n};
\left( (-1)^{2k}(\pi-\delta\psi),\frac{2\pi k}{n}\right)_{1 \le k \le n}
\right\}
\end{equation}
A comparison of the TP5 configuration and the TA5 configuration is shown in Fig.~\ref{fig:phase_diagram2} for different values of $\delta\psi$. The intersection points of the boundary curves with the $\delta\psi-$axis has been calculated by extrapolating the $(r,\delta\psi)$ data points in the range $\delta\in[0.07,\,0.8]$ with $\Delta(\delta\psi) = 2.5\pi\cdot 10^{-3}$. For small $\delta\psi$ and $r\in[3.3,\,7.5]$ the prismatic TP5 configuration is energetically favored. For $r<3.3$, however, the lattice undergoes a structural transition to the TA5 phase. For $r>7.5$ the prismatic symmetry of the TP5 configuration breaks down again. In this regime, however, the elastic energy of both configurations rapidly rises because of the lower curvature and defects disappear.

In the regime of large particle numbers, the amount of curvature required to screen the stress field of an isolated disclination in units of lattice spacing becomes too large and disclinations are unstable to grain boundary ``scars'' consisting of a linear array of tightly bound $5-7$ pairs radiating from an unpaired disclination \cite{BowickNelsonTravesset:2000,GiomiBowick:2007a}. In a manifold with variable Gaussian curvature this effects leads to a regime of coexistence of isolated disclinations (in regions of large curvature) and scars. In the case of the torus the Gaussian curvature inside ($|\psi|>\pi/2$) is always larger in magnitude than that outside ($|\psi|<\pi/2$) for any aspect ratio and so we may expect a regime in which the negative internal curvature is still large enough to support the existence of isolated $7-$fold disclinations, while on the exterior of the torus disclinations are delocalized in the form of positively charged grain boundary scars.

\begin{figure}[t]
\centering
\includegraphics[width=0.65\columnwidth]{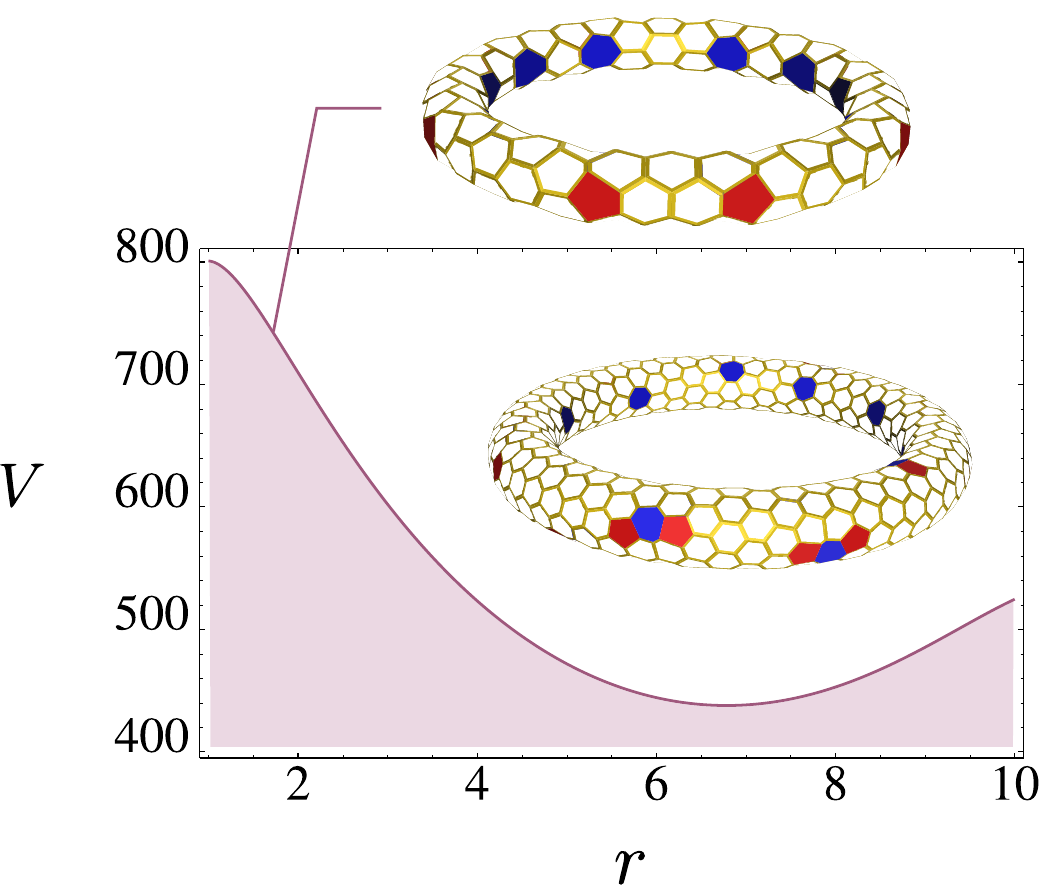}
\caption{\label{fig:phase_diagram3}(Color online) Isolated defects and scar phases in the $(r,V)$ plane. When the number of vertices $V$ increases the range of the screening curvature becomes smaller than one lattice spacing and disclinations appear delocalized in the form of a $5-7-5$ grain boundary mini-scar.}
\end{figure}

This hypothesis can be checked by comparing the energy of the TP5 lattice previously described with that of ``scarred'' configurations obtained by decorating the original toroid in such a way that each $+1$ disclination on the external equator is replaced by a $5-7-5$ mini-scar. The result of this comparison is summarized in the phase diagram of Fig. \ref{fig:phase_diagram3} in terms of $r$ and the number of vertices of the triangular lattice $V$ (the corresponding hexagonal lattice has twice the number of vertices, i.e. $V_{hex}=2V$). $V$ can be derived from the angular separation of neighboring disclinations in the same scar by approximating $V \approx A/A_{V}$, with $A_{V}=\frac{\sqrt{3}}{2}a^{2}$ the area of a hexagonal Voronoi cell and $a$ the lattice spacing. When the aspect ratio is increased from 1 to 6.8 the range of the curvature screening becomes shorter and the number of subunits required to destroy the stability of the TP5 lattice decreases. For $r>6.8$, however, the geodesic distance between the two equators of the torus becomes too small and the repulsion between like-sign defects takes over. Thus the trend is inverted.

\begin{figure}[h!]
\centering
\includegraphics[width=1\textwidth]{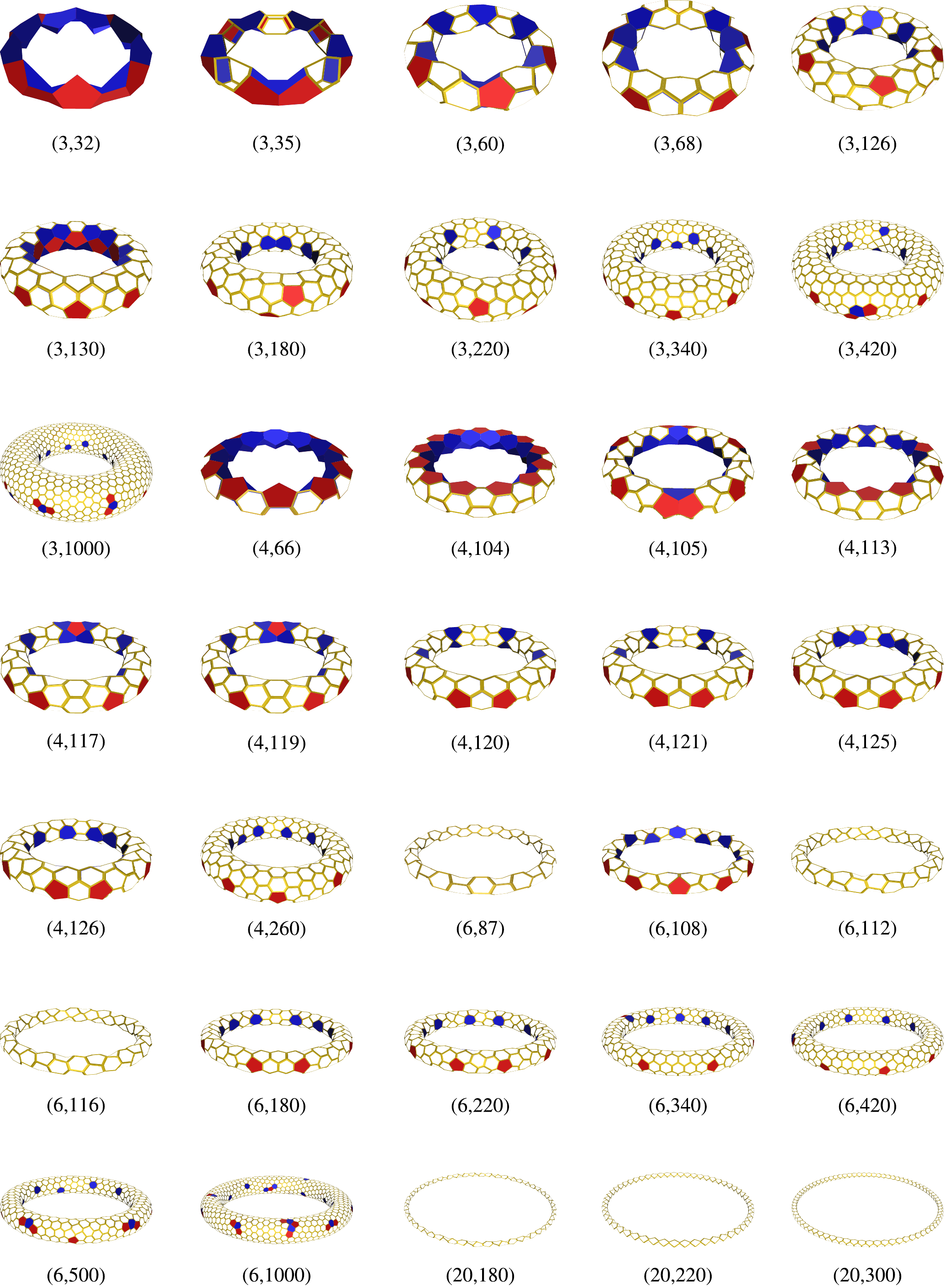}
\caption{\label{fig:numeric_tori}(Color online) Selected low energy configurations for toroidal lattices of aspect ratio $r=3,\,4,\,6$ and $20$. Lattices are labeled by $(r,V)$, with $V$ the number of particles.}
\end{figure}

Some example of toroidal lattices obtained from numerical simulations are shown in Fig. \ref{fig:numeric_tori}. The configurations are found by optimizing a system of $V$ point-like particles constrained to lie on the surface of a torus and interacting via a pair potential of the form $U_{ij}=1/|\bm{x}_{i}-\bm{x}_{j}|^{3}$ where $|\cdot|$ denotes the Euclidean distance in $\mathbb{R}^{3}$. The choice of the cubic potential is motivated here by the so called ``poppy seed bagel theorem'' \cite{HardinSaff:2005}, according to which the configuration of points that minimizes the Riesz energy $E=\sum_{i<j}1/|\bm{x}_{i}-\bm{x}_{j}|^{s}$ on a rectifiable manifold of Hausdorff dimension $d$ is uniformly distributed on the manifold for $s\ge d$. In the case of a torus of
revolution this implies that for small $s$ the points are mostly distributed on the exterior of the torus (the interior becomes completely empty in the limit $V\rightarrow\infty$). As $s$ is increased, however, the points cover a progressively larger portion of the surface. The distribution becomes uniform for $s\ge 2$. On the other hand, since the number of local minima of the Riesz energy increases with $s$, it is practical to choose a value not 
much larger than two. The choice $s=3$ has the further advantage of modelling a real physical system of neutral colloidal particles assembled at an interface \cite{Pieranski:1980} and is therefore suitable for direct comparison with experiments on colloidal suspensions. The lowest energies found, as well as the number of defects in the corresponding configuration, are reported in Table~\ref{tab:minima}. The complete set of data produced in these simulations together with a collection of interactive 3D graphics for each low energy configuration studied can be found on-line~\cite{TorusDatabase}.

\begin{table}[h!]
\tbl{Low energy configuration for a selected 
number of toroidal lattices with aspect ratios $r=3,\,4,\,6$ and $20$.
For each aspect ratio the table displays the number of particles $V$,
the lowest energy found and the number of $k-$fold vertices $V_{k}$
with $k=4$--$8$.}
{\begin{tabular}{cccccccc}
\toprule
$r$ & $V$ & $V_{4}$ & $V_{5}$ & $V_{6}$ & $V_{7}$ & $V_{8}$ & Energy \\
\colrule
\multirow{18}{*}{3} & 
  32 & 0 & 16 & 0 & 16 & 0 & 505.086593 \\
& 35 & 0 & 10 & 15 & 10 & 0 & 637.633663 \\
& 42 & 0 & 12 & 18 & 12 & 0 & 1020.466912 \\
& 120 & 0 & 14 & 92 & 14 & 0 & 14671.476332 \\
& 121 & 0 & 20 & 81 & 20 & 0 & 14981.224344 \\
& 125 & 0 & 19 & 87 & 19 & 0 & 16255.583992 \\
& 126 & 0 & 12 & 102 & 12 & 0 & 16586.793347 \\
& 130 & 0 & 20 & 90 & 20 & 0 & 17930.955152 \\
& 180 & 0 & 10 & 160 & 10 & 0 & 40623.325218 \\
& 220 & 0 & 10 & 200 & 10 & 0 & 67176.585493 \\
& 260 & 2 & 16 & 222 & 20 & 0 & 102100.926892 \\
& 300 & 1 & 10 & 277 & 12 & 0 & 146139.605664 \\
& 340 & 0 & 10 & 320 & 10 & 0 & 199812.441922 \\
& 420 & 0 & 11 & 398 & 11 & 0 & 339147.966681 \\
& 460 & 2 & 14 & 426 & 18 & 0 & 425754.968401 \\
& 500 & 2 & 16 & 462 & 20 & 0 & 524508.172150 \\
& 1000 & 1 & 17 & 963 & 19 & 0 & 2965940.674307 \\
\hline
\multirow{18}{*}{4} &
  66 & 0 & 22 & 22 & 22 & 0 & 4905.964854 \\
& 104 & 0 & 26 & 52 & 26 & 0 & 15598.534409 \\
& 105 & 0 & 15 & 75 & 15 & 0 & 15984.990289 \\
& 113 & 0 & 19 & 75 & 19 & 0 & 19237.981548 \\
& 117 & 0 & 22 & 73 & 22 & 0 & 21007.172188 \\
& 119 & 0 & 12 & 95 & 12 & 0 & 21914.283713 \\
& 120 & 0 & 10 & 100 & 10 & 0 & 22371.402771 \\
& 121 & 0 & 12 & 97 & 12 & 0 & 22859.735385 \\
& 125 & 0 & 10 & 105 & 10 & 0 & 24816.591295 \\
& 126 & 0 & 10 & 106 & 10 & 0 & 25311.298095 \\
& 180 & 0 & 10 & 160 & 10 & 0 & 62142.129092 \\
& 220 & 0 & 10 & 200 & 10 & 0 & 102919.127703 \\
& 260 & 0 & 10 & 240 & 10 & 0 & 156499.285669 \\
& 300 & 0 & 10 & 280 & 10 & 0 & 223997.341297 \\
& 340 & 0 & 10 & 320 & 10 & 0 & 306568.539431 \\
& 420 & 0 & 13 & 394 & 13 & 0 & 520431.653442 \\
& 460 & 0 & 11 & 438 & 11 & 0 & 653485.181907 \\
& 500 & 0 & 14 & 472 & 14 & 0 & 805206.972227 \\
\hline
\multirow{11}{*}{6} &
   87 & 0 & 0 & 87 & 0 & 0 &  17765.124942 \\
& 108 & 0 & 12 & 84 & 12 & 0 & 30894.374674 \\
& 112 & 0 & 0 & 112 & 0 & 0 & 33902.717714 \\
& 115 & 0 & 0 & 115 & 0 & 0 & 36254.709031 \\
& 116 & 0 & 0 & 116 & 0 & 0 & 37074.949162 \\
& 180 & 0 & 10 & 160 & 10 & 0 & 112810.451302 \\
& 220 & 0 & 10 & 200 & 10 & 0 & 187146.462245 \\
& 260 & 0 & 10 & 240 & 10 & 0 & 284907.016076\\
& 340 & 0 & 10 & 320 & 10 & 0 & 559161.546358 \\
& 420 & 0 & 10 & 400 & 10 & 0 & 950488.931696 \\
& 500 & 0 & 13 & 474 & 13 & 0 & 1471923.063515 \\
& 1000 & 0 & 30 & 940 & 30 & 0 & 8351619.696538 \\
\hline 
\multirow{6}{*}{20} &
  160 & 0 & 0 & 160 & 0 & 0 & 463967.242489 \\
& 170 & 0 & 0 & 160 & 0 & 0 & 543799.839326 \\
& 180 & 0 & 0 & 180 & 0 & 0 & 631751.371902 \\
& 220 & 0 & 0 & 220 & 0 & 0 & 1065625.748639 \\
& 260 & 0 & 0 & 260 & 0 & 0 & 1636942.532923 \\
& 300 & 0 & 0 & 300 & 0 & 0 & 2370110.403872 \\
\botrule
\end{tabular}}
\label{tab:minima}
\end{table}

\subsubsection{\label{sec:5h}The Fat Torus Limit}

We have seen that disclination defects, forbidden in the lowest energy state of a planar crystal, may be energetically favored on a substrate of  non-vanishing Gaussian curvature. It is therefore natural to ask whether large curvature can completely destroy crystalline order by driving the proliferation of a sufficiently high density of defects. The resulting state would be amorphous. The problem of generating amorphous structures by tiling a two-dimensional curved space with identical rigid subunits has drawn attention over the years, particularly through the connection to the structure of such disordered materials as supercooled liquids and metallic glasses. Since the work of Frank \cite{Frank:1952} the notion of geometrical frustration arises frequently in investigations of supercooled liquids and the glass transition. A paradigmatic example is represented by the icosahedral order in metallic liquids and glasses which, although locally favored, cannot propagate throughout all of three-dimensional Euclidean space. A two-dimensional analog, consisting of a liquid of monodisperse hard disks in a 2-manifold of constant negative Gaussian curvature (the hyperbolic plane) was first proposed by 
Nelson and coworkers in 1983 \cite{Nelson:1983,NelsonRubinstein:1983}. In such a system the impossibility of covering the entire manifold with a 6-fold coordinated array of disks mimics many aspects of the geometrical frustration of icosahedral order in three dimensions. In all these models of geometrical frustration, however, the origin of the disorder is primarily due to the short-range nature of the potential between the subunits. In a more realistic setting, part of the frustration is relieved by the fact that hexagonal unit cells can compress in order to match the underlying geometry. 
  
The embedding of a triangular lattice on an axisymmetric torus, provides a particularly suitable playground to study curvature-driven disorder. When 
$r\rightarrow 1$ the Gaussian curvature on the inside of the torus grows like $1/(r-1)$ and diverges on the internal equator at $\psi=\pi$. We thus expect 
a high density of defects in the vicinity of the curvature singularity and a resultant loss of the local $6-$fold bond orientational order. In this regime 
the system will have crystalline regions on the outside of the torus and amorphous regions near the curvature singularity.  

\begin{figure}[t]
\centering
\includegraphics[width=0.55\columnwidth]{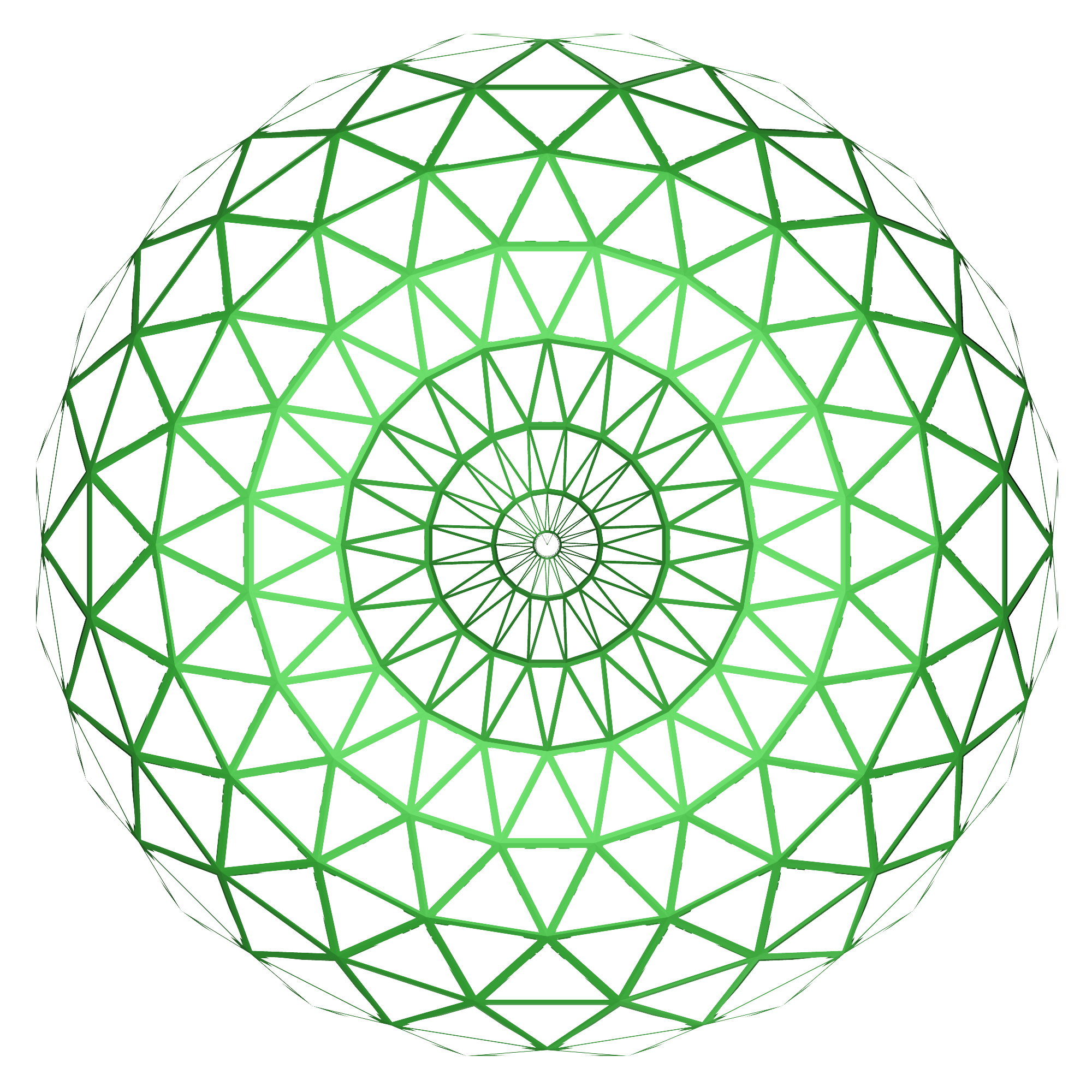}
\caption{\label{fig:fat_torus}(Color online) Top view of a defect free triangulation of a fat torus with $(n,m,l)=(10,10,20)$ and $V=400$. The corresponding
elastic energy becomes very high in the interior of the torus where the triangles are more compressed to match the reduction of surface area.}
\end{figure}

\begin{figure}[h]
\centering
\includegraphics[width=0.4\columnwidth]{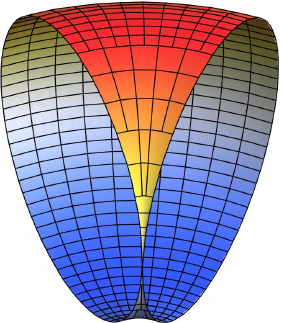}
\caption{\label{fig:dislocation}(Color online) A schematic example of a dislocation 
pile-up on a square lattice resulting from the shrinking of the area on a regular 
wedge of a fat torus.}
\end{figure}

In this section we this claim can be supported using the elastic theory of continuous distributions of edge dislocations on a ``fat'' torus.  Our argument is based on the following construction. As a consequence of the curvature singularity the surface area of an arbitrary wedge of angular width $\Delta\phi$ becomes smaller and smaller as the sectional angle $\psi$ increases and vanishes at $\psi=\pi$. If a defect-free lattice is embedded on such a wedge, Bragg rows will become closer and closer as the singularity is approached with a consequent rise in the elastic energy (see Fig.~\ref{fig:fat_torus}). An intuitive way to reduce the distortion of the lattice is to recursively remove Bragg rows as one approaches the point $\psi=\pi$ (see Fig. \ref{fig:dislocation}). This is equivalent to introducing a growing density of edge dislocations. This dislocation ``cloud'' will ultimately disorder the system by destroying the local $6-$fold bond orientational order. One might therefore view the curvature as playing the role of a local effective temperature which can drive ``melting'' by liberating disclinations and dislocations. In two-dimensional non-Euclidean crystals at $T=0$, however, the mechanism for dislocation proliferation is fundamentally different from the usual thermal melting. While the latter is governed by an entropy gain due to unbinding of dislocation pairs, the amorphization at $T=0$ is due to the adjustment of the lattice to the geometry of the embedding manifold via the proliferation of defects and the consequent release of elastic stress. A similar phenomenon occurs in the disorder-driven amorphization of vortex 
lattices in type-II and high-$T_{c}$ superconductors \cite{Kierfeld:1998,KierfeldVinokur:2000,KierfeldVinokur:2004}.

Since the shrinking area per plaquette on the inside of the torus necessitates a high density of dislocations we may approximate the dislocation cloud in this region by a continuous distribution of Burgers vector density $\bm{b}$. Minimizing the elastic energy with respect to $\bm{b}$ yields a variational equation from which the optimal dislocation density can be calculated as a function of the ratio $\epsilon_{d}/(YR^{2})$ between the dislocation core energy $\epsilon_{d}$ and elastic energy scale $YR^{2}$ with $R=R_{1}=R_{2}$. 

As a starting point, we calculate the Green function $G_{L}(\bm{x},\bm{y})$ in the fat torus limit $r\rightarrow 1$ (i.e.  $\kappa\rightarrow\infty$ and 
$\omega\rightarrow 0$, see Eq.~\eqref{eq:kappa-omega}). The conformal angle $\xi$ in this limit is
\[
\lim_{r\rightarrow 1} \xi = \tan\frac{\psi}{2}\,,
\]
and to leading order of $\kappa$ we have
\begin{gather*}
\frac{\kappa}{16\pi^{2}}\left(\psi-\frac{2}{\kappa}\xi\right)^{2}
\rightarrow \frac{\kappa}{16\pi^{2}}\,\psi^{2}-\frac{\psi}{4\pi}\tan\frac{\psi}{2}\,,\\[7pt]
\frac{\kappa}{4\pi^{2}}\Real\{\Li(\alpha e^{i\psi})\}
\rightarrow \frac{\kappa}{16\pi^{2}}\,\psi^{2}+\frac{1}{2\pi^{2}}\log\left(\cos\frac{\psi}{2}\right)\,.
\end{gather*}
To handle the limit of the Jacobi theta function we can take
$u=\Delta z/\kappa$, $q=e^{i\pi\tau}=e^{-\frac{2\pi}{\kappa}}$ and 
calculate the limit $q\rightarrow 1$. This can be done by using the modular transformation properties of Jacobi functions\cite{Mumford,Polchinski}:
\begin{equation}\label{eq:modular}
\vartheta_{1}\left(\tfrac{u}{\tau}|-\tfrac{1}{\tau}\right) = -i(-i\tau)^{\frac{1}{2}}e^{\frac{iu^{2}}{\pi\tau}}\vartheta_{1}(u|\tau)\,.
\end{equation}
Thus $\tau' = -1/\tau = i\kappa/2$, $u' = u/\tau = \Delta z/(2i)$ and $q'=e^{i\tau'}=e^{-\frac{\pi\kappa}{2}}$, where
\[
\lim_{q\rightarrow 1}\vartheta_{1}(u,q) 
= \lim_{q'\rightarrow 0} i\left(\frac{i}{\tau'}\right)^{-\frac{1}{2}} e^{\frac{iu'^{2}}{\pi\tau'}}\vartheta_{1}(u',q')\,.
\]
This is easily evaluated by means of the expansion
\[
\vartheta_{1}(u,q) = 2q^{\frac{1}{4}}\sin u + o\left(q^{\frac{9}{4}}\right)\,.
\]
Taking the logarithm and neglecting irrelevant constant terms, we obtain
\[
\log \left|\vartheta_{1}\left(\frac{z-z'}{\kappa}\Bigg|\frac{2i}{\kappa}\right)\right| \sim \log\left|\sinh\left(\frac{z-z'}{2}\right)\right|\,, 
\]
which finally leads to
\begin{equation}
G_{L}(\psi,\phi,\psi',\phi')
\sim -\frac{\psi'}{4\pi^{2}}\,\tan\frac{\psi'}{2}+\frac{1}{2\pi}\log\left|\sinh\left(\frac{z-z'}{2}\right)\right| \ ,
\end{equation}
with $z=\tan(\psi/2)+i\phi$. With the Green function in hand, we can calculate the effect of the curvature singularity at $\psi=\pi$ on the distribution of defects. Let $\bm{b}$ be the Burgers vector density of the dislocation cloud. Hereafter we work in a local frame, so that
\begin{equation}\label{eq:burger}
\bm{b}=b^{\psi}\bm{g}_{\psi}+b^{\phi}\bm{g}_{\phi}\,,
\end{equation}
with $\bm{g}_{i}=\partial_{i}\bm{R}$ a basis vector in the tangent plane of the torus. $\bm{R}$ is a three-dimensional surface vector parameterizing the torus and its Cartesian components are given in Eq. \eqref{eq:sec5-parametrization}. The quantity $\bm{b}$ has to be such that
\[
\int_{D} d^{2}x\,\bm{b}(x)=\bm{b}_{D}\,,
\]
with $\bm{b}_{D}$ the total Burger's vector in a generic domain $D$. Because on a closed manifold dislocation lines cannot terminate on the boundary, extending the integration to the whole torus we have:
\begin{equation}\label{eq:burger_neutrality}
\int d^{2}x\,\bm{b}(\bm{x})=0\,.
\end{equation}
Since the basis vectors $\bm{g}_{i}$ in Eq.~\eqref{eq:burger} have the dimension of length, contravariant coordinates $b^{i}$ have dimensions of an inverse area. Assuming all defects to be paired in the form of dislocations (i.e. $q_{i}=0$ everywhere), the total energy of the crystal reads
\begin{equation}\label{eq:cloud_energy}
F= \frac{1}{2Y} \int d^{2}x\, \Gamma^{2}(\bm{x}) + \epsilon_{d}\int d^{2}x\,|\bm{b}(\bm{x})|^{2}\,,
\end{equation}
where $\epsilon_{d}$ is the dislocation core energy. The function $\Gamma(\bm{x})$ encoding the elastic stress due to the curvature and the screening contribution of the dislocation cloud obeys
\begin{equation}\label{eq:cloud_poisson}
\frac{1}{Y}\Delta_{g}\Gamma(\bm{x})=\epsilon_{k}^{i}\nabla_{i}b^{k}(\bm{x})-K(\bm{x})\,,
\end{equation}
where $\nabla_{i}$ is the usual covariant derivative along the coordinate-direction $i$ and $\epsilon_{k}^{i}$ is the Levi-Civita antisymmetric tensor on the torus:
\[
\epsilon_{\psi\phi}=-\epsilon_{\phi\psi}=\sqrt{g}\,,
\qquad\qquad
\epsilon_{i}^{j} = g_{ik}\epsilon^{jk}\,.
\]
The stress function $\Gamma(\bm{x})$ can be expressed in the form 
$\Gamma(\bm{x}) = \Gamma_{d}(\psi,\phi)-\Gamma_{s}(\psi)$ with
\begin{subequations}
\begin{gather}
\frac{\Gamma_{s}(\psi)}{Y} = \log\left[\frac{1}{2(1+\cos\psi)}\right]+1\,,\label{eq:fat_gamma_screening}\\[7pt]
\frac{\Gamma_{s}(\psi,\phi)}{Y} = \int d^{2}y\,\epsilon_{k}^{i}\nabla_{i}b^{k}(\bm{y})\,G_{L}(\bm{x},\bm{y})\,.\label{eq:fat_gamma_defects}
\end{gather}
\end{subequations}
The integral term in Eq. \eqref{eq:fat_gamma_defects} can now be reduced to a more friendly functional of $\bm{b}$, suitable for a variational approach. Given the azimuthal symmetry we assume that all dislocations are aligned along $\bm{b}=b^{\phi}\bm{g}_{\phi}$. 
\begin{equation}\label{eq:cloud_gamma2}
\frac{\Gamma_{d}(\psi)}{Y}
= \frac{1}{2\pi}\int_{-\pi}^{\pi} d\psi'\sqrt{g}\,b^{\phi}(\psi')\left[\psi'+\sin\psi'+\pi\sgn(\psi-\psi')\right]\,.
\end{equation}
Substituting Eq.~\eqref{eq:cloud_gamma2} and \eqref{eq:fat_gamma_screening} in Eq.~\eqref{eq:cloud_energy} and minimizing with respect to $b^{\phi}$, the problem can be converted, after some algebraic manipulations, to the following Fredholm equation of the second kind:
\begin{equation}\label{eq:fredholm2}
\lambda B(\psi) - \int_{-\pi}^{\pi} d\psi'\,B(\psi')\mathcal{K}(\psi,\psi')=f(\psi)\,,
\end{equation}
where $\lambda=\epsilon_{d}/(YR^{2})$, $B(\psi)=R^{2}(1+\cos\psi)^{2}b^{\phi}(\psi)$ and the kernel $\mathcal{K}(\psi,\psi')$ is given by:
\begin{multline}
\mathcal{K}(\psi,\psi')=
\frac{1}{4(1+\cos\psi')}\Bigg\{\pi^{-1}(\psi'+\sin\psi')(\psi+\sin\psi)\\
+|\psi-\psi'|+2\cos\frac{\psi+\psi'}{2}\sin\frac{|\psi-\psi'|}{2}\Biggr\}\,.
\end{multline}
The function $f(\psi)$ on the right hand side of Eq.~\eqref{eq:fredholm2} is given by
\begin{gather}\label{eq:fredholm_known1}
f(\psi)
= -\frac{1}{4}\int_{-\psi}^{\psi} d\psi'\,(1+\cos\psi')\,\Gamma_{s}(\psi')\notag\\
= \frac{1}{2}\bigl\{[\log 2(1+\cos\psi)-2]\sin\psi-2\Cl(\psi+\pi)\bigr\}
\end{gather}
where $\Cl$ is the Clausen function (see Ref. \cite{AbramowitzStegun}, pp. 1005-1006) defined as
\[
\Cl(x)
=-\int_{0}^{x} dx\, \log\left(2\sin\frac{t}{2}\right)
= \sum_{k=1}^{\infty}\frac{\sin kx}{k^{2}}\,.
\]
\begin{figure}[t]
\centering
\includegraphics[width=0.65\columnwidth]{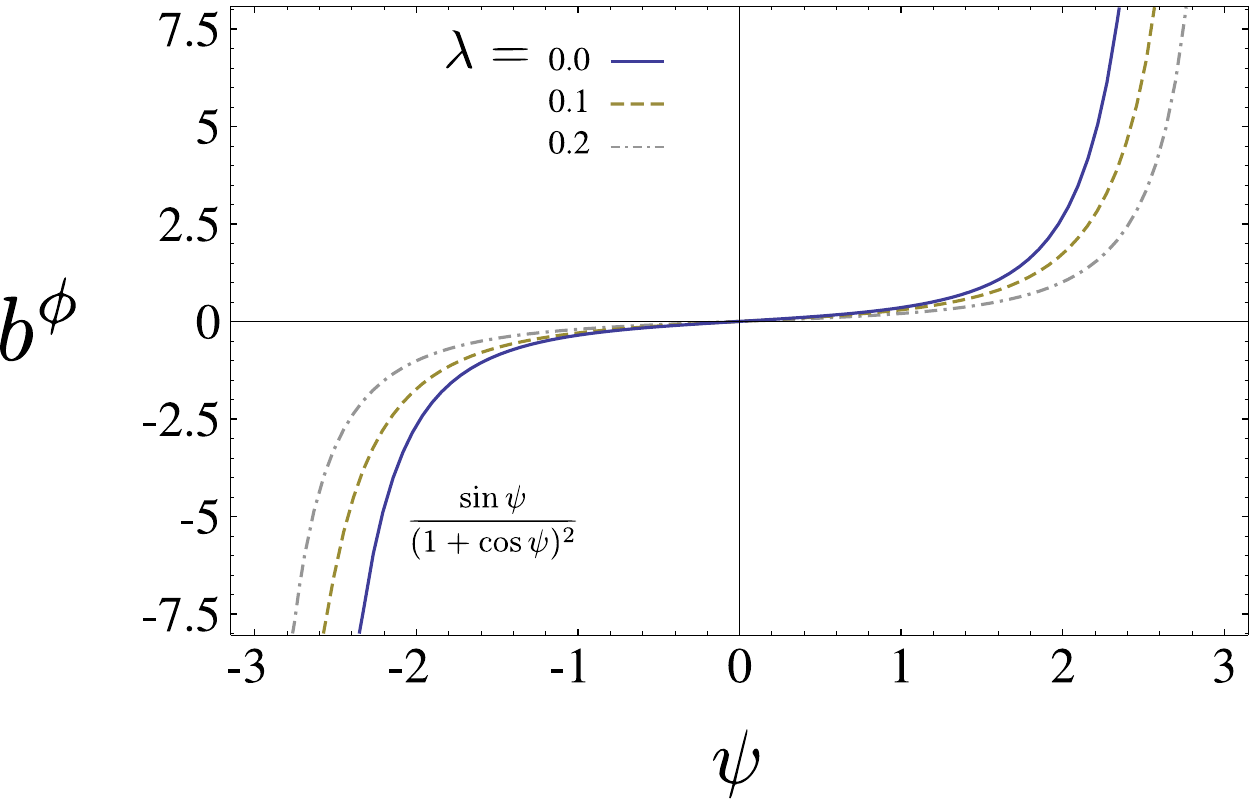}
\caption{\label{fig:burgers}(Color online) The Burgers vector component $b(\psi)$ 
for different choices of $\lambda$.}
\end{figure}

As previously noted the dislocation core energy is $\epsilon_{d}$ is much smaller than the elastic energy scale $YR^{2}$. Eq~\eqref{eq:fredholm2} is 
then suitable to be solved in powers of the dimensionless number $\lambda$:
\[
B(\psi) = B_{0}(\psi)+\lambda B_{1}(\psi)+\lambda^{2} B_{2}(\psi)+\cdots
\]
The corrections to the zero-order term $B_{0}(\psi)$ can be calculated recursively by solving a set of Fredholm equations of the first kind:
\[
B_{k-1}(\psi) = \int_{-\pi}^{\pi} d\psi'\,B_{k}(\psi)\mathcal{K}(\psi,\psi')\qquad k\ge 1\,.
\]
The function $B_{0}(\psi)$ associated with the Burgers vector density of the dislocation cloud in the limit $\lambda\rightarrow 0$, on the other hand, can be calculated directly from Eq.~\eqref{eq:cloud_poisson} by setting the effective topological charge density on the right hand side to zero:
\begin{equation}\label{eq:zero_core1}
\epsilon_{k}^{i}\nabla_{i}b^{k}(\bm{x})-K(\bm{x}) = 0\,.
\end{equation}
from which one obtains $B_{0}(\psi)=\sin\psi$ and 
\begin{equation}\label{eq:zero-order}
b^{\phi} = \frac{\sin\psi}{R^{2}(1+\cos\psi)^{2}}\,.
\end{equation}
The Burgers vector density $b^{\phi}$ obtained by a numerical solution of Eq.~\eqref{eq:fredholm2} is shown in Fig.~\ref{fig:burgers} for different values of $\lambda$. The Burgers vector density is measured in units of $R^{-2}.$ The function $b^{\phi}$ has cubic singularities at $\psi=\pm\pi$ and is approximately zero on the outside of the torus. The solid blue curve in Fig.~\ref{fig:burgers} represents the zeroth order solution of Eq.\eqref{eq:zero-order}.

\begin{figure}[h!]
\centering
\includegraphics[width=0.6\columnwidth]{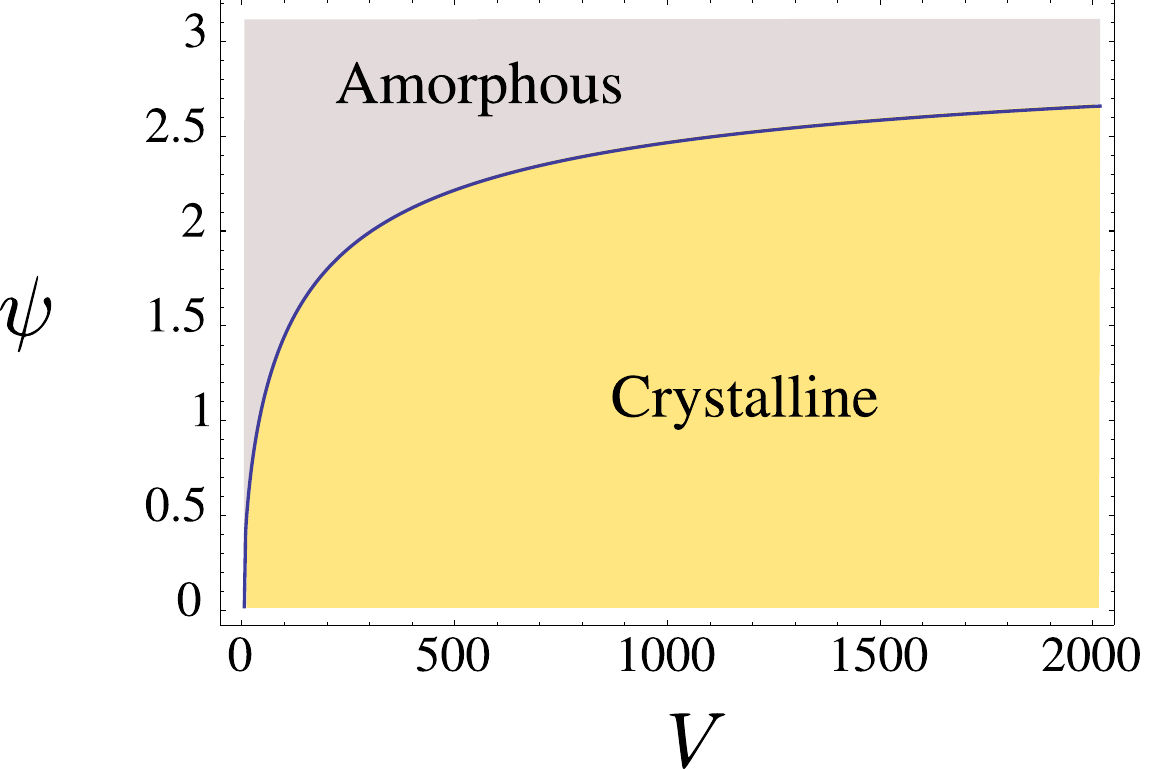}
\caption{\label{fig:melting-diagram}(Color online) Phase diagram for curvature driven amorphization. The inside of the torus contains an amorphous region whose angular size decreases with the number of vertices $V$ as a consequence of the reduction of the lattice spacing.}
\end{figure}

Now, in the theory of dislocation mediated melting a system at the solid liquid phase boundary is described as a crystalline solid saturated with dislocations. In three-dimensions, in particular, there is a strong experimental evidence of the existence of a critical dislocation density at the melting point $\rho(T_{m})\approx 0.6b^{-2}$ where $b$ is the length of the length of the smallest perfect-dislocation Burgers vector \cite{BurakovskyEtAl:2000}. Several theoretical works have motivated this evidence both for three-dimensional solids and vortex lattices in super conductors \cite{KierfeldVinokur:2000}. On the other hand, given the existence of such a critical density, its value can be empirically used to determine whether a system is in a solid or liquid-like phase in the same spirit as the Lindemann criterion. With this goal in mind we can calculate the dislocation density by requiring $|\bm{b}|=\rho a$ with $\rho$ the density of single lattice spacing dislocations. This yields:
\begin{equation}\label{eq:meltin-criterion}
\rho(\psi,V) a^{2} = 2\pi \left(\frac{\sqrt{3}}{2}\,V\right)^{-\frac{1}{2}}\left|\tan\frac{\psi}{2}\right|+o(\lambda)
\end{equation}
Solving $\rho(\psi,V)a^{2}=0.6$ as a function of $\psi$ and $V$ we obtain the diagram of Fig.~\ref{fig:melting-diagram}. As expected the inside of the torus
contains an amorphous region whose angular size decreases with the number of vertices $V$ as a consequence of the reduction of the lattice spacing.

\section{\label{sec:6}Conclusion and discussion}

In this article we have reviewed recent progress toward understanding the ground state properties of two-dimensional ordered phases on substrate of non-zero Gaussian curvature. Curved phases of matter are  found in a broad class of systems comprising materials as different as viral capsids and carbon nanotori. The geometry of the substrate gives rise to novel defective structures that would be energetically prohibitive in the ground state of conventional flat systems. Defects may appear in curved space either because they are required by the topology of the underlying substrate (as in the case of spherical crystals and nematics) or because they are favored by the curvature itself.  The purely curvature driven energetic nature of the latter phenomenon is of fundamental interest.

\begin{figure}
\centering
\includegraphics[width=0.7\textwidth]{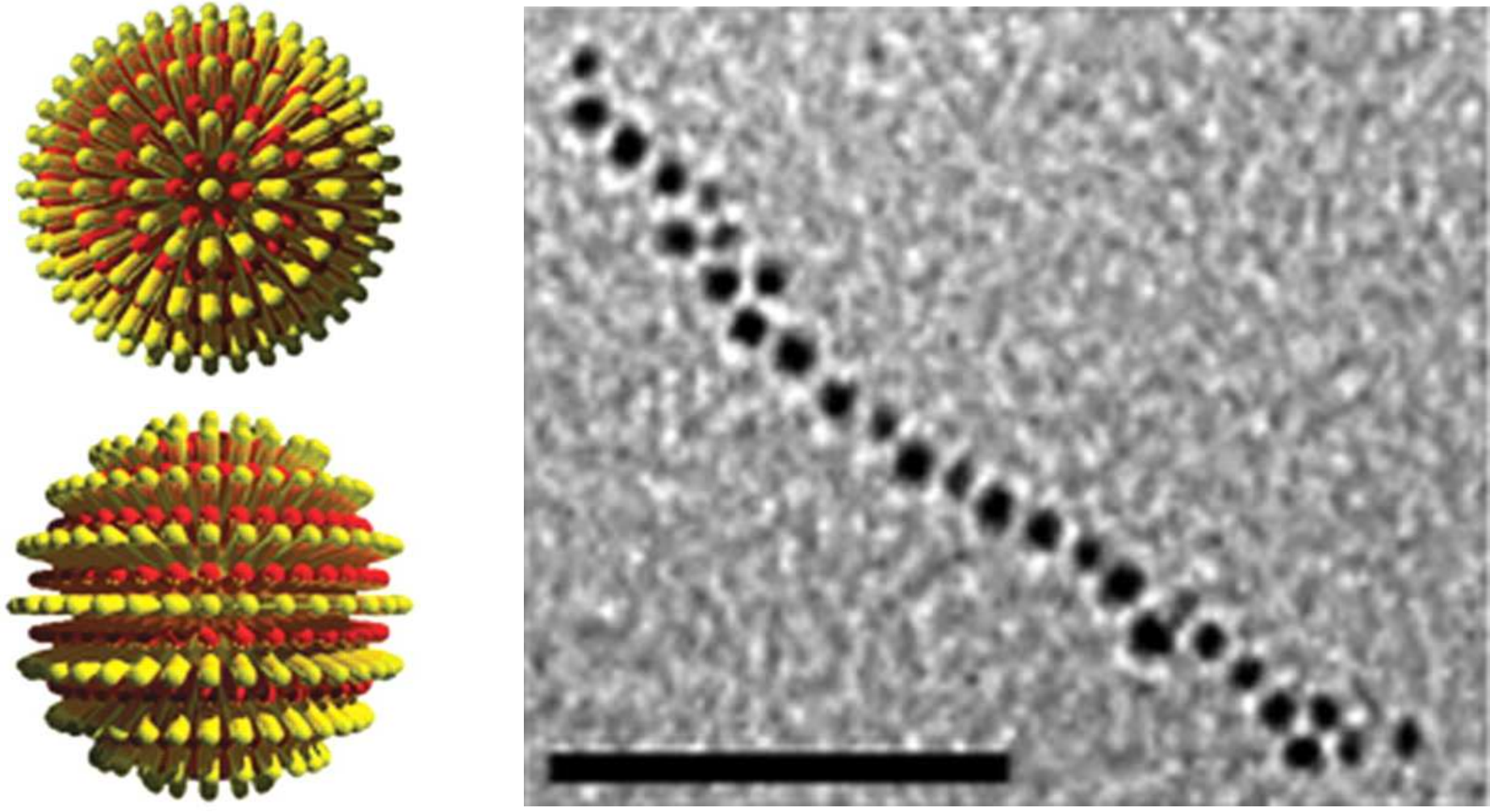}
\caption{\label{fig:sec6-stellacci}(Color online) (Left) Idealized drawing of a side and top view and of a rippled gold nanoparticle showing the two polar defects that must exist to allow the alternation of concentric rings. (Right) TEM images of a chain obtained when 11-mercaptoundecanoic acid functionalized nanoparticles are reacted with a water phase containing divalent 1,6-diaminohexane in a two-phase reaction. Scale bar 50 nm. From \cite{DeVriesEtAl:2007}.}
\end{figure}

On the material science side, non-Euclidean systems provides a promising route to constructing arrays of nanoparticles via the chemical functionalization of topological defects created on the surface of the particles by coating them with an ordered monolayer. Candidates for such coatings include triblock coplymers, gemini lipids, metallic or semiconducting nanorods and conventional liquid crystal compounds. If the induced order is vector-like or striped, then there must necessarily be two defective regions corresponding to the bald spots exhibited by a combed sphere or the source and sink of fluid flow confined to a sphere. Not only are the defects physically and mathematically distinguished, since the condensed matter order vanishes there, but it also turns out that chemical processes can detect the defects and insert linker molecules at the precise defect locations. This creates the possibility of efficiently making functionalized nanoparticles with a precise valence and corresponding directional bonding. The nanoparticles themselves can then be linked into well-defined two and three dimensional arrays. Control over the valence and the geometry of the directional bonding can be achieved by varying the nature of the ordered monolayer.

This scheme has been demonstrated recently in the work of DeVries \emph{et al}. \cite{DeVriesEtAl:2007} in which gold nanoparticles are coated with two species of naturally phase separating ligands (see Fig. \ref{fig:sec6-stellacci}). Phase separation on the spherical gold surface translates to a striped arrangement of alternating ligands. Such a spherical smectic has topological defects at the north and south pole of the nanopartcle. These may be functionalized by attaching thiol-based linker molecules at the defects via place-exchange reactions. This creates a divalent gold nanoparticle with directional bonds 180 degrees apart. These in turn can be linked to create polymers and free-standing films. 
\section*{Acknowledgements}

We thank V. Vitelli, A. Hexemer, W. Irvine, P. Chaikin, A. Lucas and A. Fonseca for sharing with us both images and data presented in this review. We thank D. R. Nelson, H. Shin and A. Travesset for long-standing collaborations and M. Forstner for discussions of amphiphilic membranes.  


\end{document}